\documentclass[nohyper,12pt,letterpaper]{JHEP3}
\usepackage{amsmath,amssymb}

\usepackage{color}

\DeclareMathOperator{\tr}{tr} 
\DeclareMathOperator{\Spin}{Spin} 
\DeclareMathOperator{\Imx}{Im} 

\DeclareMathOperator{\diag}{diag} 

\title{
Ground States of S-duality Twisted
$N=4$ Super Yang--Mills Theory
}

\author{Ori J. Ganor${}^1$, Yoon Pyo Hong${}^2$,
and Hai Siong Tan${}^1$ \\
\\
${}^1$Department of Physics,
University of California,
Berkeley, CA 94720, U.S.A.\\
${}^2$School of Physics,
Korea Institute for Advanced Study,
Seoul 130-722, Korea \\
\\
Emails:
\email{origa@socrates.berkeley.edu,
yph@kias.re.kr, haisiong\_tan@berkeley.edu}
}

\abstract{
We study the low-energy limit of
a compactification of $N=4$ $U(n)$ super Yang--Mills
theory on $S^1$ with boundary conditions
modified by an S-duality and R-symmetry twist.
This theory has $N=6$ supersymmetry in 2+1D.
We analyze the $T^2$ compactification of this 2+1D theory
by identifying a dual weakly coupled type-IIA background.
The Hilbert space of normalizable
ground states is finite-dimensional
and appears to exhibit a rich structure of sectors.
We identify most of them with Hilbert spaces
of Chern--Simons theory
(with appropriate gauge groups and levels).
We also discuss a realization of a related
twisted compactification in terms of the $(2,0)$-theory,
where the recent solution by Gaiotto and Witten of
the boundary conditions describing
D$3$-branes ending on a $(p,q)$ $5$-brane
plays a crucial role.
}

\received{July 21, 2010}

\preprint{
UCB-PTH-10/08,
KIAS-P10022
}

\begin{document}

\newcommand{\secref}[1]{\S\ref{#1}}
\newcommand{\figref}[1]{Figure~\ref{#1}}
\newcommand{\appref}[1]{Appendix~\ref{#1}}
\newcommand{\tabref}[1]{Table~\ref{#1}}

\newcommand\SUSY[1]{{${\mathcal{N}}={#1}$}}  
\newcommand\px[1]{{\partial_{#1}}}
\newcommand\qx[1]{{\partial^{#1}}}
\def\be{\begin{equation}}
\def\ee{\end{equation}}
\def\bear{\begin{eqnarray}}
\def\eear{\end{eqnarray}}
\def\nn{\nonumber}

\newcommand\rep[1]{{\mathbf{#1}}} 

\newcommand\bra[1]{{\left\langle{#1}\right\rvert}} 
\newcommand\ket[1]{{\left\lvert{#1}\right\rangle}} 

\newcommand{\C}{\mathbb{C}}
\newcommand{\R}{\mathbb{R}}
\newcommand{\Z}{\mathbb{Z}}

\def\SL{{{\mbox{\rm SL}}}} 
\def\gYM{g_{\text{YM}}} 

\def\zs{{\widehat{\mathcal{S}}}} 
\def\zr{{\widehat{\mathcal{R}}}} 

\def\xa{{\mathbf{a}}} 
\def\xb{{\mathbf{b}}} 
\def\xc{{\mathbf{c}}} 
\def\xd{{\mathbf{d}}} 

\def\zg{{\mathbf{g}}} 
\def\zt{{\mathbf{t}}} 
\def\zx{{\mathbf{x}}} 
\def\bQ{{\overline{Q}}} 
\def\bpsi{{\overline{\psi}}}
\def\Op{{\mathcal{O}}} 
\def\a{\alpha}
\def\b{\beta}
\def\dta{{\dot{\a}}}
\def\dtb{{\dot{\b}}}
\def\lvk{{k}} 
\def\pht{{\upsilon}} 
\def\phtpr{{\upsilon'}} 
\def\ord{{\mathbf{r}}} 
\def\ordpr{{\mathbf{r}'}} 
\def\gtw{{\gamma}} 
\def\ftw{{\varphi}} 
\def\Zphi{{Z}} 
\def\tA{{\widetilde{A}}} 
\def\tF{{\widetilde{F}}} 

\def\wx{{$-$}} 
\def\tx{{$\div$}} 
\def\nx{{$\times$}}

\def\Mst{M_{\text{st}}} 
\def\Mpl{M_{\text{p}}} 
\def\gst{g_{\text{st}}} 
\def\gIIA{g_{\text{IIA}}} 
\def\gIIB{g_{\text{IIB}}} 
\def\lst{\ell_{\text{st}}} 
\def\lp{\ell_{\text{P}}} 
\def\apr{{\a'}} 
\def\Area{{\mathcal{A}}} 

\def\NSols{{D_s}} 
\def\NStates{{N_s}} 
\def\Ham{{H}} 
\def\btau{{\overline{\tau}}} 

\def\MXS{{W}} 

\def\wvF{{\Psi}} 
\def\sdwvF{{\widetilde{\Psi}}} 
\def\cD{{\mathcal{D}}} 
\def\tA{{\widetilde{A}}} 

\def\gpifa{{\alpha_a}} 
\def\gpifb{{\alpha_b}} 
\def\gpifd{{\alpha_{a+b}}} 
\def\gpib{{\eta}} 
\def\ogpifa{{\overline{\alpha}_a}} 
\def\ogpifb{{\overline{\alpha}_b}} 
\def\ogpib{{\overline{\eta}}} 
\def\hfa{{\beta_a}} 
\def\hfb{{\beta_b}} 
\def\hb{{\varrho}} 
\def\wxa{{g}} 

\def\Psym{{\mathcal{U}}} 
\def\Qsym{{\mathcal{V}}} 

\def\wPsym{{\widetilde{\mathcal{U}}}} 
\def\wQsym{{\widetilde{\mathcal{V}}}} 

\def\oPsym{{\overline{\mathcal{U}}}} 
\def\oQsym{{\overline{\mathcal{V}}}} 

\def\ef{{\mathbf{e}}} 
\def\mf{{\mathbf{m}}} 
\def\tmf{{\widetilde{\mf}}} 
\def\omf{{\overline{\mf}}} 
\def\tomf{{\widetilde{\omf}}} 

\def\WilV{{\mathcal{W}}} 
\def\tHooftV{{\mathcal{M}}} 

\def\TdS{{\mathcal{S}}} 
\def\TdT{{\mathcal{T}}} 
\def\TdG{{\mathcal{G}}} 

\def\bosZ{{\phi}} 
\def\bosbZ{{\overline{\phi}}} 
\def\Vtx{{\mathcal V}} 

\def\perm{{\sigma}} 
\def\brperm{{\lbrack\perm\rbrack}} 
\def\wMa{{M_a}} 
\def\wMb{{M_b}} 
\def\wna{{\overline{M}_a}} 
\def\wnb{{\overline{M}_b}} 
\def\pKa{{K_a}} 
\def\pKb{{K_b}} 

\def\rhocp{{\rho}} 
\def\brhocp{{\overline{\rho}}} 
\def\txa{{\tilde{\mathbf{a}}}} 
\def\txb{{\tilde{\mathbf{b}}}} 
\def\txc{{\tilde{\mathbf{c}}}} 
\def\txd{{\tilde{\mathbf{d}}}} 

\def\txp{{\tilde{\mathbf{p}}}} 
\def\txq{{\tilde{\mathbf{q}}}} 

\def\xL{{L}} 
\def\xR{{R}} 
\def\xS{{S}} 

\def\Gg{{g}} 
\def\OmX{{\Upsilon}} 

\def\PQsymB{{\mathcal{K}}} 
\def\np{{\tilde{n}}} 

\def\wsig{{\sigma}} 
\def\wtau{{\eta}} 
\def\wsX{{X}} 
\def\wsY{{Y}} 
\def\wsZ{{Z}} 
\def\wsbZ{{\overline{Z}}} 
\def\oscZr{{\alpha}} 
\def\oscZl{{\tilde{\alpha}}} 
\def\oscYr{{\gamma}} 
\def\oscYl{{\tilde{\gamma}}} 
\def\oscXr{{\alpha}} 
\def\oscXl{{\tilde{\alpha}}} 
\def\fpY{{y}}
\def\pY{{P_y}}
\def\fpX{{x}}
\def\pX{{P}}
\def\fpZ{{\zeta}} 
\def\nZ{{n'}} 

\def\OrbG{{\mathcal{R}}} 

\def\wsXPsi{{\chi}} 
\def\wsZPsi{{\psi}} 
\def\twsXPsi{{\tilde{\chi}}} 
\def\twsZPsi{{\tilde{\psi}}} 
\def\oscZPsir{{\psi}}
\def\oscZPsil{{\tilde{\psi}}}
\def\oscXPsir{{\chi}}
\def\oscXPsil{{\tilde{\chi}}}
\def\xfn{{\theta}} 
\def\tL{{\tilde{L}}} 

\def\oN{{\overline{N}}} 
\def\wN{{\widetilde{N}}} 
\def\woN{{\widetilde{\overline{N}}}} 

\def\Dsp{{\rho}}
\def\tDsp{{\tilde{\rho}}}

\def\fmL{{\psi}}
\def\fmR{{\tilde{\psi}}}

\def\Ginv{G^{(\text{inv})}} 
\def\Id{{\mathbf{I}}} 
\def\Psig{{\sigma}} 
\def\Hilb{{\mathcal{H}}}
\def\HilbSun{{\widetilde{\mathcal{H}}}} 

\def\wLambda{\widetilde{\Lambda}} 
\def\wwLambda{\widetilde{\widetilde{\Lambda}}}  

\def\frl{{\nu}} 

\def\AreaForm{{\omega}} 

\def\hu{{\mathfrak{a}}} 
\def\bhu{{\overline{\hu}}} 
\def\ohu{{\hat{\hu}}} 
\def\bz{{\overline{z}}}

\def\phasezt{{\phi}} 
\def\intzt{{N}} 

\def\hFourier{{\hat{\mathfrak{f}}}} 

\def\bulkC{{\mathbf{C}'}} 
\def\lowC{{\mathbf{C}}} 
\def\lowB{{\mathbf{B}}} 

\def\Surf{{{\cal S}}} 
\def\spP{{Q}} 
\def\tauIIB{{\tau_{\text{IIB}}}} 
\def\lowD{{\mathbf{C}'''}}
\def\lowDz{\mathbf{C}''}

\def\WHV{{\mathcal{V}}} 
\def\fWHV{{\widehat{\WHV}}} 


\def\fpIIz{{
\begin{picture}(16,10)
\put(3,-2){\begin{picture}(10,12)
\thinlines
\color{blue}
\multiput(0,0)(0,10){2}{\line(1,0){10}}
\multiput(0,0)(10,0){2}{\line(0,1){10}}
\thicklines
\color{black}
\put(0,0){\circle*{2}}
\color{red}
\put(0,0){\circle{4}}
\end{picture}}
\end{picture}
}}
\def\fpIIc{{
\begin{picture}(16,10)
\put(3,-2){\begin{picture}(10,12)
\color{blue}
\thinlines
\multiput(0,0)(0,10){2}{\line(1,0){10}}
\multiput(0,0)(10,0){2}{\line(0,1){10}}
\thicklines
\color{black}
\put(5,5){\circle*{2}}
\color{red}
\put(5,5){\circle{4}}
\end{picture}}
\end{picture}
}}
\def\fpIIzz{{
\begin{picture}(16,10)
\put(3,-2){\begin{picture}(10,12)
\thinlines
\color{blue}
\multiput(0,0)(0,10){2}{\line(1,0){10}}
\multiput(0,0)(10,0){2}{\line(0,1){10}}
\thicklines
\color{black}
\put(0,0){\circle*{2}}
\color{red}
\put(0,0){\circle{4}}
\put(0,0){\circle{6}}
\end{picture}}
\end{picture}
}}

\def\fpIIcc{{
\begin{picture}(16,10)
\put(3,-2){\begin{picture}(10,12)
\color{blue}
\thinlines
\multiput(0,0)(0,10){2}{\line(1,0){10}}
\multiput(0,0)(10,0){2}{\line(0,1){10}}
\thicklines
\color{black}
\put(5,5){\circle*{2}}
\color{red}
\put(5,5){\circle{4}}
\put(5,5){\circle{6}}
\end{picture}}
\end{picture}
}}
\def\fpIIxy{{
\begin{picture}(16,10)
\put(3,-2){\begin{picture}(10,12)
\thinlines
\color{blue}
\multiput(0,0)(0,10){2}{\line(1,0){10}}
\multiput(0,0)(10,0){2}{\line(0,1){10}}
\thicklines
\color{black}
\put(0,5){\circle*{2}}
\put(5,0){\circle*{2}}
\color{red}
\put(0,5){\circle{4}}
\put(5,0){\circle{4}}
\end{picture}}
\end{picture}
}}
\def\fpIIzzz{{
\begin{picture}(16,10)
\put(3,-2){\begin{picture}(10,12)
\thinlines
\color{blue}
\multiput(0,0)(0,10){2}{\line(1,0){10}}
\multiput(0,0)(10,0){2}{\line(0,1){10}}
\thicklines
\color{black}
\put(0,0){\circle*{2}}
\color{red}
\put(0,0){\circle{4}}
\put(0,0){\circle{6}}
\put(0,0){\circle{8}}
\end{picture}}
\end{picture}
}}
\def\fpIIccc{{
\begin{picture}(16,10)
\put(3,-2){\begin{picture}(10,12)
\color{blue}
\thinlines
\multiput(0,0)(0,10){2}{\line(1,0){10}}
\multiput(0,0)(10,0){2}{\line(0,1){10}}
\thicklines
\color{black}
\put(5,5){\circle*{2}}
\color{red}
\put(5,5){\circle{4}}
\put(5,5){\circle{6}}
\put(5,5){\circle{8}}
\end{picture}}
\end{picture}
}}
\def\fpIz{{
\begin{picture}(21,14)
\put(3,-2){\begin{picture}(18,14)
\thinlines
\color{blue}
\multiput(0,0)(6,12){2}{\line(1,0){12}}
\multiput(0,0)(12,0){2}{\line(1,2){6}}
\thicklines
\color{black}
\put(0,0){\circle*{2}}
\color{red}
\put(0,0){\circle{4}}
\end{picture}}
\end{picture}
}}
\def\fpIzz{{
\begin{picture}(21,14)
\put(3,-2){\begin{picture}(18,14)
\thinlines
\color{blue}
\multiput(0,0)(6,12){2}{\line(1,0){12}}
\multiput(0,0)(12,0){2}{\line(1,2){6}}
\thicklines
\color{black}
\put(0,0){\circle*{2}}
\color{red}
\put(0,0){\circle{4}}
\put(0,0){\circle{6}}
\end{picture}}
\end{picture}
}}
\def\fpIdd{{
\begin{picture}(21,14)
\put(3,-2){\begin{picture}(18,14)
\thinlines
\color{blue}
\multiput(0,0)(6,12){2}{\line(1,0){12}}
\multiput(0,0)(12,0){2}{\line(1,2){6}}
\thicklines
\color{black}
\put(6,4){\circle*{2}}
\put(12,8){\circle*{2}}
\color{red}
\put(6,4){\circle{4}}
\put(12,8){\circle{4}}
\end{picture}}
\end{picture}
}}
\def\fpIzzz{{
\begin{picture}(21,14)
\put(3,-2){\begin{picture}(18,14)
\thinlines
\color{blue}
\multiput(0,0)(6,12){2}{\line(1,0){12}}
\multiput(0,0)(12,0){2}{\line(1,2){6}}
\thicklines
\color{black}
\put(0,0){\circle*{2}}
\color{red}
\put(0,0){\circle{4}}
\put(0,0){\circle{6}}
\put(0,0){\circle{8}}
\end{picture}}
\end{picture}
}}
\def\fpIxyd{{
\begin{picture}(21,14)
\put(3,-2){\begin{picture}(18,14)
\thinlines
\color{blue}
\multiput(0,0)(6,12){2}{\line(1,0){12}}
\multiput(0,0)(12,0){2}{\line(1,2){6}}
\thicklines
\color{black}
\put(6,0){\circle*{2}}
\put(3,6){\circle*{2}}
\put(9,6){\circle*{2}}
\color{red}
\put(6,0){\circle{4}}
\put(3,6){\circle{4}}
\put(9,6){\circle{4}}
\end{picture}}
\end{picture}
}}
\def\fpIzzzz{{
\begin{picture}(23,14)
\put(5,-2){\begin{picture}(18,14)
\thinlines
\color{blue}
\multiput(0,0)(6,12){2}{\line(1,0){12}}
\multiput(0,0)(12,0){2}{\line(1,2){6}}
\thicklines
\color{black}
\put(0,0){\circle*{2}}
\color{red}
\put(0,0){\circle{4}}
\put(0,0){\circle{6}}
\put(0,0){\circle{8}}
\put(0,0){\circle{10}}
\end{picture}}
\end{picture}
}}
\def\fpIdddd{{
\begin{picture}(21,14)
\put(3,-2){\begin{picture}(18,14)
\thinlines
\color{blue}
\multiput(0,0)(6,12){2}{\line(1,0){12}}
\multiput(0,0)(12,0){2}{\line(1,2){6}}
\thicklines
\color{black}
\put(6,4){\circle*{2}}
\put(12,8){\circle*{2}}
\color{red}
\put(6,4){\circle{4}}
\put(6,4){\circle{6}}
\put(12,8){\circle{4}}
\put(12,8){\circle{6}}
\end{picture}}
\end{picture}
}}
\def\fpIzzzzz{{
\begin{picture}(25,14)
\put(7,-2){\begin{picture}(18,14)
\thinlines
\color{blue}
\multiput(0,0)(6,12){2}{\line(1,0){12}}
\multiput(0,0)(12,0){2}{\line(1,2){6}}
\thicklines
\color{black}
\put(0,0){\circle*{2}}
\color{red}
\put(0,0){\circle{4}}
\put(0,0){\circle{6}}
\put(0,0){\circle{8}}
\put(0,0){\circle{10}}
\put(0,0){\circle{12}}
\end{picture}}
\end{picture}
}}
\def\fpIIIz{{
\begin{picture}(21,14)
\put(3,-2){\begin{picture}(18,14)
\thinlines
\color{blue}
\multiput(0,0)(6,12){2}{\line(1,0){12}}
\multiput(0,0)(12,0){2}{\line(1,2){6}}
\thicklines
\color{black}
\put(0,0){\circle*{2}}
\color{red}
\put(0,0){\circle{4}}
\end{picture}}
\end{picture}
}}
\def\fpIIIe{{
\begin{picture}(21,14)
\put(3,-2){\begin{picture}(18,14)
\thinlines
\color{blue}
\multiput(0,0)(6,12){2}{\line(1,0){12}}
\multiput(0,0)(12,0){2}{\line(1,2){6}}
\thicklines
\color{black}
\put(6,4){\circle*{2}}
\color{red}
\put(6,4){\circle{4}}
\end{picture}}
\end{picture}
}}
\def\fpIIIf{{
\begin{picture}(21,14)
\put(3,-2){\begin{picture}(18,14)
\thinlines
\color{blue}
\multiput(0,0)(6,12){2}{\line(1,0){12}}
\multiput(0,0)(12,0){2}{\line(1,2){6}}
\thicklines
\color{black}
\put(12,8){\circle*{2}}
\color{red}
\put(12,8){\circle{4}}
\end{picture}}
\end{picture}
}}
\def\fpIIIzz{{
\begin{picture}(21,14)
\put(3,-2){\begin{picture}(18,14)
\thinlines
\color{blue}
\multiput(0,0)(6,12){2}{\line(1,0){12}}
\multiput(0,0)(12,0){2}{\line(1,2){6}}
\thicklines
\color{black}
\put(0,0){\circle*{2}}
\color{red}
\put(0,0){\circle{4}}
\put(0,0){\circle{6}}
\end{picture}}
\end{picture}
}}
\def\fpIIIee{{
\begin{picture}(21,14)
\put(3,-2){\begin{picture}(18,14)
\thinlines
\color{blue}
\multiput(0,0)(6,12){2}{\line(1,0){12}}
\multiput(0,0)(12,0){2}{\line(1,2){6}}
\thicklines
\color{black}
\put(6,4){\circle*{2}}
\color{red}
\put(6,4){\circle{4}}
\put(6,4){\circle{6}}
\end{picture}}
\end{picture}
}}
\def\fpIIIff{{
\begin{picture}(21,14)
\put(3,-2){\begin{picture}(18,14)
\thinlines
\color{blue}
\multiput(0,0)(6,12){2}{\line(1,0){12}}
\multiput(0,0)(12,0){2}{\line(1,2){6}}
\thicklines
\color{black}
\put(12,8){\circle*{2}}
\color{red}
\put(12,8){\circle{4}}
\put(12,8){\circle{6}}
\end{picture}}
\end{picture}
}}


\tableofcontents

\section{Introduction}

In this paper we explore a new connection
between S-duality and pure Chern--Simons theory.
In the context of S-duality,
Chern--Simons theory has already appeared in
the work of Gaiotto and Witten \cite{Gaiotto:2008ak}
on the action of S-duality on boundary conditions.
Gaiotto and Witten studied four dimensional \SUSY{4} $U(n)$
Super Yang--Mills theory (SYM) formulated on a manifold
with a boundary.
They allowed additional degrees of freedom to be localized
on the boundary and to couple to the bulk \SUSY{4} SYM fields,
thereby generating a rich class of possibilities
for boundary conditions, generalizing
the standard Dirichlet and Neumann ones \cite{Gaiotto:2008sa}.
Chern--Simons couplings (either involving the bulk gauge
fields or boundary gauge fields) are an optional
additional ingredient
that was included in their discussion,
and S-duality can generate such couplings.

In this paper we will also study \SUSY{4} SYM with
a novel type of boundary conditions,
but these will be periodic boundary conditions
that involve S-duality at the outset.
We formulate \SUSY{4} $U(n)$ SYM on $S^1\times\R^{2,1}$
but include an S-duality and R-symmetry twist along $S^1.$
The S-duality twist is the novel feature,
which is allowed for the special value $\tau=i$
of the coupling constant
$$
\tau\equiv \frac{4\pi i}{\gYM^2} + \frac{\theta}{2\pi}\,.
$$
($\tau=i$ is the only value
invariant under $\tau\rightarrow -1/\tau.$)
The S-duality twist
is inspired by similar exotic twists
that have appeared in different contexts before
\cite{Vafa:1996xn,
Dabholkar:1998kv,
Ganor:1998ze,
Dabholkar:2002sy,
Shelton:2005cf,
Vegh:2008jn}.
(The authors of \cite{Hellerman:2006tx} coined the term
{\it monodrofolds} for such twists.)

We are interested in the low-energy limit of this setting,
as the $S^1$
shrinks to zero. In this limit, roughly speaking,
all that's left is the S-duality and R-symmetry twists.
For example, in Euclidean signature
(replacing $\R^{2,1}\rightarrow\R^3$) we can think of
the $S^1$ direction as Euclidean time and define
correlation functions of operators $\Op_1,\Op_2,\dots,$ in the
theory as
$$
\langle{\Op_1\Op_2\cdots}\rangle
\equiv
\tr((-1)^F\zs\zr e^{-2\pi\xR\Ham}\Op_1\Op_2\cdots),
$$
where $\zs$ is the S-duality operator,
$\zr$ is the R-twist operator,
$\Ham$ is the Hamiltonian of \SUSY{4} SYM,
$2\pi\xR$ is the circumference of $S^1$,
and $F$ is the fermion number [and $(-1)^F$ is a central element
of the R-symmetry group $SU(4)$].\footnote{We 
are grateful to E.~Witten for pointing out the missing $(-1)^F$
in a previous version.}
In the limit $R\rightarrow 0$
(with an appropriate treatment of zero-modes
as will be discussed later),
$\tr((-1)^F\zs\zr\Op_1\Op_2\cdots)$ is all that remains,
and the theory probes the S-duality operator through
$(-1)^F\zs\zr$.
This is the main reason why we are interested in this problem.

Since abelian S-duality is completely understood
(see, e.g., \cite{Witten:1995gf}),
the solution of our problem for $U(1)$ gauge group is
straightforward. As we explain in
\secref{sec:U(1)}, the resulting low-energy description
is a pure Chern--Simons theory with gauge group $U(1)$
at level $\lvk=2.$
(We can get other levels, $\lvk=1,3$, if we replace
the S-duality twist that realizes $\tau\rightarrow -1/\tau$
with other elements $\zg$ of the duality group $\SL(2,\Z)$
for which a self-dual coupling constant exists.)
The question that we would like to raise at this point is:
how does this statement generalize to nonabelian gauge groups?

Given the results for $U(1)$ gauge group, a na\"{i}ve conjecture would suggest that the low-energy theory is the nonabelian Chern--Simons theory at the same level as in the abelian case \cite{Ganor:2008hd}. We find, however, that the nonabelian theories present a somewhat richer picture than their abelian counterparts.

The main tool that we will use in this paper
is a weakly coupled type-IIA dual of the problem.
To arrive at this dual, we start with
type-IIB string theory, where \SUSY{4} $U(n)$ SYM is naturally
realized as the low-energy description of coincident D3-branes
\cite{Witten:1995im}, and S-duality of the gauge theory
descends from S-duality of the full string theory.
The latter can be realized as a geometrical symmetry
in a dual string theory \cite{Hull:1994ys,Witten:1995ex}.
In order to utilize this geometrical description of S-duality,
we compactify the theory on $T^2$ (replacing
$\R^{2,1}\rightarrow T^2\times\R$) and look for the
ground states.
Realizing the theory on $n$ D3-branes in type-IIB string theory,
we can map the theory to a type-IIA setting
where the question of identifying the ground states
reduces to an easily solvable geometrical problem.
With sufficient supersymmetry, the solution of the geometrical
problem after duality also solves our original problem.
This allows us to calculate the Witten Index and
analyze the space of ground states
and its symmetries in terms of the type-IIA dual background.

Based on this analysis, we argue that (for low enough rank of the gauge group) the Hilbert space of ground states decomposes into a direct sum of Hilbert spaces of Chern--Simons theories with appropriate gauge groups and levels. In particular, there exists a distinguished sector 
(which we call the $[\sigma]$-untwisted sector) that is equivalent to the Hilbert space of nonabelian Chern--Simons theory with gauge group $U(n)$ and level $(2n,2)$, where $2n$ refers to the $U(1)$ center, and $2$ refers to $SU(n)$. By ``equivalent Hilbert spaces'' we mean that their symmetry operators and their behavior under modular transformations of $T^2$ match. Our results then suggest that in the decompactification limit $T^2\times\R\rightarrow\R^{2,1}$ the Hilbert space of the low-energy theory decomposes into different superselection sectors, each described by an appropriate Chern--Simons theory.
We were also able to extend much of this picture
to the compactifications with
twists by other elements $\zg$ of the duality group
$\SL(2,\Z)$, except for a certain problematic
issue that arises for $n\ge 4$ and remains unresolved.

The paper is organized as follows.
In \secref{sec:Problem} we explain the problem in detail.
We discuss the S-duality twist, the various other $\SL(2,\Z)$
elements that can be used to construct twists,
the R-symmetry twist,
the amount of supersymmetry that is preserved,
elimination of zero-modes, and restrictions on the rank $n$
of the gauge group $U(n).$

In \secref{sec:IIA-dual} we compactify the theory on $T^2$
and find the weakly coupled type-IIA dual.
We describe in detail
the U-duality element that maps the problem
to a geometrical one, and discuss various conserved
quantum numbers that can be defined in the geometrical setting.

In \secref{sec:Ctwist}, we study as a warm-up exercise a simpler problem of compactification with charge-conjugation twist
(C-twist). This serves as an illustration of ideas developed in previous sections as well as methods that we will employ in later sections when we attack our main problem, the S-duality twist.

In \secref{sec:U(1)} we solve the
problem for $U(1)$ gauge group explicitly,
and calculate the level $\lvk$ of the low-energy
(pure) Chern--Simons theory.
We then compactify on $T^2$ and compare the Hilbert
space of ground states of abelian Chern--Simons theory
to the Hilbert space of ground states of the type-IIA dual.
We identify the type-IIA dual of Wilson loop operators as well as
other symmetries of the ground states.

In \secref{sec:U(n)-T2} we study the ground states of
the nonabelian problem [with $U(n)$ gauge group] on $T^2,$
using the type-IIA dual theory.
We show that the Hilbert space of ground states decomposes
into a direct sum of Hilbert spaces,
which in most cases we are able to identify
as the Hilbert spaces
of Chern--Simons theory with 
appropriate gauge groups [subgroups of $U(n)$]
and appropriate Chern--Simons levels.

In \secref{sec:twoz} we take another look at our problem
in terms of the $(2,0)$ theory.
We argue that the solution can be constructed
from ingredients that recently appeared in the work
of Gaiotto and Witten \cite{Gaiotto:2008ak}
in connection with the low-energy description of D$3$-branes
that end on $(p,q)$ $5$-branes.
We show how to recover the $U(1)$ result from these ingredients.

We conclude with a discussion of the results and open problems
in \secref{sec:disc}.


\section{The problem}
\label{sec:Problem}

We wish to learn new facts about the
$\SL(2,\Z)$ S-duality of
\SUSY{4} super Yang--Mills theory
by studying a circle compactification of the theory
with unconventional boundary conditions as follows.
Realizing the circle as the segment $[0,2\pi\xR]$
with endpoints $0$ and $2\pi\xR$ identified,
we require the configuration at $2\pi\xR$ to be an S-dual
of the configuration at $0.$
We will refer to this kind of boundary conditions as an {\it S-twist.}
To be specific,
we need to pick an element $\zg\in\SL(2,\Z)$,
and we need the coupling constant to be invariant under $\zg.$
There are only a small number of possibilities of this kind,
which we will list in \secref{subsec:S-twist}.

The S-twist would be easy to describe
if we knew a formulation of \SUSY{4} super Yang--Mills theory
for which S-duality is manifest. Nevertheless,
it is not hard to argue that the S-twist is consistent.
For example, in Euclidean signature we can take the direction
of the circle to be Euclidean time, and the S-twist then corresponds
to an insertion of the operator corresponding to $\zg$
on the Hilbert space of states, thus obtaining
$\tr((-1)^F\zg e^{-2\pi\xR\Ham}\cdots)$, where $\Ham$ is
the Hamiltonian, $(\cdots)$ represents additional insertions
of local operators if desired, and by a slight abuse of notation
we used the same $\zg$ to denote the action of $\zg$ on the
Hilbert space at the self-dual coupling constant.
In \secref{sec:IIA-dual} we  bring more evidence
for the consistency of the S-twist:
we present a string-theoretic
construction with an S-twist, and show that it is
dual to a conventional type-IIA string compactification.

In order to preserve some amount of supersymmetry,
we also need to pick an
appropriate nontrivial element $\gtw$
of the R-symmetry group and identify the configuration at $2\pi\xR$
with the $\gtw$-transformed
$\zg$-dual of the configuration at $0.$
For a suitable choice of $\gtw$ we can preserve $12$
supersymmetry generators, which corresponds to \SUSY{6} in
three dimensions.

Our problem is to find the effective three-dimensional
low-energy description of the theory
in the limit $R\rightarrow 0.$
We propose that for a sufficiently low rank $n$
(how low depends on $\zg$),
the requisite three-dimensional field theory is topological,
and in the next sections we will study it in special cases.

The rest of this section provides more details on the
construction above.
In \secref{subsec:notation} we introduce the notation
for the rest of this paper;
in \secref{subsec:S-twist} we discuss the various choices
for $\zg$ (there are only three) and the corresponding
self-dual coupling constants.
In \secref{subsec:R-twist} we discuss the
associated R-symmetry twist $\gtw$;
and in \secref{subsec:LE} we introduce restrictions
on the rank $n$ of the gauge group that are necessary
to eliminate unwanted low-energy moduli.
These details are a condensed version
of the discussion that can be found in \cite{Ganor:2008hd}.


\subsection{\SUSY{4} super Yang--Mills: notation}
\label{subsec:notation}
Our starting point is four-dimensional \SUSY{4}
super Yang--Mills theory with gauge group $U(n).$

We denote the complex coupling constant by
$$
\tau\equiv \frac{4\pi i}{\gYM^2} + \frac{\theta}{2\pi}\,.
$$
It transforms under an element
\be\label{eqn:xabcd}
\begin{pmatrix} \xa & \xb \\ \xc & \xd \\ \end{pmatrix}
\in\SL(2,\Z)
\ee
as
$$
\tau\rightarrow \frac{\xa\tau+\xb}{\xc\tau+\xd}\,.
$$

Our notation for the fields of $U(n)$
\SUSY{4} super Yang--Mills theory is summarized below:
\vskip 12pt
\begin{tabular}{lll}
$A_\mu$ & gauge field  &
$\mu=0,\dots ,3,$ \\
$\Phi^I$ & adjoint-valued scalars &
$I=1,\dots ,6,$ \\
$\psi^a_\a$ & adjoint-valued spinors &
$a=1,\dots ,4$ and $\a=1,2,$ \\
$\bpsi_{a\dta}$ & complex conjugate spinors &
$a=1,\dots ,4$ and
$\dta=\dot{1},\dot{2},$ \\
$Q_{a\a}$ & SUSY generators &  $a=1,\dots ,4$
and $\a=1,2,$ \\
$\bQ^a_{\dta}$ & complex conjugate generators &
$a=1,\dots ,4$  and $\dta=\dot{1},\dot{2}$. \\
\end{tabular}
\vskip 12pt
\noindent We also define the complex combinations of scalar fields
\be\label{eqn:Zphi}
\Zphi^j\equiv\Phi^j+i\Phi^{3+j}\,,
\qquad
j=1,2,3.
\ee
The $S^1$ on which we compactify is in direction $3.$


\subsection{S-duality twist}\label{subsec:S-twist}

To define the S-duality twist we need a pair $(\zg,\tau)$
comprising of an element $\zg=
\begin{pmatrix} \xa & \xb \\ \xc & \xd \\ \end{pmatrix}
\in\SL(2,\Z)$ and a self-dual coupling constant $\tau$,
satisfying
\be\label{eqn:sdtau}
\tau=\frac{\xa\tau+\xb}{\xc\tau+\xd}\,.
\ee
Assuming $\xc\neq 0$,
\eqref{eqn:sdtau} is equivalent to the quadratic equation
$\xc\tau^2 +(\xd-\xa)\tau-b=0$, and if it has solutions
away from the real axis, they must satisfy $|\tau|^2=-\xb/\xc$
and $\tau+\btau=(\xa-\xd)/\xc$, which implies that
$|\xc\tau+\xd|^2=1.$ We can therefore set
\be\label{eqn:phtdef}
\xc\tau+\xd=e^{i\pht}
\ee
for some real phase $\pht.$
It follows that $\cos\pht=\xd+\xc(\tau+\btau)/2=(\xa+\xd)/2$
can only take the values $0$ or $\pm 1/2$, and so
$\pht$ is one of $\pm \tfrac{1}{2}\pi,\pm\tfrac{1}{3}\pi,
\pm\tfrac{2}{3}\pi.$
Furthermore, it is easy to check that the eigenvalues of $\zg$
are $e^{\pm i\pht}$, and thus $\zg$ has finite order, which
can be of the three possibilities $\ord=3,4,6.$
Thus,
\be\label{eqn:def-ord}
|\pht|=\frac{2\pi}{\ord}\,.
\ee

Up to conjugation [$\zg\rightarrow \zg_0^{-1}\zg\zg_0$ for
some $\zg_0\in\SL(2,\Z)$] and inversion ($\zg\rightarrow\zg^{-1}$),
we are left with the following three choices
for $(\zg,\tau)$:
\begin{enumerate}
\item
$\tau=i$ and $\zg=\zg'\equiv\left(\begin{array}{rr}
 0 & -1 \\ 1 & 0 \\ \end{array}\right)$  of order $\ord=4$
($\pht=\tfrac{1}{2}\pi$);
\item
$\tau=e^{\pi i/3}$ and
$\zg=\zg''\equiv\left(
\begin{array}{rr} 1 & -1 \\ 1 & 0 \\ \end{array}
\right)\in\SL(2,\Z)$  of order $\ord=6$
($\pht=\tfrac{1}{3}\pi$);
\item
$\tau=e^{\pi i/3}$ and
$\zg=-\zg''{}^{-1}=\left(
\begin{array}{rr} 0 & -1 \\ 1 & -1 \\ \end{array}
\right)\in\SL(2,\Z)$ of order $\ord=3$
($\pht=\tfrac{2}{3}\pi$).
\end{enumerate}
All other possible $\zg$'s are $\SL(2,\Z)$-conjugate
to those in the list, or their inverses (which, as we will see, give
theories that are physically equivalent after a parity
transformation).

We recall \cite{Kapustin:2006pk} that $\zg$ acts nontrivially
on the supercharges
\be\label{eqn:zsQ}
\zg: Q_{a\a}\rightarrow
\left(\frac{\xc\tau+\xd}{|\xc\tau+\xd|}
\right)^{-\tfrac{1}{2}}Q_{a\a}
 = e^{-\frac{i\pht}{2}}Q_{a\a}
\,.
\ee
In order to get a supersymmetric theory,
we therefore need to supplement the S-twist with
an R-symmetry twist so that the phase in \eqref{eqn:zsQ} is cancelled.


\subsection{R-symmetry twist}\label{subsec:R-twist}

An R-symmetry twist modifies the periodic
boundary conditions by introducing additional
phases for R-charged fields.
It is a useful tool to eliminate unwanted zero
modes while preserving some amount of supersymmetry
(see for instance
\cite{Witten:1993jg,Cheung:1998te}).
In our context it also allows us
to restore some of the supersymmetry that was lost
by the S-twist.

We pick a basis of the R-symmetry group $SU(4)$
so that a diagonal element
\be\label{eqn:gtwgen}
\gtw\equiv
\begin{pmatrix}
e^{i\ftw_1} & &  & \\
& e^{i\ftw_2} &  & \\
& & e^{i\ftw_3}  & \\
& & & e^{i\ftw_4}  \\
\end{pmatrix} \in SU(4)_R\,,
\qquad
\left(\sum_a\ftw_a = 0\right)\,,
\ee
acts on the fermionic fields from \secref{subsec:notation} as
$$
\gtw(\psi^a_\a) = e^{i\ftw_a}\psi^a_\a\,,
\qquad
\gtw(\bpsi_{a\dot{\a}}) = e^{-i\ftw_a}\bpsi_{a\dot{\a}}\,,
\qquad
a=1,\dots,4,
$$
and on the bosonic fields as
$$
\gtw(A_\mu)=A_\mu\,,
\qquad
\gtw(\Zphi^j)=e^{i(\ftw_j+\ftw_4)}\Zphi^j\,,
\qquad
j=1,2,3.
$$

A $\gtw$-twist, on its own, modifies the boundary conditions to
\bear
\psi^a_\a(x_0,x_1,x_2,x_3+2\pi\xR) &=&
e^{i\ftw_a}\Lambda^{-1}\psi^a_\a(x_0,x_1,x_2,x_3)\Lambda\,,
\label{eqn:RtwistPsi}\\
\Zphi^j(x_0,x_1,x_2,x_3+2\pi\xR) &=&
e^{i(\ftw_j+\ftw_4)}\Lambda^{-1}\Zphi^j(x_0,x_1,x_2,x_3)\Lambda\,,
\label{eqn:RtwistZ}\\
A_\mu(x_0,x_1,x_2,x_3+2\pi\xR) &=&
\Lambda^{-1}A_\mu(x_0,x_1,x_2,x_3)\Lambda
+\Lambda^{-1}\px{\mu}\Lambda\,,
\label{eqn:RtwistA}
\eear
where $\Lambda$ is an arbitrary gauge transformation.
We combine the R-symmetry twist by $\gtw$ with the S-twist from
\secref{subsec:S-twist} to get an S-R-twist.
It can be formally defined by switching to Euclidean signature,
considering the direction $x_3$ as Euclidean time, and defining
correlation functions of operators,
similarly to the discussion at the top of \secref{sec:Problem},
by $\tr((-1)^F\gtw\zg e^{-2\pi\xR\Ham}\cdots)$, where $F$ 
is the fermion number, $\Ham$ is
the Hamiltonian, $(\cdots)$ represents insertions
of local operators if desired, and $\gtw$ is the R-symmetry operator
(in a slight abuse of notation we here
denote the representation of $\gtw$ on the Hilbert space by the same letter).

Combining $\gtw$ with \eqref{eqn:zsQ},
we find the action of $\gtw\zg$ on the supercharges:
$$
\gtw\zg(Q_{a\a})=
e^{i(\ftw_a-\frac{\pht}{2})}Q_{a\a}\,.
$$
Therefore, to preserve \SUSY{2} supersymmetry in three-dimensions, for example, we need to set
one of the $\ftw_a$, say $\ftw_1$, to $\frac{1}{2}\pht.$
The maximal amount of supersymmetry that we can preserve is \SUSY{6} with
\be\label{eqn:gtwN=6}
\gtw=
\begin{pmatrix}
e^{\frac{i}{2}\pht} & &  & \\
& e^{\frac{i}{2}\pht} &  & \\
& & e^{\frac{i}{2}\pht}  & \\
& & & e^{-\frac{3i}{2}\pht}  \\
\end{pmatrix} \in SU(4)_R\,.
\ee
We will work with that choice of $\gtw$ from now on.


\subsection{Low-energy limit}\label{subsec:LE}

Our goal is to study the low-energy description
of the compactification of \SUSY{4} $U(n)$ SYM on $S^1$
with a combination of
S-duality twisted boundary conditions as in
\secref{subsec:S-twist} and R-symmetry twisted boundary
conditions as in \secref{subsec:R-twist}.
With the choice of $\gtw$ as in \eqref{eqn:gtwN=6},
the theory has \SUSY{6} supersymmetry in 2+1D.
In this paper we further wish to restrict the parameters
so as to get a {\it topological} QFT in 2+1D,
for which the supersymmetry is realized trivially---all
generators are identically zero at low-energy
(which is only possible for a topological theory for which
the momentum and Hamiltonian are also zero).

This restriction requires that no massless
propagating fields shall survive at low energy.
For a $U(1)$ gauge group we will see in \secref{sec:U(1)}
that the low-energy limit is $U(1)$ Chern--Simons theory,
and indeed no low-energy propagating degrees of freedom survive;
the mass gap of our setting
is of the order of the Kaluza--Klein scale $1/\xR.$
However,
in the nonabelian case, $n>1$, the S-duality twist is poorly
understood, and it is less clear whether our setting has a mass gap
or not. In fact, we will argue in \secref{sec:U(n)-T2}
that in general our $S^1$ compactification
has several discrete choices leading
to separate superselection sectors,
each defining a different low-energy limit.
Some superselection sectors come with a mass gap,
while others do not.

In this section, however, we will introduce a necessary
requirement---that no noncompact moduli survive
the compactification to 2+1D.
This requirement seems sufficient to ensure that
the additional compactification on $T^2$ (to 0+1D),
which we will study later on, leads to a discrete spectrum.
So, we must start by eliminating the potential
zero modes arising from the dimensional reduction
of the scalar fields.

To see what that entails, let us attempt to construct
a massless degree of freedom by starting at a generic point
on the Coulomb branch of \SUSY{4} SYM in 3+1D, where
the gauge group is broken to $U(1)^n$. The 3+1D low energy physics
is described by $n$ free \SUSY{4} vector multiplets,
and the residual gauge symmetry is the permutation group $S_n$
that permutes the $n$ vector multiplets.
If the energy scale at which the $U(n)$ gauge symmetry is broken
(which is determined by the differences between the VEVs of
the scalar components of the vector multiplets) is much larger
than the compactification scale $1/R$, we can approximate
the low-energy theory by simply compactifying the $n$ free
vector multiplets on $S^1$ with the R-symmetry and S-duality twists.

Compactification of a single vector multiplet with
R-symmetry and S-duality twists leaves no massless fields in 2+1D.
To see this, consider the gauge field and the scalars and fermions
separately. Only the S-twist affects the gauge field, and only
the R-twist affects the scalars and fermions.
The gauge fields will, at best, give rise to compact moduli,
but for a single vector multiplet they do not produce
any moduli at all.
This is because for a massless mode to exist in three dimensions, we need a
solution\footnote{We will study in greater detail
the resulting low-energy limit later in \secref{sec:U(1)},
but for the purposes of the discussion in this section
it suffices to look for classical solutions.}
where the electric and magnetic fields $E_i,B_i$ (both
three-dimensional vectors with $i=1,2,3$) are independent
of $x_3$ and satisfy
\be\label{eqn:EBzm}
\begin{pmatrix} E_i \\ B_i \\ \end{pmatrix}
=
\begin{pmatrix} \xa & \xb \\ \xc & \xd \\ \end{pmatrix}
\begin{pmatrix} E_i \\ B_i \\ \end{pmatrix}\,,
\ee
as required by the S-twist.
But since $\zg=
\begin{pmatrix} \xa & \xb \\ \xc & \xd \\ \end{pmatrix}$
has nontrivial eigenvalues $e^{\pm i\pht}$,
there is no nonzero solution to \eqref{eqn:EBzm},
and no massless fields arise from the gauge fields.
For scalar zero modes we would look for solutions
to [see \eqref{eqn:RtwistZ} and \eqref{eqn:gtwN=6}]:
\be\label{eqn:Zzm}
\Zphi^j =
e^{i(\ftw_j+\ftw_4)}\Zphi^j= e^{-i\pht}\Zphi^j\,,
\qquad
j=1,2,3,
\ee
which has no nonzero solutions.
Similarly, there are no fermion zero modes,
which of course follows from supersymmetry.
So, a single vector multiplet compactified with an S-R-twist
does not have any low-energy propagating degrees of freedom.

However, as we shall now see, for $n\ge\ord$ (where $\ord$
was defined in \eqref{eqn:def-ord} as $2\pi/\pht$)
we do get massless propagating degrees of freedom.
To see this, note that the boundary conditions
in \eqref{eqn:RtwistPsi}-\eqref{eqn:RtwistA} have an
optional $U(n)$ gauge transformation $\Lambda.$
Once the gauge group is broken as $U(n)\rightarrow U(1)^n$,
we are only allowed to take $\Lambda$ in the normalizer of $U(1)^n$
in $U(n)$, which is the permutation group $S_n.$
We thus identify $\Lambda$ with some permutation $\perm\in S_n$
and modify the conditions for zero modes
\eqref{eqn:EBzm}-\eqref{eqn:Zzm} to
\be\label{eqn:EBzmP}
\begin{pmatrix}
E_i^{(\perm(l))} \\ B_i^{(\perm(l))} \\
\end{pmatrix}
=
\begin{pmatrix} \xa & \xb \\ \xc & \xd \\ \end{pmatrix}
\begin{pmatrix}
E_i^{(l)} \\ B_i^{(l)} \\
\end{pmatrix}\,,
\qquad
i=1,\dots,3,\quad l=1,\dots,n\,,
\ee
and
\be\label{eqn:ZzmP}
\Zphi^{j,\perm(l)} =
 e^{-i\pht}\Zphi^{j,l}\,,
\qquad
j=1,\dots,3,\quad l=1,\dots,n\,,
\ee
where the superscript $l$ corresponds to the $l^{th}$ $U(1)$ factor
in $U(1)^n$ and the permutation $\perm$ maps
$\{1,2,\dots,n\}$ to $\{\perm(1),\perm(2),\dots,\perm(n)\}.$
Equations \eqref{eqn:EBzmP}-\eqref{eqn:ZzmP} have nonzero solutions
if and only if $\perm$ has a cycle of length divisible by $\ord$, and
such a $\perm\in S_n$ exists if and only if $n\ge\ord.$

At the end of \secref{subsec:S-twist} we listed various possible values
of $\zg$ and $\ord.$
The corresponding restrictions on the rank $n$
of the gauge group are therefore:
$n\le 2$ for the case with $\ord=3$;
$n\le 3$ for $\ord=4$; and
$n\le 5$ for $\ord=6.$
For these cases there are no obvious zero modes,
and we are going to assume that the low-energy theory
has no noncompact moduli for $n<\ord.$


\section{Type-IIA dual}
\label{sec:IIA-dual}

The setting of \secref{sec:Problem} has a string-theoretic
realization in terms of D3-branes of type-IIB string theory.
We start with the background $\R^{9,1}$ with Cartesian coordinates
$x_0,\dots,x_9$,
and place $n$ D3-branes at $x_4=x_5=\cdots=x_9=0.$ The
type-IIB coupling constant is denoted by
$\tau=\chi+\frac{i}{\gIIB}$,
where $\gIIB$ is the string coupling constant,
and $\chi$ is the R-R scalar.
The S-duality transformation $\zg$ of \secref{subsec:S-twist} then
lifts to an S-duality transformation of the full type-IIB string theory
(that we also denote by $\zg$),
and the R-symmetry rotation $\gtw$ of \secref{subsec:R-twist}
lifts to a geometrical rotation in the $6$ directions transverse
to the D3-branes. We will now transform this background, using string dualities,
to one where S-duality is realized geometrically.

We first compactify the $x_3$-direction
on a circle of radius $2\pi \xR$ with
boundary conditions given by a simultaneous
S-duality twist $\zg$ and a $\gtw\in\Spin(6)$ geometrical twist in
directions $x_4,\dots ,x_9$,
where $\gtw$ is given by \eqref{eqn:gtwN=6} in terms of $\pht,$
and $\pht$ by \eqref{eqn:phtdef}.
This means that as we traverse
the $x_3$ circle once, we also apply a $\gtw\in\Spin(6)$
rotation in the transverse directions before
gluing $x_3=0$ to $x_3=2\pi \xR$,
similarly to the discussion in \secref{subsec:R-twist}.
We then compactify directions $x_1,x_2$,
so that $0\le x_1<2\pi \xL_1$ and
$0\le x_2<2\pi \xL_2$ are periodic.
This puts the 2+1D field theory on $T^2$
with area $4\pi^2\xL_1\xL_2$ and complex structure $i\xL_2/\xL_1.$

Now we can study different limits of the parameters
$\xL_1, \xL_2, \xR.$
First, to reproduce the field-theory
problem of \secref{sec:Problem} we need to take the limit
\be\label{eqn:LimIIB}
\xL_1, \xL_2, \xR\gg \apr^{1/2}\,,
\ee
 where $\apr^{1/2}$
is the type-IIB string scale.
In the limit \eqref{eqn:LimIIB}, we can first reduce
the description of the D3-branes to \SUSY{4} $U(n)$ SYM at low energy,
and then compactify \SUSY{4} SYM with an S-duality and R-symmetry twist.

We now consider the opposite limit $\xL_1,\xL_2\rightarrow 0$
with $\xR\rightarrow\infty$ (in the order to be specified below).
In this limit, the type-IIB description is strongly-coupled,
but we will perform a U-duality transformation,
in a series of steps described below and summarized
in \tabref{tab:Dualities}, to transform the setting to a weakly
coupled type-IIA background. This will also allow us to easily study the
ground states of the field theory.

\begin{table}[t]
\begin{tabular}{|l|l|cccccccccc|l|}
\hline\hline
Type & Brane &
$1$ & $2$ & $3$ & $4$ & $5$ & $6$ & $7$ & $8$ & $9$ & $10$ &
Apply:\\ \hline\hline
IIB  & D3 &
\wx & \wx & \tx & &&&&&&\nx &
T-duality on $1.$ \\ \hline
IIA  & D2 &
    & \wx & \tx & &&&&&&\nx &
Lift to M-theory. \\ \hline
M    & M2 &
    & \wx & \tx &     &&&&&&&
Reduction to IIA on $2.$ \\ \hline
IIA  & F1 &
     & \nx & \tx &     &&&&&&&
\\ \hline\hline
\end{tabular}
\caption{
The sequence of dualities from
$n$ D3-branes in type-IIB to $n$ fundamental strings in
type-IIA. A direction that the corresponding brane or string
wraps with periodic boundary conditions
is represented by \wx, a direction that the object wraps
with twisted boundary conditions is represented by \tx,
and a dimension that doesn't exist in the particular string
theory is represented by \nx.
All the branes in the table are at the origin
of directions $4,\dots ,9.$
}
\label{tab:Dualities}
\end{table}

The U-duality transformation proceeds as follows.
We first replace type-IIB on a circle of radius $\xL_1$
with M-theory on $T^2$ with complex structure $\tau$
and area
$\Area
=(2\pi)^2\apr^2\tau_2^{-1}\xL_1^{-2}
=(2\pi)^2\Mpl^{-3}\xL_1^{-1}$,
where $\Mpl$ is the $11$-dimensional Planck scale.
We now reduce from M-theory to type-IIA on the circle of radius
$\xL_2$ to get a theory with string coupling constant
$$
\gIIA
=(\Mpl\xL_2)^{3/2}
=\tau_2^{1/2}\xL_1^{1/2}\xL_2^{3/2}\apr^{-1}
\,,
$$
and new string scale
$$
\apr_{\text{IIA}}=\Mpl^{-3}\xL_2^{-1}
=\apr^2\tau_2^{-1}\xL_1^{-1}\xL_2^{-1}
\,.
$$

After these dualities, the D3-branes become fundamental type-IIA
strings with a total winding number $n$ in the $x_3$ direction.
The S-duality twist $\zg$ is now a diffeomorphism of
the type-IIA $T^2$, which can be realized as a rotation by an angle $\pht.$
To make this type-IIA background weakly coupled,
we assume that the limits are taken in such a way that
\be\label{eqn:LimIIA}
\Area\gg\apr_{\text{IIA}}\,,
\qquad
\gIIA\ll 1\,,
\qquad
\xR\gg\apr_{\text{IIA}}^{1/2}.
\ee
This is a different limit than \eqref{eqn:LimIIB},
but we can use the weakly coupled type-IIA background to
study the Hilbert space of (supersymmetric) ground states.
Since the type-IIA setting
is described by $n$ fundamental strings on a weakly
coupled background, the question of the Hilbert space of ground states reduces to a simple calculation in string theory.
For quick reference, we have summarized in
\tabref{tab:IIB-IIA}
the dual type-IIA
description of various charges of the original type-IIB setting.

\begin{table}[t]
\begin{tabular}{|c|c|}
\hline\hline
Type-IIB & Type-IIA \\ \hline\hline
$T^2$ is in directions $1,2$ & (Dual) $T^2$ is in directions $1,10$
\\ \hline\hline
$n$ D3-branes (directions $1,2,3$) &
$n$ F1-strings (direction $3$)\\ \hline
$\SL(2,\Z)$ diffeomorphisms of $T^2$ &
$\SL(2,\Z)$ T-duality group of (dual) $T^2$ \\ \hline
Momentum $P_1$ & D$2$-brane (wrapping directions $1,10$) charge \\ \hline
Momentum $P_2$ & D$0$-brane charge \\ \hline
String winding in direction $1$ &
Momentum $P_1$ \\ \hline
String winding in direction $2$ &
String winding in direction $10$ \\ \hline
D1 winding in direction $1$ &
Momentum $P_{10}$ \\ \hline
D1 winding in direction $2$ &
String winding in direction $1$ \\ \hline
\end{tabular}
\caption{
Mapping between the quantum numbers and other notions on the type-IIB side
to those on the type-IIA side.
}
\label{tab:IIB-IIA}
\end{table}

\subsection{The dual geometry}
\label{subsec:IIA}

After the series of dualities summarized in \tabref{tab:Dualities},
we end-up with a type-IIA string theory that we will now
describe in detail. The 9+1D
geometry is flat and free of singularities, and
the spatial part is a free orbifold of $\R^9.$
It is convenient to divide the $9$ directions in three groups
and describe the geometry as
an orbifold of $T^2\times\R\times\C^3.$
We regard the $T^2$ as the complex plane modded out by  a lattice,
$\C/(\Z+\tau\Z)$,
and take
$$
z\sim z+1\sim z+\tau\,
$$
as its coordinate.
On $\R$, the coordinate takes
$$
-\infty < x_3 < \infty\,,
$$
and on $\C^3\simeq\R^6$, we take the coordinates to be
$$
(\zeta_1,\zeta_2,\zeta_3)\,,
\qquad
\zeta_1,\zeta_2,\zeta_3\in\C\,.
$$
The orbifold is then represented by the identification
\be\label{eqn:zxzzz}
(z,x_3,\zeta_1,\zeta_2,\zeta_3)\sim
(e^{i\pht}z,x_3+2\pi\xR,
e^{i\pht}\zeta_1,e^{i\pht}\zeta_2,e^{i\pht}\zeta_3)\,.
\ee
Note that, because of the shift $x_3\rightarrow x_3+2\pi\xR$,
the orbifold has no fixed points, and the geometry is smooth.
From now until the rest of this section,
\eqref{eqn:zxzzz} will be our background.

It is also convenient to give a separate name for the
$\zeta_1=\zeta_2=\zeta_3=0$ subspace.
We will denote this smooth, flat, and compact $3$-dimensional manifold
by $\MXS.$ It is represented by the coordinates $(z,x_3)$ with identifications
\be\label{eqn:zx}
\MXS:\qquad (z,x_3)\sim (z+\tau,x_3)\sim (z+1,x_3)\sim
(e^{i\pht}z,x_3+2\pi\xR)\,.
\ee
This manifold is a $T^2$-fibration over $S^1$ with structure group $\Z_\ord.$

\subsection{Ground states}
\label{subsec:GS}

The states that are relevant to our problem are those
with a total string winding number $n$ along direction $x_3.$
A state with string winding number $n$ is
a $p$-particle (that is, $p$-string) state comprising of $1$-particle states
of winding numbers $n_1\ge n_2\ge\dots\ge n_p>0$ with
$n_1+n_2+\cdots+n_p=n.$
Thus, the Hilbert space of ground states decomposes
as a direct sum:
\be\label{eqn:HilbSumPart}
\Hilb(n,\pht)=\bigoplus_{
\begin{subarray}{c}
n_1\ge n_2\ge\dots\ge n_p > 0 \\
n_1+n_2+\cdots+n_p=n
\end{subarray}}
\Hilb_{(n_1,\dots,n_p)}(\pht)
\,.
\ee

We can recast the partition $n=n_1+\cdots+n_p$
as a conjugacy class $[\perm]$ of a permutation $\perm\in S_n$,
so that when $\perm$ is decomposed
into cycles the integers $n_1,\dots,n_p$ denote the
lengths of the cycles. So, for example $n=1+1+\dots+1$ (i.e., $p=n$)
corresponds to the identity permutation $\sigma=1$.
We therefore set
\be\label{eqn:HilbSumPerm}
\Hilb(n,\pht)=\bigoplus_{[\perm]}
\Hilb_{[\perm]}(\pht)
\,,\qquad
(\perm\in S_n)\,
.
\ee
We will refer to $\Hilb_{[1]}$ as the
{\it $[\perm]$-untwisted sector},
and to $\Hilb_{[\perm]}$ with $\perm\neq 1$ as the {\it $[\perm]$-twisted sectors.}

Understanding the multi-particle Hilbert spaces
$\Hilb_{[\perm]}(\pht)$ requires analysis of the
single-particle states of which the multi-particle states
are constructed, so
let us first discuss the single-particle string states.
The problem of superstring quantization in the flat background
\eqref{eqn:zxzzz} was studied in detail in \cite{Russo:1995ik}.\footnote{
We are grateful to Aki Hashimoto for pointing out this reference.}
In the RNS formalism, in the sector with winding
number $\np$ in direction $x_3$, the mode expansion of worldsheet fields
contains modes shifted from the standard
integer or half-integer values by  $\pm (\np/\ord).$ This is fractional
if $\np$ is not divisible by $\ord.$

For the purposes of the present paper, we do not need the details
of the worldsheet quantization or the full string
spectrum---we only need the ground states.
It turns out that (for $\np\neq 0$)
the ground states are bosonic and in the R-R sector.
In fact, the problem of finding the ground states can
be solved using essentially classical geometry:
we simply need to find classical string configurations of minimal length.
For $\np$ that is not divisible by $\ord$,
there is a basis of ground states that
are in one-to-one correspondence with loops of
minimal length and winding number $\np$
in the geometry \eqref{eqn:zxzzz}.
In the limit $\apr\rightarrow 0$,
these states reduce to the classical string configurations,
but even for finite $\apr$ these classical string configurations
are the minima of the worldsheet energy, and fluctuations around these classical
configurations correspond to massive worldsheet modes,
and there is a single ground state for each classical configuration.

To describe the classical configurations,
we can fix an $x_3$ coordinate and specify the points where the classical
string intersects the transverse coordinates $T^2\times\C^3$ in
the geometry \eqref{eqn:zxzzz}. At winding number $\np$, the string
intersects $T^2\times\C^3$ at $\np$ (not necessarily distinct) points,
and in order to be of minimal length the coordinates of
these points should be independent of $x_3.$
The classical configurations are thus characterized by a set of $\np$
points in $T^2\times\C^3$
that is invariant, as a set, under the orbifold operation
$$
(z,\zeta_1,\zeta_2,\zeta_3)\sim
(e^{i\pht}z,e^{i\pht}\zeta_1,e^{i\pht}\zeta_2,e^{i\pht}\zeta_3)\,.
$$

For $\np$ that is not divisible by $\ord$,
there is a finite number of such sets,
and they are all localized at the origin of $\C^3$, i.e.,
$\zeta_1=\zeta_2=\zeta_3=0.$
They are therefore entirely described by
the $z$-coordinates of where the string intersects $T^2$:
$z,e^{i\pht}z, e^{2i\pht},\dots,e^{i(\np-1)\pht}$,
since as we go once around the $x_3$ direction the coordinate $z$
switches to $e^{i\pht} z.$ After $\np$ loops, $z$ becomes
$e^{i\np\pht}z$ which, in order to close the loop,
should be identified with $z$, up to a shift in $\Z+\Z\tau.$
The classical string configurations are then described by
solutions $z=\fpZ_{\wMa,\wMb}$ of
\be\label{eqn:fpZsol}
e^{i\np\pht}\fpZ_{\wMa,\wMb}=\fpZ_{\wMa,\wMb}+\wMa+\wMb\tau\,,
\ee
and we consider two solutions $\fpZ_{\wMa,\wMb}$ and
$\fpZ_{\wMa',\wMb'}$ as equivalent if they differ
by a lattice element, i.e., if
$\fpZ_{\wMa,\wMb}-\fpZ_{\wMa',\wMb'}\in\Z+\Z\tau.$
In addition, $\fpZ_{\wMa,\wMb}$ and $e^{i\pht}\fpZ_{\wMa,\wMb}$
give equivalent solutions, since the intersection points of the string
with $T^2$ are unordered.
There is then only a finite number of inequivalent solutions
to \eqref{eqn:fpZsol}, and we will describe them in detail
at the end of this subsection.
We conclude that
the full single-particle string spectrum
(including excited states) decomposes into a finite sum of
distinct sectors, labeled by $\wMa,\wMb$,
and the solution $\fpZ_{\wMa,\wMb}$, which is a point on $T^2$,
describes the center of mass of the string in the directions of $T^2.$

We will denote a single-particle ground state with
winding number $\np$
by the location of the intersection of the classical string configuration
with any particular $T^2$ fiber at a constant $x_3.$
In other words, we denote a single-particle state corresponding to a solution $z$
of \eqref{eqn:fpZsol} by
\be\label{eqn:solnot}
\ket{[z, e^{i\pht}z,\dots,e^{(\np-1)i\pht}z]}\,,
\ee
where $z$ coordinates are always taken modulo the lattice $\Z+\Z\tau.$
Multi-string states are denoted by
$$
\ket{\{
[z_1, e^{i\pht}z_1,\dots,e^{(\np_1-1)i\pht}z_1],
[z_2, e^{i\pht}z_2,\dots,e^{(\np_2-1)i\pht}z_2],
\ldots,
[z_p, e^{i\pht}z_p,\dots,e^{(\np_p-1)i\pht}z_p]
\}}\,,
$$
where $z_i$ is a solution $\fpZ_{\wMa_i,\wMb_i}$ of \eqref{eqn:fpZsol}
with $\np\rightarrow \np_i,$ and $n=\sum_1^p \np_i$ is the total winding number.

The number of inequivalent solutions of \eqref{eqn:fpZsol}
for $\np=1$ will be denoted by $\lvk$.
It is a function of $\pht$ alone.
As we will see below, in our three cases we get the following three values:
\be\label{eqn:deflvk}
\left.
\begin{array}{llllll}
\lvk=1 &\text{when}&  \ord=6, & \pht=\frac{\pi}{3}, & \tau = e^{\pi i/3}, &
\zg=\left(\begin{array}{rr} 1 & -1 \\ 1 & 0 \\ \end{array}\right)\,;
\\ & & & & \\
\lvk=2 &\text{when}& \ord=4, & \pht=\frac{\pi}{2}, & \tau = i, &
\zg=\left(\begin{array}{rr}
 0 & -1 \\ 1 & 0 \\ \end{array}\right)\,;
\\ & & & & \\
\lvk=3 &\text{when}& \ord=3, & \pht=\frac{2\pi}{3}, & \tau = e^{\pi i/3}, &
\zg=\left(\begin{array}{rr} 0 & -1 \\ 1 & -1 \\ \end{array}\right)\,.
\\
\end{array}\right\}
\ee

As promised earlier, we conclude this subsection
with a full description of the single-particle ground states.
For additional clarity, we found it convenient to use
a pictorial notation.
We draw a fundamental cell of the lattice
$\C/(\Z+\Z\tau)$ as a parallelogram and explicitly mark the location
of the solutions for $z$ on it.
We denote the solutions $z$ by a dot
surrounded by a circle, and if a solution $z$ appears
with multiplicity $m$, we surround it with $m$ concentric circles.
Below, we explicitly present all the solutions.


\subsubsection*{Single-particle states for $\pht=\frac{\pi}{2}$
($\tau=i$ and $\ord=4$)}

For $n=1$ we get two fixed points:
$$
\ket{\fpIIz}=\ket{[0]}\,,\qquad
\ket{\fpIIc}=\ket{[\tfrac{1}{2}+\tfrac{1}{2}i]}\,.
$$
There are only two distinct solutions to \eqref{eqn:fpZsol},
up to a lattice element in $\Z+\Z\tau$, which can be taken as
$\fpZ_{0,0}=0$ and $\fpZ_{0,1}=\tfrac{1}{2}+\tfrac{1}{2}i.$
Two solutions $\fpZ_{\wMa,\wMb}$ and $\fpZ_{\wMa',\wMb'}$
are equivalent if $\wMa+\wMb\equiv\wMa'+\wMb'\pmod 2.$

For $n=2$ we get three fixed points:
$$
\ket{\fpIIzz}=\ket{[0,\,0]}\,,
\qquad
\ket{\fpIIxy}=\ket{[\tfrac{1}{2},\,
\tfrac{1}{2}i]}\,,
\qquad
\ket{\fpIIcc}=\ket{[\tfrac{1}{2}+\tfrac{1}{2}i,\,
\tfrac{1}{2}+\tfrac{1}{2}i]}\,.
$$
They are constructed from
$$
\fpZ_{0,0}=0\,,\quad
\fpZ_{1,0}=\tfrac{1}{2}\,,\quad
\fpZ_{0,1}=\tfrac{1}{2}i\,,\quad
\fpZ_{1,1}=\tfrac{1}{2}+\tfrac{1}{2}i\,,\qquad
\pmod{\Z+\Z i}.
$$
For $n=3$ we get two fixed points:
$$
\ket{\fpIIzzz}=\ket{[0,\,0,\,0]}\,,
\qquad
\ket{\fpIIccc}=\ket{[\tfrac{1}{2}+\tfrac{1}{2}i,\,
\tfrac{1}{2}+\tfrac{1}{2}i,\,
\tfrac{1}{2}+\tfrac{1}{2}i]}\,.
$$


\subsubsection*{Single-particle states for $\pht=\frac{\pi}{3}$
($\tau=e^{\pi i/3}$ and $\ord=6$)}

For $n=1$ we get a single fixed point:
$$
\ket{\fpIz}=\ket{[0]}\,.
$$
For $n=2$ we get two fixed points:
$$
\ket{\fpIzz}=\ket{[0,\,0]}\,,
\qquad
\ket{\fpIdd}=\ket{[\tfrac{1}{3}+\tfrac{1}{3}\tau,\,
\tfrac{2}{3}+\tfrac{2}{3}\tau]}\,.
$$
For $n=3$ we also get two fixed points:
$$
\ket{\fpIzzz}=\ket{[0,\,0,\,0]}\,,
\qquad
\ket{\fpIxyd}=\ket{[\tfrac{1}{2},\,\tfrac{1}{2}\tau,\,
\tfrac{1}{2}+\tfrac{1}{2}\tau]}\,.
$$
For $n=4$ we again get two fixed points:
$$
\ket{\fpIzzzz}=\ket{[0,\,0,\,0,\,0]}\,,
\qquad
\ket{\fpIdddd}=\ket{[
\tfrac{1}{3}+\tfrac{1}{3}\tau,\,
\tfrac{1}{3}+\tfrac{1}{3}\tau,\,
\tfrac{2}{3}+\tfrac{2}{3}\tau,\,
\tfrac{2}{3}+\tfrac{2}{3}\tau]}\,.
$$
For $n=5$ we get one fixed point:
$$
\ket{\fpIzzzzz}=\ket{[0,0,0,0,0]}\,.
$$


\subsubsection*{Single-particle states for $\pht=\frac{2\pi}{3}$
($\tau=e^{\pi i/3}$ and $\ord=3$)}

For $n=1$ we get three fixed points:
$$
\ket{\fpIIIz}=\ket{[0]}\,,\qquad
\ket{\fpIIIe}=\ket{[\tfrac{1}{3}+\tfrac{1}{3}\tau]}\,,\qquad
\ket{\fpIIIf}=\ket{[\tfrac{2}{3}+\tfrac{2}{3}\tau]}\,.
$$
For $n=2$ we also get three fixed points:
$$
\ket{\fpIIIzz}=\ket{[0,\,0]}\,,\qquad
\ket{\fpIIIee}=\ket{[\tfrac{1}{3}+\tfrac{1}{3}\tau,\,
\tfrac{1}{3}+\tfrac{1}{3}\tau]}\,,\qquad
\ket{\fpIIIff}=\ket{[\tfrac{2}{3}+\tfrac{2}{3}\tau,\,
\tfrac{2}{3}+\tfrac{2}{3}\tau]}\,.
$$


\subsubsection*{}
The single-particle states are summarized in \figref{fig:GroundStates1P}, and the complete basis
of ground states (i.e., including multi-particle states) is depicted in
\figref{fig:GroundStates}.


\FIGURE[t!]{
\begin{picture}(400,210)

\put(0,130){\begin{picture}(400,70)
\put(0,32){$\pht=\frac{\pi}{3}$}
\put(40,0){\line(0,1){70}}
\put(160,0){\line(0,1){70}}
\put(280,0){\line(0,1){70}}
\put(40,35){\line(1,0){360}}

\put(0, 0){\line(1,0){400}}
\put(0,70){\line(1,0){400}}

\put(95,45){\begin{picture}(120,40)
  \put(-50,5){$n=1$}
  \put(0,0){\begin{picture}(20,20)
    \color{blue}
    \multiput(0,0)(10,17){2}{\line(1,0){20}}
    \multiput(0,0)(20,0){2}{\qbezier(0,0)(0,0)(10,17)}
    \color{black}
    \put(0,0){\circle*{1}}
    \color{red}
    \put(0,0){\circle{4}}
    \color{black}
  \end{picture}}
\end{picture}}

\put(215,45){\begin{picture}(120,40)
  \put(-50,5){$n=2$}
  \put(0,0){\begin{picture}(20,20)
    \color{blue}
    \multiput(0,0)(10,17){2}{\line(1,0){20}}
    \multiput(0,0)(20,0){2}{\qbezier(0,0)(0,0)(10,17)}
    \color{black}
    \put(0,0){\circle*{1}}
    \color{red}
    \put(0,0){\circle{4}}
    \put(0,0){\circle{6}}
    \color{black}
  \end{picture}}

  \put(30,0){\begin{picture}(20,20)
    \color{blue}
    \multiput(0,0)(10,17){2}{\line(1,0){20}}
    \multiput(0,0)(20,0){2}{\qbezier(0,0)(0,0)(10,17)}
    \color{black}
    \put(10,6){\circle*{1}}
    \color{red}
    \put(10,6){\circle{4}}
    \color{black}
    \put(20,11){\circle*{1}}
    \color{red}
    \put(20,11){\circle{4}}
    \color{black}
  \end{picture}}
\end{picture}}

\put(335,45){\begin{picture}(120,40)
  \put(-50,5){$n=3$}
  \put(0,0){\begin{picture}(20,20)
    \color{blue}
    \multiput(0,0)(10,17){2}{\line(1,0){20}}
    \multiput(0,0)(20,0){2}{\qbezier(0,0)(0,0)(10,17)}
    \color{black}
    \put(0,0){\circle*{1}}
    \color{red}
    \put(0,0){\circle{4}}
    \put(0,0){\circle{6}}
    \put(0,0){\circle{8}}
    \color{black}
  \end{picture}}

  \put(30,0){\begin{picture}(20,20)
    \color{blue}
    \multiput(0,0)(10,17){2}{\line(1,0){20}}
    \multiput(0,0)(20,0){2}{\qbezier(0,0)(0,0)(10,17)}
    \color{black}
    \put(10,0){\circle*{1}}
    \color{red}
    \put(10,0){\circle{4}}
    \color{black}
    \put(5,8){\circle*{1}}
    \color{red}
    \put(5,8){\circle{4}}
    \color{black}
    \put(15,8){\circle*{1}}
    \color{red}
    \put(15,8){\circle{4}}
    \color{black}
  \end{picture}}
\end{picture}}

\put(95,10){\begin{picture}(120,40)
  \put(-50,5){$n=4$}
  \put(0,0){\begin{picture}(20,20)
    \color{blue}
    \multiput(0,0)(10,17){2}{\line(1,0){20}}
    \multiput(0,0)(20,0){2}{\qbezier(0,0)(0,0)(10,17)}
    \color{black}
    \put(0,0){\circle*{1}}
    \color{red}
    \put(0,0){\circle{4}}
    \put(0,0){\circle{6}}
    \put(0,0){\circle{8}}
    \put(0,0){\circle{10}}
    \color{black}
  \end{picture}}

  \put(30,0){\begin{picture}(20,20)
    \color{blue}
    \multiput(0,0)(10,17){2}{\line(1,0){20}}
    \multiput(0,0)(20,0){2}{\qbezier(0,0)(0,0)(10,17)}
    \color{black}
    \put(10,6){\circle*{1}}
    \color{red}
    \put(10,6){\circle{4}}
    \put(10,6){\circle{6}}
    \color{black}
    \put(20,11){\circle*{1}}
    \color{red}
    \put(20,11){\circle{4}}
    \put(20,11){\circle{6}}
    \color{black}
  \end{picture}}
\end{picture}}

\put(215,10){\begin{picture}(120,40)
  \put(-50,5){$n=5$}
  \put(0,0){\begin{picture}(20,20)
    \color{blue}
    \multiput(0,0)(10,17){2}{\line(1,0){20}}
    \multiput(0,0)(20,0){2}{\qbezier(0,0)(0,0)(10,17)}
    \color{black}
    \put(0,0){\circle*{1}}
    \color{red}
    \put(0,0){\circle{4}}
    \put(0,0){\circle{6}}
    \put(0,0){\circle{8}}
    \put(0,0){\circle{10}}
    \put(0,0){\circle{12}}
    \color{black}
  \end{picture}}
\end{picture}}

\end{picture}}

\put(0,45){\begin{picture}(400,80)

\put(0,37){$\pht=\frac{\pi}{2}$}
\put(0,80){\line(1,0){400}}
\put(40,0){\line(0,1){80}}
\put(40,40){\line(1,0){360}}
\put(220,0){\line(0,1){80}}
\put(0, 0){\line(1,0){400}}

\put(100,50){\begin{picture}(120,40)
  \put(-55,5){$n=1$}
  \put(0,0){\begin{picture}(20,20)
    \color{blue}
    \multiput(0,0)(0,20){2}{\line(1,0){20}}
    \multiput(0,0)(20,0){2}{\line(0,1){20}}
    \color{black}
    \put(0,0){\circle*{1}}
    \color{red}
    \put(0,0){\circle{4}}
    \color{black}
  \end{picture}}

  \put(30,0){\begin{picture}(20,20)
    \color{blue}
    \multiput(0,0)(0,20){2}{\line(1,0){20}}
    \multiput(0,0)(20,0){2}{\line(0,1){20}}
    \color{black}
    \put(10,10){\circle*{1}}
    \color{red}
    \put(10,10){\circle{4}}
    \color{black}
  \end{picture}}
\end{picture}}

\put(280,50){\begin{picture}(120,40)
  \put(-55,5){$n=2$}
  \put(0,0){\begin{picture}(20,20)
    \color{blue}
    \multiput(0,0)(0,20){2}{\line(1,0){20}}
    \multiput(0,0)(20,0){2}{\line(0,1){20}}
    \color{black}
    \put(0,0){\circle*{1}}
    \color{red}
    \put(0,0){\circle{4}}
    \put(0,0){\circle{6}}
    \color{black}
  \end{picture}}

  \put(30,0){\begin{picture}(20,20)
    \color{blue}
    \multiput(0,0)(0,20){2}{\line(1,0){20}}
    \multiput(0,0)(20,0){2}{\line(0,1){20}}
    \color{black}
    \put(10,10){\circle*{1}}
    \color{red}
    \put(10,10){\circle{4}}
    \put(10,10){\circle{6}}
    \color{black}
  \end{picture}}

  \put(60,0){\begin{picture}(20,20)
    \color{blue}
    \multiput(0,0)(0,20){2}{\line(1,0){20}}
    \multiput(0,0)(20,0){2}{\line(0,1){20}}
    \color{black}
    \put(0,10){\circle*{1}}
    \color{red}
    \put(0,10){\circle{4}}
    \color{black}
    \put(10,0){\circle*{1}}
    \color{red}
    \put(10,0){\circle{4}}
    \color{black}
  \end{picture}}
\end{picture}}

\put(100, 10){\begin{picture}(120,40)
  \put(-55,5){$n=3$}
  \put(0,0){\begin{picture}(20,20)
    \color{blue}
    \multiput(0,0)(0,20){2}{\line(1,0){20}}
    \multiput(0,0)(20,0){2}{\line(0,1){20}}
    \color{black}
    \put(0,0){\circle*{1}}
    \color{red}
    \put(0,0){\circle{4}}
    \put(0,0){\circle{6}}
    \put(0,0){\circle{8}}
    \color{black}
  \end{picture}}

  \put(30,0){\begin{picture}(20,20)
    \color{blue}
    \multiput(0,0)(0,20){2}{\line(1,0){20}}
    \multiput(0,0)(20,0){2}{\line(0,1){20}}
    \color{black}
    \put(10,10){\circle*{1}}
    \color{red}
    \put(10,10){\circle{4}}
    \put(10,10){\circle{6}}
    \put(10,10){\circle{8}}
    \color{black}
  \end{picture}}

\end{picture}}

\end{picture}}
\put(0,5){\begin{picture}(400,35)
\put(0,14){$\pht=\frac{2\pi}{3}$}
\put(40,0){\line(0,1){35}}
\put(0, 0){\line(1,0){400}}
\put(0,35){\line(1,0){400}}
\put(45,15){$n=1$}
\put(220,0){\line(0,1){35}}
\put(225,15){$n=2$}

\put(100,8){\begin{picture}(120,40)
  \put(0,0){\begin{picture}(20,20)
    \color{blue}
    \multiput(0,0)(10,17){2}{\line(1,0){20}}
    \multiput(0,0)(20,0){2}{\qbezier(0,0)(0,0)(10,17)}
    \color{black}
    \put(0,0){\circle*{1}}
    \color{red}
    \put(0,0){\circle{4}}
    \color{black}
  \end{picture}}

  \put(30,0){\begin{picture}(20,20)
    \color{blue}
    \multiput(0,0)(10,17){2}{\line(1,0){20}}
    \multiput(0,0)(20,0){2}{\qbezier(0,0)(0,0)(10,17)}
    \color{black}
    \put(10,6){\circle*{1}}
    \color{red}
    \put(10,6){\circle{4}}
    \color{black}
  \end{picture}}

  \put(60,0){\begin{picture}(20,20)
    \color{blue}
    \multiput(0,0)(10,17){2}{\line(1,0){20}}
    \multiput(0,0)(20,0){2}{\qbezier(0,0)(0,0)(10,17)}
    \color{black}
    \put(20,11){\circle*{1}}
    \color{red}
    \put(20,11){\circle{4}}
    \color{black}
  \end{picture}}

\end{picture}}

\put(280,8){\begin{picture}(120,40)
  \put(0,0){\begin{picture}(20,20)
    \color{blue}
    \multiput(0,0)(10,17){2}{\line(1,0){20}}
    \multiput(0,0)(20,0){2}{\qbezier(0,0)(0,0)(10,17)}
    \color{black}
    \put(0,0){\circle*{1}}
    \color{red}
    \put(0,0){\circle{4}}
    \put(0,0){\circle{6}}
    \color{black}
  \end{picture}}

  \put(30,0){\begin{picture}(20,20)
    \color{blue}
    \multiput(0,0)(10,17){2}{\line(1,0){20}}
    \multiput(0,0)(20,0){2}{\qbezier(0,0)(0,0)(10,17)}
    \color{black}
    \put(10,6){\circle*{1}}
    \color{red}
    \put(10,6){\circle{4}}
    \put(10,6){\circle{6}}
    \color{black}
  \end{picture}}

  \put(60,0){\begin{picture}(20,20)
    \color{blue}
    \multiput(0,0)(10,17){2}{\line(1,0){20}}
    \multiput(0,0)(20,0){2}{\qbezier(0,0)(0,0)(10,17)}
    \color{black}
    \put(20,11){\circle*{1}}
    \color{red}
    \put(20,11){\circle{4}}
    \put(20,11){\circle{6}}
    \color{black}
  \end{picture}}

\end{picture}}

\end{picture}}
\end{picture}
\caption{
Single-particle ground states.
Each ground state is depicted by the intersection of the strings
with the $T^2$ fiber at a fixed $x_3.$
The string can intersect the fiber more than once at the same point,
and the number of small circles surrounding the intersection point
represents the number of times that the
string intersects the fiber at that point.
}
\label{fig:GroundStates1P}
}


\FIGURE[t!]{
\begin{picture}(400,400)

\put(0,205){\begin{picture}(400,150)
\put(0,88){$\pht=\frac{\pi}{3}$}
\put(40,0){\line(0,1){175}}
\put(0, 0){\line(1,0){400}}
\put(0,175){\line(1,0){400}}
\put(40,140){\line(1,0){360}}
\put(40,105){\line(1,0){360}}
\put(40,70){\line(1,0){360}}
\put(40,35){\line(1,0){360}}

\put(100,145){\begin{picture}(120,40)
  \put(-55,10){$n=1$}
  \put(0,0){\begin{picture}(20,20)
    \color{blue}
    \multiput(0,0)(10,17){2}{\line(1,0){20}}
    \multiput(0,0)(20,0){2}{\qbezier(0,0)(0,0)(10,17)}
    \color{black}
    \put(0,0){\circle*{1}}
    \color{red}
    \put(0,0){\circle{4}}
    \color{black}
    \put(10,18){$1$}
  \end{picture}}
\put(200,10){$\NStates=1$}
\end{picture}}

\put(100,110){\begin{picture}(120,40)
  \put(-55,10){$n=2$}
  \put(0,0){\begin{picture}(20,20)
    \color{blue}
    \multiput(0,0)(10,17){2}{\line(1,0){20}}
    \multiput(0,0)(20,0){2}{\qbezier(0,0)(0,0)(10,17)}
    \color{black}
    \put(0,0){\circle*{1}}
    \color{red}
    \put(0,0){\circle{4}}
    \put(0,0){\circle{6}}
    \color{black}
    \put(10,18){$2$}
  \end{picture}}

  \put(30,0){\begin{picture}(20,20)
    \color{blue}
    \multiput(0,0)(10,17){2}{\line(1,0){20}}
    \multiput(0,0)(20,0){2}{\qbezier(0,0)(0,0)(10,17)}
    \color{black}
    \put(10,6){\circle*{1}}
    \color{red}
    \put(10,6){\circle{4}}
    \color{black}
    \put(20,11){\circle*{1}}
    \color{red}
    \put(20,11){\circle{4}}
    \color{black}
    \put(10,18){$1$}
  \end{picture}}
\put(200,10){$\NStates=3$}
\end{picture}}

\put(100,75){\begin{picture}(120,40)
  \put(-55,10){$n=3$}
  \put(0,0){\begin{picture}(20,20)
    \color{blue}
    \multiput(0,0)(10,17){2}{\line(1,0){20}}
    \multiput(0,0)(20,0){2}{\qbezier(0,0)(0,0)(10,17)}
    \color{black}
    \put(0,0){\circle*{1}}
    \color{red}
    \put(0,0){\circle{4}}
    \put(0,0){\circle{6}}
    \put(0,0){\circle{8}}
    \color{black}
    \put(10,18){$3$}
  \end{picture}}

  \put(30,0){\begin{picture}(20,20)
    \color{blue}
    \multiput(0,0)(10,17){2}{\line(1,0){20}}
    \multiput(0,0)(20,0){2}{\qbezier(0,0)(0,0)(10,17)}
    \color{black}
    \put(0,0){\circle*{1}}
    \color{red}
    \put(0,0){\circle{4}}
    \color{black}
    \put(10,6){\circle*{1}}
    \color{red}
    \put(10,6){\circle{4}}
    \color{black}
    \put(20,11){\circle*{1}}
    \color{red}
    \put(20,11){\circle{4}}
    \color{black}
    \put(10,18){$1$}

  \end{picture}}

  \put(60,0){\begin{picture}(20,20)
    \color{blue}
    \multiput(0,0)(10,17){2}{\line(1,0){20}}
    \multiput(0,0)(20,0){2}{\qbezier(0,0)(0,0)(10,17)}
    \color{black}
    \put(10,0){\circle*{1}}
    \color{red}
    \put(10,0){\circle{4}}
    \color{black}
    \put(5,8){\circle*{1}}
    \color{red}
    \put(5,8){\circle{4}}
    \color{black}
    \put(15,8){\circle*{1}}
    \color{red}
    \put(15,8){\circle{4}}
    \color{black}
    \put(10,18){$1$}
  \end{picture}}
\put(200,10){$\NStates=5$}
\end{picture}}

\put(100,40){\begin{picture}(120,40)
  \put(-55,10){$n=4$}
  \put(0,0){\begin{picture}(20,20)
    \color{blue}
    \multiput(0,0)(10,17){2}{\line(1,0){20}}
    \multiput(0,0)(20,0){2}{\qbezier(0,0)(0,0)(10,17)}
    \color{black}
    \put(0,0){\circle*{1}}
    \color{red}
    \put(0,0){\circle{4}}
    \put(0,0){\circle{6}}
    \put(0,0){\circle{8}}
    \put(0,0){\circle{10}}
    \color{black}
    \put(10,18){$5$}
  \end{picture}}

  \put(30,0){\begin{picture}(20,20)
    \color{blue}
    \multiput(0,0)(10,17){2}{\line(1,0){20}}
    \multiput(0,0)(20,0){2}{\qbezier(0,0)(0,0)(10,17)}
    \color{black}
    \put(0,0){\circle*{1}}
    \color{red}
    \put(0,0){\circle{4}}
    \put(0,0){\circle{6}}
    \color{black}
    \put(10,6){\circle*{1}}
    \color{red}
    \put(10,6){\circle{4}}
    \color{black}
    \put(20,11){\circle*{1}}
    \color{red}
    \put(20,11){\circle{4}}
    \color{black}
    \put(10,18){$2$}
  \end{picture}}

  \put(60,0){\begin{picture}(20,20)
    \color{blue}
    \multiput(0,0)(10,17){2}{\line(1,0){20}}
    \multiput(0,0)(20,0){2}{\qbezier(0,0)(0,0)(10,17)}
    \color{black}
    \put(0,0){\circle*{1}}
    \color{red}
    \put(0,0){\circle{4}}
    \color{black}
    \put(10,0){\circle*{1}}
    \color{red}
    \put(10,0){\circle{4}}
    \color{black}
    \put(5,8){\circle*{1}}
    \color{red}
    \put(5,8){\circle{4}}
    \color{black}
    \put(15,8){\circle*{1}}
    \color{red}
    \put(15,8){\circle{4}}
    \color{black}
    \put(10,18){$1$}
  \end{picture}}

  \put(90,0){\begin{picture}(20,20)
    \color{blue}
    \multiput(0,0)(10,17){2}{\line(1,0){20}}
    \multiput(0,0)(20,0){2}{\qbezier(0,0)(0,0)(10,17)}
    \color{black}
    \put(10,6){\circle*{1}}
    \color{red}
    \put(10,6){\circle{4}}
    \put(10,6){\circle{6}}
    \color{black}
    \put(20,11){\circle*{1}}
    \color{red}
    \put(20,11){\circle{4}}
    \put(20,11){\circle{6}}
    \color{black}
    \put(10,18){$2$}
  \end{picture}}
\put(200,10){$\NStates=10$}
\end{picture}}

\put(100,5){\begin{picture}(120,40)
  \put(-55,10){$n=5$}
  \put(0,0){\begin{picture}(20,20)
    \color{blue}
    \multiput(0,0)(10,17){2}{\line(1,0){20}}
    \multiput(0,0)(20,0){2}{\qbezier(0,0)(0,0)(10,17)}
    \color{black}
    \put(0,0){\circle*{1}}
    \color{red}
    \put(0,0){\circle{4}}
    \put(0,0){\circle{6}}
    \put(0,0){\circle{8}}
    \put(0,0){\circle{10}}
    \put(0,0){\circle{12}}
    \color{black}
    \put(10,18){$7$}
  \end{picture}}

  \put(30,0){\begin{picture}(20,20)
    \color{blue}
    \multiput(0,0)(10,17){2}{\line(1,0){20}}
    \multiput(0,0)(20,0){2}{\qbezier(0,0)(0,0)(10,17)}
    \color{black}
    \put(0,0){\circle*{1}}
    \color{red}
    \put(0,0){\circle{4}}
    \put(0,0){\circle{6}}
    \put(0,0){\circle{8}}
    \color{black}
    \put(10,6){\circle*{1}}
    \color{red}
    \put(10,6){\circle{4}}
    \color{black}
    \put(20,11){\circle*{1}}
    \color{red}
    \put(20,11){\circle{4}}
    \color{black}
    \put(10,18){$3$}
  \end{picture}}

  \put(60,0){\begin{picture}(20,20)
    \color{blue}
    \multiput(0,0)(10,17){2}{\line(1,0){20}}
    \multiput(0,0)(20,0){2}{\qbezier(0,0)(0,0)(10,17)}
    \color{black}
    \put(0,0){\circle*{1}}
    \color{red}
    \put(0,0){\circle{4}}
    \put(0,0){\circle{6}}
    \color{black}
    \put(10,0){\circle*{1}}
    \color{red}
    \put(10,0){\circle{4}}
    \color{black}
    \put(5,8){\circle*{1}}
    \color{red}
    \put(5,8){\circle{4}}
    \color{black}
    \put(15,8){\circle*{1}}
    \color{red}
    \put(15,8){\circle{4}}
    \color{black}
    \put(10,18){$2$}
  \end{picture}}

  \put(90,0){\begin{picture}(20,20)
    \color{blue}
    \multiput(0,0)(10,17){2}{\line(1,0){20}}
    \multiput(0,0)(20,0){2}{\qbezier(0,0)(0,0)(10,17)}
    \color{black}
    \put(10,6){\circle*{1}}
    \color{red}
    \put(0,0){\circle{4}}
    \color{black}
    \put(10,6){\circle*{1}}
    \color{red}
    \put(10,6){\circle{4}}
    \put(10,6){\circle{6}}
    \color{black}
    \put(20,11){\circle*{1}}
    \color{red}
    \put(20,11){\circle{4}}
    \put(20,11){\circle{6}}
    \color{black}
    \put(10,18){$2$}
  \end{picture}}

  \put(120,0){\begin{picture}(20,20)
    \color{blue}
    \multiput(0,0)(10,17){2}{\line(1,0){20}}
    \multiput(0,0)(20,0){2}{\qbezier(0,0)(0,0)(10,17)}
    \color{black}
    \put(10,6){\circle*{1}}
    \color{red}
    \put(10,6){\circle{4}}
    \color{black}
    \put(20,11){\circle*{1}}
    \color{red}
    \put(20,11){\circle{4}}
    \color{black}
    \put(10,0){\circle*{1}}
    \color{red}
    \put(10,0){\circle{4}}
    \color{black}
    \put(5,8){\circle*{1}}
    \color{red}
    \put(5,8){\circle{4}}
    \color{black}
    \put(15,8){\circle*{1}}
    \color{red}
    \put(15,8){\circle{4}}
    \color{black}
    \put(10,18){$1$}
  \end{picture}}

\put(200,10){$\NStates=15$}
\end{picture}}

\end{picture}}

\put(0,80){\begin{picture}(400,90)

\put(0,60){$\pht=\frac{\pi}{2}$}
\put(0,120){\line(1,0){400}}
\put(40,0){\line(0,1){120}}
\put(45,90){$n=1$}
\put(40,40){\line(1,0){360}}
\put(45,50){$n=2$}
\put(40,80){\line(1,0){360}}
\put(45,10){$n=3$}
\put(0, 0){\line(1,0){400}}

\put(100,85){\begin{picture}(120,40)
  \put(0,0){\begin{picture}(20,20)
    \color{blue}
    \multiput(0,0)(0,20){2}{\line(1,0){20}}
    \multiput(0,0)(20,0){2}{\line(0,1){20}}
    \color{black}
    \put(0,0){\circle*{1}}
    \color{red}
    \put(0,0){\circle{4}}
    \color{black}
    \put(0,21){$1$}
  \end{picture}}

  \put(30,0){\begin{picture}(20,20)
    \color{blue}
    \multiput(0,0)(0,20){2}{\line(1,0){20}}
    \multiput(0,0)(20,0){2}{\line(0,1){20}}
    \color{black}
    \put(10,10){\circle*{1}}
    \color{red}
    \put(10,10){\circle{4}}
    \color{black}
    \put(0,21){$1$}
  \end{picture}}
\put(200,10){$\NStates=2$}
\end{picture}}

\put(100,45){\begin{picture}(120,40)
  \put(0,0){\begin{picture}(20,20)
    \color{blue}
    \multiput(0,0)(0,20){2}{\line(1,0){20}}
    \multiput(0,0)(20,0){2}{\line(0,1){20}}
    \color{black}
    \put(0,0){\circle*{1}}
    \color{red}
    \put(0,0){\circle{4}}
    \put(0,0){\circle{6}}
    \color{black}
    \put(0,21){$2$}
  \end{picture}}

  \put(30,0){\begin{picture}(20,20)
    \color{blue}
    \multiput(0,0)(0,20){2}{\line(1,0){20}}
    \multiput(0,0)(20,0){2}{\line(0,1){20}}
    \color{black}
    \put(10,10){\circle*{1}}
    \color{red}
    \put(10,10){\circle{4}}
    \put(10,10){\circle{6}}
    \color{black}
    \put(0,21){$2$}
  \end{picture}}

  \put(60,0){\begin{picture}(20,20)
    \color{blue}
    \multiput(0,0)(0,20){2}{\line(1,0){20}}
    \multiput(0,0)(20,0){2}{\line(0,1){20}}
    \color{black}
    \put(0,0){\circle*{1}}
    \color{red}
    \put(0,0){\circle{4}}
    \color{black}
    \put(10,10){\circle*{1}}
    \color{red}
    \put(10,10){\circle{4}}
    \color{black}
    \put(0,21){$1$}
  \end{picture}}

  \put(90,0){\begin{picture}(20,20)
    \color{blue}
    \multiput(0,0)(0,20){2}{\line(1,0){20}}
    \multiput(0,0)(20,0){2}{\line(0,1){20}}
    \color{black}
    \put(0,10){\circle*{1}}
    \color{red}
    \put(0,10){\circle{4}}
    \color{black}
    \put(10,0){\circle*{1}}
    \color{red}
    \put(10,0){\circle{4}}
    \color{black}
    \put(0,21){$1$}
  \end{picture}}
\put(200,10){$\NStates=6$}
\end{picture}}

\put(100, 5){\begin{picture}(120,40)
  \put(0,0){\begin{picture}(20,20)
    \color{blue}
    \multiput(0,0)(0,20){2}{\line(1,0){20}}
    \multiput(0,0)(20,0){2}{\line(0,1){20}}
    \color{black}
    \put(0,0){\circle*{1}}
    \color{red}
    \put(0,0){\circle{4}}
    \put(0,0){\circle{6}}
    \put(0,0){\circle{8}}
    \color{black}
    \put(0,21){$3$}
  \end{picture}}

  \put(30,0){\begin{picture}(20,20)
    \color{blue}
    \multiput(0,0)(0,20){2}{\line(1,0){20}}
    \multiput(0,0)(20,0){2}{\line(0,1){20}}
    \color{black}
    \put(0,0){\circle*{1}}
    \color{red}
    \put(0,0){\circle{4}}
    \put(0,0){\circle{6}}
    \color{black}
    \put(10,10){\circle*{1}}
    \color{red}
    \put(10,10){\circle{4}}
    \color{black}
    \put(0,21){$2$}
  \end{picture}}

  \put(60,0){\begin{picture}(20,20)
    \color{blue}
    \multiput(0,0)(0,20){2}{\line(1,0){20}}
    \multiput(0,0)(20,0){2}{\line(0,1){20}}
    \color{black}
    \put(0,0){\circle*{1}}
    \color{red}
    \put(0,0){\circle{4}}
    \color{black}
    \put(10,10){\circle*{1}}
    \color{red}
    \put(10,10){\circle{4}}
    \put(10,10){\circle{6}}
    \color{black}
    \put(0,21){$2$}
  \end{picture}}

  \put(90,0){\begin{picture}(20,20)
    \color{blue}
    \multiput(0,0)(0,20){2}{\line(1,0){20}}
    \multiput(0,0)(20,0){2}{\line(0,1){20}}
    \color{black}
    \put(10,10){\circle*{1}}
    \color{red}
    \put(10,10){\circle{4}}
    \put(10,10){\circle{6}}
    \put(10,10){\circle{8}}
    \color{black}
    \put(0,21){$3$}
  \end{picture}}

  \put(120,0){\begin{picture}(20,20)
    \color{blue}
    \multiput(0,0)(0,20){2}{\line(1,0){20}}
    \multiput(0,0)(20,0){2}{\line(0,1){20}}
    \color{black}
    \put(10,0){\circle*{1}}
    \color{red}
    \put(10,0){\circle{4}}
    \color{black}
    \put(0,10){\circle*{1}}
    \color{red}
    \put(0,10){\circle{4}}
    \color{black}
    \put(0,0){\circle*{1}}
    \color{red}
    \put(0,0){\circle{4}}
    \color{black}
    \put(0,21){$1$}
  \end{picture}}

  \put(150,0){\begin{picture}(20,20)
    \color{blue}
    \multiput(0,0)(0,20){2}{\line(1,0){20}}
    \multiput(0,0)(20,0){2}{\line(0,1){20}}
    \color{black}
    \put(10,0){\circle*{1}}
    \color{red}
    \put(10,0){\circle{4}}
    \color{black}
    \put(0,10){\circle*{1}}
    \color{red}
    \put(0,10){\circle{4}}
    \color{black}
    \put(10,10){\circle*{1}}
    \color{red}
    \put(10,10){\circle{4}}
    \color{black}
    \put(0,21){$1$}
  \end{picture}}
\put(200,10){$\NStates=12$}
\end{picture}}

\end{picture}}
\put(0,5){\begin{picture}(400,150)
\put(0,38){$\pht=\frac{2\pi}{3}$}
\put(40,0){\line(0,1){70}}
\put(0, 0){\line(1,0){400}}
\put(0,70){\line(1,0){400}}
\put(40,70){\line(1,0){360}}
\put(45,50){$n=1$}
\put(40,35){\line(1,0){360}}
\put(45,15){$n=2$}

\put(100,40){\begin{picture}(120,40)
  \put(0,0){\begin{picture}(20,20)
    \color{blue}
    \multiput(0,0)(10,17){2}{\line(1,0){20}}
    \multiput(0,0)(20,0){2}{\qbezier(0,0)(0,0)(10,17)}
    \color{black}
    \put(0,0){\circle*{1}}
    \color{red}
    \put(0,0){\circle{4}}
    \color{black}
    \put(10,18){$1$}
  \end{picture}}

  \put(30,0){\begin{picture}(20,20)
    \color{blue}
    \multiput(0,0)(10,17){2}{\line(1,0){20}}
    \multiput(0,0)(20,0){2}{\qbezier(0,0)(0,0)(10,17)}
    \color{black}
    \put(10,6){\circle*{1}}
    \color{red}
    \put(10,6){\circle{4}}
    \color{black}
    \put(10,18){$1$}
  \end{picture}}

  \put(60,0){\begin{picture}(20,20)
    \color{blue}
    \multiput(0,0)(10,17){2}{\line(1,0){20}}
    \multiput(0,0)(20,0){2}{\qbezier(0,0)(0,0)(10,17)}
    \color{black}
    \put(20,11){\circle*{1}}
    \color{red}
    \put(20,11){\circle{4}}
    \color{black}
    \put(10,18){$1$}
  \end{picture}}

\put(200,10){$\NStates=3$}
\end{picture}}

\put(100,5){\begin{picture}(120,40)
  \put(0,0){\begin{picture}(20,20)
    \color{blue}
    \multiput(0,0)(10,17){2}{\line(1,0){20}}
    \multiput(0,0)(20,0){2}{\qbezier(0,0)(0,0)(10,17)}
    \color{black}
    \put(0,0){\circle*{1}}
    \color{red}
    \put(0,0){\circle{4}}
    \put(0,0){\circle{6}}
    \color{black}
    \put(10,18){$2$}
  \end{picture}}

  \put(30,0){\begin{picture}(20,20)
    \color{blue}
    \multiput(0,0)(10,17){2}{\line(1,0){20}}
    \multiput(0,0)(20,0){2}{\qbezier(0,0)(0,0)(10,17)}
    \color{black}
    \put(10,6){\circle*{1}}
    \color{red}
    \put(10,6){\circle{4}}
    \put(10,6){\circle{6}}
    \color{black}
    \put(10,18){$2$}
  \end{picture}}

  \put(60,0){\begin{picture}(20,20)
    \color{blue}
    \multiput(0,0)(10,17){2}{\line(1,0){20}}
    \multiput(0,0)(20,0){2}{\qbezier(0,0)(0,0)(10,17)}
    \color{black}
    \put(20,11){\circle*{1}}
    \color{red}
    \put(20,11){\circle{4}}
    \put(20,11){\circle{6}}
    \color{black}
    \put(10,18){$2$}
  \end{picture}}

  \put(90,0){\begin{picture}(20,20)
    \color{blue}
    \multiput(0,0)(10,17){2}{\line(1,0){20}}
    \multiput(0,0)(20,0){2}{\qbezier(0,0)(0,0)(10,17)}
    \color{black}
    \put(0,0){\circle*{1}}
    \color{red}
    \put(0,0){\circle{4}}
    \color{black}
    \put(10,6){\circle*{1}}
    \color{red}
    \put(10,6){\circle{4}}
    \color{black}
    \put(10,18){$1$}
  \end{picture}}

  \put(120,0){\begin{picture}(20,20)
    \color{blue}
    \multiput(0,0)(10,17){2}{\line(1,0){20}}
    \multiput(0,0)(20,0){2}{\qbezier(0,0)(0,0)(10,17)}
    \color{black}
    \put(0,0){\circle*{1}}
    \color{red}
    \put(0,0){\circle{4}}
    \color{black}
    \put(20,11){\circle*{1}}
    \color{red}
    \put(20,11){\circle{4}}
    \color{black}
    \put(10,18){$1$}
  \end{picture}}

  \put(150,0){\begin{picture}(20,20)
    \color{blue}
    \multiput(0,0)(10,17){2}{\line(1,0){20}}
    \multiput(0,0)(20,0){2}{\qbezier(0,0)(0,0)(10,17)}
    \color{black}
    \put(10,6){\circle*{1}}
    \color{red}
    \put(10,6){\circle{4}}
    \color{black}
    \put(20,11){\circle*{1}}
    \color{red}
    \put(20,11){\circle{4}}
    \color{black}
    \put(10,18){$1$}
  \end{picture}}
\put(200,10){$\NStates=9$}
\end{picture}}

\end{picture}}
\end{picture}
\caption{
The complete basis of ground states.
A ground state in this basis
comprises of one or more single-particle states from
\figref{fig:GroundStates1P}.
As in the previous figure,
each ground state is depicted by the intersection of the strings
with the fiber at $x_3=0.$ Several different states could have
the same depiction (if they decompose as $n=n_1+\cdots + n_p$ in different ways),
and the numbers on top of each cell indicate the multiplicity.
$N_s$ is the total number of states.
}
\label{fig:GroundStates}
}




\subsubsection*{Notation for multi-particle states}

The multi-particle states are states in the Fock space of
identical bosons.
We denote multi-particle states by combining inside a single ket
the pictorial representations of the individual single-particle
states which make up the multi-particle state.
For example, for $\pht=\frac{\pi}{2}$
($\tau=i$ and $\ord=4$) and $n=2$
we get the following $2$-particle states:
$$
\ket{\fpIIz\fpIIz}
\,,\quad
\ket{\fpIIz\fpIIc}
\,,\quad
\ket{\fpIIc\fpIIc}
\,.
$$
In the middle state, the two particles occupy different
single-particle states, while in the leftmost and rightmost states
the two particles occupy the same single-particle state.
Note that, by definition, the corresponding wavefunctions
are symmetric, so for example:
$\ket{\fpIIz\fpIIc}\equiv\ket{\fpIIc\fpIIz}.$


Next, we will discuss symmetries of the string background.
We will identify two $\Z_\lvk$ symmetries, which act on the full spectrum,
but in what follows we will only need their action on ground states.


\subsection{$\Z_\lvk$ momentum}
\label{subsec:P}

The space $\MXS$ defined in \eqref{eqn:zx} possesses an isometry
\be\label{eqn:MXSisom}
\Psym:(z,x_3)\mapsto (z+\tfrac{1}{\lvk}+\tfrac{1}{\lvk}\tau,x_3)\,,
\ee
where $\lvk$ is the number of ground states of the
$\np=1$ problem, listed in \eqref{eqn:deflvk}.
It is not hard to check that the isometry is compatible with
the structure group of the fibration since, for all three cases $\lvk=1,2,3$,
the $T^2$ point with coordinate
$z=\frac{1}{\lvk}(1+\tau)$ is a solution to the $\np=1$ version of \eqref{eqn:fpZsol},
and so
$\frac{1}{\lvk}(1+\tau)e^{i\pht}$ and $\frac{1}{\lvk}(1+\tau)$
differ by an element of the lattice $\Z+\Z\tau.$
Thus, $\Psym$ defines an operator on the Hilbert space of states,
and since $\Psym^\lvk=1$ it follows that the eigenvalues
of $\Psym$ take the form $e^{2\pi i j/\lvk}$ with $j\in\Z_k.$
We interpret this $j$ as a discrete $\Z_\lvk$ momentum.

The operator $\Psym$ takes single-particle states to single-particle states
with the same winding number $\np$, and its action on any ground state
can be computed from its action on the single-particle states.
For future reference, we list the action explicitly below.


\subsubsection*{Action of $\Psym$ on
single-particle states for $\pht=\frac{\pi}{2}$
($\tau=i$ and $\ord=4$)}

In this case $\lvk=2.$
For $\np=1$, $\Psym$ acts as
$$
\Psym\ket{[0]}=\ket{[\tfrac{1}{2}+\tfrac{1}{2}i]}\,,
\qquad
\Psym\ket{[\tfrac{1}{2}+\tfrac{1}{2}i]}=\ket{[0]}\,,
$$
or in pictorial notation,
$$
\Psym\ket{\fpIIz}=\ket{\fpIIc}\,,\qquad
\Psym\ket{\fpIIc}=\ket{\fpIIz}\,.
$$

For $\np=2$, $\Psym$ acts as:
$$
\Psym\ket{[0,0]}=\ket{[\tfrac{1}{2}+\tfrac{1}{2}i,\tfrac{1}{2}+\tfrac{1}{2}i]}\,,
\quad
\Psym\ket{[\tfrac{1}{2},\tfrac{1}{2}i]}=
\ket{[\tfrac{1}{2},\tfrac{1}{2}i]}\,,
\quad
\Psym\ket{[\tfrac{1}{2}+\tfrac{1}{2}i,\tfrac{1}{2}+\tfrac{1}{2}i]}
=\ket{[0,0]}\,.
$$
$$
\Psym\ket{\fpIIzz}=\ket{\fpIIcc}\,,\qquad
\Psym\ket{\fpIIcc}=\ket{\fpIIzz}\,,\qquad
\Psym\ket{\fpIIxy}=\ket{\fpIIxy}\,.
$$

For $\np=3$, $\Psym$ acts as:
$$
\Psym\ket{[0,0,0]}=\ket{[\tfrac{1}{2}+\tfrac{1}{2}i,
\tfrac{1}{2}+\tfrac{1}{2}i,\tfrac{1}{2}+\tfrac{1}{2}i]}
\,,\quad
\Psym{[\tfrac{1}{2}+\tfrac{1}{2}i,
\tfrac{1}{2}+\tfrac{1}{2}i,\tfrac{1}{2}+\tfrac{1}{2}i]}=
\ket{[0,0,0]}\,.
$$
$$
\Psym\ket{\fpIIzzz}=\ket{\fpIIccc}\,,
\qquad
\Psym\ket{\fpIIccc}=\ket{\fpIIzzz}\,.
$$


\subsubsection*{Action of $\Psym$ on single-particle states
for $\pht=\frac{\pi}{3}$ ($\tau=e^{\pi i/3}$ and $\ord=6$)}

In this case $\lvk=1$ and $\Psym$ is the identity.


\subsubsection*{Action of $\Psym$ on
single-particle states for $\pht=\frac{2\pi}{3}$
($\tau=e^{\pi i/3}$ and $\ord=3$)}

In this case $\lvk=3.$
For $\np=1$, $\Psym$ acts as:
$$
\Psym\ket{[0]}=\ket{[\tfrac{1}{3}+\tfrac{1}{3}\tau]}\,,\qquad
\Psym\ket{[\tfrac{1}{3}+\tfrac{1}{3}\tau]}=
\ket{[\tfrac{2}{3}+\tfrac{2}{3}\tau]}\,,\qquad
\Psym\ket{[\tfrac{2}{3}+\tfrac{2}{3}\tau]}=
\ket{[0]}\,.
$$
$$
\Psym\ket{\fpIIIz}=\ket{\fpIIIe}\,,\qquad
\Psym\ket{\fpIIIe}=\ket{\fpIIIf}\,,\qquad
\Psym\ket{\fpIIIf}=\ket{\fpIIIz}\,.
$$

For $\np=2$, $\Psym$ acts as:
$$
\left.
\begin{array}{lll}
\Psym\ket{[0,0]} &=\ket{[\tfrac{1}{3}+\tfrac{1}{3}\tau,
\tfrac{1}{3}+\tfrac{1}{3}\tau]}
\\
\Psym\ket{[\tfrac{1}{3}+\tfrac{1}{3}\tau,
\tfrac{1}{3}+\tfrac{1}{3}\tau]} &=
\ket{[\tfrac{2}{3}+\tfrac{2}{3}\tau,
\tfrac{2}{3}+\tfrac{2}{3}\tau]}
\\
\Psym\ket{[\tfrac{2}{3}+\tfrac{2}{3}\tau,
\tfrac{2}{3}+\tfrac{2}{3}\tau]} &=
\ket{[0,0]}\\
\end{array}\right\}
$$
$$
\Psym\ket{\fpIIIzz}=\ket{\fpIIIee}\,,\qquad
\Psym\ket{\fpIIIee}=\ket{\fpIIIff}\,,\qquad
\Psym\ket{\fpIIIff}=\ket{\fpIIIzz}\,.
$$


\subsection{$\Z_\lvk$ winding number}
\label{subsec:W}

Our problem has a second conserved $\Z_\lvk$ quantum number.
This one is defined by the winding number of the string in the fiber direction.
The winding number takes values in the first homology group of the space,
which in our case is homotopically equivalent to the space $\MXS$ defined in
\eqref{eqn:zx}. As we will check below, the first homology group is
$H_1(\MXS,\Z)\simeq\Z\oplus\Z_\lvk.$

To see this, let us pick the origin $(z=0,x_3=0)$ as a marked point,
and let us define three elements of the fundamental group
$\pi_1(\MXS)$ as the equivalence classes of the following three loops:
\be\label{eqn:gpibfafb}
\left.
\begin{array}{ll}
\gpib &= [t\mapsto(z=0,x_3=2\pi R t)]
\\
\gpifa &= [t\mapsto(z=t,x_3=0)]
\\
\gpifb&=[t\mapsto(z=t \tau,x_3=0)]
\\
\end{array}
\right\}
\qquad
0\le t<1.
\ee
The loops that define $\gpifa,\gpifb$ run along the $T^2$ fiber,
while $\gpib$ is defined by a loop that runs along the $S^1$ base.
Note that the fundamental group $\pi_1(\MXS)$ is generated
by $\gpib,\gpifa,\gpifb$ with the relations
$$
\gpifa\gpifb=\gpifb\gpifa\,,\qquad
\gpib^{-1}\gpifa\gpib=\gpifa^\xa\gpifb^\xb\,,\qquad
\gpib^{-1}\gpifb\gpib=\gpifa^\xc\gpifb^\xd\,,
$$
where $\xa,\xb,\xc,\xd$ are the elements of the $\SL(2,\Z)$ matrix $\zg$
defined in \secref{subsec:S-twist}.
The homology group $H_1(\MXS,\Z)$ is homeomorphic to
the abelianization of $\pi_1(\MXS).$
The abelianization of $\pi_1(\MXS)$ ignores the order of operators
in the relations, and so we get an abelian group generated
by $\ogpib$ and $\ogpifa,\ogpifb$ with the relations (over the ring $\Z$)
$$
\ogpifa=\ogpifa^\xa\ogpifb^\xb\,,
\qquad
\ogpifb=\ogpifa^\xc\ogpifb^\xd\,.
$$

Since it is more appropriate to denote the group operation
in $H_1(\MXS,\Z)$ by a sum instead of a product,
we switch notation from  $\ogpib,\ogpifa,\ogpifb$ to
$\hb,\hfa,\hfb.$
The result is that $H_1(\MXS,\Z)$ is generated
by $\hb,\hfa,\hfb,$ with the relations
$$
\hfa=\xa\hfa+\xb\hfb\,,
\qquad
\hfb=\xc\hfa+\xd\hfb\,.
$$
It is useful at this point to separate the cases:
\begin{itemize}
\item
For $\tau=i$, $\pht = \frac{\pi}{2},$
$\zg=\zg'\equiv
\left(\begin{array}{rr}
 0 & -1 \\ 1 & 0 \\ \end{array}\right),$ 
we have $\lvk=2$ and $\hfa=-\hfb=-\hfa$,
so $H_1(\MXS)=\Z\oplus\Z_2,$ generated by $\hb\in\Z$
and $\hfa\in\Z_2.$

\item
For $\tau=e^{\pi i/3}$, $\pht= \frac{\pi}{3},$
$\zg=\zg''\equiv\left(
\begin{array}{rr} 1 & -1 \\ 1 & 0 \\ \end{array}
\right),$ 
we have $\lvk=1$ and $\hfb=\hfa$ and
$\hfa=\hfa-\hfb$
so $\hfa=\hfb=0,$ and $H_1(\MXS)=\Z,$ generated by $\hb.$
\item
For $\tau=e^{\pi i/3}$, $\pht= \frac{2\pi}{3},$
$\zg=-\zg''{}^{-1}=\left(
\begin{array}{rr}  0 & -1 \\ 1 & -1 \\ \end{array}
\right),$ 
we have $\lvk=3$ and $\hfb=-\hfa$ and $\hfa=\hfb-\hfa.$
So $H_1(\MXS)=\Z\oplus\Z_3,$
generated by $\hb\in\Z$ and $\hfa\in\Z_3.$

\end{itemize}

To a string configuration that winds $n$ times
around the base and has the homology class $n\hb+\wxa\hfa$
(with $\wxa\in\Z_\lvk$), we assign a $\Z_\lvk$ charge of $\wxa.$
We now define the quantum operator $\Qsym$ to take the eigenvalue
$e^{2\pi i \wxa/\lvk}$ on such a state.

There is some arbitrariness in the definition
of $\wxa$ because of our arbitrary choice
of the origin $z=0$ of the $T^2$ fiber of $\MXS.$
Consider, for example, the case $\tau=i.$
The loop $[t\mapsto(z=0,x_3=2\pi t R)]$ was defined to have
homology class $n\hb$ (with $\wxa=0$) and the loop
$[t\mapsto(z=\tfrac{1}{2}+\tfrac{1}{2}i,x_3=2\pi t R)]$ then has
homology class $n\hb+\hfa.$ But we could have just as well
chosen the origin at $z=\tfrac{1}{2}+\tfrac{1}{2}i$, thereby switching
the eigenvalues of $\Qsym.$
In general, replacing
\be\label{eqn:Qphi}
\Qsym\rightarrow e^{i\phi}\Qsym
\ee
for some arbitrary (constant) phase $\phi$
results in an equally reasonable definition of $\Qsym.$
We will, nevertheless, stick to the definition of $\Qsym$ with the origin
set at $z=0.$


At this point we have found two operators $\Qsym,\Psym,$
acting on the Hilbert space of ground states (and, in fact, on the
full Hilbert space). Each operator defines
a conserved $\Z_\lvk$ quantum number, and by definition they satisfy
\be\label{eqn:QkPk}
\Qsym^\lvk=\Psym^\lvk=1\,.
\ee
They obey the commutation relation
\be\label{eqn:QPQP}
\Qsym\Psym\Qsym^{-1}\Psym^{-1}=e^{\frac{2\pi i n}{\lvk}}
\,,
\ee
which can be verified as follows.
Consider a (classical) string configuration with $n=1$
given by a section $z=f(x_3)$ of the fiber bundle $W$, where $f$ is some continuous
function on the interval $0\le x_3<2\pi\xR$ with
$f(2\pi\xR)=e^{i\pht}f(0).$
Now, translate the section by a small $c$ to $z=f(x_3)+c.$ This no longer
satisfies the boundary conditions, and so to close the string we need
to add a piece of string that connects $(z=f(0)+c,x_3=0)$ to
$(z=f(0)+e^{i\pht}c,x_3=0).$ As $c$ is increased from $0$ to $\frac{1}{\lvk}(1+\tau)$,
the extra piece of string increases until it becomes a piece of string
that stretches from start-point to end-point
by the complex vector $\frac{1}{\lvk}(1+\tau)(1-e^{i\pht}).$
It is easy to check that for the cases $\lvk=2,3$ this vector is in the homology
class $\hfa.$ We have thus arrived at the following conclusion:
acting with $\Psym$  on a string state with homology class $n\hb+\wxa\hfa$ produces
a string state with homology class $n\hb+(\wxa+n)\hfa.$
The commutation relation \eqref{eqn:QPQP} follows immediately from that observation.

For future reference, we list the action of $\Qsym$
on single-particle states explicitly below.


\subsubsection*{Action of $\Qsym$ on
single-particle states for $\pht=\frac{\pi}{2}$
($\tau=i$ and $\ord=4$)}

In this case $\lvk=2.$
For $\np=1$, $\Qsym$ acts as:
$$
\Qsym\ket{[0]}=\ket{[0]}
\,,\quad
\Qsym\ket{[\tfrac{1}{2}+\tfrac{1}{2}i]}=
 -\ket{[\tfrac{1}{2}+\tfrac{1}{2}i]}\,.
$$
\be\label{eqn:Qsym,k=2,n=1}
\Qsym\ket{\fpIIz}=\ket{\fpIIz}\,,\qquad
\Qsym\ket{\fpIIc}=-\ket{\fpIIc}\,.
\ee

For $\np=2$, $\Qsym$ acts as:
$$
\Qsym\ket{[0,0]}=\ket{[0,0]}
\,,\quad
\Qsym\ket{[\tfrac{1}{2},\tfrac{1}{2}i]}=
 -\ket{[\tfrac{1}{2},\tfrac{1}{2}i]}
\,,\quad
\Qsym\ket{[\tfrac{1}{2}+\tfrac{1}{2}i,\tfrac{1}{2}+\tfrac{1}{2}i]}
=\ket{[\tfrac{1}{2}+\tfrac{1}{2}i,\tfrac{1}{2}+\tfrac{1}{2}i]}\,.
$$
$$
\Qsym\ket{\fpIIzz}=\ket{\fpIIzz}\,,\qquad
\Qsym\ket{\fpIIcc}=\ket{\fpIIcc}\,,\qquad
\Qsym\ket{\fpIIxy}=-\ket{\fpIIxy}\,.
$$

For $\np=3$, $\Qsym$ acts as:
$$
\Qsym\ket{[0,0,0]}=\ket{[0,0,0]}
\,,\quad
\Qsym\ket{[\tfrac{1}{2}+\tfrac{1}{2}i,\tfrac{1}{2}+\tfrac{1}{2}i,
\tfrac{1}{2}+\tfrac{1}{2}i]}=
-\ket{[\tfrac{1}{2}+\tfrac{1}{2}i,
\tfrac{1}{2}+\tfrac{1}{2}i,\tfrac{1}{2}+\tfrac{1}{2}i]}\,.
$$
$$
\Qsym\ket{\fpIIzzz}=\ket{\fpIIzzz}\,,
\qquad
\Qsym\ket{\fpIIccc}=-\ket{\fpIIccc}\,.
$$


\subsubsection*{Action on single-particle states
for $\pht=\frac{\pi}{3}$ ($\tau=e^{\pi i/3}$ and $\ord=6$)}

In this case $\lvk=1$ and $\Qsym$ is the identity.


\subsubsection*{Action on
single-particle states for $\pht=\frac{2\pi}{3}$
($\tau=e^{\pi i/3}$ and $\ord=3$)}

In this case $\lvk=3.$
For $\np=1$, $\Qsym$ acts as:
$$
\Qsym\ket{[0]}=\ket{[0]}
\,,\quad
\Qsym\ket{[\tfrac{1}{3}+\tfrac{1}{3}\tau]}=
e^{\frac{2\pi i}{3}}\ket{[\tfrac{1}{3}+\tfrac{1}{3}\tau]}
\,,\quad
\Qsym\ket{[\tfrac{2}{3}+\tfrac{2}{3}\tau]}=
e^{-\frac{2\pi i}{3}}\ket{[\tfrac{2}{3}+\tfrac{2}{3}\tau]}
\,.
$$
$$
\Qsym\ket{\fpIIIz}=\ket{\fpIIIz}\,,\qquad
\Qsym\ket{\fpIIIe}=e^{\frac{2\pi i}{3}}\ket{\fpIIIe}\,,\qquad
\Qsym\ket{\fpIIIf}=e^{-\frac{2\pi i}{3}}\ket{\fpIIIf}\,.
$$

For $\np=2$, $\Qsym$ acts as:
\bear
&&\Qsym\ket{[0,0]} =\ket{[0,0]}
\,,\qquad
\Qsym\ket{[\tfrac{1}{3}+\tfrac{1}{3}\tau,
\tfrac{1}{3}+\tfrac{1}{3}\tau]}=
e^{-\frac{2\pi i}{3}}
\ket{[\tfrac{1}{3}+\tfrac{1}{3}\tau,
\tfrac{1}{3}+\tfrac{1}{3}\tau]}
\nn\\
&&\Qsym\ket{[\tfrac{2}{3}+\tfrac{2}{3}\tau,
\tfrac{2}{3}+\tfrac{2}{3}\tau]}
=
e^{\frac{2\pi i}{3}}
\ket{[\tfrac{2}{3}+\tfrac{2}{3}\tau,
\tfrac{2}{3}+\tfrac{2}{3}\tau]}\,.
\nn
\eear
$$
\Qsym\ket{\fpIIIzz}=\ket{\fpIIIzz}\,,\qquad
\Qsym\ket{\fpIIIee}=e^{-\frac{2\pi i}{3}}\ket{\fpIIIee}\,,\qquad
\Qsym\ket{\fpIIIff}=e^{\frac{2\pi i}{3}}\ket{\fpIIIff}\,.
$$

\subsection{Worldsheet symmetries}
\label{subsec:ws}

We will now discuss additional symmetries of the worldsheet CFT
that do not directly correspond to symmetries of the Hilbert space
of ground states, but will later be used as building blocks to construct
operators that do act on the Hilbert space of ground states.
Furthermore, these extra worldsheet symmetries will be useful in
\secref{subsec:T} when we study T-duality.

The worldsheet theory
can be regarded as a $\Z_\ord$ orbifold of a compactification of type-IIA theory
on $S^1\times T^2$, where $S^1$ has radius $2\pi\ord\xR$,
and $\Z_\ord$ is generated by an isometry as in \eqref{eqn:zxzzz}.
The sector of the Hilbert space that corresponds to strings of
winding number $\np$ is a twisted sector of this orbifold theory.
The rules of orbifolds \cite{Dixon:1985jw} dictate
that the Hilbert space $\Hilb$ of such one-particle states
is the $\Z_\ord$-invariant subspace of the Hilbert space
$\Hilb'\supseteq\Hilb$
of the CFT on a circle with twisted boundary conditions given by
the identification
\be\label{eqn:zxzzzn}
(z,x_3,\zeta_1,\zeta_2,\zeta_3)\sim
(e^{i\np\pht}z,x_3+2\pi\np\xR,
e^{i\np\pht}\zeta_1,e^{i\np\pht}\zeta_2,e^{i\np\pht}\zeta_3)\,,
\qquad
\pht\equiv\frac{2\pi}{\ord}\,.
\ee
If $\np$ and $\ord$ are not relatively prime (i.e.,
$\np=2,3,4$ for $\ord=6$, and $\np=2$ for $\ord=4$),
the Hilbert space $\Hilb'$
possesses discrete symmetries in addition to $\Psym,\Qsym.$
This is because
the identification \eqref{eqn:zxzzzn} generates a
$\Z_{\ord/\gcd(\ord,\np)}\subset\Z_{\ord}$ subgroup of the orbifold group,
and thus, as far as $\Hilb'$ is concerned, the effective geometry
is a $\Z_{\ord/\gcd(\ord,\np)}$-orbifold, which can have a larger
group of symmetries than the $\Z_\ord$-orbifold.

To explain this in more detail, we need a worldsheet realization of
the type-IIA background,
but it will be sufficient to consider only the bosons and
only in the directions of $\MXS.$
We represent $\MXS$ as a $\Z_\ord$ orbifold of $T^2\times S^1$,
where $T^2$ is parameterized by $z\sim z+1\sim z+\tau$ as above,
and $S^1$ is parameterized by $0 \le y < 2\pi \ord R$,
and the orbifold group is generated by
$$
(z,y)\mapsto (e^{i\pht}z,y+2\pi\xR).
$$
We define worldsheet coordinates $(\wsig,\wtau)$
and worldsheet bosons $\wsZ(\wsig,\wtau),\wsY(\wsig,\wtau)$
corresponding to the coordinates $z$ and $y$,
so that $\wsY$ is real and $\wsZ$ is complex.
We work in a twisted sector for which
$$
\wsZ(\wsig+2\pi,\wtau)=
e^{i\np\pht}\wsZ(\wsig,\wtau)+\wMa+\wMb\tau\,,\qquad
\wsY(\wsig+2\pi,\wtau)=\wsY(\wsig,\wtau)+2\pi\np R\,.
$$
In this sector, the worldsheet fields have an expansion
\bear
\wsY &=& \fpY_0 + \pY\wtau + \np\wsig R
+\sum_{\nZ\neq 0}\frac{i}{\nZ}\oscYr_{-\nZ}
e^{i \nZ(\wtau-\wsig)}
+\sum_{\nZ\neq 0}\frac{i}{\nZ}\oscYl_{-\nZ}
e^{i \nZ(\wtau+\wsig)}\,,
\\
\wsZ &=& \fpZ_{\wMa,\wMb}
+\sum_{\nZ\in\Z}\frac{i}{\nZ-\frac{\np}{\ord}}
\oscZr_{-\nZ+\frac{\np}{\ord}}
e^{i (\nZ-\frac{\np}{\ord})(\wtau-\wsig)}
+\sum_{\nZ\in\Z}\frac{i}{\nZ+\frac{\np}{\ord}}
\oscZl_{-\nZ-\frac{\np}{\ord}}
e^{i (\nZ+\frac{\np}{\ord})(\wtau+\wsig)}\,,
\nn\\ &&
\eear
where $\pY$ is the $\wsY$-momentum,
$\oscYr_{-\nZ}$ and $\oscYl_{-\nZ}$ are the integer-moded
right and left moving oscillators for $\wsY$,
$\oscZr_{-\nZ-\frac{\np}{\ord}}$ and $\oscZl_{-\nZ-\frac{\np}{\ord}}$
are the fractionally moded right and left moving oscillators
for $\wsZ$,
and $\fpZ_{\wMa,\wMb}$ is the solution of \eqref{eqn:fpZsol}
and is a fixed point of rotation of the
$T^2$ fiber by an angle $\np\pht.$
We define two solutions $\fpZ_{\wMa,\wMb}$ and $\fpZ_{\wMa',\wMb'}$
as equivalent if they differ by a lattice vector, i.e.,
$\fpZ_{\wMa,\wMb}-\fpZ_{\wMa',\wMb'}\in\Z+\Z\tau.$
In the ground states of the CFT
the oscillators are not excited and $\pY=0.$
The states fall into a finite number of sectors labeled
by the inequivalent solutions $\fpZ_{\wMa,\wMb}.$
We will denote these states by $\ket{\fpZ_{\wMa,\wMb}}.$

Note that if $\fpZ_{\wMa,\wMb}$ is a solution,
then so is
$$
e^{i\pht}\fpZ_{\wMa,\wMb}=\fpZ_{\xd\wMa+\xb\wMb,\,\xc\wMa+\xa\wMb}\,,
$$
where we have used \eqref{eqn:sdtau}-\eqref{eqn:phtdef}.
But if $e^{i\pht}\fpZ_{\wMa,\wMb}-\fpZ_{\wMa,\wMb}$ is not in $\Z+\Z\tau$,
then
as far as the worldsheet CFT is concerned,
$\ket{\fpZ_{\wMa,\wMb}}$
and $\ket{e^{i\pht}\fpZ_{\wMa,\wMb}}$ are different states.
However, to get the {\it string} ground states, we need to impose
(i) invariance under translations in $\wsig$, and
(ii) invariance under the orbifold group $\Z_\ord.$
Now, define the operator $\OrbG$ by
$$
\OrbG\ket{\fpZ_{\wMa,\wMb}}=\ket{e^{i\pht}\fpZ_{\wMa,\wMb}}.
$$
On ground states, it is equivalent to a combination of the $\Z_\ord$
orbifold generator and worldsheet translation in $\wsig.$
The CFT ground states
that correspond to string ground states must therefore
be invariant under $\OrbG$ and thus are linear combinations of
states of the form
\be\label{eqn:HpBasis}
\sum_{j=0}^{\np-1}\ket{e^{i j\pht}\fpZ_{\wMa,\wMb}}\,.
\ee
(Note that
$\ket{e^{i\np\pht}\fpZ_{\wMa,\wMb}}
=\ket{\fpZ_{\wMa,\wMb}}$, by virtue of \eqref{eqn:fpZsol}.)
Since we are not concerned with excited states,
we will take $\Hilb'$ as the Hilbert space spanned by
the states $\ket{\fpZ_{\wMa,\wMb}}$,
and $\Hilb\subseteq\Hilb'$ as the subspace spanned by
the states \eqref{eqn:HpBasis}.

Now, let us assume that $\gcd(\np,\ord)>1.$
We can then find additional symmetries of $\Hilb'$ that
do not commute with $\OrbG$ as follows.
For any $\fpZ\in\C/\Z+\Z\tau$ that satisfies
\be\label{eqn:fpZmfpZ}
e^{i\np\pht}\fpZ-\fpZ\in\Z+\Z\tau
\,,
\ee
we define two operators $\wPsym(\fpZ),\wQsym(\fpZ)$
on $\Hilb'$ by
\be\label{eqn:defwPQ}
\wPsym(\fpZ)\ket{\fpZ_{\wMa,\wMb}}=\ket{\fpZ_{\wMa,\wMb}+\fpZ}
\,,\qquad
\wQsym(\fpZ)\ket{\fpZ_{\wMa,\wMb}}=
e^{4\pi i\text{Re}(\fpZ^*\fpZ_{\wMa,\wMb})}\ket{\fpZ_{\wMa,\wMb}}
\,.
\ee
We have the commutation relations
$$
\wPsym(\fpZ)\wPsym(\fpZ')=\wPsym(\fpZ')\wPsym(\fpZ)
\,,\qquad
\wQsym(\fpZ)\wQsym(\fpZ')=\wQsym(\fpZ')\wQsym(\fpZ)
\,,
$$
$$
\wPsym(\fpZ)\wQsym(\fpZ')=
e^{-4\pi i\text{Re}(\fpZ^*\fpZ')}\wQsym(\fpZ')\wPsym(\fpZ)
\,.
$$
For example,
the symmetry operators $\Psym,\Qsym$ defined in \secref{subsec:P}-\secref{subsec:W}
can be written as
\be\label{eqn:PQsumr}
\Psym=\wPsym (\tfrac{1}{\lvk}+\tfrac{1}{\lvk}\tau )\,,\qquad
\Qsym=\wQsym (\tfrac{1}{\lvk}+\tfrac{1}{\lvk}\tau )\,.
\ee

But in general, the operators $\wPsym(\zeta),\wQsym(\zeta)$
do not preserve
the subspace of physical string states $\Hilb\subset\Hilb'$,
because in general $\wPsym(\zeta),\wQsym(\zeta)$ do not commute with $\OrbG.$
However, we can form $\OrbG$-invariant combinations such as
$$
\sum_{j=0}^{\np-1}\wPsym(e^{i j\pht}\fpZ)\,,
\quad
\sum_{j=0}^{\np-1}\wQsym(e^{i j\pht}\fpZ)\,,
\quad
\prod_{j=0}^{\np-1}
\wPsym(e^{i j\pht}\fpZ)\wQsym(e^{i j\pht}\fpZ)\,,
\cdots
$$
that do preserve $\Hilb$ and therefore define operators that act
on the Hilbert space of string ground states. We will 
return to these constructions in \secref{subsec:Wilson}.

Let us proceed to examples.
For the first example, consider the case
$k=2$ ($\tau=i$ and $\pht=\frac{\pi}{2}$, $\ord=4$) and $\np=2$.
We have four inequivalent solutions to \eqref{eqn:fpZsol}:
\be\label{eqn:fpZ,n=2,r=4}
\fpZ_{0,0}=0,\quad
\fpZ_{1,0}=\tfrac{1}{2},\quad
\fpZ_{0,1}=\tfrac{1}{2}i,\quad
\fpZ_{1,1}=\tfrac{1}{2}(1+i)\pmod{\Z+\Z\tau}.
\ee
If we define
\be\label{eqn:defPQab}
\wQsym_a\equiv\wQsym(\fpZ_{1,0})
\,,\quad
\wPsym_a\equiv\wPsym(\fpZ_{1,0})
\,,\quad
\wQsym_b\equiv\wQsym(\fpZ_{0,1})
\,,\quad
\wPsym_b\equiv\wPsym(\fpZ_{0,1})
\,,
\ee
then they act on the four-dimensional Hilbert space $\Hilb'$ as
\be\label{eqn:PQabAction}
\left.
\begin{array}{llll}
\wQsym_a\ket{\fpZ_{\wMa,\wMb}}&=(-1)^{\wMa}\ket{\fpZ_{\wMa,\wMb}}
\,,&
\wQsym_b\ket{\fpZ_{\wMa,\wMb}}&=(-1)^{\wMb}\ket{\fpZ_{\wMa,\wMb}}
\,,
\\
\wPsym_a\ket{\fpZ_{\wMa,\wMb}}&=\ket{\fpZ_{\wMa+1,\wMb}}
\,,&
\wPsym_b\ket{\fpZ_{\wMa,\wMb}}&=\ket{\fpZ_{\wMa,\wMb+1}}\,,
\\
\end{array}\right\}
\ee
where $(\wMa+1)$ and $(\wMb+1)$ are understood to
be additions in $\Z_2.$
The states $\ket{\fpZ_{\wMa,\wMb}}$ are eigenstates of
$\wQsym_a,\wQsym_b$,
while
\be\label{eqn:KK}
\ket{\pKa,\pKb}\equiv
\frac{1}{2}\sum_{\wMa,\wMb\in\Z_2}(-1)^{\pKa\wMa+\pKb\wMb}
\ket{\fpZ_{\wMa,\wMb}}
\ee
are eigenstates of $\wPsym_a,\wPsym_b.$
The operators $\Psym$ and $\Qsym$ are related to $\wPsym_a,\wPsym_b$
and $\wQsym_a,\wQsym_b$ by
$$
\Psym\equiv\wPsym(\fpZ_{1,1})=\wPsym_a\wPsym_b
\,,\qquad
\Qsym\equiv\wQsym(\fpZ_{1,1})=\wQsym_a\wQsym_b
\,.
$$

We can regard $\wQsym_a,\wQsym_b$ as associated with
two independent winding numbers $\wna,\wnb\in\Z_2$,
which characterize the topology of the map $\wsZ=\wsZ(\sigma,\eta)$
from the worldsheet to $T^2.$
We define them by
$$
\wna\equiv\wMa\pmod 2
\,,\qquad
\wnb\equiv\wMb\pmod 2
\,.
$$
The winding number $\wna$ is associated with the $\hfa$ cycle
(a loop along a straight path $\wsZ\rightarrow\wsZ+1$)
and $\wnb$
is associated with the $\hfb$ cycle ($\wsZ\rightarrow\wsZ+i$).
Beyond the CFT, in the full theory,
the cycles $\hfa,\hfb$ were identified in homology,
and only one $\Z_2$ winding number remained as an independent
quantum number.
But in the worldsheet CFT sector with $\np=2$, we have a larger symmetry.
The only identification is $\wsZ\sim -\wsZ$ (together with
$\wsY\sim\wsY+4\pi R$).
Thus, for each of the two cycles of $T^2$ we
end up with a separate $\Z_2$ winding number in the
worldsheet theory.

Now, let us recover the Hilbert subspace
$\Hilb\subset\Hilb'$ of string ground states.
In the string theory, we need to keep only the states that
are invariant under the entire $\Z_\ord=\Z_4$ orbifold group, generated by
$\wsZ\rightarrow i\wsZ$ together
with $\wsY\rightarrow\wsY+2\pi\xR.$
The resulting $\Z_\ord$-invariant states span
a $3$-dimensional subspace $\Hilb$ of the $4$-dimensional $\Hilb'$, and
the $\Z_\ord$-invariant combinations that correspond to the states
in \secref{subsec:GS} are:
\be\label{eqn:fpIIfpZ}
\ket{\fpZ_{0,0}}
\rightarrow
\ket{\fpIIzz}
\,,\qquad
\ket{\fpZ_{1,1}}
\rightarrow
\ket{\fpIIcc}
\,,\qquad
\frac{1}{\sqrt{2}}(\ket{\fpZ_{0,1}}+\ket{\fpZ_{1,0}})
\rightarrow
\ket{\fpIIxy}
\,.
\ee
The operators
$$
\wPsym_a+\wPsym_b\,,
\quad
\wQsym_a+\wQsym_b\,,
\quad
\wQsym_a\wPsym_a+\wQsym_b\wPsym_b\,,
\ldots
$$
preserve the $3$-dimensional Hilbert space
spanned by $\ket{\fpIIzz},\ket{\fpIIxy},\ket{\fpIIcc}.$
They act as
\be\label{eqn:UaUb-VaVb}
\left.
\begin{array}{ll}
(\wQsym_a+\wQsym_b)\ket{\fpIIzz}=2\ket{\fpIIzz}
\,,\qquad &
(\wPsym_a+\wPsym_b)\ket{\fpIIzz}=\sqrt{2}\ket{\fpIIxy}\,,
\\
(\wQsym_a+\wQsym_b)\ket{\fpIIxy}=0
\,,\qquad &
(\wPsym_a+\wPsym_b)\ket{\fpIIxy}=
\sqrt{2}(\ket{\fpIIzz}+\ket{\fpIIcc})\,,
\\
(\wQsym_a+\wQsym_b)\ket{\fpIIcc}=-2\ket{\fpIIcc}
\,,\qquad &
(\wPsym_a+\wPsym_b)\ket{\fpIIcc}=
\sqrt{2}\ket{\fpIIxy}\,,
\\
\end{array}\right\}
\ee
\be\label{eqn:UaVaUbVb}
\left.
\begin{array}{l}
(\wQsym_a\wPsym_a+\wQsym_b\wPsym_b)\ket{\fpIIzz} =
-\sqrt{2}\ket{\fpIIxy}\,,
\\
(\wQsym_a\wPsym_a+\wQsym_b\wPsym_b)\ket{\fpIIxy}=
\sqrt{2}(\ket{\fpIIzz}-\ket{\fpIIcc})\,,
\\
(\wQsym_a\wPsym_a+\wQsym_b\wPsym_b)\ket{\fpIIcc}=
\sqrt{2}\ket{\fpIIxy}\,,
\\
\end{array}\right\}
\ee
and so on.

As another example of this technique,
consider the case $k=1$ ($\tau=e^{\pi i/3}$ and $\pht=\frac{\pi}{3}$, $\ord=6$) and $\np=2$.
The relevant solutions to \eqref{eqn:fpZsol} are:
$$
\fpZ_{0,0}=0
\,,\quad
\fpZ_{1,0}=\frac{i}{\sqrt{3}}
\,,\quad
\fpZ_{0,1}= -\frac{i}{\sqrt{3}}
\quad
\pmod{\Z+\Z\tau}.
$$
(But also note the equivalent solutions   $\fpZ_{0,0}\simeq\fpZ_{1,1}\simeq\fpZ_{2,2}$,
$\fpZ_{1,0}\simeq
\fpZ_{2,1}\simeq
\fpZ_{0,2}$, and $\fpZ_{0,1}\simeq
\fpZ_{2,0}\simeq
\fpZ_{1,2}$.)
The orbifold generator
acts on these fixed points $\fpZ_{\wMa,\wMb}$ as multiplication
by $e^{\pi i/3}$, and the invariant combinations are
$$
\ket{\fpZ_{0,0}}
\,,\qquad
\frac{1}{\sqrt{2}}\left(\ket{\fpZ_{1,0}}+\ket{\fpZ_{0,1}}\right)
\,.
$$
They correspond to the string ground states
$$
\ket{\fpZ_{0,0}}
\rightarrow\ket{[0,0]}=\ket{\fpIzz}
\,,\qquad
\frac{1}{\sqrt{2}}\left(\ket{\fpZ_{1,0}}+\ket{\fpZ_{0,1}}\right)
\rightarrow\ket{[\tfrac{1}{3}+\tfrac{1}{3}\tau,
\tfrac{2}{3}+\tfrac{2}{3}\tau]}
=\ket{\fpIdd}
\,.
$$

The group of additional worldsheet symmetries is generated by
\be\label{eqn:eqn:PQa,r=6,n=2}
\wQsym_a\equiv\wQsym(\fpZ_{1,0})
\,,\qquad
\wPsym_a\equiv\wPsym(\fpZ_{1,0})\,.
\ee
They satisfy
$$
\wQsym_a^3=\wPsym_a^3=1,
$$
and can be regarded as related to $\Z_3$ winding number and momentum.
They act on states as
\be\label{eqn:wVawUa}
\wQsym_a\ket{\fpZ_{\wMa,\wMb}}
\equiv e^{\frac{2\pi i}{3}(\wMb-\wMa)}\ket{\fpZ_{\wMa,\wMb}}
\,,\qquad
\wPsym_a\ket{\fpZ_{\wMa,\wMb}}
=\ket{\fpZ_{\wMa+1,\wMb}}\,.
\ee
The operators
$$
\wPsym_a+\wPsym_a^{-1}\,,
\quad
\wQsym_a+\wQsym_a^{-1}\,,
\quad
\wQsym_a\wPsym_a+\wQsym_a^{-1}\wPsym_a^{-1}\,,
\ldots
$$
preserve the $2$-dimensional Hilbert space
spanned by $\ket{\fpIzz},\ket{\fpIdd}$,
and act as
\be
\left.
\begin{array}{ll}
(\wQsym_a+\wQsym_a^{-1})\ket{\fpIdd}
=-\ket{\fpIdd}
\,,\qquad &
(\wPsym_a+\wPsym_a^{-1})\ket{\fpIdd}
=\ket{\fpIdd}+\sqrt{2}\ket{\fpIzz}
\,, \\
(\wQsym_a+\wQsym_a^{-1})\ket{\fpIzz}
=2\ket{\fpIzz}
\,,\qquad &
(\wPsym_a+\wPsym_a^{-1})\ket{\fpIzz}
=\sqrt{2}\ket{\fpIdd}
\,,\\
\end{array}
\right\}
\ee
\be
\left.
\begin{array}{l}
(\wQsym_a\wPsym_a+\wQsym_a^{-1}\wPsym_a^{-1})\ket{\fpIzz}
=\sqrt{2}e^{-\frac{2\pi i}{3}}\ket{\fpIdd}
\,,\\
(\wQsym_a\wPsym_a+\wQsym_a^{-1}\wPsym_a^{-1})\ket{\fpIdd}
=e^{\frac{2\pi i}{3}}\ket{\fpIdd}+\sqrt{2}\ket{\fpIzz}
\,,
\end{array}
\right\}
\ee
and so on.

\subsection{Dependence on complex structure}
\label{subsec:rhocp}

At the beginning of this section,
we compactified the field theory on $T^2$ (on the type-IIB side)
with periodic coordinates $0\le x_1<2\pi \xL_1$ and
$0\le x_2<2\pi \xL_2.$
For simplicity we took the metric to be $ds^2=dx_1^2+dx_2^2,$
which sets the complex structure of $T^2$
to be $\rhocp=i\xL_1/\xL_2.$
For this metric $\rhocp$ is purely imaginary, but
we can easily
allow a more general flat metric with a complex structure
that has a nonzero real part.
We can then define the action of a group $\SL(2,\Z)$
of large diffeomorphisms on $T^2$ by
\be\label{eqn:defTdG}
\begin{pmatrix} x_1 \\ x_2 \\ \end{pmatrix}
\mapsto
\TdG
\begin{pmatrix} x_1 \\ x_2 \\ \end{pmatrix}
\,,\qquad
\TdG\equiv
\begin{pmatrix} \txa & \txb \\ \txc & \txd \\ \end{pmatrix}
\in\SL(2,\Z)
\,,
\ee
which acts on the complex structure $\rhocp$ as
\be\label{eqn:TdGi}
\rhocp\rightarrow \frac{\txa\rhocp+\txb}{\txc\rhocp+\txd}\,.
\qquad
\ee
(This $\SL(2,\Z)$ is, of course,
not related to the S-duality group of
\secref{subsec:S-twist}.
In the following, we hope that the context makes
it clear which $\SL(2,\Z)$ we are referring to.)

The full Hilbert space is fibered over the
moduli space of $\rhocp$'s,
which is
$$\SL(2,\Z)\backslash\SL(2,\R)/SO(2)\,,$$
and two subgroups of $\SL(2,\Z)$ become symmetries
at two special values of $\rhocp$:
$\Z_4\subset\SL(2,\Z)$ generated by
$\begin{pmatrix} 0 & -1 \\ 1 & 0 \\ \end{pmatrix}$
is a symmetry at $\rhocp=i$, and
$\Z_6\subset\SL(2,\Z)$ generated by
$\begin{pmatrix} 1 & -1 \\ 1 & 0 \\ \end{pmatrix}$
is a symmetry at $\rhocp=e^{\pi i /3}$.

If we are only interested in the finite-dimensional
Hilbert space of supersymmetric ground states, as is the case here,
we can say more.
This finite-dimensional Hilbert space is the fiber of
a flat vector bundle over the moduli space of $\rhocp.$
Thus, the fibers at different complex
structures $\rhocp$ can be naturally identified,
and the action of $\SL(2,\Z)$, which is the holonomy group of
the vector bundle, can be naturally defined on the fiber.
In this way we get a full $\SL(2,\Z)$ symmetry acting
on the Hilbert space of ground states.
Unlike the operators $\Psym,\Qsym$,
this $\SL(2,\Z)$ group is not a symmetry of the full theory,
but only a low-energy symmetry.
Let us now identify the action of this $\SL(2,\Z)$ on the
type-IIA side.

\subsection{T-duality}
\label{subsec:T}

Following the sequence of dualities in \tabref{tab:Dualities},
we find that on the type-IIA side (the last row in the table)
we can identify $\rhocp$ (defined in \secref{subsec:rhocp})
 as the complexified area modulus
of the $T^2$ fiber of $\MXS$:
\be\label{eqn:rhocp}
\rhocp =
\frac{i}{\alpha'_{\text{IIA}}}
\text{Area}(T^2)
+\frac{1}{2\pi}\int_{T^2}B\,.
\ee
Here, $B$ is the NS-NS two-form potential.
The $\SL(2,\Z)$ group from \secref{subsec:rhocp}
becomes T-duality, and is generated by
$$
\TdS\rightarrow
\begin{pmatrix} 0 & -1 \\ 1 & 0 \\ \end{pmatrix}
\in\SL(2,\Z)\,,
\qquad
\TdS:\qquad\rhocp\rightarrow-\frac{1}{\rhocp}\,,
$$
and
$$
\TdT\rightarrow
\begin{pmatrix} 1 & 1 \\ 0 & 1 \\ \end{pmatrix}
\in\SL(2,\Z)\,,
\qquad
\TdT:\qquad\rhocp\rightarrow\rhocp+1\,.
$$

At $\rhocp=i$, $\TdS$ generates a $\Z_4$ symmetry of the full theory.
However,
$\TdT$ can never be extended to a symmetry of the
full spectrum, while $\TdT\TdS$ has order $6$ and
is a symmetry of the full spectrum for $\rhocp=e^{\pi i/3}.$
Let us now determine the action of $\TdT$ and $\TdS$ on the ground states.

Partial information can be gleaned from the commutation
relations of $\TdS,\TdT$ with $\Psym,\Qsym$ defined in
\secref{subsec:P}-\secref{subsec:W}.
Since we associated $\Psym$ with $\Z_\lvk$ momentum,
and $\Qsym$ with $\Z_\lvk$ winding number,
and since T-duality exchanges
these two quantum numbers, we set:
\be\label{eqn:SPQ}
\TdS^{-1}\Qsym\TdS=\Psym\,,\qquad
\TdS^{-1}\Psym\TdS=\Qsym^{-1}
\,.
\ee
We also expect that a general T-duality element
$\TdG\in\SL(2,\Z)$ [defined in \eqref{eqn:defTdG}] acts as
\be\label{eqn:GPQ}
\TdG^{-1}\Qsym\TdG=e^{i\phi_1}\Qsym^\txd\Psym^{-\txc}\,,\qquad
\TdG^{-1}\Psym\TdG=e^{i\phi_2}\Qsym^{-\txb}\Psym^\txa\,.
\ee
We included undetermined phases $\phi_1,\phi_2$
in the commutation relations, because $\Psym,\Qsym$ do not commute
and their order in the expressions
on the right-hand side of the equations in \eqref{eqn:GPQ} is important.
Part of this phase ambiguity
can be absorbed by a redefinition
\be\label{eqn:TdGredef}
\TdG\rightarrow\Psym^\txp\Qsym^\txq\TdG\,,
\ee
under which
$$
\phi_1\rightarrow\phi_1+\frac{2\pi n}{\lvk}\txp
\,,\qquad
\phi_2\rightarrow\phi_2-\frac{2\pi n}{\lvk}\txq
\,.
$$
But in general $\phi_1,\phi_2$ need to be nonzero,
so that the eigenvalues of the left- and right-hand sides
of the equations in \eqref{eqn:GPQ} will agree.

Now, let us specialize to $\TdT.$
The generator $\TdT$ of $\SL(2,\Z)$ commutes with
the winding number operator $\Qsym,$
so we can choose $\phi_1=0$ in \eqref{eqn:GPQ}.
We can also take $\phi_2 = \pm\pi n(k-1)/\lvk$,
so that the eigenvalues of the left- and right-hand sides
of the rightmost equation of \eqref{eqn:GPQ} will agree. (We take $-$ sign for $k=2$ and $+$ sign for $k=3$.)
Thus we get
\be\label{eqn:TPQ}
\TdT^{-1}\Qsym\TdT=\Qsym\,,\qquad
\TdT^{-1}\,\Psym\TdT
=e^{\frac{i\pi n (-1)^{\lvk+1}(\lvk-1)}{\lvk}}\Psym\Qsym^{-1}\,.
\ee

For single-particle ground states of winding number $\np=n$
that is relatively prime to $\ord$,
equations \eqref{eqn:SPQ} and \eqref{eqn:TPQ}
are sufficient to determine $\TdS$ and $\TdT$, up to
multiplication by an overall phase and the freedom
\eqref{eqn:TdGredef}.
In principle, these ambiguities can be further
restricted by requiring the $\SL(2,\Z)$ relations
$\TdS^2=(\TdS\TdT)^3=-1$, but this will not be required
for our present purposes.
The results for $\TdS,\TdT$ are listed in \appref{app:SL(2,Z)}.

If $\gcd(n,\ord)>1$, \eqref{eqn:SPQ} and \eqref{eqn:TPQ}
are insufficient to completely determine $\TdS$ and $\TdT$,
and we need to study the worldsheet theory
more carefully.
In this case, the worldsheet theory,
as we saw in \secref{subsec:ws},
possesses additional discrete symmetries that can be regarded as
additional components of $\Z_2$ or $\Z_3$ winding and momentum.
These symmetries
do not commute with the $\Z_\ord$-orbifold symmetry generator
$\OrbG$ and therefore do not lead
to symmetries of the Hilbert space of string ground states.
However, since T-duality is a duality at the level of CFT,
we can use the additional discrete symmetries to glean
additional information about the action of $\TdS,\TdT.$
We will demonstrate how this works below.

As a first example, consider the case $\pht=\frac{\pi}{2}$ ($\tau=i$).
We will start with the $n=1$,
for which of course $\gcd(n,\ord)=1$ and we do not get
additional worldsheet symmetries; but it is still instructive
to start with this case. Referring to the notation of
\secref{subsec:ws}, we have only two inequivalent solutions
to \eqref{eqn:fpZsol}:
$$
\fpZ_{1,1}\simeq\fpZ_{0,0}=0,\quad\fpZ_{1,0}\simeq\fpZ_{0,1}
\simeq\tfrac{1}{2}(1+i)\pmod{\Z+\Z\tau}.
$$
The states $\ket{\fpZ_{0,0}}$ and $\ket{\fpZ_{1,1}}$
are eigenstates of winding, while
$\frac{1}{\sqrt{2}}(\ket{\fpZ_{0,0}}\pm\ket{\fpZ_{1,1}})$
are eigenstates of the translation
$\wsZ\rightarrow\wsZ+\tfrac{1}{2}(1+i).$
Hence, $\TdS$ maps $\ket{\fpZ_{0,0}}$ to
$\frac{1}{\sqrt{2}}(\ket{\fpZ_{0,0}}+\ket{\fpZ_{1,1}})$,
and maps $\ket{\fpZ_{1,1}}$ to
$\frac{1}{\sqrt{2}}(\ket{\fpZ_{0,0}}-\ket{\fpZ_{1,1}}).$

Now consider the case $\pht=\frac{\pi}{2}$ and
$n=2$,
for which $\gcd(n,\ord)=2,$ and
we do get additional worldsheet symmetries,
$\wQsym_a,\wQsym_b,\wPsym_a,\wPsym_b,$
as explained in \secref{subsec:ws}.
The T-duality generators $\TdS,\TdT$ are required to satisfy commutation
relations similar to \eqref{eqn:SPQ}-\eqref{eqn:TPQ}:
\be\label{eqn:SPQab(k=2)}
\TdS^{-1}\wQsym_a\TdS=\wPsym_b\,,\quad
\TdS^{-1}\wQsym_b\TdS=\wPsym_a^{-1}\,,\quad
\TdS^{-1}\wPsym_a\TdS=\wQsym_b\,,\quad
\TdS^{-1}\wPsym_b\TdS=\wQsym_a^{-1}\,,
\ee
\be\label{eqn:TPQab(k=2)}
\TdT^{-1}\wQsym_a\TdT=\wQsym_a\,,\quad
\TdT^{-1}\wQsym_b\TdT=\wQsym_b\,,
\quad
\TdT^{-1}\wPsym_a\TdT=\wPsym_a\wQsym_b^{-1}\,,\quad
\TdT^{-1}\wPsym_b\TdT=\wPsym_b\wQsym_a\,.
\ee
Solving \eqref{eqn:SPQab(k=2)}-\eqref{eqn:TPQab(k=2)},
we find the explicit expressions:
\be\label{eqn:TdSfpZ}
\TdS\ket{\fpZ_{\pKa,\pKb}}=
\frac{1}{2}\sum_{\wMa,\wMb\in\Z_2}(-1)^{\pKa\wMa+\pKb\wMb}
\ket{\fpZ_{\wMa,\wMb}}
\equiv
\ket{\pKa,\pKb}
\,,
\ee
and
\be\label{eqn:TdTfpZ}
\TdT\ket{\fpZ_{\pKa,\pKb}}=(-1)^{\pKa\pKb}\ket{\fpZ_{\pKa,\pKb}}\,.
\ee
The action of $\TdS,\TdT$ on the subspace of string ground states
can be deduced from \eqref{eqn:fpIIfpZ} and
\eqref{eqn:TdSfpZ}-\eqref{eqn:TdTfpZ}.
The complete expressions are listed in \appref{app:SL(2,Z)}.

As another example of this technique,
consider the case $\pht=\frac{\pi}{3}$ ($\tau=e^{\pi i/3}$) and $n=2$.
The commutation relations of $\TdS,\TdT$
with the extra symmetry generators \eqref{eqn:eqn:PQa,r=6,n=2} are:
\be\label{eqn:TSPQ(k=1)}
\TdS^{-1}\wQsym_a\TdS=\wPsym_a\,,\quad
\TdS^{-1}\wPsym_a\TdS=\wQsym_a^{-1}
\,,\quad
\TdT^{-1}\wQsym_a\TdT=\wQsym_a\,,\quad
\TdT^{-1}\wPsym_a\TdT
=e^{\frac{2\pi i}{3}}\wPsym_a\wQsym_a\,,
\ee
where we have chosen the phase in the rightmost equation
so that (i) the eigenvalues of the left- and right-hand sides
will agree, and (ii) so that
the subspace of string states
will be invariant under $\TdT.$
The solutions for $\TdS,\TdT$ are listed in
\appref{app:SL(2,Z)}.

\subsubsection*{Worldsheet derivation of the action of $\TdT$}
We will end this section by checking the formulas for $\TdT$
directly from the worldsheet description.
Denote
$$
\AreaForm_F\equiv\frac{1}{2i\Imx\tau}
dz\wedge d\overline{z}\,.
$$
The integral of $\AreaForm_F$ on any $T^2$
fiber of $\MXS$ is $1.$
The operator $\TdT$ acts by shifting the NS-NS $2$-form
$B$-field of type-IIA by
$$
B\rightarrow B + \AreaForm_F\,.
$$

We will now check how this shift affects the phase
of scattering amplitudes of the string ground states.
A string ground state, as discussed in \secref{subsec:GS},
corresponds to a curve $\gamma$ in target space,
which we can take for the present discussion to be $\MXS.$
Consider a worldsheet configuration that contributes
to a scattering amplitude taking string ground states that
correspond to the curves $\gamma_1,\gamma_2,\cdots,\gamma_p$
into another ground states corresponding to curves
$\gamma_1',\gamma_2',\cdots,\gamma_q'.$
The image of this worldsheet configuration in target
space is a surface $\Sigma$ whose boundary is
\be\label{eqn:SigmaBoundary}
\partial\Sigma =
\left(\bigcup_{i=1}^p\gamma_i^{-1}\right)
\bigcup
\left(\bigcup_{j=1}^q\gamma_j'\right)
\,,
\ee
where $\gamma_i^{-1}$ is the curve with oposite orientation
of $\gamma_i.$
Define the phase factor
$$
e^{i\Phi(\gamma_1^{-1},\gamma_2^{-1},\cdots,\gamma_p^{-1},
\gamma_1',\gamma_2',\cdots,\gamma_q')}
\equiv
\exp\left(i\int_\Sigma\AreaForm_F\right).
$$
This phase is clearly independent of which $\Sigma$
we choose, as long as it satisfies \eqref{eqn:SigmaBoundary},
because $\AreaForm_F/2\pi$ is an integral cohomology class,
whose integral over any closed surface is an integer.
Thus, the phase $\exp(i\Phi)$ only depends on
the curves $\gamma_1^{-1},\dots,\gamma_q'.$

Consider, for example, the case $\pht=\frac{\pi}{2}$ and $n=1$.
We show in \appref{app:SL(2,Z)} that
$$
\TdT\ket{\fpIIz}=\ket{\fpIIz}
\,,\qquad
\TdT\ket{\fpIIc}=e^{\frac{i\pi}{2}}\ket{\fpIIc}
\,.
$$
We would like to verify this phase difference of
$e^{\pi i/2}=i$ using the explicit
worldsheet considerations as above.
So we study the action of $\TdT$
on a scattering amplitude with initial state $\ket{\fpIIz}$
and final state $\ket{\fpIIc}.$
But because of $\Qsym$-conservation
(see \eqref{eqn:Qsym,k=2,n=1}), we have to have an even number
of $\ket{\fpIIc}$ in the final state.
So, we consider the scattering amplitude
of two $\ket{\fpIIz}$ states into two $\ket{\fpIIc}$ states.
(See \figref{fig:Scattering}.)
$\TdT$ acts as multiplication
by $i^2 = -1$ on this $4$-point scattering amplitude,
and this is what we wish to verify.

\FIGURE[t!]{
\begin{picture}(410,100)

\put(100,50){
\begin{picture}(200,100)

\put(-8,40){(a)}

\qbezier(-70,-45)(0,-20)(70,-45)
\qbezier(-75,-35)(30,0)(-75,35)
\qbezier(-70,45)(0,20)(70,45)
\qbezier(75,-35)(-30,0)(75,35)

\qbezier(-70,-45)(-75,-45)(-75,-35)
\qbezier(70,45)(75,45)(75,35)
\qbezier(-70,45)(-75,45)(-75,35)
\qbezier(70,-45)(75,-45)(75,-35)

\qbezier(-70,-45)(-70,-35)(-75,-35)
\qbezier(70,45)(70,35)(75,35)
\qbezier(-70,45)(-70,35)(-75,35)
\qbezier(70,-45)(70,-35)(75,-35)

\put(73,-47){\fpIIc}
\put(73,40){\fpIIc}
\put(-95,-47){\fpIIz}
\put(-95,40){\fpIIz}

\put(-10,-8){\Huge $\Sigma$}
\put(-45,-30){$\gamma_1$}
\put(-45,-30){\vector(-3,-1){24}}
\put(-45,28){$\gamma_1$}
\put(-45,30){\vector(-3,1){24}}
\put(35,28){$\gamma_2$}
\put(45,30){\vector(3,1){24}}
\put(35,-32){$\gamma_2$}
\put(45,-30){\vector(3,-1){24}}
\end{picture}}

\put(200,0){\line(0,1){100}}

\put(300,50){

\put(-8,30){(b)}

\begin{picture}(200,100)
\multiput(-70,-10)(0,20){2}{\line(1,0){140}}
\multiput(-70,0)(140,0){2}{
\qbezier(0,-10)(-10,5)(0,10)
\qbezier(0,-10)(10,-5)(0,10)
}
\put(-36,0){\vector(-1,0){25}}
\put(-34,-2){$\gamma_1^2$}
\put(36,0){\vector(1,0){25}}
\put(24,-2){$\gamma_2^2$}

\put(-95,-3){\fpIIzz}
\put(76,-3){\fpIIcc}
\put(-10,-5){$\Sigma'$}
\end{picture}}

\end{picture}
\caption{
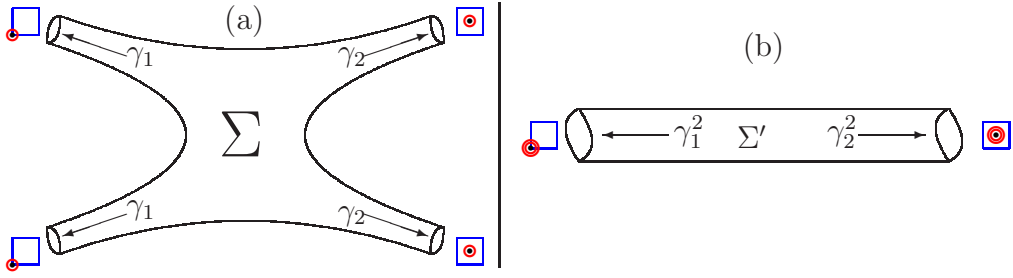
The image in target space of
string worldsheets representing
a scattering amplitude of string ground states.
(a) Scattering of two identical string ground states
into two other identical string ground states.
The boundary of the image of the worldsheet is
the union of four loops, corresponding to the
four string states; the phase acquired under
$B\rightarrow B+\frac{1}{2i}dz\wedge d\overline{z}$ is equal
to the phase that the same transformation induces in
(b) a worldsheet diagram for a $2$-point function
of string ground states of winding number $2$ .
}
\label{fig:Scattering}
}

With the parameterization
$$
0\le t<1,
$$
define the loops
$$
\gamma_1 = [t\mapsto(z=0,x_3=2\pi R t)]
\,,\qquad
\gamma_2 = [t\mapsto(z=\frac{1}{2}(1+\tau),x_3=2\pi R t)]
\,.
$$
We also use the standard loop-space product to define
the double-wound loops:
$$
\gamma_1^2 = [t\mapsto(z=0,x_3=4\pi R t)]
\,,\qquad
\gamma_2^2 = [t\mapsto(z=\frac{1}{2}(1+\tau),x_3=4\pi R t)]
\,.
$$
In addition define the loops
$$
\gpifa=[t\mapsto(z=t,x_3=0)]
\,,\qquad
\gpifb=[t\mapsto(z=t \tau,x_3=0)]
\,,
$$
$$
\gpifd=[t\mapsto(z=t (\tau+1),x_3=0)]
\,.
$$
(See \eqref{eqn:gpibfafb} for similar definitions.)
The phase $\Phi$ is clearly additive, so
$$
\Phi(\gamma_1^{-1},\gamma_1^{-1},\gamma_2,\gamma_2)
\equiv
\Phi(\gamma_1^{-2},\gamma_2^2)
\pmod{2\pi}
\,.
$$
We calculate the latter as follows.
First note that the following $3$-point phase vanishes:
$$
\Phi(\gamma_1^{-2},\gamma_2^2,\gpifd)\equiv 0\pmod{2\pi}
\,.
$$
To quickly see this, take $\Sigma$ to be the following surface:
$$
\Sigma=[(\wsig,\wtau)\mapsto
(z=\tfrac{1}{2}\wsig(1+\tau),x_3=4\pi R\wtau)]
\,,\qquad
0\le\wsig,\wtau<1\,,
$$
for which $\partial\Sigma=\gamma_1^{-2}\cup\gamma_2^2\cup\gpifd$
and $\int_\Sigma\AreaForm_F=0.$
We also have
$$
\Phi(\gpifa^{-1},\gpifb^{-1})\equiv 0\pmod{2\pi}
\,,
$$
which can be verified by taking
$$
\Sigma=[(\wsig,\wtau)\mapsto
(z=\wsig,x_3=2\pi R\wtau)]
\,,\qquad
0\le\wsig,\wtau<1\,.
$$
Finally, note that
$$
\Phi(\gpifa^{-1},\gpifb^{-1},\gpifd)\equiv \pi\pmod{2\pi}
\,.
$$
To see this, consider $\Sigma$ that is confined to one fiber
at $x_3=0$ and is bounded by the three cycles
$\gpifa^{-1},\gpifb^{-1},\gpifd.$ This $\Sigma$ is a triangle
and integrating $\AreaForm_F$ on it gives $\pi.$
The above results imply that
$$
\Phi(\gamma_1^{-2},\gamma_2^2)\equiv
\Phi(\gpifd)\equiv\pi
\pmod{2\pi}\,,
$$
as claimed.

As another example, consider
a $3$-string scattering amplitude
$\ket{\fpIIxy}\rightarrow\ket{\fpIIz\fpIIc}$
(which preserves both $\Z_2$ momentum and winding).
The action of $\TdT$ on this amplitude will tell
us the phase difference between the $\TdT$-eigenvalue
of $\ket{\fpIIxy}$ and the $\TdT$-eigenvalue of
$\ket{\fpIIz\fpIIc}.$
(This nontrivial phase will have an important consequence
in \secref{subsec:other-sectors}.)
{}From \eqref{eqn:TdT-II1}
and \eqref{eqn:TdT-II2}, we know that this phase difference
is $e^{-\pi i/2}$.
To verify it, define the loop,
$$
\gamma_3 = [t\mapsto(z=\tfrac{1}{2},x_3=4\pi R t)]
\,.
$$
The loop $\gamma_3$ corresponds to the state
$\ket{\fpIIxy}$, since at $t=\tfrac{1}{2}$ we have
$(z=\tfrac{1}{2},x_3=2\pi R)\simeq (z=\tfrac{1}{2}\tau,x_3=0)$
by \eqref{eqn:zx}.
What we need then is the phase $\Phi(\gamma_3,\gamma_1^{-1},\gamma_2^{-1}).$
To calculate it, consider the following two surfaces
(here $\tau=i$):
$$
\Sigma_1=[(\wsig,\wtau)\mapsto
(z=\tfrac{1}{2}\wsig,x_3=2\pi R\wtau)]
\,,\qquad
0\le\wsig,\wtau<1\,,
$$
and
$$
\Sigma_2=[(\wsig,\wtau)\mapsto
(z=\tfrac{1}{2}\wsig\tau,x_3=2\pi R\wtau)]
\,,\qquad
0\le\wsig,\wtau<1\,.
$$
Also, define the curve:
$$
\delta = [t\mapsto\left\{
\begin{array}{ll}
(z=2t,x_3=0) & 0\le t\le\frac{1}{4} \\
(z=\tfrac{1}{2}+2(t-\frac{1}{4})\tau,x_3=0) &
\frac{1}{4}\le t\le\frac{1}{2} \\
(z=2(\tfrac{3}{4}-t)+\tfrac{1}{2}\tau,x_3=0) &
\frac{1}{2}\le t\le\frac{3}{4} \\
(z=2(1-t)\tau,x_3=0) &
\frac{3}{4}\le t\le 1 \\
\end{array}\right\}]
\,.
$$
Note that $\delta$ traces a square
with vertices
$z=0,\tfrac{1}{2},\tfrac{1}{2}(1+\tau),\tfrac{1}{2}\tau$
inside the fiber over $x_3=0$, and the area bounded by it
is $\tfrac{1}{4}.$
To complete the calculation of the phase, we note that
$$
\partial(\Sigma_1\cup\Sigma_2)=
\gamma_1^{-1}\cup\gamma_2^{-1}\cup\gamma_3\cup\delta
\,,
\quad
\text{and}\quad \int_{\Sigma_1\cup\Sigma_2}\AreaForm_F=0\,.
$$
Thus,
$$
\Phi(\gamma_1^{-1},\gamma_2^{-1},\gamma_3)
\equiv-\Phi(\delta)\equiv -\frac{\pi}{2}
\pmod{2\pi}\,.
$$


\def\fpCz{{
\begin{picture}(26,10)
\put(3,-2){\begin{picture}(20,12)
\thinlines
\color{blue}
\multiput(0,0)(0,10){2}{\line(1,0){20}}
\multiput(0,0)(20,0){2}{\line(0,1){10}}
\thicklines
\color{black}
\put(0,0){\circle*{2}}
\color{red}
\put(0,0){\circle{4}}
\end{picture}}
\end{picture}
}}
\def\fpCx{{
\begin{picture}(26,10)
\put(3,-2){\begin{picture}(20,12)
\thinlines
\color{blue}
\multiput(0,0)(0,10){2}{\line(1,0){20}}
\multiput(0,0)(20,0){2}{\line(0,1){10}}
\thicklines
\color{black}
\put(10,0){\circle*{2}}
\color{red}
\put(10,0){\circle{4}}
\end{picture}}
\end{picture}
}}
\def\fpCy{{
\begin{picture}(26,10)
\put(3,-2){\begin{picture}(20,12)
\thinlines
\color{blue}
\multiput(0,0)(0,10){2}{\line(1,0){20}}
\multiput(0,0)(20,0){2}{\line(0,1){10}}
\thicklines
\color{black}
\put(0,5){\circle*{2}}
\color{red}
\put(0,5){\circle{4}}
\end{picture}}
\end{picture}
}}
\def\fpCc{{
\begin{picture}(26,10)
\put(3,-2){\begin{picture}(20,12)
\color{blue}
\thinlines
\multiput(0,0)(0,10){2}{\line(1,0){20}}
\multiput(0,0)(20,0){2}{\line(0,1){10}}
\thicklines
\color{black}
\put(10,5){\circle*{2}}
\color{red}
\put(10,5){\circle{4}}
\end{picture}}
\end{picture}
}}
\def\fpCzy{{
\begin{picture}(26,10)
\put(3,-2){\begin{picture}(20,12)
\thinlines
\color{blue}
\multiput(0,0)(0,10){2}{\line(1,0){20}}
\multiput(0,0)(20,0){2}{\line(0,1){10}}
\thicklines
\color{black}
\put(0,5){\circle*{2}}
\put(0,0){\circle*{2}}
\color{red}
\put(0,5){\circle{4}}
\put(0,0){\circle{4}}
\end{picture}}
\end{picture}
}}
\def\fpCxy{{
\begin{picture}(26,10)
\put(3,-2){\begin{picture}(20,12)
\thinlines
\color{blue}
\multiput(0,0)(0,10){2}{\line(1,0){20}}
\multiput(0,0)(20,0){2}{\line(0,1){10}}
\thicklines
\color{black}
\put(10,0){\circle*{2}}
\put(0,5){\circle*{2}}
\color{red}
\put(10,0){\circle{4}}
\put(0,5){\circle{4}}
\end{picture}}
\end{picture}
}}
\def\fpCzc{{
\begin{picture}(26,10)
\put(3,-2){\begin{picture}(20,12)
\thinlines
\color{blue}
\multiput(0,0)(0,10){2}{\line(1,0){20}}
\multiput(0,0)(20,0){2}{\line(0,1){10}}
\thicklines
\color{black}
\put(10,5){\circle*{2}}
\put(0,0){\circle*{2}}
\color{red}
\put(10,5){\circle{4}}
\put(0,0){\circle{4}}
\end{picture}}
\end{picture}
}}
\def\fpCxc{{
\begin{picture}(26,10)
\put(3,-2){\begin{picture}(20,12)
\thinlines
\color{blue}
\multiput(0,0)(0,10){2}{\line(1,0){20}}
\multiput(0,0)(20,0){2}{\line(0,1){10}}
\thicklines
\color{black}
\put(10,5){\circle*{2}}
\put(10,0){\circle*{2}}
\color{red}
\put(10,5){\circle{4}}
\put(10,0){\circle{4}}
\end{picture}}
\end{picture}
}}



\section{Warm-up: C-twist}
\label{sec:Ctwist}

In \secref{subsec:S-twist} we were interested only
in $\zg\in\SL(2,\Z)$ that act nonperturbatively
and fix a strongly-coupled value of $\tau.$
But there is another element $\zg$ that we can consider:
$\zg=\begin{pmatrix} -1 & 0\\0 & -1 \\ \end{pmatrix}.$
It preserves every $\tau$, and acts on the theory as
charge conjugation. It corresponds to $\pht=\pi$ and
has order $\ord=2.$
We will refer to this twist as a {\it C-twist}.

Compactification with C-twist actually preserves the full
\SUSY{8} supersymmetry in three dimensions.
In addition, we can keep $\Imx\tau\gg 1$,
so as to have a weakly-coupled theory.
We will now study the C-twist and demonstrate
some of the ideas in the previous section explicitly in this setting.
We will study only the cases of $U(1)$ and $U(2)$ gauge group.
The case of $U(n)$ with $n\ge 3$ is more involved and will
not be addressed here.


\subsection{Group theory}
\label{subsec:C-GT}

Combining the C-twist with the appropriate R-twist,
and adjusting \eqref{eqn:RtwistPsi}-\eqref{eqn:RtwistA}
 to include a charge
conjugation, we get the following boundary conditions
\bear
\left\lbrack\bpsi_a^\dta(x_0,x_1,x_2,x_3+2\pi\xR)
\right\rbrack^* &=&
i\Lambda^{-1}\psi^a_\a(x_0,x_1,x_2,x_3)\Lambda\,,
\qquad a=1,\dots,4\,,
\label{eqn:CtwistPsi}\\
\left\lbrack\Phi^I(x_0,x_1,x_2,x_3+2\pi\xR)\right\rbrack^* &=&
-\Lambda^{-1}\Phi^I(x_0,x_1,x_2,x_3)\Lambda\,,
\qquad I=1\dots 6\,,
\label{eqn:CtwistPhi}\\
-A^*_\mu(x_0,x_1,x_2,x_3+2\pi\xR) &=&
\Lambda^{-1}A_\mu(x_0,x_1,x_2,x_3)\Lambda
-i\Lambda^{-1}\px{\mu}\Lambda\,,
\label{eqn:CtwistA}
\eear
where $[\cdots]^*$ is the complex conjugate $n\times n$ matrix
(not the adjoint matrix),
and $\Lambda$ is an arbitrary gauge transformation.

Now consider a closed path $C$ at a constant $x_3=0$
that starts and ends  at the origin,
and consider the $U(n)$-holonomy $\Gg=P \exp (i\oint_C A).$
Set $\Omega=\Lambda(0,0,0,0).$
The combined charge conjugation and gauge transformation
act on $\Gg$ as
$$
\Gg \mapsto [\Omega^{-1}\Gg\Omega]^*.
$$
We will need the invariant subgroup of $U(n)$, which is the
subgroup of solutions to
$$
\Gg=[\Omega^{-1}\Gg\Omega]^*.
$$
We denote it by $\Ginv_\Omega\subset U(n)$, since
it generally depends on $\Omega.$
We now proceed to study the $U(1)$ and $U(2)$ cases in more detail.


\subsection{$U(1)$ gauge group}
\label{subsec:U(1)-C}

In this case $\Ginv_\Omega=O(1)\simeq\Z_2.$
At low-energy,
no propagating degrees of freedom survive
the twist \eqref{eqn:CtwistPsi}-\eqref{eqn:CtwistA}.
The low-energy gauge group is $O(1)\simeq\Z_2$,
which means that when we compactify the 2D space
on $T^2$ we can have nontrivial $\Z_2$ Wilson lines
around the two independent $1$-cycles of $T^2.$
Let $w_a\in\Z_2\simeq\{1,-1\}$ be the $\Z_2$ Wilson
line along $1$-cycle $a$ ($a=1,2$).
(For convenience, we take the $\Z_2$ group to be
multiplicative instead of additive.)
The four vacua are then labeled by $\ket{w_1,w_2}$,
and we have a mass gap of $1/(2\xR).$

Now consider the type-IIA dual description of the vacua,
as in \secref{subsec:GS}. The effect of charge conjugation here is that it rotates the (dual) $T^2$ by $\pht=\pi$. The four vacua are therefore
$$
\ket{\fpCz}=\ket{\{0\}}\,,\quad
\ket{\fpCx}=\ket{\{\tfrac{1}{2}\}}\,,\quad
\ket{\fpCy}=\ket{\{\tfrac{1}{2}\tau\}}\,,\quad
\ket{\fpCc}=\ket{\{\tfrac{1}{2}(1+\tau)\}}\,.
$$
In order to match these type-IIA states with the
field theory vacua $\ket{w_1,w_2}$, we define, as
in \secref{subsec:P},
the $\Z_2$ momentum operators
\be\label{eqn:PP-C-U(1)}
\left.
\begin{array}{l}
\Psym_1\ket{[\tfrac{1}{2}M+\tfrac{1}{2}N\tau]}
=\ket{[\tfrac{1}{2}(1-M)+\tfrac{1}{2}N\tau]}\,,
\\
\Psym_2\ket{[\tfrac{1}{2}M+\tfrac{1}{2}N\tau]}
=\ket{[\tfrac{1}{2}M+\tfrac{1}{2}(1-N)\tau]}\,,
\end{array}\right\}
\ee
and the $\Z_2$ winding number operators,
as in \secref{subsec:W},
\be\label{eqn:QQ-C-U(1)}
\left.
\begin{array}{l}
\Qsym_1\ket{[\tfrac{1}{2}M+\tfrac{1}{2}N\tau]}
=(-1)^M\ket{[\tfrac{1}{2}M+\tfrac{1}{2}N\tau]}\,,
\\
\Qsym_2\ket{[\tfrac{1}{2}M+\tfrac{1}{2}N\tau]}
=(-1)^N\ket{[\tfrac{1}{2}M+\tfrac{1}{2}N\tau]}\,.
\end{array}\right\}
\ee
We will now argue that
\be\label{eqn:PQsymC}
\Psym_1 = (-1)^{\mf_1}
\,,\quad
\Psym_2 = (-1)^{\ef_1}
\,,\quad
\Qsym_1 = (-1)^{\ef_2}
\,,\quad
\Qsym_2 = (-1)^{\mf_2}
\,,
\ee
where $\ef_1,\ef_2$ are the electric flux operators in directions $1,2$ respectively,
and $\mf_1,\mf_2$ are the magnetic flux operators.

Equations \eqref{eqn:PQsymC} can be derived
by following the chain of dualities of
\tabref{tab:Dualities} backwards.
Starting
on the type-IIA side (the last row of \tabref{tab:Dualities}),
take a state with Kaluza--Klein momenta $p_1,p_{10}\in\Z$
 in directions $x_1, x_{10}.$ (We can assume that the state
is localized in the $x_3$ direction.)
The unitary operator $\Psym_1$ acts as a translation in
the direction of $x_{10}$
and therefore multiplies the state by the phase
$e^{\pi i p_{10}}$.
Similarly, $\Psym_2$ multiplies the state
by $e^{\pi i p_1}$.
Following the chain of dualities backwards in \tabref{tab:Dualities},
we find that on the type-IIB side $p_1$ becomes
fundamental string (F1) winding number in direction $x_1$,
while $p_{10}$ becomes D1 winding number in direction $x_1.$ (See \tabref{tab:IIB-IIA}.)
The Kaluza--Klein state on the type-IIA side therefore becomes
a $(p,q)$-string, with $p=p_1$ and $q=p_{10}.$
Bound to $n$ D3-branes, these quantum numbers become
\cite{Witten:1995im,Douglas:1995bn}
$\ef_1=p_1$ units of electric flux in direction $1$
and $\mf_1=p_{10}$ units of magnetic flux in the same direction.
Similarly,
$\Qsym_1$ corresponds to the exponential of string winding number in direction $10$, and
$\Qsym_2$ to the exponential of string winding number in direction $1.$
On the type-IIB side, these become fundamental string winding number
and D1-brane winding number in direction $2.$

Now, let's relate the $\ket{w_1,w_2}$ basis (on the field theory/type-IIB side)
to the $\ket{[\tfrac{1}{2}M+\tfrac{1}{2}N\tau]}$ basis on the type-IIA side.
On the field theory side,
$\Qsym_1=(-1)^{\ef_2}$
can be interpreted as the operator of a large gauge transformation
acting on the components of the gauge field as
$$
A_1\rightarrow A_1
\,,\qquad
A_2\rightarrow A_2 +\frac{1}{2\xL_2}
\,.
$$
Similarly,
$\Psym_2=(-1)^{\ef_1}$
can be interpreted as the operator of a large gauge transformation
$$
A_1\rightarrow A_1 +\frac{1}{2\xL_1}
\,,\qquad
A_2\rightarrow A_2
\,.
$$
We can therefore identify the action on eigenstates of Wilson lines as
\be\label{eqn:PQonWW}
\Qsym_1\ket{w_1,w_2}=\ket{w_1,-w_2}\,,\qquad
\Psym_2\ket{w_1,w_2}=\ket{-w_1,w_2}\,.
\ee
Comparing \eqref{eqn:PQonWW} to
\eqref{eqn:PP-C-U(1)}-\eqref{eqn:QQ-C-U(1)}
we find
\be\label{eqn:MNww}
\ket{[\tfrac{1}{2}M+\tfrac{1}{2}N\tau]}_{\text{IIA}}
=\frac{1}{\sqrt{2}}\sum_{M'=0,1} (-1)^{M' M}
\ket{(-1)^N,(-1)^{M'}}_{\text{IIB}}\,.
\ee


Now consider the operators $\Psym_1,\Qsym_2$,
which according to \eqref{eqn:PQsymC} are related to magnetic flux.
Using \eqref{eqn:PP-C-U(1)}-\eqref{eqn:QQ-C-U(1)} and \eqref{eqn:MNww},
we find
\be\label{eqn:Psym1Qsym2}
\Psym_1\ket{w_1,w_2}=w_2\ket{w_1,w_2}
\,,\qquad
\Qsym_2\ket{w_1,w_2}=w_1\ket{w_1,w_2}
\,.
\ee
So $w_1$ is the eigenvalue of magnetic flux $(-1)^{\mf_2}$,
and $w_2$ is the eigenvalue of magnetic flux $(-1)^{\mf_1}$.

The connection between the discrete $\Z_2$ Wilson line $w_1$ and the
magnetic flux $\mf_2$ can be understood as follows.
Let us pick a uniform gauge field with Wilson line $w_1=-1$:
$A = \frac{1}{2\xL_1}dx_1.$
The charge conjugate field is $-A$, so we have to pick a nonzero
$\Lambda$ in \eqref{eqn:CtwistA}.
Specifically, the gauge transformation that converts
$A$ to $-A$ is $\Lambda=\exp(-i x_1/\xL_1).$
This gauge transformation accompanies the coordinate transformation
$x_3\rightarrow x_3+2\pi\xR$, and for an ordinary $T^3$ compactification
it would be interpreted \cite{'tHooft:1977hy}
as one unit of magnetic flux in direction $2$, i.e., $\mf_2=1.$
The connection between $w_2$ and $\mf_1$ is similar.

We can now understand the action of $\Qsym_1$ and $\Psym_2$ as follows.
According to \eqref{eqn:PQsymC}, $\Psym_2=(-1)^{\ef_1}$
and therefore acts as a discontinuous gauge transformation
with gauge parameter $\wLambda(x_1,x_2)=\exp(-i x_1/2\xL_1).$
Such a gauge transformation does not preserve the boundary
conditions \eqref{eqn:CtwistA}, because charge conjugation
converts $\wLambda$ to $\wLambda^{-1}=\exp(i x_1/2\xL_1)$,
but this can be fixed by modifying
the gauge transformation $\Lambda$
that appears in \eqref{eqn:CtwistA} to
$$
\Lambda\rightarrow\Lambda e^{\frac{i x_1}{\xL_1}}
\,.
$$
This implies that
$\Psym_2$ changes the magnetic flux $\mf_2$
by one unit (modulo $2$).
Similarly, $\Qsym_1=(-1)^{\ef_2}$ changes the magnetic flux $\mf_1$
by one unit. Since, as we have seen in \eqref{eqn:Psym1Qsym2}
[combined with \eqref{eqn:PQsymC}], $w_1,w_2$ can be identified
with the magnetic fluxes $(-1)^{\mf_2},(-1)^{\mf_1}$, we recover
the expressions \eqref{eqn:PQonWW}
for the action of $\Qsym_1,\Psym_2$ on states.
We have therefore completely mapped the field theory
ground states to the type-IIA ground states.

Let us conclude this subsection with a few additional comments.
First we note that
the magnetic flux $\mf_3$ has to vanish, because charge conjugation
acts on it as $\mf_3\rightarrow -\mf_3$, and this cannot be fixed
by any gauge transformation $\Lambda$ in \eqref{eqn:CtwistA}.
The electric flux $\ef_3$ therefore also vanishes by S-duality.
Finally, let us also write down the T-duality
transformations $\TdS,\TdT.$
On the type-IIB (field theory) side they act geometrically,
so we have
$$
\TdS\ket{w_1,w_2}=\ket{w_2,w_1}\,,\qquad
\TdT\ket{w_1,w_2}=\ket{w_1,w_1 w_2}\,.
$$


\subsection{$U(2)$ gauge group}
\label{subsec:U(2)-C}

Let's now study the case of gauge group $U(2).$
On the type-IIA side, a basis state is of one of two types:
(i) a single string with winding number $2$;
or (ii) two strings with winding number $1$.

The single-particle string states of winding number $2$ are
built from one of the four types of states
$$
\ket{[z,-z]}\,,\qquad
\ket{[z,\tfrac{1}{2}-z]}\,,\qquad
\ket{[z,\tfrac{1}{2}\tau-z]}\,,\qquad
\ket{[z,\tfrac{1}{2}+\tfrac{1}{2}\tau-z]}\,,
$$
where $z$ is a free parameter on $T^2/\Z_2$, which needs to be quantized.
In addition, the location of the strings in the $\R^6$ transverse
directions is free and needs to be quantized, too.
This results in a continuous spectrum.

The two-particle states are given by combining two strings of winding number $1.$
Each of these strings can be at any of the four locations
studied in \secref{subsec:U(1)-C}, and since they are identical bosons,
the order is not important. We denote the states by
$\ket{\{[z],[z']\}}$, where $z,z'\in\C/(\Z+\Z\tau)$ are any one of
$0,\tfrac{1}{2},\tfrac{1}{2}\tau,\tfrac{1}{2}+\tfrac{1}{2}\tau.$
Altogether we get $10$ states (two identical bosons
with $4$ single-particle states).
Similarly to \eqref{eqn:PP-C-U(1)},
we define the symmetry operators $\oPsym_1,\oPsym_2$ by
\be\label{eqn:PP-C-U(2)}
\left.\begin{array}{l}
\oPsym_1\ket{\{[\tfrac{1}{2}M+\tfrac{1}{2}N\tau],[\tfrac{1}{2}M'+\tfrac{1}{2}N'\tau]\}}
=\ket{[\tfrac{1}{2}(1-M)+\tfrac{1}{2}N\tau],[\tfrac{1}{2}(1-M')+\tfrac{1}{2}N'\tau]}
\\
\\
\oPsym_2\ket{\{[\tfrac{1}{2}M+\tfrac{1}{2}N\tau],[\tfrac{1}{2}M'+\tfrac{1}{2}N'\tau]\}}
=\ket{[\tfrac{1}{2}M+\tfrac{1}{2}(1-N)\tau],[\tfrac{1}{2}M'+\tfrac{1}{2}(1-N')\tau]}
\\
\end{array}\right\}\,,
\ee
and similarly to \eqref{eqn:QQ-C-U(1)},
we define $\oQsym_1,\oQsym_2$ by
\be\label{eqn:QQ-C-U(2)}
\left.\begin{array}{l}
\oQsym_1\ket{\{[\tfrac{1}{2}M+\tfrac{1}{2}N\tau],[\tfrac{1}{2}M'+\tfrac{1}{2}N'\tau]\}}
=(-1)^{M+M'}
\ket{\{[\tfrac{1}{2}M+\tfrac{1}{2}N\tau],[\tfrac{1}{2}M'+\tfrac{1}{2}N'\tau]\}}
\\
\\
\oQsym_2\ket{\{[\tfrac{1}{2}M+\tfrac{1}{2}N\tau],[\tfrac{1}{2}M'+\tfrac{1}{2}N'\tau]\}}
=(-1)^{N+N'}
\ket{\{[\tfrac{1}{2}M+\tfrac{1}{2}N\tau],[\tfrac{1}{2}M'+\tfrac{1}{2}N'\tau]\}}
\\
\end{array}\right\}
\,.
\ee
Note that the $4$ operators
$\oPsym_1,\oPsym_2,\oQsym_1,\oQsym_2$ are mutually commuting.

Next, let us see how to get this spectrum from
the field theory (type-IIB) side.
Since $U(2)=[SU(2)\times U(1)]/\Z_2$,
we can start by separately discussing the $SU(2)$ and $U(1)$
degrees of freedom, and then consider how they
combine to form states of the full $U(2)$ theory.

We begin with the $SU(2)$ degrees of freedom.
Since $SU(2)$ is pseudo-real, charge conjugation
is equivalent to a gauge transformation.
Explicitly, the gauge transformation is realized by
the matrix $i\Psig_2\in SU(2)$
(we denote the Pauli matrices by $\Psig_1,\Psig_2,\Psig_3$),
and for an adjoint-valued field $\phi$ we have
$$
-\phi^* = (i\Psig_2)^{-1}\phi (i\Psig_2)\,.
$$
Thus, combining the extra gauge parameter $i\Psig_2$ with $\Lambda$
in \eqref{eqn:CtwistPsi}-\eqref{eqn:CtwistA}, we find that
the C-twist has no effect on the $SU(2)$ degrees of freedom.
As far as the $SU(2)$ degrees of freedom are concerned,
we therefore have a standard compactification of \SUSY{4} $SU(2)$ SYM
on $T^3$, preserving $16$ supersymmetries, and
we are interested in the normalizable ground states.

Let $\ef'_1,\ef'_2,\ef'_3$ be the $\Z_2$ 't Hooft electric fluxes
of a state of the $SU(2)$ theory, and
$\mf'_1,\mf'_2,\mf'_3$ the $\Z_2$ 't Hooft magnetic fluxes.\footnote{
Here we do not restrict the magnetic or electric fluxes, since we need
to combine the $SU(2)$ degrees of freedom with the $U(1)$ later on.
Of course, if we had just the $SU(2)$ degrees of freedom, we would have
had to set all magnetic fluxes to zero, and if we had just $SO(3)\simeq SU(2)/\Z_2$
we would have had to set all electric fluxes to zero.}
It turns out \cite{Henningson:2007dq}
that there is one supersymmetric ground state
for every combination of 't Hooft fluxes that satisfies
\be\label{eqn:effmffz}
\ef'_1\mf'_2-\ef'_2\mf'_1 =
\ef'_1\mf'_3-\ef'_3\mf'_1 =
\ef'_2\mf'_3-\ef'_3\mf'_2 = 0\,.
\ee
We denote the corresponding ground state by
$\ket{\ef'_1,\ef'_2,\ef'_3,\mf'_1,\mf'_2,\mf'_3}_{SU(2)}$,
and by convention this state is identically zero if \eqref{eqn:effmffz}
is not satisfied. We will soon require $\ef'_3=\mf'_3=0$,
and then the only nontrivial condition in \eqref{eqn:effmffz} is
\be\label{eqn:efmfz}
\ef'_1\mf'_2-\ef'_2\mf'_1 = 0\,.
\ee
There are exactly $10$
combinations of the $\Z_2$ fluxes
$\ef'_1,\ef'_2,\mf'_1,\mf'_2$ that satisfy \eqref{eqn:efmfz}.

Let us comment that the result on the number of $SU(2)$ ground
states can be obtained in several ways.
One way is to count the supersymmetric bound states of
$2$ D$3$-branes on $T^3.$ This system is described at low-energy
by $U(2)$ super Yang--Mills theory, and its Hilbert space
is a tensor product of sectors of $U(1)$ and $SU(2)$ Hilbert spaces
with the $SU(2)$ electric and magnetic fluxes determined
by the modulo $2$ residue of the $U(1)$ electric and magnetic fluxes.
The $U(1)$ electric and magnetic fluxes correspond to
fundamental string (F$1$) and D$1$-charge.
The result, which can be established by T-duality
on the three directions of $T^3$,
is that there is one supersymmetric bound state for each
combination of the electric and magnetic fluxes.
It is a ``bound state at threshold'' if all magnetic fluxes vanish,
and not at threshold otherwise. The condition
\eqref{eqn:effmffz} ensures that the total momentum
carried by the $U(1)$ flux is an integer.
Alternatively, the result can be established entirely in field theory
(with the assumption that the Witten index
is identical to the number of ground states)
\cite{Henningson:2007dq},
using results on the number of normalizable ground states
in theories with $16$ supersymmetries
\cite{Claudson:1984th,Witten:1995im,Yi:1997eg,Sethi:1997pa}.

Next, let us discuss the $U(1)$ degrees of freedom.
In \secref{subsec:U(1)-C} we showed that the $U(1)$ theory has
$4$ ground states, $\ket{w_1,w_2}$ (with $w_1,w_2=\pm 1$).
However, the discussion of \secref{subsec:U(1)-C}
needs to be modified in order to be
applicable to the $U(2)$ theory, as we shall now explain.
Generally speaking, the problem is that
the pure $U(1)$ theory is invariant under certain
large gauge transformations that can no longer be considered
good gauge transformations in the $U(2)$ theory.
To explain this in detail, we need to discuss the electric
and magnetic fluxes more thoroughly.

Consider a $U(2)$ gauge configuration $A^{U(2)}$ which we regard
locally as a $2\times 2$ matrix of $1$-forms.
{}From this matrix we construct a $U(1)$ gauge field by taking the trace,
$A^{U(1)}=\tr A^{U(2)}.$
This normalization is actually a matter of convention.
For example, for a standard toroidal compactification of $U(n)$
gauge theory on $T^3$, we can choose to define
$A^{U(1)}=\tr A^{U(n)}$, which corresponds to a surjective map
$U(n)\xrightarrow{\det}U(1)$,
or we can choose to define $A^{U(1)}=\tfrac{1}{n}\tr A^{U(n)}$,
which corresponds to an injective map
$U(1)\xrightarrow{\cdot I} U(n).$
Neither choice is optimal, however, because with the second choice
we are forced to include sectors with fractional magnetic flux
(like $\tfrac{1}{n}$), and with the first choice
we are forced to include sectors with fractional electric flux.
We will see a manifestation of this below \eqref{eqn:wwLam}
where we will 
have to include sectors for which 
a proper gauge transformation ($\det\wLambda$) does not 
act as the identity operator. In section \secref{subsec:U(1)subset}
we will choose to work with the second convention
but for the present section we proceed with 
$A^{U(1)}=\tr A^{U(2)}.$

We now introduce $U(1)$ magnetic fluxes $\mf_1,\mf_2,\mf_3.$
We have already seen in \secref{subsec:U(1)-C} that the C-twist requires
$\mf_3=0.$
States with integer $U(1)$ magnetic fluxes $\mf_1,\mf_2$
can be realized by the following classical solution to \eqref{eqn:CtwistA}:
$$
A^{U(2)} = \begin{pmatrix}
\frac{\mf_2}{2\xL_1}dx_1+\frac{\mf_1}{2\xL_2}dx_2 & 0 \\ 0 & 0 \\
\end{pmatrix}\,,
\qquad
\Lambda =  \begin{pmatrix}
e^{-\frac{i\mf_2 x_1}{\xL_1}-\frac{i\mf_1 x_2}{\xL_2}} & 0 \\ 0 & 1 \\
\end{pmatrix}\,.
$$
Following \eqref{eqn:PQsymC} we define
\be\label{eqn:PQsymC-U(2)-a}
\Psym_1 \equiv (-1)^{\mf_1}
\,,\qquad
\Qsym_2 \equiv (-1)^{\mf_2}
\,.
\ee
Now, let's find the $SU(2)$ magnetic fluxes $\mf_1',\mf_2'\in\Z_2$
of this configuration.
Locally, we can split $\Lambda$ into  $U(1)$ and $SU(2)$ parts as
$$
\Lambda =
e^{-\frac{i\mf_2 x_1}{2\xL_1}-\frac{i\mf_1 x_2}{2\xL_2}}
\begin{pmatrix}
e^{-\frac{i\mf_2 x_1}{2\xL_1}-\frac{i\mf_1 x_2}{2\xL_2}} & 0 \\ 0 &
e^{\frac{i\mf_2 x_1}{2\xL_1}+\frac{i\mf_1 x_2}{2\xL_2}} \\
\end{pmatrix}\,.
$$
We can read off the $SU(2)$ 't Hooft magnetic fluxes
$\mf_1',\mf_2'$ from the $SU(2)$ matrix on the right-hand side.
This shows that, as in ordinary toroidal compactifications,
the $SU(2)$ 't Hooft magnetic flux is determined by
the $U(1)$ magnetic flux according to
\be\label{eqn:Lam-mm}
0 = \mf_1 + \mf_1' = \mf_2 + \mf_2'
{\pmod 2}\,.
\ee
Similarly, $\mf_3'=\mf_3\pmod 2$, and since $\mf_3=0$ we get $\mf_3'=0.$

Let us now turn to electric fluxes.
Consider the large gauge transformation
\be\label{eqn:wwLam}
\wLambda = \begin{pmatrix}
e^{\frac{i x_1}{\xL_1}} & 0 \\ 0 & 1 \\
\end{pmatrix}
=
e^{\frac{i x_1}{2\xL_1}}
\begin{pmatrix}
e^{\frac{i x_1}{2\xL_1}} & 0 \\ 0 & e^{-\frac{i x_1}{2\xL_1}} \\
\end{pmatrix}
\,.
\ee
All states of the $U(2)$ theory must be invariant under $\wLambda.$
On the right-hand side of \eqref{eqn:wwLam} we decomposed
$\wLambda$ locally
into a $U(1)$ gauge transformation and an $SU(2)$ gauge transformation.
However, note that
the latter actually
generates a discontinuous gauge
transformation of $SU(2)$---applying this
gauge transformation locally is
equivalent to acting with the operator $(-1)^{\ef_1'}$,
according to 't Hooft's definition \cite{'tHooft:1977hy}.
On the $U(1)$ degrees of freedom,
with our normalization, $\wLambda$ acts as
$\det\wLambda=\exp(i x_1/\xL_1)$, which is a proper gauge
transformation.

In \secref{subsec:U(1)-C} we defined the operators
$\Psym_2,\Qsym_1$
which correspond to gauge transformations by
discontinuous gauge parameters
$\exp(i x_1/2\xL_1)$ and $\exp(i x_2/2\xL_2)$, respectively.
The $U(1)$ part of the gauge transformation $\wLambda$
can therefore be identified with $\Psym_2^2$, and we conclude that
in the $U(2)$ theory all states must satisfy
\be\label{eqn:QPef}
\Psym_2^2(-1)^{\ef_1'} = \Qsym_1^2(-1)^{\ef_2'} = 1.
\ee
In the realization \eqref{eqn:PP-C-U(1)}-\eqref{eqn:QQ-C-U(1)}
of $\Psym_1,\Psym_2,\Qsym_1,\Qsym_2$,
we had $\Qsym_1^2=\Psym_2^2=1$ identically,
but in the present context of the $U(2)$ theory, this is too restrictive.
For example, $\Psym_2^2$ corresponds to
the gauge parameter
$$
\begin{pmatrix}
e^{\frac{i x_1}{2\xL_1}} & 0 \\
0 & e^{\frac{i x_1}{2\xL_1}} \\
\end{pmatrix}
\,,
$$
which is discontinuous in $U(2)$
and therefore not required to be the identity on physical states.

Thus, the Hilbert space of the $U(1)$ theory has to be a representation of
the algebra generated by $\Psym_1,\Psym_2,\Qsym_1,\Qsym_2$
with the relations
\be\label{eqn:PQalgI}
\begin{split}
&\Psym_1^2=\Qsym_2^2=1
\,,\quad
\Psym_1\Psym_2 = \Psym_2\Psym_1
\,,\quad
\Qsym_1\Qsym_2 = \Qsym_2\Qsym_1
\,,\\
&\Psym_i\Qsym_j = (-1)^{\delta_{ij}}\Qsym_j\Psym_i
\,,\qquad i,j=1,2,
\end{split}
\ee
and we can add the conditions
\be\label{eqn:PQalgII}
\Qsym_1^4=\Psym_2^4=1
\,,
\ee
since $\Psym_2^4$ and $\Qsym_1^4$ are generated by the continuous
large gauge transformations $e^{\frac{i x_1}{\xL_1}}$ and
$e^{\frac{i x_2}{\xL_2}}$, and those do have to be the identity
on physical states.

Note that $\Qsym_1^2$ and $\Psym_2^2$ are central elements of this algebra,
and all irreducible representations with $\Qsym_1^2=\Psym_2^2=1$
are equivalent to the one we studied in \secref{subsec:U(1)-C}.
But we can find other irreducible representations
by allowing one or both of $\Qsym_1^2$ and $\Psym_2^2$
to be $(-1).$
In light of \eqref{eqn:QPef}, we can identify
$$
\Psym_2^2 = (-1)^{\ef_1'}\,,\qquad
\Qsym_1^2 = (-1)^{\ef_2'}\,.
$$
The algebra \eqref{eqn:PQalgI}-\eqref{eqn:PQalgII} then has
the following $4$-dimensional irreducible representation
with states $\ket{w_1,w_2,\ef_1,\ef_2}$, where $\ef_1,\ef_2\in\Z_2$
are fixed (and we dropped the primes now),
and $w_1,w_2\in\{-1,1\}$ take all possible values:
\be\label{eqn:PPQQwwee}
\left\{
\begin{array}{l}
\Psym_1\ket{w_1,w_2,\ef_1,\ef_2} = w_2\ket{w_1,w_2,\ef_1,\ef_2}
\,,\\
\Psym_2\ket{w_1,w_2,\ef_1,\ef_2} =
i^{\ef_1}\ket{-w_1,w_2,\ef_1,\ef_2}
\,,\\
\Qsym_1\ket{w_1,w_2,\ef_1,\ef_2} =
i^{\ef_2}\ket{w_1,-w_2,\ef_1,\ef_2}
\,,\\
\Qsym_2\ket{w_1,w_2,\ef_1,\ef_2} = w_1\ket{w_1,w_2,\ef_1,\ef_2}
\,.\\
\end{array}\right.
\ee
(Note that replacing $i^{\ef_j}$ by $(-i)^{\ef_j}$ can be absorbed
by a change of basis, so we picked one choice of square-root of
$(-1)^{\ef_j}$ at random.)

We can now combine the $U(1)$ and $SU(2)$ parts to form physical $U(2)$ states.
Incorporating the conditions
\eqref{eqn:Lam-mm} and \eqref{eqn:QPef}
with the identifications
\eqref{eqn:PQsymC-U(2)-a} and \eqref{eqn:PPQQwwee},
we can construct
a basis of physical states of the form
\bear
\lefteqn{
\ket{\ef'_1,\ef'_2,\mf'_1,\mf'_2}_{U(2)}\equiv
}\nn\\ &&
\ket{w_1=(-1)^{\mf_2'},w_2=(-1)^{\mf_1'},\ef_1',\ef_2'}_{U(1)}
\otimes
\ket{\ef'_1,\ef'_2,0,\mf'_1,\mf'_2,0}_{SU(2)}
\,,
\label{eqn:phys-U(2)-C}
\eear
with
$$
\ef_1',\ef_2',\mf_1',\mf_2'\in\Z_2
\,,
\qquad
\ef_3'=\mf_3'=0.
$$
These are a total of $2^4=16$ states, but they are
reduced to $10$ states when we implement the condition
\eqref{eqn:efmfz}.

Let us now map the basis of
states \eqref{eqn:phys-U(2)-C}
to the basis of states
$\ket{\{[\tfrac{1}{2}M+\tfrac{1}{2}N\tau],
[\tfrac{1}{2}M'+\tfrac{1}{2}N'\tau]\}}$
discussed at the beginning of this subsection.
The discussion above motivates us to postulate the following
relations between the type-IIA and field-theory symmetry operators:
$$
\oPsym_1 = \Psym_1
\,,
\qquad
\oPsym_2 = \Psym_2^2
\,,
\qquad
\oQsym_1 = \Qsym_1^2
\,,
\qquad
\oQsym_2 = \Qsym_2
\,.
$$
With these identifications, the commutation relations agree.
Comparing \eqref{eqn:PP-C-U(2)}-\eqref{eqn:QQ-C-U(2)}
with \eqref{eqn:PPQQwwee}, using \eqref{eqn:phys-U(2)-C},
we find the relation between
the type-IIA and $U(2)$ field-theory states:
\bear
\lefteqn{
\ket{\{[\tfrac{1}{2}M +\tfrac{1}{2}N\tau],
       [\tfrac{1}{2}M'+\tfrac{1}{2}N'\tau]\}}_{\text{IIA}}
=
}\nn\\ &&
\frac{1}{2}
\sum_{K=0}^1
\sum_{L=0}^1
(-1)^{M K + N L}
\ket{\ef'_1=L,\ef'_2=M-M',\mf'_1=K,\mf'_2=N-N'}_{U(2)}
\,,
\label{eqn:U(2)-IIA}
\eear
where $M-M',$ and $N-N'$ are understood to be mod $2$,
and of course, we have used $M-M'\equiv M+M'\pmod 2$
and $N-N'\equiv N+N'\pmod 2.$
We have also fixed an arbitrary phase in the definition
of the states $\ket{\ef'_1,\ef'_2,\mf'_1,\mf'_2}_{U(2)}$.

Where does the condition \eqref{eqn:efmfz} come from?
It comes from the fact that the type-IIA strings are identical bosons.
To see this,
note that the expression \eqref{eqn:U(2)-IIA}
is a priori not symmetric under the interchange
$(M,N)\leftrightarrow (M',N')$.
This exchange does not affect $(M-M')$ and $(N-N')$,
since they are $\Z_2$-valued, but it
replaces $(-1)^{M K + N L}$ by $(-1)^{M' K + N' L}$.
The operator that exchanges $(M,N)\leftrightarrow (M',N')$
therefore acts as multiplication by
$(-1)^{(M-M')K + (N-N')L}$
and can be identified with
$(-1)^{\ef'_1\mf'_2-\ef'_2\mf'_1}$ acting on
the states $\ket{\ef'_1,\ef'_2,\mf'_1,\mf'_2}_{U(2)}$.
Requiring the states to be invariant under the exchange
$(M,N)\leftrightarrow (M',N')$ is therefore
equivalent to \eqref{eqn:efmfz}.

We conclude the discussion of the $U(2)$
C-twist by writing down the reverse
transformation from the type-IIA states to the field theory states:
\bear\label{eqn:IIA-U(2)}
\lefteqn{
\ket{\ef'_1,\ef'_2,\mf'_1,\mf'_2}_{U(2)}
=
}\nn\\ &&
\frac{1}{2}
\sum_{M=0}^1
\sum_{N=0}^1
(-1)^{M \mf'_1 + N \ef'_1}
\ket{\{[\tfrac{1}{2}M +\tfrac{1}{2}N\tau],
       [\tfrac{1}{2}(M+\ef'_2)+\tfrac{1}{2}(N+\mf'_2)\tau]
     \}}_{\text{IIA}}
\,.
\eear


\section{Solution for $U(1)$ gauge theory}
\label{sec:U(1)}

We will now present in detail the solution of the problem
presented in \secref{sec:Problem} for $U(1)$ gauge group,
in which case the action of the three-dimensional field theory
can be written down exactly.


\subsection{The field theory side}
\label{subsec:U(1)-FT}

The scalars, fermions, and gauge fields decouple from each other,
and the scalars and fermions are described by a free field
theory with R-twisted boundary conditions
as in \secref{subsec:R-twist}:
\bear
\psi^a_\a(x_0,x_1,x_2,x_3+2\pi\xR) &=&
e^{i\ftw_a}\psi^a_\a(x_0,x_1,x_2,x_3)\,,
\qquad
a=1,\dots,4.
\nn\\
\Zphi^j(x_0,x_1,x_2,x_3+2\pi\xR) &=&
e^{i(\ftw_j+\ftw_4)}\Zphi^j(x_0,x_1,x_2,x_3)\,,
\qquad
j=1,2,3.
\nn
\eear
Here we take the \SUSY{6} twist \eqref{eqn:gtwN=6},
for which $\ftw_1=\ftw_2=\ftw_3=\frac{1}{2}\pht$
and $\ftw_4=-\frac{3}{2}\pht$ and there are no zero modes.
The scalars and fermions therefore do not give rise
to any 2+1D low-energy fields.

The vector field is a bit more involved.
The action for the vector field contains two terms:
a 3+1D bulk term in the coordinate range $0<x_3<2\pi\xR$,
and a 2+1D ``boundary'' term at $x_3=0$
(or $x_3=2\pi\xR$) associated with the S-twist.
The bulk term is a standard
$U(1)$ Yang--Mills action on the interval
$0< x_3< 2\pi\xR$, but instead of identifying the two
endpoints, we allow the gauge
fields at $x_3=0$ and $x_3=2\pi\xR$  to be independent,
and define the 2+1D fields
$$
A(0)\equiv \sum_{\mu=0}^2 A_\mu(x_0,x_1,x_2,x_3=0) dx^\mu\,,
\quad
A(2\pi\xR)\equiv \sum_{\mu=0}^2 A_\mu(x_0,x_1,x_2,x_3=2\pi\xR) dx^\mu\,.
$$
The gauge transformations are also not required to be periodic
in $x_3.$

The additional boundary term depends on the specific
element $\zg\in\SL(2,\Z)$ used in the twist.
For $\zg=\zg'\equiv\left(\begin{array}{rr}
 0 & -1 \\ 1 & 0 \\ \end{array}\right)$
the S-twist is incorporated by adding
the following 2+1D term
to the action \cite{Lozano:1995aq,Ganor:1996pe,Gaiotto:2008ak}:
\be\label{eqn:IS}
I_S(\zg') =\frac{1}{2\pi}\int A(0)\wedge dA(2\pi\xR)\,.
\ee
Here, the exterior derivative $d$ in $dA(2\pi\xR)$
is a 2+1D derivative.
One way to see that this term realizes the S-twist is to switch
to Euclidean signature and think of $x_3$  as a Euclidean time
coordinate (instead of $x_0$).
Then, in the Hamiltonian formalism, $e^{i I_S}$ represents the kernel of the S-duality operator
which acts on wavefunctions (in the $A_3=0$ gauge) as
a kind of ``Fourier transform'':
\be\label{eqn:Kernel}
\wvF(A)\rightarrow\sdwvF(\tA)=
\int[\cD A]
\exp\left\{
\frac{i}{2\pi}\int A\wedge d\tA
\right\}\wvF(A)\,,
\ee
and the last expression can be seen by requiring
the $\zg'$-duality to act on operators as
\be\label{eqn:EB}
E_i\rightarrow B_i\,,
\quad
B_i\rightarrow -E_i\,.
\ee
If instead of $\zg=\zg'$ we had picked
$\zg=-\zg'=(\zg')^{-1}$, we would have ended up with
the boundary action
\be\label{eqn:ISinv}
I_S(-\zg') =-\frac{1}{2\pi}\int A(0)\wedge dA(2\pi\xR)\,.
\ee

For $\zg=\zg''\equiv\left(
\begin{array}{rr} 1 & -1 \\ 1 & 0 \\ \end{array}
\right)=
\left(
\begin{array}{rr} 1 & 1 \\0  & 1 \\ \end{array}
\right)\zg'$ we get
\be\label{eqn:ITS}
I_S(\zg'') =\frac{1}{4\pi}\int \left\{-
A(2\pi\xR)\wedge dA(2\pi\xR)
+2A(0)\wedge dA(2\pi\xR)
\right\}\,,
\ee
since the effect of the $\SL(2,\Z)$ transformation $\left(
\begin{array}{rr} 1 & 1 \\0  & 1 \\ \end{array}
\right)$ on wavefunctions is multiplication by a Chern--Simons
term at level $\lvk=-1$:
$$
\wvF(A)\rightarrow
\exp\left\{-
\frac{i}{4\pi}\int A\wedge dA
\right\}\wvF(A)\,.
$$
Similarly, the action of $\zg=-\zg''
=\left(
\begin{array}{rr} 1 & 1 \\0  & 1 \\ \end{array}
\right)(\zg')^{-1}$ is realized by the boundary term
\be\label{eqn:ITSinv}
I_S(-\zg'') =\frac{1}{4\pi}\int \left\{-
A(2\pi\xR)\wedge dA(2\pi\xR)
-2A(0)\wedge dA(2\pi\xR)
\right\}\,.
\ee

Now, given the various expressions for the boundary
terms $I_S$ in \eqref{eqn:IS},\eqref{eqn:ITS},\eqref{eqn:ITSinv},
it is easy to take the low-energy limit.
We simply set $A(0)=A(2\pi\xR)$ (up to a gauge transformation),
and find that $I_S(\pm\zg')$ reduces to a Chern--Simons
action at level $\lvk=\pm 2$, and $I_S(\pm\zg'')$
reduces to a Chern--Simons action at level $\lvk=-1\pm 2.$
This can be summarized in the formula
$$
I_S\rightarrow
\frac{2-\xa-\xd}{4\pi\xc}\int A\wedge dA\,,
$$
which is a $U(1)$ Chern--Simons theory at level
\be\label{eqn:deflvkadc}
\lvk\equiv(2-\xa-\xd)/\xc\,.
\ee
Note that the gauge transformation parameter $\Lambda$,
appearing in the gauge transformation $A\rightarrow A+d\Lambda$,
is not required to be periodic in $x_3.$
Therefore, unlike an ordinary $S^1$ compactification,
the Wilson line $\int_0^{2\pi\xR} A_3 dx_3$
(at fixed $x_0,x_1,x_2$) can be gauged away,
and there is no additional massless mode arising from
the dimensional reduction of $A_3.$

Let us summarize the results in the following list:
\begin{itemize}
\item
for $\tau=i$, $\pht = \frac{\pi}{2},$
$\zg=\zg'\equiv
\left(\begin{array}{rr}
 0 & -1 \\ 1 & 0 \\ \end{array}\right),$ 
we have $\lvk=2$;
\item
for $\tau=e^{\pi i/3}$, $\pht= \frac{\pi}{3},$
$\zg=\zg''\equiv\left(
\begin{array}{rr} 1 & -1 \\ 1 & 0 \\ \end{array}
\right),$ 
we have $\lvk=1$;
\item
for $\tau=e^{\pi i/3}$, $\pht= \frac{2\pi}{3},$
$\zg=-\zg''{}^{-1}=\left(
\begin{array}{rr} 0 & -1 \\ 1 & -1 \\ \end{array}
\right),$ 
we have $\lvk=3$.
\end{itemize}
Note that the value of the Chern--Simons level $\lvk$ is the same as the number of ground states with winding number $n=1$ in type-IIA theory, given in \eqref{eqn:deflvk}. This is consistent with the fact that $U(1)$ Chern--Simons theory at level $\lvk$ has $\lvk$ ground states, as we will see in \secref{subsec:U(1)-CS}.


\subsection{Toroidal compactification}
\label{subsec:U(1)-T3}

The expressions \eqref{eqn:IS},\eqref{eqn:ISinv},
and \eqref{eqn:ITS} that we found in \secref{subsec:U(1)-FT}
assume that the theory is formulated on $\R^{2,1}$ with
certain boundary conditions that will become evident shortly.
When these boundary conditions are relaxed, or
when the theory is compactified on $T^2$,
additional terms need to be added
due to the possibility of electric or magnetic fluxes,
as we will now explain.
For the sake of the discussion, let us assume
Euclidean signature, and let us compactify
all directions $0,1,2$ on $T^3$, so that
$$
0\le x_0<2\pi\xL_0
\,,\qquad
0\le x_1<2\pi\xL_1
\,,\qquad
0\le x_2<2\pi\xL_2
\,.
$$

In \eqref{eqn:Kernel} we wrote down the
transformation from a wavefunction $\wvF(A)$
to the dual wavefunction $\sdwvF(\tA).$
The expression contained the integral
$\int A\wedge d\tA.$
On $T^3$, this expression is not well-defined,
because $A$ is not globally well-defined
in sectors with nonzero magnetic flux.
To obtain correct expression, consider a sector with magnetic flux
$(\omf_0,\omf_1,\omf_2)$,
where $\omf_0,\omf_1,\omf_2$ are integers.
We define the associated gauge field
$$
A =
\frac{\omf_0 x_1}{2\pi\xL_1\xL_2}dx_2
+\frac{\omf_1 x_2}{2\pi\xL_2\xL_0}dx_0
+\frac{\omf_2 x_0}{2\pi\xL_0\xL_1}dx_1
+A'\,,
$$
where $A'$ is a globally defined 1-form.
Similarly, set
$$
\tA=
\frac{\tomf_0 x_1}{2\pi\xL_1\xL_2}dx_2
+\frac{\tomf_1 x_2}{2\pi\xL_2\xL_0}dx_0
+\frac{\tomf_2 x_0}{2\pi\xL_0\xL_1}dx_1
+\tA'\,.
$$
Then \eqref{eqn:Kernel} should read
\bear
\lefteqn{
\wvF(A'; \omf_0,\omf_1,\omf_2)\rightarrow
}\nn\\ &&
\sdwvF(\tA;\tomf_0,\tomf_1,\tomf_2)=
\sum_{\omf_0\in\Z}
\sum_{\omf_1\in\Z}
\sum_{\omf_2\in\Z}
\int[\cD A']
\exp\Bigl\{
\nn\\ &&
i\int A'\wedge
(
\frac{\tomf_0}{4\pi^2\xL_1\xL_2}dx_1\wedge dx_2
+\frac{\tomf_1}{4\pi^2\xL_2\xL_0}dx_2\wedge dx_0
+\frac{\tomf_2}{4\pi^2\xL_0\xL_1}dx_0\wedge dx_1
)
\nn\\ &&
-i\int\tA'\wedge
(
\frac{\omf_0}{4\pi^2\xL_1\xL_2}dx_1\wedge dx_2
+\frac{\omf_1}{4\pi^2\xL_2\xL_0}dx_2\wedge dx_0
+\frac{\omf_2}{4\pi^2\xL_0\xL_1}dx_0\wedge dx_1
)
\nn\\ &&
+\frac{i}{2\pi}\int A'\wedge d\tA'
\Bigr\}\wvF(A'; \omf_0, \omf_1,\omf_2)\,.
\label{eqn:KernelT3}
\eear
The expression within the curly brackets $\{\cdots\}$ can be written
in terms of the discontinuous $A$ and $\tA$ fields as
\bear
\frac{i}{2\pi}\int A\wedge d\tA
+i\pi(\omf_0\tomf_1
+\omf_1\tomf_2
+\omf_2\tomf_0)
+\frac{i\omf_0}{\xL_2} \int\tA_0(x_0,0,x_2)dx_0 dx_2
\nn\\
+\frac{i\omf_1}{\xL_0} \int\tA_1(x_0,x_1,0)dx_0 dx_1
+\frac{i\omf_2}{\xL_1} \int\tA_2(0,x_1,x_2)dx_1 dx_2
\,.
\label{eqn:KernelT3b}
\eear
The last three terms may seem a little odd,
especially since they are evaluated at arbitrary locations
($x_1=0$, $x_2=0$, and $x_0=0$), but they are required because
of the discontinuity in $A$ at those locations.

We can now Wick-rotate the expression \eqref{eqn:KernelT3b}
by setting
\be\label{eqn:omfi}
\omf_0= \int dA\wedge\delta(x_0)dx_0
\,,
\ee
and
\be\label{eqn:omfii}
\begin{split}
\omf_1&=\int dA\wedge\delta(x_1)dx_1
=\frac{1}{2\pi\xL_1}\int dA\wedge dx_1
\,,\\
\omf_2&=\int dA\wedge\delta(x_2)dx_2
=\frac{1}{2\pi\xL_2}\int dA\wedge dx_2\,,
\end{split}
\ee
and similarly,
\be\label{eqn:tomfi}
\tomf_0=\int d\tA\wedge\delta(x_0)dx_0
\,,
\ee
and
\be\label{eqn:tomfii}
\begin{split}
\tomf_1&=
\int d\tA\wedge\delta(x_1)dx_1
=\frac{1}{2\pi\xL_1}\int d\tA\wedge dx_1
\,,\\
\tomf_2&=\int d\tA\wedge\delta(x_2)dx_2
=\frac{1}{2\pi\xL_2}\int d\tA\wedge dx_2\,.
\end{split}
\ee
We now find the correction to \eqref{eqn:IS} by
combining \eqref{eqn:KernelT3b}
with \eqref{eqn:omfi}-\eqref{eqn:tomfii},
and setting $A\equiv A(0)$ and $\tA\equiv A(2\pi R).$

For most purposes, \eqref{eqn:IS} will be sufficient.
In particular, the low-energy limit is simply
the compactification on $T^2$ of the Chern--Simons theory
found in \secref{subsec:U(1)-FT}.
This can be seen by setting $A=\tA$ (i.e.,
$A'=\tA'$ and $\omf_j=\tomf_j$)
in \eqref{eqn:KernelT3b}.
However, one place where we should be careful is
when we consider electric fluxes.
For example, let us discuss the electric flux
$\ef_1$ in the direction of $x_1.$
First, note that $\ef_1$ by itself is not S-duality
invariant and so is ill-defined in our setting.
However, if we add the magnetic flux $\mf_1$ we find that
the combination $\ef_1+\mf_1\pmod 2$ is S-duality invariant.
Thus,
$$
(-1)^{\ef_1+\mf_1}
$$
is a well-defined $\Z_2$ quantum number.
More generally, for other $\SL(2,\Z)$ elements,
$\ef_j+\mf_j\pmod\lvk$, where $\lvk$ is the Chern--Simons level
defined in \eqref{eqn:deflvkadc}, is invariant under the S-duality
twist, and
$$
e^{\frac{2\pi i}{\lvk}(\ef_j+\mf_j)}
$$
is a well-defined $\Z_\lvk$ quantum number.
Let us see how to interpret this statement from the action.

The operator $(-1)^{\ef_1}$ acts as the
discontinuous gauge transformation
$$
A\rightarrow A+\frac{dx_1}{2\xL_1}\,.
$$
In the action \eqref{eqn:KernelT3} this translates to
\be\label{eqn:AptAp}
A'\rightarrow A'+\frac{dx_1}{2\xL_1}\,,\qquad
\tA'\rightarrow \tA'+\frac{dx_1}{2\xL_1}\,,
\ee
while keeping $\omf_j$ and $\tomf_j$ ($j=0,1,2$) unchanged.
Under \eqref{eqn:AptAp},
the action \eqref{eqn:KernelT3b} then picks up an extra term
$i\pi (\tomf_1-\omf_1)$,
which, using \eqref{eqn:omfi}-\eqref{eqn:tomfii},
can be written as
$$
\frac{i}{2}\int [F_{02}(x_3=0)-F_{02}(x_3=2\pi R)]dx_0 dx_2
\,.
$$
Assuming that $A_0$ is periodic in $x_2$,
this becomes the difference of Wilson lines:
\bear
\lefteqn{
i\pi\int [A_2(x_3=2\pi R,x_0=\infty)-A_2(x_3=0,x_0=\infty)] dx^2
}\nn\\ &&
-i\pi\int [A_2(x_3=2\pi R,x_0=-\infty)-A_2(x_3=0,x_0=-\infty)] dx^2\,.
\eear

Now, what is the operator $(-1)^{\mf_1}$?
Acting on quantum states, it would multiply the wavefunction by
$$
\exp\left\{i\pi
\int F_{23} dx_2 dx_3
\right\}
\stackrel{A_3=0}{\longrightarrow}
\exp\left\{i\pi\left(
\int [A_2(x_3=2\pi R)-A_2(x_3=0)] dx^2
\right)\right\}.
$$
If a symmetry multiplies
quantum states by $e^{i\phi(x_0)}$, where $\phi(x_0)$ is a
time-dependent phase, then it multiplies the path-integral by
$\exp\{i(\phi(x_0=\infty)-\phi(x_0=-\infty))\}.$
Thus, altogether
$(-1)^{\mf_1+\ef_1}$ keeps the action invariant.
We also see that this operator reduces to $(-1)^{\ef_1}$
in the low-energy Chern--Simons theory,
because the low-energy Chern--Simons theory action
depends only on $A'=\tA'$ on which \eqref{eqn:AptAp}
acts as $(-1)^{\ef_1}.$


\subsection{$U(1)$ Chern--Simons theory on $T^2$}
\label{subsec:U(1)-CS}

Now we describe in detail the theory upon compactification of the
two spatial directions on $T^2$
(parameterized by $0\le x_j\le 2\pi\xL_j$ for $j=1,2$).
As we have seen  in \secref{subsec:U(1)-FT},
the low-energy theory is a $U(1)$ Chern--Simons theory
at level $\lvk$ with action
$$
I = \frac{\lvk}{4\pi}\int A\wedge dA
\,.
$$
We are interested in the Hilbert space of
states of this theory on $T^2$
(all are ground states since the theory is topological),
and we will now take a few paragraphs to review it
(see \cite{Elitzur:1989nr} for more details).

We denote the $k^{th}$ root of unity by
$$
\omega\equiv e^{\frac{2\pi i}{\lvk}}\,.
$$
We also define two independent Wilson loop operators
in terms of the integrals of the gauge fields on two
independent $1$-cycles of $T^2$:
\be\label{eqn:WilV12}
\WilV_1 = \exp\left\{\int_0^{2\pi\xL_1} A_1(t,0)dt\right\}\,,
\qquad
\WilV_2 = \exp\left\{\int_0^{2\pi\xL_2} A_2(0,t)dt\right\}\,.
\ee
The Hilbert space of Chern--Simons theory on $T^2$ at level $\lvk$
has $\lvk$ states, and we can pick a basis where
$\WilV_1,\WilV_2$ are represented by
\be\label{eqn:WilVs}
\WilV_1 = \begin{pmatrix}
 1 & & & & \\
 & \omega  & & & \\
 & & \ddots & & \\
 & & & & \omega^{\lvk-2} & \\
 & & & & & \omega^{\lvk-1} \\
\end{pmatrix}\,,
\qquad
\WilV_2 = \begin{pmatrix}
 &  & & & 1 \\
1 & &  & & \\
 & 1 & & & \\
 & & \ddots &  &  \\
 & & &  1&  \\
\end{pmatrix}\,.
\ee
These operators satisfy the relations
\be\label{eqn:U(1)-WW}
\WilV_1\WilV_2 = \omega\WilV_2\WilV_1\,,
\qquad
\WilV_1^{\lvk}=\WilV_2^{\lvk}=1\,.
\ee
We denote the states in the basis in which
\eqref{eqn:WilVs} holds by
\be\label{eqn:CSstates}
\ket{p}\,,\qquad
p=0,\dots,\lvk-1\in\Z/\lvk\Z\,,
\ee
so that
$$
\WilV_1\ket{p}=\omega^p\ket{p}\,,\qquad
\WilV_2\ket{p}=\ket{p+1}\,.
$$

The Chern--Simons theory is topological and therefore independent
of the metric on $T^2.$
There is an $\SL(2,\Z)$ group of large diffeomorphisms,
introduced in \secref{subsec:rhocp},
that acts on $T^2$ as
\be\label{eqn:zt}
\begin{pmatrix} x_1 \\ x_2 \\ \end{pmatrix}
\mapsto
\TdG
\begin{pmatrix} x_1 \\ x_2 \\ \end{pmatrix}
\,,\qquad
\TdG\equiv
\begin{pmatrix} \txa & \txb \\ \txc & \txd \\ \end{pmatrix}
\in\SL(2,\Z)
\,,
\ee
(We stress again that this $\SL(2,\Z)$ should not be confused
with the S-duality group.)
For the same reason as in \secref{subsec:T},
it is represented projectively on the Hilbert space
(i.e., commutation relations close up to a phase).
That is, we can require an element $\TdG\in\SL(2,\Z)$ to satisfy
\be\label{eqn:zsWilVV}
\TdG^{-1}\WilV_1\TdG=\WilV_2^{-\txb}\WilV_1^{\txa}\,,
\qquad
\TdG^{-1}\WilV_2\TdG=\WilV_2^{\txd}\WilV_1^{-\txc}\,,
\ee
but the order of the operators on the right-hand side of each equation
is arbitrary, since
the Wilson operators $\WilV_1$ and $\WilV_2$ do not commute,
and this is why we get only a projective representation.
Another ordering would correspond to replacing
$\TdG$ by $\WilV_1^l\WilV_2^m\TdG$ for some integers $l,m.$
For our purposes we will only need to realize two
elements of $\SL(2,\Z)$, and the projective nature of
the representation will not be important to us.
The elements that we need are listed below.

The element $\TdS\equiv\begin{pmatrix}0 & -1 \\ 1 & 0\\ \end{pmatrix}
\in\SL(2,\Z)$ acts as
\be\label{eqn:zspr}
\TdS\ket{p}=\frac{1}{\sqrt{\lvk}}\sum_q\omega^{p q}\ket{q},
\ee
and is a special case of the Verlinde matrix
\cite{Verlinde:1988sn,Witten:1988hf}.
It is easy to verify that
$$
\TdS^{-1}\WilV_1\TdS = \WilV_2\,,\qquad
\TdS^{-1}\WilV_2\TdS = \WilV_1^{-1}\,.
$$

The other generator of $\SL(2,\Z)$
is $\TdT\equiv\begin{pmatrix}
1 & 1 \\ 0 & 1 \\ \end{pmatrix}$.
For even $\lvk$ it acts as
\be\label{eqn:zteven}
\TdT\ket{p}=
e^{\frac{\pi i}{\lvk}p^2}\ket{p}\,,
\qquad
\lvk\equiv 0\pmod 2\,.
\ee
For odd $\lvk$, \eqref{eqn:zteven}
is ill-defined since the phase
$\exp(\frac{\pi i}{\lvk}p^2)$ depends on $p$ and not just
on $p$ mod $\lvk.$
In fact,
as we will discuss in \secref{subsec:U(1)-wf},
things get a little more complicated for odd $\lvk$,
but with the proper modification of the definition of $\TdT$,
it turns out that we can use the expression
\be\label{eqn:ztoddk}
\TdT\ket{p}=e^{\frac{\pi i}{\lvk}p(p+\lvk)}\ket{p}\,.
\ee
In general, there is some freedom in the expressions
that we have given above for $\TdT.$
For odd as well as even $\lvk$,
we can replace
$\TdT\rightarrow e^{i\phasezt}\WilV_1^{\intzt'}\TdT$
for some arbitrary integer $\intzt'$
and phase $\phasezt$,
and this will only introduce
an inconsequential phase in the commutation relation
$\TdT^{-1}\WilV_2\TdT$.
In principle, $\phasezt$ can be determined
if we wish to preserve the relation $(\TdT\TdS)^3=-1$,
not just up to a phase.

Now, let us discuss electric flux.
Consider large discontinuous $U(1)$ gauge transformations of the form
\be\label{eqn:LargeG}
\Lambda_1(x_1,x_2)=e^{\frac{i \frl x_1}{\xL_1}}\,,
\qquad
\Lambda_2(x_1,x_2)=e^{\frac{i \frl x_2}{\xL_2}}\,,
\ee
where $0<\frl<1$ is arbitrary, for the time being.
Let $\Omega_1,\Omega_2$ be the corresponding operators
on the Hilbert space, which we can identify with exponentials
of the electric fluxes $\ef_1,\ef_2$:
\be\label{eqn:Omef}
\Omega_1 = e^{2\pi i\frl\ef_1}
\,,
\qquad
\Omega_2 = e^{2\pi i\frl\ef_2}
\,.
\ee
They act by conjugation on the Wilson operators:
$$
\Omega_1^{-1}\WilV_1\Omega_1 = e^{2\pi i\frl}\WilV_1\,,\quad
\Omega_2^{-1}\WilV_1\Omega_2 = \WilV_1\,,\quad
\Omega_1^{-1}\WilV_2\Omega_1 = \WilV_2\,,\quad
\Omega_2^{-1}\WilV_2\Omega_2 = e^{2\pi i\frl}\WilV_2\,.
$$

These equations are solvable only if $\frl$ is an integer
multiple of $1/\lvk.$
Setting
\be\label{eqn:LargeGk}
\frl=\frac{1}{\lvk}\,,\qquad
\Lambda_1(x_1,x_2)=e^{\frac{i x_1}{\lvk\xL_1}}\,,
\qquad
\Lambda_2(x_1,x_2)=e^{\frac{i x_2}{\lvk\xL_2}}\,,
\ee
we can identify
\be\label{eqn:OmW}
\Omega_1\equiv\WilV_2\,,\qquad
\Omega_2\equiv\WilV_1^{-1}\,,
\ee
and they act on states as:
\be\label{eqn:OmegaU(1)}
\Omega_1\ket{p} = \ket{p+1}
\,,\qquad
\Omega_2\ket{p} = \omega^{-p}\ket{p}\,.
\ee
Thus, electric flux is defined only modulo $\lvk.$
The state $\ket{p}$ is an eigenstate of $\Omega_2$
with eigenvalue $\omega^{-p}$, and hence has $\ef_2=-p$,
and the state
$$
\frac{1}{\sqrt{\lvk}}\sum_{j=0}^{\lvk-1}
\omega^{-q j}\ket{j}
$$
is an eigenstate of $\Omega_1$ with eigenvalue $\omega^q$,
and hence has $\ef_1=q.$


\subsection{Wavefunctions}
\label{subsec:U(1)-wf}

An explicit description of the
Hilbert space of Chern--Simons theory on $T^2$ can be
given in terms of $\theta$-functions
\cite{Elitzur:1989nr,Bos:1989kn,Labastida:1989dv,Isidro:1991fp}.

In general,
the states of Chern--Simons theory on a Riemann surface
can be obtained by quantization of the space of flat connections
on the Riemann surface \cite{Witten:1988hf}.
For $U(1)$ gauge group on $T^2$ of complex structure
$\rhocp=\rhocp_1 + i\rhocp_2$,
a flat connection corresponds
to a constant gauge field $A\equiv A_1 dx_1 + A_2 dx_2$.
We parameterize it by the complex combination
$$
\hu\equiv -\frac{i\rhocp_2}{\pi} A_\bz
= \frac{1}{2\pi}(-\rhocp A_1 + A_2)
\,,
$$
where we used the complex coordinate $z\equiv x_1 + \rhocp x_2.$
Because gauge equivalent configurations are identified,
we find that $\hu$ lives on a $T^2$ of complex structure $\rhocp$,
with $\hu\simeq\hu+1\simeq\hu+\rhocp.$

The Chern--Simons action implies a nonzero
commutation relation between the operator $\ohu$
that represents $\hu$
and its conjugate $\ohu^\dagger$.
These commutation relations can be represented
by the following operators,
\be\label{eqn:ohudg}
\ohu^\dagger=\frac{\rhocp_2}{\pi\lvk}\frac{\partial}{\partial\hu}
\,,\qquad
\ohu = \hu
\,,
\ee
acting on analytic functions $\psi(\hu).$
The formulas above for $\ohu$ and its complex conjugate
$\ohu^\dagger$
are compatible with an inner product
\cite{Bos:1989kn} given by
\be\label{eqn:psipsi}
\langle\psi|\psi\rangle
=\int e^{-\frac{\pi\lvk}{\rhocp_2}|\hu|^2}|\psi|^2 d^2\hu
\,.
\ee
Imposing the periodicity conditions
$\hu\simeq\hu+1\simeq\hu+\rhocp$,
we get a $\lvk$-dimensional Hilbert space with a basis
\be\label{eqn:psiptheta}
\psi_p(\hu) =
\theta(\lvk\hu + p\rhocp;\lvk\rhocp)
e^{\frac{\pi\lvk}{2\rhocp_2}\hu^2+
\frac{1}{\lvk}\pi i \rhocp p^2 + 2\pi i p\hu}
\,,
\ee
where the $\theta$-function is given by
$$
\theta(\hu;\rhocp)\equiv
\sum_{n=-\infty}^\infty
e^{\pi i \rhocp n^2 + 2\pi i n\hu}
\,.
$$

The operators $\WilV_1,\WilV_2$ act on a
generic wavefunction $\psi(\hu)$ [which is understood
to be a linear combination of the $\psi_p(\hu)$'s] as
\be\label{eqn:WilV12onpsi}
\WilV_1\psi(\hu) = e^{-\frac{\pi}{\rhocp_2}\hu-\frac{\pi}{2\lvk\rhocp_2}}
\psi(\hu+\tfrac{1}{\lvk})
\,,\qquad
\WilV_2\psi(\hu) =
e^{-\frac{\pi\brhocp}{\rhocp_2}\hu
-\frac{\pi|\rhocp|^2}{2\lvk\rhocp_2}}
\psi(\hu+\tfrac{\rhocp}{\lvk})
\,.
\ee
The factors
$e^{-\frac{\pi}{\rhocp_2}\hu-\frac{\pi}{2\lvk\rhocp_2}}$
and $e^{-\frac{\pi\brhocp}{\rhocp_2}\hu
-\frac{\pi|\rhocp|^2}{2\lvk\rhocp_2}}$
are required in order to preserve unitarity,
which can be understood as follows:
the transformation $\hu\rightarrow\hu+\frac{1}{\lvk}$,
for example, needs to be accompanied by
$\hu^\dagger\rightarrow\hu^\dagger + \frac{1}{\lvk}$,
and the latter is generated by $e^{-\frac{\pi}{\rhocp_2}\hu}.$

In this representation it is easy to check
the $\TdT$-transformation
\be\label{eqn:ztzt}
\TdT\ket{p}=e^{\frac{\pi i}{\lvk}p^2}\ket{p}
\ee
for even $\lvk$ (up to an unimportant overall phase),
and odd $\lvk$ will be discussed below.


\subsubsection*{Odd $\lvk$}

For odd $k$, the definition of Chern--Simons theory
requires a spin structure on the three-manifold
\cite{Witten:2003ya}. For that reason,
theories with odd $\lvk$
are sometimes referred to as {\it spin Chern--Simons}
theories \cite{Belov:2005ze}.

In our case, we can see a manifestation of this in
the behavior of the wavefunctions \eqref{eqn:psiptheta}
under $\SL(2,\Z).$
There are $4$ distinct spin structures on $T^2$, and
the wavefunction \eqref{eqn:psiptheta} corresponds to one particular
spin structure. The transformation
$\rhocp\rightarrow\rhocp+1$ does not preserve this spin structure,
and indeed, for odd $\lvk$
the Hilbert space is not closed under it.
(It is only closed under its square $\rhocp\rightarrow\rhocp+2$.)
We can see this by a direct calculation:
$$
\psi_p(\hu,\rhocp+1) =
e^{\frac{1}{\lvk}\pi i p^2+\pi i p}
e^{-\frac{\pi\lvk}{2\rhocp_2}\hu}
e^{-\frac{\pi\lvk}{8\rhocp_2}}
\psi_p(\hu+\tfrac{1}{2},\rhocp).
$$

However, if we define
\be\label{eqn:modifiedTdT}
\TdT\psi(\hu,\rhocp)
\equiv
e^{-\frac{\pi\lvk}{2\rhocp_2}\hu-\frac{\pi\lvk}{8\rhocp_2}}
\psi(\hu+\tfrac{1}{2},\rhocp+1)
\,,
\ee
then we find closure:
$$
\TdT\psi_p = e^{\frac{1}{\lvk}\pi i p^2+\pi i p}\psi_p
\,.
$$
Note that the factor $e^{-\frac{\pi\lvk}{2\rhocp_2}\hu}$
in \eqref{eqn:modifiedTdT} can be understood as
realizing $\hu^\dagger\rightarrow\hu^\dagger+\tfrac{1}{2}$,
in agreement with \eqref{eqn:ztzt}.
Thus, \eqref{eqn:modifiedTdT} represents the
large diffeomorphism $\rho\rightarrow\rho+1$
augmented by a change of coordinates
that represents translation by $1/2$ of the $T^2$.

The dependence of the theory on the spin structure of $T^2$
is related to the dependence of the partition function
of a 5+1D (anti-)self-dual free $2$-form on spin structure
\cite{Witten:1996hc}.
The connection arises because our setting is related to
a compactification of the $(2,0)$-theory on
$\MXS\times T^2$ (see \secref{sec:twoz}).


\subsection{Connecting to the type-IIA picture}
\label{subsec:U(1)-IIA}

We would like to match the states $\ket{p}$ ($p=0,\dots,\lvk-1$)
of Chern--Simons theory with linear combinations of the $\lvk$
ground states of the type-IIA theory that we found in \secref{subsec:GS}.
Our strategy is to identify the
symmetry operators $\Psym,\Qsym$ that we defined
in \secref{subsec:P}-\secref{subsec:W}
with operators on the Chern--Simons Hilbert space.

We believe that the correct identification is
\be\label{eqn:PQW}
\Qsym=\WilV_1
\,,
\qquad
\Psym=\WilV_2
\,.
\ee
As a check, note that with this identification
the commutation relations
\eqref{eqn:QkPk}-\eqref{eqn:QPQP}
agree with \eqref{eqn:U(1)-WW}.

To motivate \eqref{eqn:PQW} further,
we can compare the connection between
the operators above and electric flux.
Combining \eqref{eqn:Omef} (with $\frl=1/\lvk$)
and \eqref{eqn:OmW}
we get
\be\label{eqn:WWef}
\WilV_1=\omega^{-\ef_2}\,,\qquad
\WilV_2=\omega^{\ef_1}\,.
\ee
Now, similarly to what we did in \secref{subsec:U(1)-C},
we can follow the
chain of dualities of \secref{sec:IIA-dual} backwards,
starting with $\Psym,\Qsym$ on the type-IIA side,
to find out what they do on the type-IIB side.
On the type-IIA side (the last row of \tabref{tab:Dualities}),
take an eigenstate with Kaluza--Klein momenta $p_1,p_{10}\in\Z$
in directions $x_1, x_{10}.$ (We can assume that the eigenstate
is localized in the $x_3$ direction.)
The unitary operator $\Psym$ acts as a translation,
and therefore multiplies its eigenstate by the phase
$e^{2\pi i (p_1 + p_{10})/\lvk}.$
Following the chain of dualities backwards in \tabref{tab:Dualities},
we saw in \secref{subsec:U(1)-C} that
the Kaluza--Klein state on the type-IIA side becomes
a $(p,q)$-string with $p=p_1$ and $q=p_{10}$ on the type-IIB side, and that
bound to $n$ D3-branes, these quantum numbers become
\cite{Witten:1995im,Douglas:1995bn}
$\ef_1=p_1$ units of electric flux
and $\mf_1=p_{10}$ units of magnetic flux in direction 1.
We conclude that on the gauge theory side $\Psym$ acts as
\be\label{eqn:PsymEM}
\Psym = e^{\frac{2\pi i}{\lvk}(\ef_1+\mf_1)}\,.
\ee
Similarly, $\Qsym$ is related to fundamental string
winding number, and we find
\be\label{eqn:QsymEM}
\Qsym = e^{-\frac{2\pi i}{\lvk}(\ef_2+\mf_2)}\,.
\ee
To see this, we note, for example, that the operator that has the
eigenvalue $e^{2\pi i (p_1 + p_{10})/\lvk}$ under the
adjoint action of $\Qsym$ is simply $\WilV_1.$

To interpret \eqref{eqn:PsymEM}-\eqref{eqn:QsymEM} correctly,
we need to discuss how to define the electric and magnetic fluxes
in the presence of the S-duality twist.
First, note that $\ef_j+\mf_j$ ($j=1,2$) is generally
not invariant under S-duality,
but it is not hard to check that if $\lvk$ is determined
by $\zg$ as in \eqref{eqn:deflvk}, then $(\ef_j+\mf_j)$
is invariant mod $\lvk.$
As we have argued in \secref{subsec:U(1)-T3},
the operator $\exp[\frac{2\pi i}{\lvk}(\ef_j+\mf_j)]$
in the full 3+1D theory
reduces to $\exp[\frac{2\pi i}{\lvk}\ef_j]$
in the low-energy Chern--Simons theory.
We can therefore identify
$\Qsym$ and $\Psym$ as
\be\label{eqn:PQef}
\Psym=\omega^{\ef_1}
\,,
\qquad
\Qsym=\omega^{-\ef_2}
\,,
\ee
which together with \eqref{eqn:WWef} leads to
\eqref{eqn:PQW}.

The basis states defined at the end of \secref{subsec:GS}
are eigenstates of $\Qsym$ with eigenvalues $\omega^p.$
Up to an unimportant phase, they can be identified
as eigenstates $\ket{p}$ of $\WilV_1$ defined in \eqref{eqn:CSstates}.
We conclude with a list of identifications of these states.


\subsubsection*{Single-particle states for $\pht=\frac{\pi}{2}$
($\tau=i$ and $\lvk=2$)}
$$
\ket{\fpIIz}=\ket{0}
\,,\qquad
\ket{\fpIIc}=\ket{1}\,.
$$


\subsubsection*{Single-particle states for $\pht=\frac{\pi}{3}$
($\tau=e^{\pi i/3}$ and $\lvk=1$)}

$$
\ket{\fpIz}=\ket{0}\,.
$$


\subsubsection*{Single-particle states for $\pht=\frac{2\pi}{3}$
($\tau=e^{\pi i/3}$ and $\lvk=3$)}

\be
\ket{\fpIIIz} =
\ket{0}
\,,\qquad
\ket{\fpIIIe} =
\ket{1}
\,,\qquad
\ket{\fpIIIf} =
\ket{2}
\,.
\ee


\section{$U(n)$ gauge group on $T^2$}
\label{sec:U(n)-T2}

In \secref{sec:U(1)} we have seen that for $U(1)$ gauge group,
the solution to our problem (as posed in \secref{sec:Problem})
is a Chern--Simons theory at one of the levels $\lvk=1,2,3.$
The level is determined by the choice of coupling constant $\tau$
and the $\SL(2,\Z)$ element $\zg.$
We now turn to the nonabelian case of $U(n)$ gauge group.
A natural question, then, is whether the solution to the $U(n)$
problem (with the restrictions on $n$
as given in \secref{subsec:LE})
is also a Chern--Simons theory.
And if not, what is it?
To explore this question we will use
the dual type-IIA description of the Hilbert space
of ground states of the $U(n)$ theory
on $T^2$ that we found in \secref{sec:IIA-dual},
and compare it to the Hilbert space of Chern--Simons
theory at the appropriate level.

As we saw in \eqref{eqn:HilbSumPerm}, the type-IIA Hilbert space $\Hilb(n,\pht)$ can be decomposed
into subspaces $\Hilb_{(n_1,n_2,\dots,n_p)}(\pht)$
by specifying the winding numbers $n_1,\dots,n_p$
of the individual strings.
We will analyze these subspaces separately, using the following three tools:
\begin{enumerate}
\item
The T-duality group $\SL(2,\Z)$ generated by $\TdT,\TdS$;
\item
The $\Z_\lvk\times\Z_\lvk$ symmetry generated
by $\Psym,\Qsym$ (which is useful for $\lvk>1$);
\item
The decomposition of the gauge group
$U(n)=[U(1)\times SU(n)]/\Z_n$.
\end{enumerate}

Together with the known solution for $U(1)$,
the last point allows us to construct from each of the
$\Hilb_{(n_1,\dots,n_p)}$'s another Hilbert space
$\HilbSun_{(n_1,\dots,n_p)}$
of states that
we can associate with the $SU(n)$ degrees of freedom only.
This will be done in \secref{subsec:U(1)subset}.
The coupling to the $U(1)$ degrees of freedom is
encoded in the action of large gauge transformations
related to the $\Z_n\subset SU(n)$ center.
They form a $\Z_n\times\Z_n$ symmetry group that is
generated by a pair
of large gauge transformations.
In this way, we will end up with Hilbert subspaces
$\HilbSun_{(n_1,\dots,n_p)}(\pht)$ on which an action
of the semidirect product of
$\SL(2,\Z)$ and $\Z_n\times\Z_n$ is given.
This will also be described in detail in
\secref{subsec:U(1)subset} [see \eqref{eqn:Omjpp}].

In general the analysis depends on the total winding number $n$ and
the angle $\pht$,
but there is one observation that we can make independently
of them.
For every $n$ and $\pht$,
there is unique sector $\HilbSun_{(1,1,\dots,1)}(\pht)$
with $n$ particles of winding number $1.$
In \secref{subsec:winding-1} we will argue that this sector
corresponds to the Hilbert space of
$SU(n)$ Chern--Simons theory at level $\lvk.$
The interpretation of other sectors is more mysterious
and we will defer the
discussion of them to \secref{subsec:other-sectors}.

We take the $T^2$ to be in directions $x_1,x_2$, and
as in \secref{sec:Problem},
the $S^1$ is of radius $\xR$ in direction $x_3.$


\subsection{The center $U(1)\subset U(n)$}
\label{subsec:U(1)subset}

Except for global issues related to electric and magnetic
fluxes, the $U(1)$ center of the gauge group $U(n)$ decouples.
Arguments similar to those in \secref{sec:U(1)}
lead to the conclusion that at low-energy (below the
compactification scale $1/\xR$) it gives rise
to a decoupled
sector of $U(1)$ Chern--Simons theory.
(The global issues will be discussed below.) The level is
$\lvk'=\lvk n$, where $\lvk$ is that of the $U(1)$ problem
given in \secref{sec:U(1)}. This can be seen as follows.

A gauge field $A$ of the $U(1)$ center
is diagonally embedded in $U(n)$ as
$$
A\rightarrow \begin{pmatrix}
A & & \\ & A & \\ & & \ddots \\
\end{pmatrix}\,.
$$
Inserted into the $U(n)$ Yang--Mills action,
this normalization of $A$
gives rise to a $U(1)$ action with
coupling constant $n\tau.$
Then the S-dual $U(1)$ action,
expressed in terms of a dual $U(1)$ gauge field $\tA$,
has coupling constant $-\frac{1}{n\tau}.$
Now consider the S-twist by $\zg'$
(the case with $\pht=\frac{\pi}{2}$
and $\lvk=2$), which
requires the theory to be self-dual.
To achieve self-duality,we need to rescale the dual gauge field
by defining $\tA=n\tA'$.
We can then set the S-twisted boundary
conditions to be, roughly speaking, $\tA'(2\pi\xR)=A(0).$
Inspecting \eqref{eqn:Kernel},
and repeating the arguments of \secref{sec:U(1)},
then shows that the proper Chern--Simons level is $\lvk'=2n.$
Similarly, it can be checked that the effective $U(1)$
Chern--Simons level is $\lvk'=\lvk n$ for the other values
of $\lvk$, where $\lvk$ is the function of
$\pht$ defined in \eqref{eqn:deflvk}. (When checking this,
note that the shift of the $U(n)$
$\theta$-angle that corresponds to $\tau\rightarrow\tau+1$
induces $\tau\rightarrow\tau+n$ for the $U(1)$ variables.)

Now compactify the remaining
two spatial directions of the theory on $T^2$
(parameterized by $0\le x_j\le 2\pi\xL_j$ for $j=1,2$).
The Hilbert space of states of $U(1)$ Chern--Simons theory
at level $\lvk'=\lvk n$ on $T^2$ has $\lvk'$ states,
which we denote by
$$
\ket{p}_{U(1)}\,,\qquad p=0,\dots,\lvk'-1\,.
$$
We pick a basis of the Hilbert space so that
these states are eigenstates of the $U(1)$ Wilson line
operator $\WilV_1$
corresponding to the $1$-cycle around the $x_1$-axis,
\be\label{eqn:WilV1}
\WilV_1 = \exp\left\{\int_0^{2\pi\xL_1} A_1(x_1,0)dx_1\right\}\,,
\ee
so that
$$
\WilV_1\ket{p}_{U(1)}=e^{\frac{2\pi i p}{\lvk n}}\ket{p}_{U(1)}\,.
$$
(For a quick review
of $U(1)$ Chern--Simons theory on $T^2$,
see \secref{subsec:U(1)-CS}, replacing $\lvk$
that appeared there with $\lvk'$.)
A general
state $\ket{\psi}_{U(n)}$ of the Hilbert space of the $U(n)$
theory can then be decomposed as
\be\label{eqn:decpsi}
\ket{\psi}_{U(n)} =
\sum_{p=0}^{\lvk n-1}\ket{\psi;p}_{SU(n)}\otimes\ket{p}_{U(1)}\,,
\ee
where $\ket{\psi;p}_{SU(n)}$ are the ``coefficients''
which can be interpreted as wavefunctions of the $SU(n)$
degrees of freedom only.

Let us now discuss the global issues that arise
because $U(n)$ is not $U(1)\times SU(n)$ but rather
$[U(1)\times SU(n)]/\Z_n.$
When compactifying a $U(n)$ gauge theory on $T^2$,
we require the Hilbert space of states to be invariant
under large $U(n)$ gauge transformations.
In particular, we need to consider the two gauge
transformations $\Omega_j$ ($j=1,2$) with gauge parameters
$$
\Omega_j(x_1,x_2) = \diag(e^{\frac{i x_j}{\xL_j}},1,\cdots,1),
$$
which are continuous in $U(n)$,
but cannot be lifted to continuous
gauge transformations in $U(1)\times SU(n).$
Indeed, they can be written as $\Omega_j=\Omega_j'\Omega_j''$
with
\bear
\Omega_j'(x_1,x_2) &=& \diag(e^{\frac{i x_j}{n\xL_j}},
e^{\frac{i x_j}{n\xL_j}},\cdots,e^{\frac{i x_j}{n\xL_j}})\in U(1),
\label{eqn:Omjpdef}\\
\Omega_j''(x_1,x_2) &=& \diag(e^{\frac{(n-1)i x_j}{n\xL_j}},
e^{-\frac{i x_j}{n\xL_j}},\cdots,e^{-\frac{i x_j}{n\xL_j}})
\in SU(n),
\label{eqn:Omjppdef}
\eear
but $\Omega_j'$ and $\Omega_j''$ have a discontinuity
at $x_j=0\equiv 2\pi\xL_j.$

Nevertheless, $\Omega_1'$ and $\Omega_2'$
define unitary operators
on the Hilbert space of $U(1)$ Chern--Simons theory on $T^2$
which act on the states $\ket{p}_{U(1)}$ as
(see \secref{subsec:U(1)-CS})
$$
\Omega_1'\ket{p}_{U(1)}=\ket{p+\lvk}_{U(1)}\,,
\qquad
\Omega_2'\ket{p}_{U(1)}=e^{-\frac{2\pi i p}{n}}\ket{p}_{U(1)}\,.
$$
The decomposition \eqref{eqn:decpsi} then implies that
\be\label{eqn:Omjpp}
\Omega_1''\ket{\psi;p}_{SU(n)}
=\ket{\psi;p+\lvk}_{SU(n)}\,,
\qquad
\Omega_2''\ket{\psi;p}_{SU(n)}
=e^{\frac{2\pi i p}{n}}\ket{\psi;p}_{SU(n)}\,.
\qquad
\ee
We conclude that a state of the form
\eqref{eqn:decpsi} is in the Hilbert space of the $U(n)$ theory
provided that the $SU(n)$ states in the decomposition
satisfy \eqref{eqn:Omjpp}.


\subsubsection*{Identifying the $\Z_\lvk$ momentum and winding number operators}
In \secref{subsec:P}-\secref{subsec:W}
we defined the symmetry operators $\Psym,\Qsym$
on the type-IIA dual. Let us identify these
operators on the gauge theory side.
{}From the definition it is clear that $\Psym,\Qsym$
act only on the $U(1)\subset U(n)$ degrees of freedom,
since on the type-IIA side they are defined in terms
of the ``center-of-mass'' of the strings.
We therefore turn to the analysis of
large gauge transformations that
act only on the $U(1)$ degrees of freedom.

Define the discontinuous $U(1)$ gauge transformations
$$
\OmX_j^{(\alpha)}(x_1,x_2) =
\diag(e^{\frac{i \alpha x_j}{\lvk'\xL_j}},
\cdots,e^{\frac{i \alpha x_j}{\lvk'\xL_j}})\in U(1)\,,
\qquad
j=1,2,
$$
where $\alpha$ is a real parameter.
We will see momentarily that it has to be an integer.
Let $\WilV_1$ be the $U(1)$ Wilson line as in
\eqref{eqn:WilV1}, and define the Wilson line $\WilV_2$
in direction $2$ similarly.
Then, by definition, $\OmX_1^{(\alpha)}$ has the following
commutation relations with the Wilson lines:
\be\label{eqn:OmXComm}
(\OmX_1^{(\alpha)})^{-1}\WilV_1\OmX_1^{(\alpha)}
=e^{\frac{2\pi i\alpha}{\lvk'}}\WilV_1
\,,\qquad
(\OmX_1^{(\alpha)})^{-1}\WilV_2\OmX_1^{(\alpha)}
=\WilV_2\,,
\ee
and similarly for $\OmX_2^{(\alpha)}.$
Given the explicit $\lvk'$-dimensional representation
of $\WilV_1,\WilV_2$ (see \secref{subsec:U(1)-CS}),
it is not hard to check that a solution to \eqref{eqn:OmXComm}
exists only for integer $\alpha$, in which case we can take
$$
\OmX_1^{(\alpha)}=\WilV_2^\alpha,
\qquad
\OmX_2^{(\alpha)}=\WilV_1^{-\alpha}\,.
$$

But even with $\alpha\in\Z$, the operators $\OmX_j^{(\alpha)}$
might not preserve the Hilbert space of $U(n)$ Chern--Simons theory.
For example, acting on \eqref{eqn:decpsi} we get
$$
\OmX_2^{(\alpha)}\ket{\psi}_{U(n)} =
\sum_{p=0}^{\lvk n-1}
e^{-\frac{2\pi i\alpha p}{\lvk n}}\ket{\psi;p}_{SU(n)}
\otimes\ket{p}_{U(1)}\,,
$$
and assuming the ``coefficients''
$\ket{\psi;p}_{SU(n)}$
satisfy  the condition \eqref{eqn:Omjpp}, we find that
the new ``coefficients''
$e^{\frac{2\pi i\alpha p}{\lvk n}}\ket{\psi;p}_{SU(n)}$
do not satisfy the rightmost equation of \eqref{eqn:Omjpp}
unless $\alpha\in n\Z.$
Similarly,
$$
\OmX_1^{(\alpha)}\ket{\psi}_{U(n)} =
\sum_{p=0}^{\lvk n-1}
\ket{\psi;p-\alpha}_{SU(n)}
\otimes\ket{p}_{U(1)}\,,
$$
and the leftmost condition of \eqref{eqn:Omjpp}
is not satisfied unless $\alpha\in n\Z.$
In order to preserve the Hilbert space, we therefore
require $\alpha$ to be an integer multiple of $n.$
We therefore define
\be\label{eqn:PQBW}
\PQsymB_1\equiv\OmX_1^{(n)}=\WilV_2^n,
\qquad
\PQsymB_2\equiv\OmX_2^{(n)}=\WilV_1^{-n}\,.
\ee
They generate a $\lvk^2$-dimensional group that preserves
the $U(n)$ Hilbert space, and they act as
\be\label{eqn:PQBdecpsi}
\begin{split}
\PQsymB_1\ket{\psi}_{U(n)} &=
\sum_{p=0}^{\lvk n-1}
\ket{\psi;p-n}_{SU(n)}
\otimes\ket{p}_{U(1)}\,,
\\
\PQsymB_2\ket{\psi}_{U(n)} &=
\sum_{p=0}^{\lvk n-1}
e^{-\frac{2\pi i p}{\lvk}}\ket{\psi;p}_{SU(n)}
\otimes\ket{p}_{U(1)}\,.
\end{split}
\ee
The operators $\PQsymB_1,\PQsymB_2$
satisfy the clock-and-shift relations
$$
\PQsymB_1\PQsymB_2=
e^{\frac{2\pi i n}{\lvk}}
\PQsymB_2\PQsymB_1\,,
\qquad
(\PQsymB_1)^\lvk=(\PQsymB_2)^\lvk=1\,.
$$

We can now connect the type-IIA operators
$\Psym,\Qsym$ to the gauge theory by identifying
\be\label{eqn:PQPQB}
\Psym=\PQsymB_1
\,,\qquad
\Qsym=\PQsymB_2^{-1}
\,,
\ee
in analogy with \eqref{eqn:PQW}.
We therefore get
\be\label{eqn:PQonpsi}
\begin{split}
\Qsym\ket{\psi}_{U(n)} &=
\sum_{p=0}^{\lvk n-1}
e^{\frac{2\pi i p}{\lvk}}\ket{\psi;p}_{SU(n)}
\otimes\ket{p}_{U(1)}\,,
\\
\Psym\ket{\psi}_{U(n)} &=
\sum_{p=0}^{\lvk n-1}
\ket{\psi;p-n}_{SU(n)}
\otimes\ket{p}_{U(1)}\,.
\end{split}
\ee

\subsubsection*{Action of $\SL(2,\Z)$}
There are two more operators that we find useful
to define on the Hilbert space of the $U(n)$ theory on $T^2.$
In \secref{subsec:rhocp} we discussed the $\SL(2,\Z)$
action of large diffeomorphisms of $T^2$,
and we mentioned that it induces an action on the Hilbert space
of ground states.
The action of $\SL(2,\Z)$ on the $U(1)$ states is well-known.
Setting the generators of $\SL(2,\Z)$ to be
$$
\TdT=\begin{pmatrix} 1 & 1 \\ 0 & 1 \\ \end{pmatrix}\,,
\qquad
\TdS=\begin{pmatrix} 0 & -1 \\ 1 & 0 \\ \end{pmatrix}\,,
$$
we have (see \secref{subsec:U(1)-CS})
\be\label{eqn:TSonU(1)}
\begin{split}
\TdS\ket{p}_{U(1)} &=
\frac{1}{\sqrt{\lvk n}}\sum_{q=0}^{\lvk n-1}
e^{\frac{2\pi i}{\lvk n}p q}\ket{q}_{U(1)}\,,
\\
\TdT\ket{p}_{U(1)} &=
\left\{
\begin{array}{ll}
e^{\frac{i\pi}{\lvk n}p^2}\ket{p}_{U(1)}
& \text{for even $\lvk n$,} \\
e^{\frac{i\pi}{\lvk n}p(p+\lvk n)}\ket{p}_{U(1)}
& \text{for odd $\lvk n$.} \\
\end{array}\right.
\end{split}
\ee

There are restrictions on the action of
$\TdT,\TdS$ on the $SU(n)$ states
since they have to preserve
the $U(n)$ Hilbert space \eqref{eqn:decpsi}
with the conditions \eqref{eqn:Omjpp}.
The restrictions have certain implications
for the commutation relations among $\TdT,\TdS$ and
$\Omega_1'',\Omega_2''.$
We have
$$
\TdT\ket{\psi}_{U(n)} =
\left\{
\begin{array}{ll}
\sum_{p=0}^{\lvk n-1}
e^{\frac{\pi i p^2}{\lvk n}}\TdT\ket{\psi;p}_{SU(n)}
\otimes\ket{p}_{U(1)}
& \text{for even $\lvk n$,} \\
\sum_{p=0}^{\lvk n-1}
e^{\frac{\pi i p(p+\lvk n)}{\lvk n}}\TdT\ket{\psi;p}_{SU(n)}
\otimes\ket{p}_{U(1)}
& \text{for odd $\lvk n$,} \\
\end{array}\right.
$$
and therefore \eqref{eqn:Omjpp} implies
\be\label{eqn:TdTOmjppA}
\TdT^{-1}\Omega_2''\TdT\ket{\psi;p}_{SU(n)}
=e^{\frac{2\pi i p}{n}}\ket{\psi;p}_{SU(n)}\,,
\ee
and
\be\label{eqn:TdTOmjppB}
\TdT^{-1}\Omega_1''\TdT\ket{\psi;p}_{SU(n)}
=
\left\{
\begin{array}{ll}
e^{\frac{2\pi i p}{n}
+\frac{\pi i \lvk}{n}}\ket{\psi;p+\lvk}_{SU(n)}
& \text{for even $\lvk n$,} \\
e^{\frac{2\pi i p}{n}
+\frac{\pi i \lvk(n+1)}{n}}
\ket{\psi;p+\lvk}_{SU(n)}
& \text{for odd $\lvk n$.} \\
\end{array}\right.
\ee
Using \eqref{eqn:Omjpp} again, we can rewrite these relations
as
\be\label{eqn:TdTOmTdT}
\TdT^{-1}\Omega_2''\TdT=\Omega_2''\,,
\qquad
\TdT^{-1}\Omega_1''\TdT
=
\left\{
\begin{array}{ll}
e^{-\frac{\pi i\lvk}{n}}\Omega_2''\Omega_1''
& \text{for even $\lvk n$} \\
e^{\frac{\pi i\lvk(n-1)}{n}}\Omega_2''\Omega_1''
& \text{for odd $\lvk n$} \\
\end{array}\right.
\qquad
\text{(on ground states).}
\ee
Similarly, we find
\be\label{eqn:TdSOmTdS}
\TdS^{-1}\Omega_1''\TdS=\Omega_2''\,,
\qquad
\TdS^{-1}\Omega_2''\TdS=(\Omega_1'')^{-1}\,
\qquad
\text{(on ground states).}
\ee
So, in order for $\TdT$ and $\TdS$ to preserve
the Hilbert space with the conditions \eqref{eqn:Omjpp},
they must obey the commutation relations
\eqref{eqn:TdTOmTdT}-\eqref{eqn:TdSOmTdS}.


\subsection{$U(n)$ Chern--Simons Hilbert space
as a symmetric product}
\label{subsec:SU(n)review}

The states of $SU(n)$ Chern--Simons theory on $T^2$
at level $\lvk$ are in one-to-one correspondence
with irreducible representations of $SU(n)$ that correspond
to Young diagrams with at most $\lvk$ columns
\cite{Witten:1988hf}.
For example, for $SU(2)$ the states are labeled by an irreducible
representation of $SU(2)$ with spin $j$ at most $\lvk/2$,
and so we can label the states by an integer $m=2j=0,\ldots,\lvk.$

We are interested in how the $\SL(2,\Z)$ generators $\TdT,\TdS$
act on the states, as well as in the large gauge transformations
$\Omega''_1,\Omega''_2$, which are
defined similarly to \eqref{eqn:Omjppdef}
and generate two $\Z_n$
symmetries of the Hilbert space.
For this purpose,
we work with a particularly convenient representation
of the Hilbert space that is derived from the Hilbert space
of $U(n)$ Chern--Simons theory, as we shall now explain.

To describe Chern--Simons theory with $U(n)$ gauge group
requires two levels---a level $\lvk'$ for $U(1)$ and a level
$\lvk$ for $SU(n)$. The theory is then denoted by
$$
[U(1)_{\lvk'}\times SU(n)_{\lvk}]/\Z_n\,,
$$
where the $\Z_n$ quotient refers to modding out by
large gauge transformations $\Omega_j=\Omega_j'\Omega_j''$
defined similarly to \eqref{eqn:Omjpdef}-\eqref{eqn:Omjppdef}.
The level $\lvk'$ can in principle be any integer multiple of $n$,
and can be changed by adding to the Chern--Simons action
a $U(1)$ Chern--Simons term for the trace of the gauge field.
One customary choice is $\lvk'=n(\lvk+n)$,
for which the (bare) Lagrangian doesn't have a separate
Chern--Simons term for the trace of the gauge field.
Another choice is $\lvk'=\lvk n.$
This choice is particularly convenient, because we have
the equivalence of Hilbert spaces
\cite{Elitzur:1989nr,Bos:1989kn}:
\be\label{eqn:U(n)asU(1)n}
\Hilb([U(1)_{\lvk n}\times SU(n)_\lvk]/\Z_n)
\simeq
\Hilb(U(1)_{\lvk})^{\otimes n}/S_n
\ee
where $[\cdots]/S_n$
denotes the symmetric part of the tensor product.

Equation \eqref{eqn:U(n)asU(1)n} is to be understood as follows:
both sides are equivalent representations of $\TdT,\TdS$,
as well as $\Omega_1,\Omega_2.$
In particular, the dimensions of both sides are equal:
\be\label{eqn:dimHilb1}
\begin{split}
\dim\Hilb([U(1)_{\lvk n}\times SU(n)_\lvk]/\Z_n) &=
\lvk n
\begin{pmatrix}n+\lvk-1\\ \lvk\\ \end{pmatrix}
\frac{1}{n^2}
=\begin{pmatrix}n+\lvk-1\\ \lvk-1\\ \end{pmatrix}\\
&=\dim\left\lbrack\Hilb(U(1)_{\lvk})^{\otimes n}\right\rbrack_S
\,.
\end{split}
\ee

In fact, \eqref{eqn:U(n)asU(1)n} can be understood
in terms of wavefunctions as well.
To explain this, we need to first discuss the wavefunctions
of $U(n)$ Chern--Simons theory on $T^2.$
The states of Chern--Simons theory on a Riemann surface
can be obtained by quantization of the space of flat connections
on the Riemann surface \cite{Witten:1988hf}.
For $T^2$, the flat connections can be encoded
in the conjugacy class
of the two commuting
holonomies of the gauge field around two independent
cycles of $T^2.$
The resulting wavefunctions have been explicitly described in
\cite{Elitzur:1989nr,Bos:1989kn,Murayama:1989we,Labastida:1989dv,Isidro:1991fp}.
For $U(1)$ gauge group, the two holonomies can be combined
into a complex variable that takes values on a dual $T^2.$
This dual $T^2$ also has complex structure $\rhocp$,
and the wavefunctions are related to $\theta$-functions,
as we reviewed in \secref{subsec:U(1)-wf}.
For $U(n)$ gauge group,
with the help of a gauge transformation,
the two commuting holonomies can be reduced to
a maximal torus $U(1)^n\subset U(n).$
The two holonomies
associated with the $i^{th}$
$U(1)$ factor ($i=1,\dots,n$)  can be combined
into a complex variable $\hu_i$ which takes values in $T^2$,
and so is subject to the identifications
$\hu_i\sim\hu_i+1\sim\hu_i+\rhocp.$
The wavefunctions $\psi(\hu_1,\dots,\hu_n)$
are required to be symmetric in the $n$ variables
(because of the Weyl group $S_n$) and can be expressed
in terms of partition functions
of $U(n)$ WZW models at level $\lvk$
(which are characters of the corresponding
affine Lie algebra \cite{Bardakci:1970nb,Kac:1990gs}).
Explicit expressions can be found in
\cite{Elitzur:1989nr,Bos:1989kn,Labastida:1989dv,Isidro:1991fp}.

On the other hand, the wavefunctions of
$\Hilb(U(1)_{\lvk})^{\otimes n}$
are proportional to symmetrized products of the
wavefunctions $\psi_{p_i}(\hu_i)$ described in
\secref{subsec:U(1)-wf}:
\be\label{eqn:Psippp}
\Psi_{p_1,\dots,p_n}(\hu_1,\dots,\hu_n)\equiv
\sum_{\sigma\in S_n}\prod_{i=1}^n\psi_{p_i}(\hu_{\sigma(i)})
\,.
\ee
Here, again, the $U(1)$
wavefunctions can be expressed in terms of $\theta$-functions
as reviewed in \secref{subsec:U(1)-wf}.
Now, the main point is that these symmetrized products
of $\theta$-functions \eqref{eqn:Psippp}
span the same Hilbert space as the characters of
$U(n)$ WZW at level $\lvk$, as explained in
\cite{Elitzur:1989nr,Bos:1989kn}!

Using \eqref{eqn:U(n)asU(1)n} it is easy to
calculate the action of $\TdT,\TdS,\Omega_1'',\Omega_2''$
on $\Hilb(SU(n)_\lvk)$ by calculating the action
of $\TdT,\TdS,\Omega_1,\Omega_2$ on
the right-hand side using the formulas of \secref{sec:U(1)},
and then extracting the $SU(n)$ degrees of freedom
from the left-hand side using the $U(1)$ results
in \secref{subsec:U(1)subset}. In doing so, it is very useful,
using \eqref{eqn:PQPQB},
to compare the action of $\Psym,\Qsym$
on type-IIA states and on the expansion \eqref{eqn:decpsi}
[as given in \eqref{eqn:PQonpsi}],
and derive restrictions on the the $SU(n)$ states
that appear in the expansion \eqref{eqn:decpsi} as coefficients.
Once we have these restrictions, we can derive the action
of the $\SL(2,\Z)$ operators $\TdT$ and $\TdS$ on
the $SU(n)$ states. We illustrate this procedure
with an example in \secref{subsec:ExU(1)inU(2)}.
The general case is described in \appref{app:SL(2,Z)-CS}.


\subsection{Example: $U(2)$}
\label{subsec:ExU(1)inU(2)}

We will demonstrate the decomposition into $U(1)$ and
$SU(n)$ degrees of freedom
in the case $\lvk=2$ and $n=2$.

Consider the sector of $2$-particle states on the type-IIA side
where each string has winding number 1. Since we identified
the single-particle states with those of $U(1)$ Chern--Simons theory
in \secref{sec:U(1)},
\eqref{eqn:U(n)asU(1)n} implies that states in this sector can be
identified with those of $U(2)$ Chern--Simons theory.
We expand the basis states
given in \secref{subsec:GS} using \eqref{eqn:PQonpsi}:
\bear
\ket{\fpIIz\fpIIz} &=&
\sum_{p=0}^{3}\left(
\ket{\fpIIz\fpIIz;p}_{SU(n)}\otimes\ket{p}_{U(1)}\right)\,,
\nn\\
\ket{\fpIIz\fpIIc} &=&
\sum_{p=0}^{3}\left(
\ket{\fpIIz\fpIIc;p}_{SU(n)}\otimes\ket{p}_{U(1)}\right)\,,
\nn\\
\ket{\fpIIc\fpIIc} &=&
\sum_{p=0}^{3}\left(
\ket{\fpIIc\fpIIc;p}_{SU(n)}\otimes\ket{p}_{U(1)}\right)\,.
\nn
\eear
Now let's compare the eigenvalues of $\Qsym$ on both sides.
Since $\ket{\fpIIc\fpIIc}$ and $\ket{\fpIIz\fpIIz}$
have $\Qsym$-eigenvalue $+1$, only even $p$'s can appear
in their expansions [see \eqref{eqn:PQonpsi}].
Similarly, $\ket{\fpIIz\fpIIz}$ has $\Qsym$-eigenvalue $-1$,
and therefore only odd $p$'s can appear in its expansion.
Next, we note that $\ket{\fpIIz\fpIIz}+\ket{\fpIIc\fpIIc}$
and $\ket{\fpIIz\fpIIc}$ have $\Psym$-eigenvalue $+1$
while $\ket{\fpIIz\fpIIz}-\ket{\fpIIc\fpIIc}$
has $\Psym$-eigenvalue $-1.$
We conclude that the expansion of $U(2)$ states
in terms of $SU(2)$ and $U(1)$ states must take the form
\bear
\ket{\fpIIz\fpIIz} &=&
\ket{a}_{SU(2)}\otimes\ket{0}_{U(1)}
+\ket{c}_{SU(2)}\otimes\ket{2}_{U(1)}
\label{eqn:decoSU(2)zz}\\
\ket{\fpIIz\fpIIc} &=&
\ket{b}_{SU(2)}\otimes\left(\ket{1}_{U(1)}+\ket{3}_{U(1)}\right)
\,,
\label{eqn:decoSU(2)zc}\\
\ket{\fpIIc\fpIIc} &=&
\ket{c}_{SU(2)}\otimes\ket{0}_{U(1)}
+\ket{a}_{SU(2)}\otimes\ket{2}_{U(1)}
\,,
\label{eqn:decoSU(2)cc}
\eear
where $\ket{a},\ket{b},\ket{c}$ are $3$ states
of the $SU(2)$ degrees of freedom.

We now note that the $\SL(2,\Z)$ action of large diffeomorphisms
becomes T-duality on the type-IIA side.
Using the action of $\TdT,\TdS$ on the single-particle states
as listed in \appref{app:SL(2,Z)},
and the action of $\TdT,\TdS$
on the $U(1)$ variables as given in \eqref{eqn:TSonU(1)},
we can find the action of $\TdT,\TdS$ on the
$SU(2)$ basis of states $\ket{a},\ket{b},\ket{c}$:
\be\label{eqn:TSabc}
\TdS=
\begin{pmatrix}
\frac{1}{2} &  \frac{1}{\sqrt{2}} & \frac{1}{2} \\
\frac{1}{\sqrt{2}} & 0  & -\frac{1}{\sqrt{2}} \\
\frac{1}{2} & -\frac{1}{\sqrt{2}} & \frac{1}{2} \\
\end{pmatrix}
\,,\qquad
\TdT=
\begin{pmatrix}
1 &  0 & 0 \\
0 &  e^{\frac{\pi i}{4}} & 0 \\
0 &  0 & -1 \\
\end{pmatrix}\,.
\ee
We also find, using \eqref{eqn:Omjpp}, the action of large gauge
transformations:
\be\label{eqn:omeabc}
\Omega_1''=
\begin{pmatrix}
0 &0&1 \\ 0&1&0 \\ 1&0&0
\end{pmatrix}\,,\qquad
\Omega_2''=
\begin{pmatrix}
1&0&0 \\ 0&-1&0 \\ 0&0&1
\end{pmatrix}
\,.
\ee

We will now present explicit expressions for
the wavefunctions that realize the decomposition
\eqref{eqn:decoSU(2)zz}-\eqref{eqn:decoSU(2)cc}.
The single-particle states $\ket{\fpIIz}$ and $\ket{\fpIIc}$
can be identified with wavefunctions of $U(1)$ Chern--Simons
theory. The latter can be explicitly represented in terms
of $\theta$-functions, as we recalled in \eqref{eqn:psiptheta}.
In order to realize \eqref{eqn:decoSU(2)zz}-\eqref{eqn:decoSU(2)cc},
we need to recast the product of two such wavefunctions
in a way that separates the ``center of mass'' $U(1)$ variable.

We denote the wavefunctions of Chern--Simons theory for any $\lvk$
as\footnote{For the purposes of this discussion,
we can actually be more general, and do not need
to restrict ourselves to $\lvk=2.$}
(see \secref{subsec:U(1)-wf})
\be\label{eqn:psipk}
\psi_{p,\,\lvk}(\hu) =
\theta(\lvk\hu + p\rhocp;\lvk\rhocp)
e^{\frac{\pi\lvk}{2\rhocp_2}\hu^2+
\frac{1}{\lvk}\pi i \rhocp p^2 + 2\pi i p\hu}
\,,\qquad
p=0,\dots,\lvk-1
\,.
\ee
The correspondence between the single-particle states and their wavefunctions
can be found using  \eqref{eqn:WilV12onpsi} and \eqref{eqn:PQW}; for example, for $k=2$, we get
$$
\ket{\fpIIz}\rightarrow\psi_{0,\,2}
\,,\qquad
\ket{\fpIIc}\rightarrow\psi_{1,\,2}
\,.
$$
Next, we use the identity
$$
\theta(z_1;\tau)\theta(z_2;\tau)
=
\theta(z_1+z_2;2\tau)
\theta(z_1-z_2;2\tau)
+e^{\pi i (\tau +2 z_2)}\theta(z_1+z_2+\tau;2\tau)
\theta(z_1-z_2-\tau;2\tau)
$$
to rewrite the $2$-particle wavefunctions as
\bear
\lefteqn{
\psi_{p_1,\,\lvk}(\hu_1)
\psi_{p_2,\,\lvk}(\hu_2)
+
\psi_{p_1,\,\lvk}(\hu_2)
\psi_{p_2,\,\lvk}(\hu_1)
}\nn\\
&=&
\psi_{p_1+p_2,\,2\lvk}\left(\frac{\hu_1+\hu_2}{2}\right)
\left\lbrack
\psi_{p_1-p_2,\,2\lvk}\left(\frac{\hu_1-\hu_2}{2}\right)
+
\psi_{p_2-p_1,\,2\lvk}\left(\frac{\hu_1-\hu_2}{2}\right)
\right\rbrack
\nn\\ &&+
\psi_{p_1+p_2+\lvk,\,2\lvk}\left(\frac{\hu_1+\hu_2}{2}\right)
\left\lbrack
\psi_{p_1-p_2-\lvk,\,2\lvk}\left(\frac{\hu_1-\hu_2}{2}\right)
+
\psi_{p_2-p_1+\lvk,\,2\lvk}\left(\frac{\hu_1-\hu_2}{2}\right)
\right\rbrack
\,.
\label{eqn:psipp2k}
\eear
We interpret the functions of $(\hu_1+\hu_2)$ as the $U(1)$ parts,
$$
\ket{p}_{U(1)}\rightarrow
\psi_{p,\,2\lvk}\left(\frac{\hu_1+\hu_2}{2}\right)
\,,
$$
and the factors in the square brackets $\lbrack\cdots\rbrack$ 
in \eqref{eqn:psipp2k} as the $SU(2)$ parts.
Specializing to $\lvk=2$ again,
we see that the decomposition
\eqref{eqn:decoSU(2)zz}-\eqref{eqn:decoSU(2)cc}
is consistent with:
\bear
\ket{a}_{SU(2)} &\rightarrow&
\psi_{0,\,4}\left(\frac{\hu_1-\hu_2}{2}\right)
\,,\nn\\
\ket{b}_{SU(2)} &\rightarrow&
\frac{1}{\sqrt{2}}\left[
\psi_{1,\,4}\left(\frac{\hu_1-\hu_2}{2}\right)
+\psi_{3,\,4}\left(\frac{\hu_1-\hu_2}{2}\right)
\right]
\,,\label{eqn:abc-psi}\\
\ket{c}_{SU(2)} &\rightarrow&
\psi_{2,\,4}\left(\frac{\hu_1-\hu_2}{2}\right)
\,.\nn
\eear
In \secref{subsec:winding-1} we will interpret these
as wavefunctions of $SU(2)$ Chern--Simons theory at level $2.$


\subsubsection*{$\SL(2,\Z)$ action on $SU(2)$ Chern--Simons theory}
For future reference, we list here the action of $\TdT,\TdS,\Omega_1'',\Omega_2''$ on the Hilbert space of $SU(2)$
Chern--Simons theory at level $\lvk=1,2,3$.

For $\lvk=1$, the basis states $\ket{a},\ket{b}$ are defined, using \eqref{eqn:U(n)asU(1)n}, by
$$
\ket{\fpIz\fpIz}=\ket{a}_{SU(2)}\otimes\ket{0}_{U(1)}+\ket{b}_{SU(2)}\otimes\ket{1}_{U(1)}\,.
$$
In this basis, we have
\be\label{eqn:tsk=1}
\TdT=
\begin{pmatrix}
1&0 \\ 0&e^{-\pi i/2}
\end{pmatrix}
\,,\qquad
\TdS=\tfrac{1}{\sqrt{2}}
\begin{pmatrix}
1&1 \\ 1&-1
\end{pmatrix}
\,,
\ee
\be
\Omega_1''=
\begin{pmatrix}
0&1 \\ 1 &0
\end{pmatrix}
\,,\qquad
\Omega_2''=
\begin{pmatrix}
1&0 \\ 0&-1
\end{pmatrix}
\,.
\ee

For $\lvk=2$, the results are in \eqref{eqn:TSabc}-\eqref{eqn:omeabc}.

For $\lvk=3$, the basis states $\ket{a},\ket{b},\ket{c},\ket{d}$ are defined by the decomposition
\begin{equation*}
\begin{split}
\ket{\fpIIIz\fpIIIz}&=\ket{a}_{SU(2)}\otimes\ket{0}_{U(1)}+\ket{b}_{SU(2)}\otimes\ket{3}_{U(1)}\,,\\
\ket{\fpIIIe\fpIIIe}&=\ket{a}_{SU(2)}\otimes\ket{2}_{U(1)}+\ket{b}_{SU(2)}\otimes\ket{5}_{U(1)}\,,\\
\ket{\fpIIIf\fpIIIf}&=\ket{a}_{SU(2)}\otimes\ket{4}_{U(1)}+\ket{b}_{SU(2)}\otimes\ket{1}_{U(1)}\,,\\
\ket{\fpIIIz\fpIIIe}&=\ket{c}_{SU(2)}\otimes\ket{1}_{U(1)}+\ket{d}_{SU(2)}\otimes\ket{4}_{U(1)}\,,\\
\ket{\fpIIIe\fpIIIf}&=\ket{c}_{SU(2)}\otimes\ket{3}_{U(1)}+\ket{d}_{SU(2)}\otimes\ket{0}_{U(1)}\,,\\
\ket{\fpIIIz\fpIIIf}&=\ket{c}_{SU(2)}\otimes\ket{5}_{U(1)}+\ket{d}_{SU(2)}\otimes\ket{2}_{U(1)}\,,
\end{split}
\end{equation*}
where we have used the same argument as in the paragraph preceding \eqref{eqn:decoSU(2)zz}-\eqref{eqn:decoSU(2)cc} to simplify the decomposition into $U(1)$ and $SU(n)$ degrees of freedom. We get
\be\label{eqn:tsk=3}
\TdT=
\begin{pmatrix}
1&0 & 0 &0 \\ 0&e^{\pi i/2}&0&0 \\ 0&0& e^{-5\pi i/6} &0 \\
0&0&0& e^{2\pi i/3}
\end{pmatrix}
\,,\qquad
\TdS=\tfrac{1}{\sqrt{6}}
\begin{pmatrix}
1&1 &\sqrt{2} & \sqrt{2} \\ 1&-1 & -\sqrt{2} & \sqrt{2} \\
\sqrt{2} & -\sqrt{2} & 1 & -1 \\ \sqrt{2} & \sqrt{2} & -1 & -1
\end{pmatrix}
\,,
\ee
\be\label{eqn:Z2inSU(2)}
\Omega_1''=
\begin{pmatrix}
0&1 & 0 & 0\\ 1 &0 &0&0 \\ 0&0&0&1 \\ 0&0&1&0
\end{pmatrix}
\,,\qquad
\Omega_2''=
\begin{pmatrix}
1&0&0&0 \\ 0&-1&0&0 \\ 0&0&-1&0 \\ 0&0&0&1
\end{pmatrix}
\,.
\ee


Note that $\TdS,\TdT$, as given above, are not the same as
the realization of $\SL(2,\Z)$ on Chern--Simons states as
given by \cite{Verlinde:1988sn,Witten:1988hf}.
In \cite{Verlinde:1988sn,Witten:1988hf}, $\TdS$ is realized
by the Verlinde matrix:
\be\label{eqn:TdSV}
\TdS_V\ket{m}=
\sqrt{\tfrac{2}{\lvk+2}}
\sum_{m'}\sin\frac{\pi (m+1)(m'+1)}{\lvk+2}\ket{m'}
\,,
\qquad m=0,\dots,\lvk
\,,
\ee
where $\ket{m}$ is the state corresponding to spin $m/2$,
and $\TdT$ is realized diagonally by
\be\label{eqn:TdTV}
\TdT_V\ket{m}= e^{\frac{\pi i m(m+2)}{2(\lvk+2)}}\ket{m}
\qquad m=0,\dots,\lvk
\,.
\ee
Formulas \eqref{eqn:TdSV}-\eqref{eqn:TdTV} were derived
from the equivalence between states of Chern--Simons theory
on $T^2$ and characters of the WZW model,
while formulas \eqref{eqn:TSabc} and \eqref{eqn:tsk=1},\eqref{eqn:tsk=3}
were derived by realizing the Hilbert space
of $SU(2)$ Chern--Simons theory
as a subspace of $\Hilb(U(1)_\lvk)^2/S_2.$
The discrepancy between \eqref{eqn:TdSV}-\eqref{eqn:TdTV}
and \eqref{eqn:TSabc}, \eqref{eqn:tsk=1},\eqref{eqn:tsk=3}
is because they are written in difference bases, and
the transformation from one basis to the other involves
nontrivial coefficients that are
functions of the complex structure $\rhocp$.
These coefficients
transform nontrivially themselves
under $\TdS$ and $\TdT$, and hence the resulting formulas
are different. This point will be demonstrated explicitly in
an example in \secref{subsec:winding-1}.


\subsection{Chern--Simons theory and the
$\brperm$-untwisted sector
$\Hilb_{[1]}(\pht)$}
\label{subsec:winding-1}

Each of the three cases $\lvk=1,2,3$ considered in \secref{subsec:GS}
has a special sector $\Hilb_{[1]}(\pht)=\Hilb_{(1,1,\dots,1)}(\pht)$
comprising of $n$ string states,
all of which have winding number $1.$
The location of these strings can be any one of $\lvk$ choices,
so altogether the Hilbert space is the symmetric product of single-particle
Hilbert spaces:
\be\label{eqn:Hilb1S}
\Hilb_{(1,1,\dots,1)}(\pht)\simeq
\Hilb(U(1)_{\lvk})^{\otimes n}/S_n
\,.
\ee
In fact, for the present discussion
the restrictions on $n$ from \secref{subsec:LE}
can be relaxed, and we can allow any $n\ge 1$,
because even the cases $n\ge\ord$
still have a finite-dimensional
subspace of normalizable ground states
(even though there is no mass gap now).
The finite-dimensional
Hilbert space $\Hilb_{[1]}(\pht)$
is therefore well-defined for all $n.$
This is the sector we
referred to in \secref{subsec:GS}
as the {\it $\brperm$-untwisted sector.}

We can now state our main observation:
{\bf
$\Hilb_{[1]}$ is equivalent to
the Hilbert space of
$[U(1)_{\lvk n}\times SU(n)_\lvk]/\Z_n$
Chern--Simons theory at level $\lvk.$
}
This follows immediately from \eqref{eqn:U(n)asU(1)n}.

As an example, take the case $n=2.$
In \secref{subsec:ExU(1)inU(2)} we studied the
basis of symmetric $2$-particle states of
$\Hilb(U(1)_\lvk)^{\otimes 2}$
with wavefunctions of the form
$\psi_{p_1}(\hu_1)\psi_{p_2}(\hu_2)
+\psi_{p_1}(\hu_2)\psi_{p_2}(\hu_1).
$
In \eqref{eqn:psipp2k} we expressed these products
as a linear combination of products of wavefunctions of
$(\hu_1+\hu_2)$ and wavefunctions of $(\hu_1-\hu_2).$
According to \eqref{eqn:U(n)asU(1)n},
the symmetric part of the
space $\Hilb(U(1)_\lvk)^{\otimes 2}$
is the Hilbert space of $U(2)$ Chern--Simons theory
where the $U(1)$ center is at level $2\lvk$
and the $SU(2)$ is at level $\lvk.$ Indeed,
the functions of $(\hu_1+\hu_2)$ correspond to the $U(1)$
factor and are wavefunctions of $U(1)$ theory at level $2\lvk$,
while the functions of $(\hu_1-\hu_2)$
can be recast as wavefunctions of $SU(2)$ theory at level $\lvk$
using the connection
\cite{Verlinde:1988sn,Witten:1988hf,Elitzur:1989nr,Bos:1989kn}
between the latter and characters of affine Lie algebras.
Let us demonstrate how this is done.

The characters of the $SU(2)$ affine Lie algebra at level $\lvk$
are related to the wavefunctions we found in
\eqref{eqn:psipp2k} as follows.
Consider the Weyl--Kac characters
\be\label{eqn:weyl-kac characters}
\text{ch}_\lvk^\lambda(\hu) =
\sum_{m=1-\lvk}^{\lvk}C_m^\lambda (\rhocp) \Theta_{m,k},
\qquad\Theta_{m,k}\equiv \sum_{n\in \Z+m/2\lvk}
e^{2\pi i(n^2 \rhocp - n\hu)}
\,,
\ee
where $C_m^\lambda(\rho)$
are the ``string functions,'' which satisfy the following relations \cite{Kac:1984mq}:
\be\label{string fn constraint}
C_m^\lambda = 0\quad \forall\, \lambda \neq m \pmod{2},\qquad C_m^\lambda=C_{-m}^\lambda,
\qquad C_m^\lambda=C_{\lvk+m}^{\lvk-\lambda}
\,.
\ee
Here $\lambda=0,\dots,\lvk$ corresponds to twice the ``spin''
of the highest weight of the representation.
For $\lvk=2$, for example,
these constraints yield $3$ independent $C_m^\lambda$'s
whose exact forms are
\be\label{eqn:exact string fn}
C_+\equiv C_0^0 + C_2^0=
\frac{\eta(q)}{\eta(\sqrt{q})\eta(q^2)}
\,,\quad
C_{-}\equiv C_0^0 - C_2^0=\frac{\eta(\sqrt{q})}{[\eta(q)]^2}
\,,\quad C_1^1=\frac{\eta(q^2)}{[\eta(q)]^2}\,,
\ee
where $\eta(q)$ is the Dedekind function of
$q\equiv e^{2\pi i\rhocp}$.
Relating them back to our wavefunctions in \eqref{eqn:abc-psi},
we find the relations:
\bear
\ket{a}_{SU(2)} &=&
\frac{e^{\pi\lvk\hu_{-}^2/2\rhocp_2}}{C_+C_-}\left(C_0^0\,
\text{ch}_2^0(\hu_{-}) - C_2^0\,\text{ch}_2^2(\hu_{-}) \right)
\,,\\
\ket{c}_{SU(2)} &=& \frac{e^{\pi\lvk\hu_{-}^2/2\rhocp_2}}{C_+C_-}
\left(-C_2^0\,\text{ch}_2^0(\hu_{-}) + C_0^0
\,\text{ch}_2^2(\hu_{-}) \right)
\,,\\
\ket{b}_{SU(2)} &=&
 \frac{e^{\pi\lvk\hu_{-}^2/2\rhocp_2}}{C_1^1}\text{ch}_2^1 (\hu_{-})
\,,
\eear
where $\hu_{-}\equiv (\hu_2-\hu_1)/2$.

Thus, we see explicitly that the wavefunctions of
$\Hilb_{(1,1)}(\pht)$ correspond to a basis of the wavefunctions
of $U(2)$ Chern--Simons theory at level $\lvk.$
In particular, the states $\ket{a},\ket{b},\ket{c}$ of
\eqref{eqn:abc-psi}
correspond to a linear combination of the states of $SU(2)$ at level
$\lvk=2$ with highest weight $j=0,1,2$, respectively.
The subtle point about the linear coefficients being functions of
$\rhocp$ is that in the language of holomorphic quantization,
the basis furnished by the string theory is not yet normalized.
A straightforward computation reveals that the modular transformation properties of the string functions explain
the discrepancy between our formulae for $\TdT,\TdS$ and those found in standard literature for Chern--Simons theory.
For example, under $\rhocp\rightarrow\rhocp+1$ we find
$$
C_1^1\rightarrow C_1^1
\,,\qquad
C_{-}\rightarrow e^{-\frac{\pi i}{8}}C_{+}
\,,\qquad
C_{+}\rightarrow e^{-\frac{\pi i}{8}}C_{-}
\,.
$$
Thus, if one further orthonormalizes the string theory states,
then, as shown above, the states are those
of nonabelian Chern--Simons theory.


\subsection{$\brperm$-twisted sectors}
\label{subsec:other-sectors}
We now turn to the sectors of the Hilbert space where some strings
have winding number greater than $1.$
These are the sectors $\Hilb_{(n_1,\dots,n_p)}(\pht)$
that we called $\brperm$-twisted
in \secref{subsec:GS}.
We distinguish two kinds of sectors:
\begin{enumerate}
\item
{\it Irreducible sectors}---those
sectors for which all individual strings
have the same winding number, i.e.,
$n_1=\dots = n_p.$
Those Hilbert spaces cannot be written as a product
of two Hilbert spaces with a smaller number of strings
and same value of $\pht$.
In particular, all single-particle sectors ($p=1$ and $n_1=n$)
are irreducible.

\item
{\it Reducible sectors}---those
sectors for which at least two individual strings
have different winding numbers, i.e.,
$n_1>n_p.$
Those Hilbert spaces can always be written as a product
of at least two Hilbert spaces with a smaller number of strings
and same value of $\pht$.
\end{enumerate}
The Hilbert spaces of reducible sectors can always be written
as tensor products of Hilbert spaces of irreducible sectors,
and are therefore equivalent to Hilbert spaces of
a sum of decoupled Chern--Simons theories with gauge groups of
lower rank. For example, for $\pht=\frac{\pi}{2}$ and $n=3$
we have
$$
\Hilb_{(2,1)}(\tfrac{\pi}{2})
\simeq
\Hilb_{(1)}(\tfrac{\pi}{2})
\otimes
\Hilb_{(2)}(\tfrac{\pi}{2})\,.
$$
We have already identified $\Hilb_{(1)}(\frac{\pi}{2})$
as equivalent to the Hilbert space of $U(1)_2$ Chern--Simons theory,
and below we will identify $\Hilb_{(2)}(\frac{\pi}{2})$
as the Hilbert space of $[U(1)_4\times SU(2)_{-2}]/\Z_2$
Chern--Simons theory (the gauge group here is
$U(2)\simeq [U(1)\times SU(2)]/\Z_2$), so altogether
we can identify $\Hilb_{(2,1)}(\frac{\pi}{2})$
as equivalent to the Hilbert space of a $U(1)\times U(2)$
Chern--Simons theory.
It therefore suffices to study the irreducible sectors,
which we shall undertake below.

Let us here begin by outlining the plan.
In each case we will decompose
an irreducible space $\Hilb_{(n_1,\dots,n_p)}(\pht)$
into irreducible representations of $\SL(2,\Z)$ and present
the action of $\TdT,\TdS.$ We then extract the $SU(n)$ degrees
of freedom as in \eqref{eqn:decpsi} and calculate
the action of $\TdT,\TdS$ on the
resulting states $\ket{\psi;p}_{SU(n)}.$
Next, we will attempt to map the states $\ket{\psi;p}_{SU(n)}$
to ground states of $SU(n)$ Chern--Simons theory on $T^2$
at some level $\lvk''.$
We will find that this is possible in all
single-particle cases, and we will identify the level.
A useful tool is the action of the $\Z_n\times\Z_n$
symmetry group generated by the large gauge transformations
$\Omega_1'',\Omega_2''$ as in \eqref{eqn:Omjpp}.

We should make it clear that
at this point we are not claiming that the theory
{\it is} Chern--Simons theory at level $\lvk''$
(although it is very likely),
but only that the single-particle
Hilbert space $\Hilb_{(n)}(\pht)$
is equivalent as a representation of $\SL(2,\Z)$ and
$\Psym,\Qsym$ to the Hilbert space of
$[U(1)_{\lvk n}\times SU(n)_{\lvk''}]/\Z_n.$
However,
whether the low-energy theory really is
Chern--Simons theory or not,
the $SU(n)$ states $\ket{\psi;p}_{SU(n)}$ have to form
a representation of the $\Z_n\times\Z_n$ group,
and in this sense we can say that at the very least the
low-energy theory is a $\Z_n$ gauge theory (but
of course $\Z_n$
might be a subgroup of a bigger gauge group).

Accepting the equivalence
between the single-particle sectors and their corresponding
$U(n)$ Chern--Simons Hilbert spaces,
and given the
equivalence between the untwisted sector
$\Hilb_{(1,1,\dots,1)}(\pht)$ and the corresponding
$U(n)$ Chern--Simons Hilbert space that we established
in \secref{subsec:winding-1}, it is straightforward
to construct the Chern--Simons Hilbert space equivalent
of reducible sectors $\Hilb_{(n_1,\dots,n_p)}(\pht)$,
as long as no $n_j\ge 2$ appears more than once in the sequence
$(n_1,\dots,n_p).$
Thus, assuming
\be\label{eqn:norder}
n_1 > \cdots > n_{p-q} > n_{p-q+1} = \cdots = n_p = 1
\,,
\ee
and denoting, for brevity,
$$
U(n')_{\lvk n',\lvk''} \equiv
\Hilb([U(1)_{\lvk n'}\times SU(n')_{\lvk''}]/\Z_{n'})
\,,
$$
we can identify
\be\label{eqn:HilbSumU}
\Hilb_{(n_1,\dots,n_p)}(\pht) \simeq
U(n_1)_{\lvk n_1,k_1} \otimes \cdots \otimes
U(n_{p-q})_{\lvk n_{p-q},k_{p-q}}
\otimes U(q)_{\lvk q,\lvk}
\,,
\ee
where $U(n_j)_{\lvk n_j,\lvk_j}$ is our proposal, to
be developed below,
for the single-particle sector $\Hilb_{(n_j)}(\pht)$,
and $\lvk_j$ ($j=1,\dots,p-q$) depends on $n_j$ and $\lvk.$

Note that \eqref{eqn:HilbSumU} is not explicitly
in the form $[U(1)_{\lvk n}\times (\cdots)]/\Z_n$.
To reconcile \eqref{eqn:HilbSumU} with our discussion
on the level $\lvk n$ of the $U(1)$ center
in \secref{subsec:U(1)subset}, we can consider
a  wavefunction in the right-hand side of \eqref{eqn:HilbSumU}.
It is a product of wavefunctions of the component
$U(n')_{\lvk n',\lvk''}$ Hilbert spaces.
As we have explained in \secref{subsec:ExU(1)inU(2)},
these wavefunctions are
products of $\theta$-functions in variables
$\hu_1,\hu_2,\dots,$ which take values on $T^2.$
To address the
question of the $U(1)$ center, we fix
$\hu_1,\hu_2,\dots,$ and translate all variables
by $\zeta$, which we take to be some holomorphic
coordinate on $T^2$:
$$
\hu_1\rightarrow\hu_1+\zeta,
\quad
\hu_2\rightarrow\hu_2+\zeta,
\quad\dots\,.
$$
The main point is that a wavefunction in
$U(n')_{\lvk n',\lvk''}$ is a linear combination
of level-$\lvk n'$ $\theta$-functions in $\zeta.$
In other words, it is a section of a holomorphic
line bundle over $T^2$ with first Chern class $c_1 = \lvk n'.$
As a function of $\zeta,$
the product of the wavefunctions in all the component
Hilbert spaces on the right-hand side of \eqref{eqn:HilbSumU}
is a linear combination of $\theta$-functions of level
$\lvk(q+\sum_{j=1}^{p-q}n_j) = \lvk n$, as it should be.

The condition \eqref{eqn:norder} is satisfied
by all irreducible sectors except
$\Hilb_{(2,2)}(\frac{\pi}{3}).$
There are therefore two sectors that are not covered by
our results, both for $\lvk=1.$ The first is
$\Hilb_{(2,2)}(\frac{\pi}{3})$ itself for $n=4$,
and the second is the reducible sector
$\Hilb_{(2,2,1)}(\frac{\pi}{3})$ for $n=5$,
which decomposes as
$U(1)_1\times\Hilb_{(2,2)}(\frac{\pi}{3}).$
We discuss the sector $\Hilb_{(2,2)}(\frac{\pi}{3})$
in some detail in \appref{app:HilbDecomp}, but it
generally remains a mystery to us.

We now turn to a case-by-case analysis of
the single-particle irreducible sectors.


\subsubsection{$\pht=\frac{\pi}{2}$ ($\lvk=2$)}
\label{subsubsec:sigma-twisted(k=2)}

For $\lvk=2$ and $n=2$ we have,
on the type-IIA side,
$3$ single-particle states
$$
\ket{\fpIIzz}\,,\quad\ket{\fpIIxy}\,,\quad\ket{\fpIIcc}\,,
$$
which are a basis for a subspace we denote by
$\Hilb_{(2)}(\frac{\pi}{2}).$
Using the same argument as in the paragraph preceding \eqref{eqn:decoSU(2)zz}-\eqref{eqn:decoSU(2)cc},
we can separate the $SU(2)$ degrees of freedom as follows:
\bear
\ket{\fpIIzz} &=&
\ket{a}_{SU(2)}\otimes\ket{0}_{U(1)}
+\ket{c}_{SU(2)}\otimes\ket{2}_{U(1)}
\label{eqn:decoSU(2)zz2}\\
\ket{\fpIIxy} &=&
\ket{b}_{SU(2)}\otimes\left(\ket{1}_{U(1)}+\ket{3}_{U(1)}\right)
\,,
\label{eqn:decoSU(2)xy2}\\
\ket{\fpIIcc} &=&
\ket{c}_{SU(2)}\otimes\ket{0}_{U(1)}
+\ket{a}_{SU(2)}\otimes\ket{2}_{U(1)}
\,.
\label{eqn:decoSU(2)cc2}
\eear
[We used the same notation $\ket{a},\ket{b},\ket{c}$ as in
\eqref{eqn:decoSU(2)zz}-\eqref{eqn:decoSU(2)cc},
but these states are, of course, unrelated to the
states $\ket{a},\ket{b},\ket{c}$ of
\eqref{eqn:decoSU(2)zz2}-\eqref{eqn:decoSU(2)cc2}.]
Then, we can read off the action of the $\SL(2,\Z)$
generators $\TdT,\TdS,$ and
the large $\Z_2$ gauge transformations $\Omega_1'',\Omega_2''$
on $\ket{a},\ket{b},\ket{c}$:
\be\label{eqn:su2at-2}
\TdT\ket{a}_{SU(2)}=\ket{a}_{SU(2)}\,,\quad
\TdT\ket{b}_{SU(2)}=e^{-\frac{\pi i}{4}}\ket{b}_{SU(2)}\,,\quad
\TdT\ket{c}_{SU(2)}=-\ket{c}_{SU(2)}\,,
\ee
\be
\left.\begin{array}{llll}
\TdS\ket{a}_{SU(2)}&=\tfrac{1}{2}\ket{a}_{SU(2)} &+\tfrac{1}{\sqrt{2}}\ket{b}_{SU(2)} & +\tfrac12\ket{c}_{SU(2)}\,,\\
\\
\TdS\ket{b}_{SU(2)}&=\tfrac{1}{\sqrt{2}}\ket{a}_{SU(2)} & & -\tfrac{1}{\sqrt{2}}\ket{c}_{SU(2)}\,,\\
\\
\TdS\ket{c}_{SU(2)}&=\tfrac{1}{2}\ket{a}_{SU(2)} &-\tfrac{1}{\sqrt{2}}\ket{b}_{SU(2)} & +\tfrac12\ket{c}_{SU(2)}\,,
\end{array}\right\}
\ee
and
\be
\Omega_1''\ket{a}_{SU(2)} = \ket{c}_{SU(2)}\,,\quad
\Omega_1''\ket{b}_{SU(2)} = \ket{b}_{SU(2)}\,,\quad
\Omega_1''\ket{c}_{SU(2)} = \ket{a}_{SU(2)}\,,
\ee
\be
\Omega_2''\ket{a}_{SU(2)}= \ket{a}_{SU(2)}\,,\quad
\Omega_2''\ket{b}_{SU(2)}= -\ket{b}_{SU(2)}\,,\quad
\Omega_2''\ket{c}_{SU(2)}= \ket{c}_{SU(2)}\,.
\ee
Comparing the above with \eqref{eqn:TSabc}-\eqref{eqn:omeabc}, we see that
the action of $\TdT,\TdS,\Omega_1'',\Omega_2''$ agrees with that on the Hilbert space of $SU(2)$ Chern--Simons theory at level $\lvk=-2$. [To see this, note that the eigenvalues of $\TdT$ above are, up to an overall phase, conjugates of those in \eqref{eqn:TSabc}.] Chern--Simons theories with
negative levels $\lvk<0$ are equivalent to the
theories with positive levels $(-\lvk)$ but with the opposite
orientation of spacetime.
Thus, if we wish to keep the same spacetime orientation for all
the sectors of the theory,
we have to include negative Chern--Simons levels. We conclude that $\Hilb_{(2)}(\frac{\pi}{2})$  is equivalent to the Hilbert space of $[U(1)_4\times SU(2)_{-2}]/\Z_2$ Chern--Simons theory.

For $n=3$ we get two $\brperm$-twisted sectors.
The first, corresponding to $[\sigma]=(3)$, is $2$-dimensional and spanned by
$$
\ket{\fpIIzzz},\qquad
\ket{\fpIIccc}.
$$
We denote it by $\Hilb_{(3)}(\frac{\pi}{2}).$
Let us first separate the $U(1)_{6}$ center, as in
\eqref{eqn:decpsi}:
\bear
\ket{\fpIIzzz} &=&
 \ket{a}_{SU(3)}\otimes\ket{0}_{U(1)}
+\ket{b}_{SU(3)}\otimes\ket{2}_{U(1)}
+\ket{c}_{SU(3)}\otimes\ket{4}_{U(1)}
\,,\nn\\
\ket{\fpIIccc} &=&
 \ket{a}_{SU(3)}\otimes\ket{3}_{U(1)}
+\ket{b}_{SU(3)}\otimes\ket{5}_{U(1)}
+\ket{c}_{SU(3)}\otimes\ket{1}_{U(1)}
\,,\nn
\eear
where $\ket{p}_{U(1)}$ ($p=0,\dots,5$)
are states of $U(1)$ Chern--Simons theory at level $\lvk n=6$,
and we have used the known
action of $\Psym,\Qsym$ to simplify the decomposition.
$\ket{a}_{SU(3)},\ket{b}_{SU(3)},\ket{c}_{SU(3)}$
are unspecified states associated with the $SU(3)$
degrees of freedom only.
Using \eqref{eqn:TdTfpII} and \eqref{eqn:TSonU(1)} we calculate
(up to an overall phase):
$$
\TdT\ket{a}_{SU(3)} = \ket{a}_{SU(3)}
\,,\qquad
\TdT\ket{b}_{SU(3)} =
e^{-\frac{2\pi i}{3}}\ket{b}_{SU(3)}
\,,\qquad
\TdT\ket{c}_{SU(3)} =
e^{-\frac{2\pi i}{3}}\ket{c}_{SU(3)}
\,,
$$
and
\be\label{eqn:TdSabc(n=3)}
\left.
\begin{array}{lr}
\TdS\ket{a}_{SU(3)} &=
\frac{1}{\sqrt{3}}
\Bigl(
\ket{a}_{SU(3)}+\ket{b}_{SU(3)}+\ket{c}_{SU(3)}
\Bigr)
\\ & \\
\TdS\ket{b}_{SU(3)} &=
\frac{1}{\sqrt{3}}
\Bigl(\ket{a}_{SU(3)}
+e^{\frac{2\pi i}{3}}\ket{b}_{SU(3)}
+e^{-\frac{2\pi i}{3}}\ket{c}_{SU(3)}
\Bigr)
\\ & \\
\TdS\ket{c}_{SU(3)} &=
\frac{1}{\sqrt{3}}
\Bigl(\ket{a}_{SU(3)}
+e^{-\frac{2\pi i}{3}}\ket{b}_{SU(3)}
+e^{\frac{2\pi i}{3}}\ket{c}_{SU(3)}
\Bigr)
\\
\end{array}\right\}\,.
\ee
The $\Z_3\subset SU(3)$ center acts, according to
\eqref{eqn:Omjpp}, as
$$
\Omega_2''\ket{a}_{SU(3)} = \ket{a}_{SU(3)}
\,,\qquad
\Omega_2''\ket{b}_{SU(3)} =
e^{-\frac{2\pi i}{3}}\ket{b}_{SU(3)}
\,,\qquad
\Omega_2''\ket{c}_{SU(3)} =
e^{\frac{2\pi i}{3}}\ket{c}_{SU(3)}
\,,
$$
and
$$
\Omega_1''\ket{a}_{SU(3)} = \ket{b}_{SU(3)}
\,,\qquad
\Omega_1''\ket{b}_{SU(3)} =\ket{c}_{SU(3)}
\,,\qquad
\Omega_1''\ket{c}_{SU(3)} =\ket{a}_{SU(3)}
\,.
$$
These formulas for $\TdT,\TdS,\Omega_1'',\Omega_2''$,
are consistent with the Hilbert space of $SU(3)$
Chern--Simons theory
at level $\lvk=-1$.
To check that $\ket{a}_{SU(3)},\ket{b}_{SU(3)},\ket{c}_{SU(3)}$
agree with the states of $SU(3)_{-1}$ Chern--Simons theory,
we note that $U(3)_{-1}=[U(1)_{-3}\times SU(3)_{-1}]/\Z_3$ has
a one-dimensional Hilbert space, spanned by a state of the form
$$
\ket{0}_{U(1)_{-3}}\otimes\ket{a'}_{SU(3)_{-1}}
+
\ket{1}_{U(1)_{-3}}\otimes\ket{b'}_{SU(3)_{-1}}
+
\ket{2}_{U(1)_{-3}}\otimes\ket{c'}_{SU(3)_{-1}}
$$
where, as the notation suggests,
$\ket{p}_{U(1)_{-3}}$ ($p=0,1,2$) are the states
of $U(1)_{-3}$ Chern--Simons theory.
The $\TdT,\TdS$ transformations of
$\ket{a'}_{SU(3)_{-1}},
\ket{b'}_{SU(3)_{-1}},\ket{c'}_{SU(3)_{-1}}$
can then be recovered from \eqref{eqn:zspr} and \eqref{eqn:ztoddk}.

The second $\brperm$-twisted sector for $n=3$ corresponds to $[\sigma]=(2,1)$ and is spanned by
\be\label{eqn:H3(3,2)states}
\ket{\fpIIz\fpIIzz},\quad
\ket{\fpIIz\fpIIxy},\quad
\ket{\fpIIz\fpIIcc},\quad
\ket{\fpIIc\fpIIzz},\quad
\ket{\fpIIc\fpIIxy},\quad
\ket{\fpIIc\fpIIcc}.
\ee
We denote it by $\Hilb_{(2,1)}(\frac{\pi}{2}).$
As explained at the top of \secref{subsec:other-sectors},
this sector is reducible, and equivalent to the Hilbert space
of $U(1)_2\times [U(1)_4\times SU(2)_{-2}]/\Z_2$ Chern--Simons theory.

So, altogether,
in the case $n=3$ we found that the Hilbert space
is a direct sum of three Hilbert spaces:
$$
U(3)_{6,2}\oplus U(3)_{6,-1}\oplus [U(1)_2\otimes U(2)_{4,-2}]\,.
$$
The first two have gauge group $U(3)$, and the third
has gauge group $U(1)\times U(2).$


\subsubsection{$\pht=\tfrac{\pi}{3}$ ($\lvk=1$)
and $\pht=\tfrac{2\pi}{3}$ ($\lvk=3$)}
\label{subsubsec:k=1,3}
Except for the mysterious
$\Hilb_{(2,2)}(\frac{\pi}{3})$ sector
mentioned above, we again find that each
single-particle sector
$\Hilb_{(n)}(\pht)$
is equivalent to a Chern--Simons Hilbert space.
The derivations are presented in \appref{app:HilbDecomp},
and the results are summarized in
\tabref{tab:lvkpp} in the concluding section.


\subsection{Wilson loop operators}
\label{subsec:Wilson}

So far we have found a
correspondence between the Hilbert space of
ground states
of the S-duality twisted compactification of
\SUSY{4} SYM of \secref{sec:Problem}
and
the Hilbert space of ground states
of the type-IIA background
of \secref{sec:IIA-dual}.
The next step is to extend this correspondence to operators.
The natural operators to start with on the type-IIB
(gauge theory) side are Wilson loops
at a constant $x_3$ and along a curve $C\subset\R^{2,1}$
where $\R^{2,1}$ corresponds to directions $0,1,2.$
Let us denote a Wilson loop operator in the fundamental
representation of $U(n)$ by $\WilV(C,x_3).$
We assume that the curvature of $C$ is much smaller than
the compactification scale $1/\xR.$
We can also consider supersymmetric extensions
of Wilson loops, as constructed in 
\cite{Rey:1998ik,Maldacena:1998im}. These include
additional terms that depend on the scalar fields of
\SUSY{4} SYM, but since we have eliminated all the zero modes
of scalar fields in \secref{subsec:R-twist}
we can expect that at low-energy
the scalar fields are effectively zero, and
it is likely that the difference between ordinary
and supersymmetric Wilson loops disappears.
In any case, we ask to what operator $\WilV(C,x_3)$
flows to at low-energy.
This question will be addressed in more detail
in an upcoming paper \cite{InPreparation},
but we will make a few preliminary remarks in the present
subsection. For simplicity, we restrict to the case $\lvk=2.$

Note that because of the S-duality twist,
the operator $\WilV(C,x_3)$ satisfies 
the boundary conditions
\be\label{eqn:WiltHooft}
\WilV(C,x_3) = \tHooftV(C,x_3+2\pi\xR)
= \WilV(C,x_3+4\pi\xR)^\dagger
= \tHooftV(C,x_3+6\pi\xR)^\dagger
= \WilV(C,x_3+8\pi\xR)
\,,
\ee
where $\tHooftV$ is the magnetic dual `t Hooft loop operator,
and $\WilV^\dagger$ is the charge-conjugate Wilson loop
operator in the anti-fundamental representation of $U(n).$
We now define linear combinations which diagonalize
the boundary conditions \eqref{eqn:WiltHooft}:
\be\label{eqn:WHdef}
\WHV^{(p)}(C,x_3) \equiv
\WilV(C,x_3)
+i^p\tHooftV(C,x_3)
+(-1)^p\WilV(C,x_3)^\dagger
+(-i)^p\tHooftV(C,x_3)^\dagger
\,,\qquad
p=0,1,2,3.
\ee
Their Fourier transforms along $x_3$ are
\be\label{eqn:FWH}
\WHV^{(p)}(C,x_3) = \sum_{m\in\Z}
\fWHV_{m+\frac{p}{4}}^{(p)}(C)e^{(m+\frac{p}{4})\frac{i x_3}{\xR}}
\,.
\ee
For $p\neq 0$,
when acting on the ground states all 
the modes $\fWHV_{m+\frac{p}{4}}^{(p)}(C)$ 
create linear combinations of states with 
nonzero fractional Kaluza-Klein momentum, and therefore
have energy at least $\frac{1}{4\xR}$.
Thus, when we project these operators to the 
Hilbert space of ground states 
they all vanish except $\WHV^{(0)}.$
We can therefore surmise that 
the operators $\WilV(C,x_3)$,
$\tHooftV(C,x_3)$, $\WilV(C,x_3)^\dagger$,
and $\tHooftV(C,x_3)^\dagger$, all flow at low-energy
to the same operator:
\be\label{eqn:WMflow}
\WilV(C,x_3),\,
\tHooftV(C,x_3),\, 
\WilV(C,x_3)^\dagger,\,
\tHooftV(C,x_3)^\dagger
\stackrel{\text{IR}}{\longrightarrow}
\tfrac{1}{4}\fWHV^{(0)}_0
\,.
\ee
In other words, at low-energy only the S-duality
invariant combination
$$
\WHV^{(0)}(C,x_3) \equiv
\WilV(C,x_3)
+\tHooftV(C,x_3)
+\WilV(C,x_3)^\dagger
+\tHooftV(C,x_3)^\dagger
$$
is relevant.
And in particular we note that even though the gauge group
is complex, $\WHV^{(0)}(C,x_3)$
is real and gives rise to a self-adjoint operator
on the Hilbert space of ground states.
For example, for $U(1)$ gauge group we saw that
the Wilson loops
$\WilV_1,\WilV_2$ defined in \eqref{eqn:WilVs}
are self-adjoint for $\lvk=2$,
even though they are not self-adjoint in Chern-Simons theory
at level $\lvk>2.$
A similar phenomenon occurs for the C-twist
that we studied in \secref{sec:Ctwist}.
This time the combinations that survive the low-energy
limit are $\WilV+\WilV^\dagger.$
In \secref{subsec:U(1)-C} we saw that 
starting with $U(2)$ gauge group, with the help of 
a C-twist, we get a low-energy $SU(2)$ gauge theory.
So, while Wilson loop operators in $U(2)$ gauge theory
are not
self-adjoint, they are in $SU(2)$ since its fundamental
and anti-fundamental representations are equivalent!

The action of $\WHV^{(0)}(C,x_3)$ on ground states
can be studied using the type-IIA dual by introducing
probe strings, but this is beyond the scope of the present
paper and will be discussed in detail in \cite{InPreparation}.
We will only mention that the operators
defined in \eqref{eqn:UaUb-VaVb} play a role in the construction.


\section{Realization via the $(2,0)$-theory}
\label{sec:twoz}

S-duality is geometrically realized in terms of
the six-dimensional $(2,0)$-theory.
In this section we will discuss a geometrical construction
in terms of the $(2,0)$-theory of a setting
similar to that of \secref{sec:Problem}.
The $(2,0)$-theory that was proposed by Witten
in \cite{Witten:1995zh} is still poorly understood,
but there are at least two proposals for a definition:
one as a M(atrix)-model \cite{Aharony:1997th}
and another in terms of deconstruction \cite{ArkaniHamed:2001ie}.
(For some attempts in other directions see
\cite{Ganor:1996nf}-\cite{Rey:2010uz}.)
In this section we will actually not have to use
any of the fundamental definitions, however,
because known results about the low-energy description
of the theory will suffice.
The $(2,0)$-theory has an $SO(5)$ R-symmetry, so
we cannot reproduce the identical setting of \secref{sec:Problem},
because the full $SO(6)$ R-symmetry twist
cannot be realized in terms of the
$(2,0)$-theory.
Instead, we will produce a closely related setting as follows.

As Witten proposed \cite{Witten:1995zh},
\SUSY{4} $U(n)$ super Yang--Mills theory with coupling constant
$\tau$ is the low-energy limit of a six-dimensional theory
compactified on $T^2,$ with $\tau$ being the complex structure
parameter of the torus, so that
S-duality $\tau\rightarrow (\xa\tau+\xb)/(\xc\tau+\xd)$
is realized as an element of the mapping class group of the $T^2.$
This immediately leads to a realization of the S-duality
twisted compactification defined in \secref{subsec:S-twist}:
we simply take the $(2,0)$-theory (for the appropriate $n$)
and compactify it on the space $\MXS$ defined in
\secref{subsec:IIA}.
Recall that $\MXS\simeq (T^2\times S^1)/\Z_\ord$,
where $S^1$ has radius $2\pi\xR\ord.$
The torus $T^2$ has complex structure $\tau$, and
we denote its area by $\Area$,
so that in the limit
\be\label{eqn:AreaxRlim1}
\Area\ll\xR^2\,,
\ee
we recover
the S-duality twisted compactification of \secref{subsec:S-twist}.

The R-symmetry twist of \secref{subsec:R-twist},
however, is more difficult to realize because the $(2,0)$-theory
only has an $SO(5)$ global R-symmetry, not $SO(6).$
The enhanced $SO(6)$ R-symmetry of \SUSY{4} SYM only arises
as an effective low-energy symmetry.
To get around this obstacle,
we note that the R-symmetry twist we
used in \secref{subsec:R-twist} can be continuously deformed
while preserving \SUSY{4} SUSY in 2+1D by replacing
\eqref{eqn:gtwN=6} with
\be\label{eqn:gtwN=4}
\gtw'=
\begin{pmatrix}
e^{\frac{i}{2}\pht} & &  & \\
& e^{\frac{i}{2}\pht} &  & \\
& & e^{i\epsilon-\frac{i}{2}\pht}  & \\
& & & e^{-i\epsilon-\frac{i}{2}\pht}  \\
\end{pmatrix} \in SU(4)_R\,.
\ee
For $\epsilon=\pht$ we recover \eqref{eqn:gtwN=6}, while
for $\epsilon=0$ we get an R-symmetry twist in a subgroup
$SO(2)\subset SO(5)\subset SO(6)$, and thus it can be realized
inside the $SO(5)$ R-symmetry of the $(2,0)$-theory.
For $\epsilon=0$ we also get additional bosonic
and fermionic zero-modes from the scalars and gluinos,
but for $0<\epsilon\le\pht$ they are absent.
Presumably, the low-energy description for $0<\epsilon\le\pht$
is independent of $\epsilon$
(by supersymmetry, or if the theory is indeed topological),
and so we can study the theory at $\epsilon=0$ first,
and then deform by a small $\epsilon$, provided we
can understand that deformation in the low-energy description
of the $\epsilon=0$ setting. In fact,
for the specific purpose
of understanding some of the $\brperm$-twisted sectors
in \secref{subsec:twozero-U(2)}, it will suffice to
study the $\epsilon=0$ case.

To better understand the low-energy limit of the $\epsilon=0$
theory, which is a 2+1D theory
with \SUSY{4} supersymmetry, we will make the
plausible assumption that
the low-energy theory is independent of the dimensionless parameter
$\Area/\xR^2$ and take the limit opposite to
\eqref{eqn:AreaxRlim1}, namely
\be\label{eqn:AreaLarger}
\frac{\Area}{\xR^2}\rightarrow\infty.
\ee
To analyze this limit, it is convenient to describe
$\MXS$ as an $S^1$ fibration over a base $T^2/\Z_\ord.$
The fibers are constructed as follows.
Fix a point on $T^2$ that corresponds to coordinate
$z$ (with the
identification $z\sim z+1\sim z+\tau$),
and consider the set of all points with coordinates
$(z,x_3)$, where $x_3$ is arbitary.
The generic fiber is an $S^1$ of circumference
$2\pi\xR\ord.$
Because of the $\Z_\ord$ action, the fibers that we get
for $z$ and $e^{2\pi i/\ord}z$ are identical,
so the base is $T^2/\Z_\ord$, as stated above.

This fibration is not quite a circle bundle, however,
because there exist special points on the $T^2/\Z_\ord$ base
where the fiber is smaller than the generic one.
This happens
if $z$ is invariant (up to $\Z+\Z\tau$)
under some nontrivial element of the orbifold group $\Z_\ord.$
For $\tau=i$ this is the case for three
inequivalent $z$'s: $z=0,\frac{1}{2},$ and
$\frac{1}{2}(1+i).$
The $T^2$ points $0$ and $\frac{1}{2}(1+i)$ are fixed
by the entire $\Z_4$, and the fiber over those points
is of size $2\pi\xR$, i.e., $\frac{1}{4}$
of the generic fiber. The point $\frac{1}{2}$
is fixed by a $\Z_2\subset\Z_4$ subgroup and the fiber
over it is of size $4\pi\xR$, i.e., $\frac{1}{2}$
of the generic fiber.
We can choose
the fundamental domain of the $\Z_4$ action
on $T^2$ to be a triangle with vertices
$z=0,\frac{1}{2}(1+i),1$ and with extra identifications
on the boundary of the triangle which are
induced by the identification
$z\simeq 1+i z$ and $z\simeq 1-z.$
The result is depicted in the $\lvk=2$ portion of
\figref{fig:FundamentalDomains}.
The situation is similar for $\tau=e^{\pi i/3}.$
Here again there are three special points of $T^2/\Z_\ord$
which are invariant under subgroups of $\Z_\ord$, and
the fibers there are smaller than the generic fiber.


\FIGURE[t!]{
\begin{picture}(430,135)

\put(10,15){\begin{picture}(150,110)
\put(150,-10){\line(0,1){120}}
\put(30,110){($\lvk=1$)}
\put(90,110){($\ord=6$)}

\color{green}
\thinlines
\qbezier(3,0)(3,0)(3,2)
\qbezier(6,0)(6,0)(6,3)
\qbezier(9,0)(9,0)(9,5)
\qbezier(12,0)(12,0)(12,7)
\qbezier(15,0)(15,0)(15,9)
\qbezier(18,0)(18,0)(18,10)
\qbezier(21,0)(21,0)(21,12)
\qbezier(24,0)(24,0)(24,14)
\qbezier(27,0)(27,0)(27,16)
\qbezier(30,0)(30,0)(30,17)
\qbezier(33,0)(33,0)(33,19)
\qbezier(36,0)(36,0)(36,21)
\qbezier(39,0)(39,0)(39,23)
\qbezier(42,0)(42,0)(42,24)
\qbezier(45,0)(45,0)(45,26)

\qbezier(87,0)(87,0)(87,2)
\qbezier(84,0)(84,0)(84,3)
\qbezier(81,0)(81,0)(81,5)
\qbezier(78,0)(78,0)(78,7)
\qbezier(75,0)(75,0)(75,9)
\qbezier(72,0)(72,0)(72,10)
\qbezier(69,0)(69,0)(69,12)
\qbezier(66,0)(66,0)(66,14)
\qbezier(63,0)(63,0)(63,16)
\qbezier(60,0)(60,0)(60,17)
\qbezier(57,0)(57,0)(57,19)
\qbezier(54,0)(54,0)(54,21)
\qbezier(51,0)(51,0)(51,23)
\qbezier(48,0)(48,0)(48,24)
\qbezier(45,0)(45,0)(45,26)

\color{blue}
\thicklines

\multiput(0,0)(45,78){2}{\qbezier(0,0)(0,0)(90,0)}
\multiput(0,0)(90,0){2}{\qbezier(0,0)(0,0)(45,78)}

\thinlines
\qbezier(90,0)(90,0)(45,78)
\qbezier(45,26)(45,26)(0,0)
\qbezier(45,26)(45,26)(90,0)
\qbezier(45,26)(45,26)(45,78)

\qbezier(90,52)(90,52)(135,78)
\qbezier(90,52)(90,52)(90,0)
\qbezier(90,52)(90,52)(45,78)

\thicklines
\color{red}
\put(0,0){\circle*{4}}
\put(45,26){\circle*{4}}
\put(45,0){\circle*{4}}

\thicklines
\color{black}
\multiput(21,-5)(3,0){2}{\line(0,1){10}}
\multiput(66,-5)(3,0){2}{\line(0,1){10}}


\qbezier(26,8)(26,8)(20,18)
\qbezier(64,8)(64,8)(70,18)

\color{black}
\put(30,32){$(\tfrac{1}{3})$}
\put(37,-15){$(\tfrac{1}{2})$}
\put(-18,-3){$(\tfrac{1}{6})$}

\color{black}
\end{picture}}

\put(180,15){\begin{picture}(100,110)
\put(120,-10){\line(0,1){120}}
\put(0,110){($\lvk=2$)}
\put(60,110){($\ord=4$)}

\color{blue}
\thicklines
\multiput(0,0)(90,0){2}{\line(0,1){90}}
\multiput(0,0)(0,90){2}{\line(1,0){90}}

\thinlines

\color{green}

\put(5,5){\line(1,0){80}}
\put(10,10){\line(1,0){70}}
\put(15,15){\line(1,0){60}}
\put(20,20){\line(1,0){50}}
\put(25,25){\line(1,0){40}}
\put(30,30){\line(1,0){30}}
\put(35,35){\line(1,0){20}}
\put(40,40){\line(1,0){10}}


\thinlines
\color{blue}
\put(45,45){\line(1,1){45}}
\put(45,45){\line(1,-1){45}}
\put(45,45){\line(-1,1){45}}
\put(45,45){\line(-1,-1){45}}

\thicklines
\color{black}
\multiput(26,16)(3,3){2}{\line(-1,1){10}}
\multiput(64,16)(-3,3){2}{\line(1,1){10}}

\put(22,-5){\line(0,1){10}}
\put(68,-5){\line(0,1){10}}

\thicklines
\color{red}
\put(0,0){\circle*{4}}
\put(45,0){\circle*{4}}
\put(45,45){\circle*{4}}

\color{black}
\put(37,-14){$(\tfrac{1}{2})$}
\put(37,53){$(\tfrac{1}{4})$}
\put(-18,-3){$(\tfrac{1}{4})$}

\color{black}
\end{picture}}

\put(310,15){\begin{picture}(140,110)
\put(0,110){($\lvk=3$)}
\put(60,110){($\ord=3$)}

\color{green}
\thinlines
\qbezier(90,4)(90,4)(87,2)	
\qbezier(90,8)(90,8)(83,4)	
\qbezier(90,12)(90,12)(80,6)	
\qbezier(90,16)(90,16)(76,8)	
\qbezier(90,20)(90,20)(73,10)	
\qbezier(90,24)(90,24)(69,12)	
\qbezier(90,28)(90,28)(66,14)	
\qbezier(90,32)(90,32)(62,16)	
\qbezier(90,36)(90,36)(59,18)	
\qbezier(90,40)(90,40)(55,20)	
\qbezier(90,44)(90,44)(52,22)	
\qbezier(90,48)(90,48)(48,24)	
\qbezier(90,52)(90,52)(45,26)	

\qbezier(45,74)(45,74)(48,76)
\qbezier(45,70)(45,70)(52,74)
\qbezier(45,66)(45,66)(55,72)
\qbezier(45,62)(45,62)(59,70)
\qbezier(45,58)(45,58)(62,68)
\qbezier(45,54)(45,54)(66,66)
\qbezier(45,50)(45,50)(69,64)
\qbezier(45,46)(45,46)(73,62)
\qbezier(45,42)(45,42)(76,60)
\qbezier(45,38)(45,38)(80,58)
\qbezier(45,34)(45,34)(83,56)
\qbezier(45,30)(45,30)(87,54)

\color{blue}
\thicklines

\multiput(0,0)(45,78){2}{\qbezier(0,0)(0,0)(90,0)}
\multiput(0,0)(90,0){2}{\qbezier(0,0)(0,0)(45,78)}

\thinlines
\qbezier(45,26)(45,26)(0,0)
\qbezier(45,26)(45,26)(90,0)
\qbezier(45,26)(45,26)(45,78)

\qbezier(90,52)(90,52)(135,78)
\qbezier(90,52)(90,52)(90,0)
\qbezier(90,52)(90,52)(45,78)

\thicklines
\color{red}
\put(45,78){\circle*{4}}
\put(45,26){\circle*{4}}
\put(90,52){\circle*{4}}

\color{black}
\put(38,86){$(\tfrac{1}{3})$}
\put(38,13){$(\tfrac{1}{3})$}
\put(83,60){$(\tfrac{1}{3})$}

\thicklines
\qbezier(63,9)(66,14)(69,19)
\qbezier(67,7)(70,12)(73,17)

\qbezier(84,26)(90,26)(96,26)
\qbezier(39,50)(45,50)(51,50)
\qbezier(39,54)(45,54)(51,54)

\qbezier(65,60)(68,65)(71,70)

\color{black}
\end{picture}}

\end{picture}
\caption{
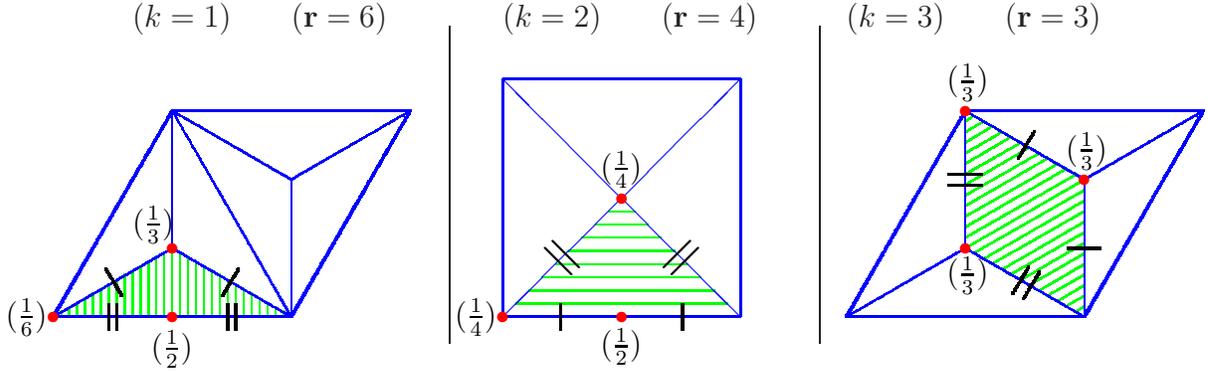
Fundamental domains of the $\Z_\ord$ action on $T^2.$
Fundamental domains are triangles for $\lvk=1,2$
or a rhombus for $\lvk=3$ with edges identified as
indicated by the markings.
The special points are fixed points of $\Z_\ord$ or a proper
subgroup of it, and the fractions indicate the size
of the $S^1$ fiber. These fractions
are the inverses of the orders of the fixed-point subgroup.
}
\label{fig:FundamentalDomains}
}


For $i=1,2,3$, we denote the $i^{th}$ special point by
$\spP_i\in T^2/\Z_\ord.$
We denote the order of the subgroup of $\Z_\ord$
that fixes the special point $\spP_i$ by $p_i.$
The generic fiber has circumference $2\pi\ord\xR$,
and so the fiber at $\spP_i$ has circumference $2\pi\ord\xR/p_i.$
We find the following values of $p_i$:
\be\label{eqn:pivals}
(p_1, p_2, p_3) = \left\{\begin{array}{lll}
(3,3,3) & \text{for $\ord=3$ ($\tau=e^{\pi i/3}$),} \\
(4,4,2) & \text{for $\ord=4$ ($\tau=i$),} \\
(6,3,2) & \text{for $\ord=6$ ($\tau=e^{\pi i/3}$).} \\
\end{array}\right.
\ee
\figref{fig:FundamentalDomains} shows convenient
reresentations of $T^2/\Z_\ord$ with the special points
marked by the fraction $1/p_i.$
Note that in all three cases
\be\label{eqn:sumpi}
1 = \sum_{i=1}^3\frac{1}{p_i}
\,.
\ee


\subsection{Reduction to 4+1D and 2+1D}
\label{subsec:Reduction}

When the $(2,0)$-theory is compactified on $\MXS$
in the limit \eqref{eqn:AreaLarger},
we can ``dimensionally reduce'' the theory on the generic
$S^1$ fiber to get, away from the three singular points
$\spP_1,\spP_2,\spP_3,$
a low-energy 4+1D \SUSY{2}
super Yang--Mills theory (with $16$ supersymmetry generators).
The theory is formulated on $\R^{2,1}\times (T^2/\Z_\ord)$,
and has a coupling constant
$$
\gYM^{\text{(5D)}} = 2\pi(2\xR\ord)^{\frac{1}{2}}.
$$
The space $T^2/\Z_\ord$ is locally flat, except for
curvature singularities
at the special points $\spP_1,\spP_2,\spP_3.$

We denote the bulk 4+1D $U(n)$ gauge field by $\bulkC$,
and for simplicity of the discussion ignore the superpartners.
The resulting low-energy description
is constructed by combining the bulk 4+1D action for $\bulkC$
with additional localized interactions at the special points
$\spP_1,\spP_2,\spP_3.$

What is the contribution of the special point $\spP_i$
to the action?
Near $\spP_i$ the base looks like $\R^2/\Z_{p_i}$,
which is a cone.
The total space looks like $(S^1\times\R^2)/\Z_{p_i}$
where $\Z_{p_i}$ acts as rotation by $2\pi/p_i$ on $\R^2$
and translation by $2\pi\xR\ord/p_i$ on $S^1.$
To proceed, we switch to the M-theory realization
where we have $n$ M5-branes on $(S^1\times\R^2)/\Z_{p_i}.$
We also need to realize the R-symmetry twist.
For \eqref{eqn:gtwN=4} with $\epsilon=0$,
this twist can be expressed as a $2\pi/p_i$ rotation
in an additional $\R^2$ plane transverse to the
M$5$-branes. Altogether,
combining this transverse $\R^2\simeq\C$ with the
$\R^2\simeq\C$ that appears in $(S^1\times\R^2)/\Z_{p_i}$,
we get M-theory on $(S^1\times\C^2)/\Z_{p_i}$
where the generator of
$\Z_{p_i}$ acts on a point of $(S^1\times\C^2)$
with coordinates $(x_3,\zeta_1,\zeta_2)$ as:
\be\label{eqn:Zpi}
\Z_{p_i}:
(x_3,\zeta_1,\zeta_2)\mapsto
(x_3+\tfrac{2\pi\xR\ord}{p_i},
e^{\frac{2\pi i}{p_i}}\zeta_1,
e^{-\frac{2\pi i}{p_i}}\zeta_2)
\,.
\ee

We are now ready to describe the low-energy contribution
of $\spP_i$ to the action.
We can arrive at the answer by combining a
thirteen-year-old result of Witten
\cite{Witten:1997kz} with a fairly recent
result of Gaiotto and Witten \cite{Gaiotto:2008ak}.
In \cite{Witten:1997kz}, Witten showed that
M-theory on $(S^1\times\C^2)/\Z_{p_i}$
in the $\xR\rightarrow 0$ limit and in the
region near the origin ($\zeta_1=\zeta_2=0$)
is dual to a $(1,p_i)$ $5$-brane
(an object with $p_i$ units of NS$5$-brane
charge and $1$ unit of D$5$-brane charge) of type-IIB 
string theory.
We will review Witten's arguments below, and see
that under the duality the $n$ M$5$-branes
are transformed into $n$ D$3$-branes that end on
the $(1,p_i)$ $5$-brane.
Luckily, in the last section of \cite{Gaiotto:2008ak},
Gaiotto and Witten described the boundary interaction
of $n$ D$3$-branes ending on a $(1,p_i)$ $5$-brane,
and so we can use that interaction to describe
the vicinity of our special point $\spP_i.$

Before we proceed to the details of the interaction,
let us review the part of
Witten's arguments from \cite{Witten:1997kz} that
apply to our case.
Starting with $(S^1\times\C^2)/\Z_{p_i}$,
we first replace $\C^2$ with a Taub-NUT space,
whose metric can be written as
\be\label{eqn:dsTN}
ds^2 =
\left(1+\frac{\xS}{2r}\right)^{-1} (dy +\cos\theta\,d\phi)^2
+\left(1+\frac{\xS}{2r}\right)
 (dr^2 + r^2 (d\theta^2+\sin^2\theta\,d\phi^2))
\,,
\ee
where $y$ is a periodic coordinate with range
$0\le y<2\pi\xS.$
The origin $r=0$ is a smooth point,
and the isometry that acts as $y\rightarrow y+2\pi\xS/p_i$
(keeping the other coordinates unchanged)
rotates the tangent plane at the origin
in exactly the same way that the
$\C^2$ parameterized by
$(\zeta_1,\zeta_2)$ is rotated in \eqref{eqn:Zpi}.
We then replace $\C^2$ in the space
$(S^1\times\C^2)/\Z_{p_i}$ with the Taub-NUT space
\eqref{eqn:dsTN} and take the limit of large $\xS.$

Next, we take $r\rightarrow\infty$ at constant $\theta,\phi,$
and focus on the $T^2$ in the $(x_3,y)$ directions.
The periodicities and the $\Z_{p_i}$ orbifold
induce the identifications
$$
(x_3,y)\simeq
(x_3,y+2\pi\xS)\simeq
(x_3+\tfrac{2\pi\xR\ord}{p_i},y-\tfrac{2\pi\xS}{p_i})\,.
$$
Changing coordinates to a complex variable
$$
w = \frac{1}{2\pi\xR\ord}(x_3 + i y)\,,
$$
we find the identifications
$$
w\simeq w+1\simeq w-\frac{1}{p_i}+i\frac{\xS}{p_i\xR\ord}
\,.
$$
We now reduce M-theory on
the $T^2$ that is in the $(x_3,y)$
directions to type-IIB with complex coupling constant
\be\label{eqn:tauIIB}
\tauIIB=-\frac{1}{p_i}+i\frac{\xS}{p_i\xR\ord}\,.
\ee

To identify which $(p,q)$ $5$-brane
we get in type-IIB, we have to find the Taub-NUT charge
of the metric in terms of $w.$
More explicitly, for fixed and large $r$, the $T^2$
is fibered over the $S^2$ (parameterized by the $\theta,\phi$
coordinates), and the structure group of the fibration
is generated by translations in $y.$ (See \figref{fig:pq}.)
In terms of $w$, the translation
$(x_3,y)\rightarrow (x_3,y+\epsilon)$
is equivalent to $w\rightarrow w + \epsilon(p_i\tauIIB+1).$
The combination $p_i\tauIIB+1$ identifies the Taub-NUT charge
as the one that reduces to the $(1,p_i)$ $5$-brane.

\FIGURE[t!]{
\begin{picture}(430,135)

\thinlines
\color{blue}

\thinlines
\color{blue}

\multiput(130,20)(120,0){3}{\line(-2,1){40}}
\multiput(90,40)(120,0){3}{\line(-2,1){40}}
\put(410,60){\line(2,-1){20}}
\multiput(50,60)(120,0){4}{\line(-2,1){40}}
\multiput(130,80)(120,0){3}{\line(-2,1){40}}

\multiput(10,20)(0,20){5}{\line(1,0){420}}

\color{red}
\thicklines
\put(130,20){\vector(-2,1){40}}
\put(130,20){\vector(0,1){60}}
\multiput(90,41)(0,12){5}{\line(0,1){10}}
\multiput(130,80)(-10,5){4}{\qbezier(0,0)(0,0)(-6,3)}
\put(132,65){$\vec{a}$}
\put(86,26){$\vec{b}$}

\color{black}
\put(10,15){\vector(1,0){60}}
\put(72,12){$y$}

\put(10,20){\vector(0,1){20}}
\put(12,28){$x_3$}

\put(250,15){\vector(1,0){120}}
\put(370,15){\vector(-1,0){120}}
\put(250,13){\line(0,1){4}}
\put(370,13){\line(0,1){4}}
\put(302,2){$2\pi\xS$}

\put(298,20){\line(1,0){4}}
\put(298,40){\line(1,0){4}}
\put(300,20){\vector(0,1){20}}
\put(300,40){\vector(0,-1){20}}
\put(302,27){$2\pi\xR\ord/p_i$}

\end{picture}
\caption{
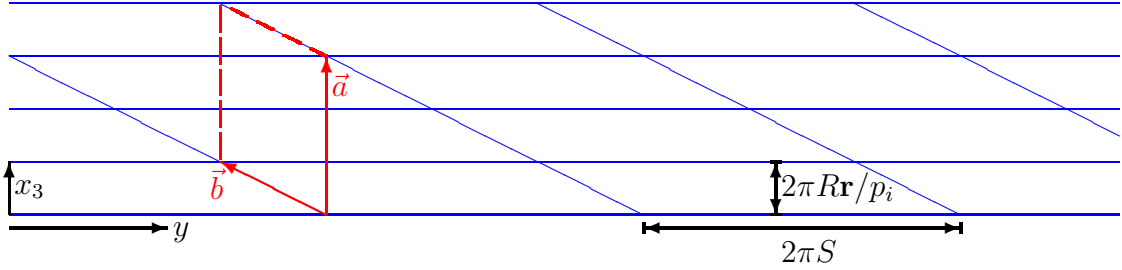
The fiber of the space $(S^1\times \text{Taub-NUT})/\Z_{p_i}$
at $r\rightarrow\infty$ and constant $\theta,\phi$
is a $T^2.$
The $S^1$ is in direction $x_3$,
and the Taub-NUT fiber is in direction $y.$
The $T^2$ is represented here
as a fundamental cell of a lattice,
and we picked the fundamental cell generated by the vectors
$\vec{a},\vec{b}.$ The Taub-NUT direction is then
$\vec{a}-p_i\vec{b}.$
In this example $p_i=3.$
}
\label{fig:pq}
}
So, after reduction to type-IIB,
we get $n$ D$3$-branes ending on a $(1,p_i)$ $5$-brane.
Let $\lowD$ be the 3+1D $U(n)$ gauge
field on the D$3$-branes, and let $\lowC$
be the 2+1D boundary value of the gauge field
at the endpoint where the D$3$-branes meet the $(1,p_i)$
$5$-brane. From the 4+1D perspective, $\lowC$ can be
identified with the restriction
of the 2+1D components of the bulk gauge field
$\bulkC$ to the special point $\spP_i.$
In the discussion that follows we will suppress
the superpartners for simplicity.

The description that Gaiotto and Witten provide for $n$
D$3$-branes ending on a $(1,p_i)$ $5$-brane
was derived
as the S-dual of the description of
$n$ D$3$-branes ending on a $(p_i,1)$ $5$-brane.
The latter configuration
is described simply by adding a Chern--Simons coupling
for the boundary gauge field.
The Chern--Simons level is $p_i$,
and this can be derived
by a standard $\SL(2,\Z)$
transformation that maps a $(p_i,1)$ $5$-brane
to a $(0,1)$ $5$-brane while changing the type-IIB
coupling constant as $\tauIIB\rightarrow\tauIIB+p_i$
(see \cite{Kitao:1998mf}).
We denote the $U(n)$ gauge field of this Chern--Simons theory
by $\lowB_i.$

Following Gaiotto and Witten, S-duality is realized by
coupling $\lowB_i$
to $\lowC$ through additional degrees of freedom
with global $U(n)\times U(n)$
symmetry (the ``$T(U(n))$'' theory of \cite{Gaiotto:2008ak})
and gauging one $U(n)$ factor with $\lowB_i$ and the other with
$\lowC.$ (See \secref{subsec:twozero-U(1)}
for an example of how this works for $U(1)$ gauge theory.)
This description is valid if $\text{Re}\tauIIB=0.$
But in our case,
the type-IIB coupling constant \eqref{eqn:tauIIB}
has a nonzero real part $\text{Re}\tauIIB = -1/p_i.$
This adds an additional interaction in terms of
$F=d\lowD+\lowD\wedge\lowD$:
\be\label{eqn:FFpi}
-\frac{1}{4\pi p_i}\int_{D3} \text{tr}(F\wedge F)
=\frac{1}{4\pi p_i}\int
(\lowC\wedge d\lowC+\tfrac{2}{3}\lowC\wedge\lowC\wedge\lowC)
\,,
\ee
where we have integrated $\text{tr}(F\wedge F)$
to obtain a Chern--Simons coupling at level $1/p_i$
at the boundary. In principle there is an equal and opposite
term at the other end of the D$3$-branes, wherever
it may be, but this is of no concern to us
since we are only interested in the interactions near
the end of the $(1,p_i)$ $5$-brane.

The final step is to reduce to the 2+1D low-energy theory.
At low energy the 2+1D components
of the bulk gauge field $\bulkC$ can be assumed to be
constant along $T^2/\Z_\ord$, and in particular
we can identify the three $\lowC$ gauge fields
as one and the same.
Adding up the $3$ fractional Chern--Simons interactions
\eqref{eqn:FFpi} at levels $1/p_i$, and using
\eqref{eqn:sumpi}, we find that the low-energy
effective action has a Chern--Simons interaction
at level $1$ for $\lowC$:
$$
\frac{1}{4\pi}\tr\int (\lowC\wedge d\lowC
+\tfrac{2}{3}\lowC\wedge\lowC\wedge\lowC)\,.
$$
In addition, the Lagrangian has
$3$ Chern--Simons interactions --
an interaction at level $p_i$ for $\lowB_i$ ($i=1,2,3$):
$$
\sum_{i=1}^3\frac{p_i}{4\pi}\tr\int (\lowB_i\wedge d\lowB_i
+\tfrac{2}{3}\lowB_i\wedge\lowB_i\wedge\lowB_i)\,,
$$
and three copies ($i=1,2,3$)
of the $T(U(n))$ theories described in
\cite{Gaiotto:2008ak}, each coupled to $\lowB_i$ and $\lowC.$
At this point 
we point out again that superpartners
of the gauge fields have been suppressed.


\subsection{Recovering the $U(1)$ result}
\label{subsec:twozero-U(1)}

For $U(1)$ gauge theory, we construct the low-energy
2+1D interactions as follows.
First, we have a low-energy gauge field $\lowC$ that descends
from the bulk field $\bulkC.$ It has a Chern--Simons
interaction at level $1$, i.e.,
$\frac{1}{4\pi}\int\lowC\wedge d\lowC.$
Then, we have additional degrees of freedom from
the three special points. These are equivalent to
the degrees of freedom of a D$3$-brane that ends on
a $(1,p)$ $5$-brane.
The description of that system
was given in \cite{Gaiotto:2008ak} in terms
of the action
\be\label{eqn:GWBC}
\frac{1}{4\pi}\int\left(
p\lowB\wedge d\lowB + 2\lowB\wedge d\lowC_b
\right),
\ee
where $\lowC_b$ is the bulk D3-brane gauge field, restricted
to the boundary. We can identify it with our low-energy
field $\lowC.$

Let us briefly comment on how
the expression \eqref{eqn:GWBC} was derived.
It is the S-dual of the boundary interaction of a D$3$-brane
ending on a $(p,1)$ $5$-brane, the latter being given by
a level $p$ Chern--Simons interaction of the boundary gauge field $\lowC_b.$
As explained in \cite{Gaiotto:2009we}, the
$2\lowB\wedge d\lowC_b$ term realizes the S-duality
[see \eqref{eqn:IS}].

For each of the cases listed in \eqref{eqn:pivals}, we have three special points, so
we need to include three interactions of the type \eqref{eqn:GWBC},
with the appropriate values of $p.$
We denote the $3$ localized gauge fields by
$\lowB_1, \lowB_2, \lowB_3.$
The various values of $p$ are the denominators of the fractions
appearing in \figref{fig:FundamentalDomains}.
Thus, we have
\be\label{eqn:lowCS}
I = I_{\text{sp}} + \frac{1}{2\pi}\int
(\lowB_1 + \lowB_2 + \lowB_3)\wedge d\lowC
\,,
\ee
with
\be\label{eqn:lowCSsp}
\left.
\begin{array}{ll}
I_{\text{sp}} = \frac{1}{4\pi}\int\left(
6\lowB_1\wedge d\lowB_1
+3\lowB_2\wedge d\lowB_2
+2\lowB_3\wedge d\lowB_3
\right)\,,
& \quad(\lvk = 1) \\
I_{\text{sp}} = \frac{1}{4\pi}\int\left(
4\lowB_1\wedge d\lowB_1
+4\lowB_2\wedge d\lowB_2
+2\lowB_3\wedge d\lowB_3
\right)\,,
& \quad(\lvk = 2) \\
I_{\text{sp}} = \frac{1}{4\pi}\int\left(
3\lowB_1\wedge d\lowB_1
+3\lowB_2\wedge d\lowB_2
+3\lowB_3\wedge d\lowB_3
\right)\,,
& \quad(\lvk = 3) \\
\end{array}
\right\}.
\ee
The general form of the interaction
\eqref{eqn:lowCS}-\eqref{eqn:lowCSsp} is therefore
\be\label{eqn:lowCSgen}
I = \frac{1}{4\pi}\int\left(
\sum_{i=1}^3 p_i \lowB_i\wedge d\lowB_i
+2d\lowC\wedge\sum_{i=1}^3\lowB_i
\right)
\,,
\ee
where $p_1,p_2,p_3$ are integers determined by the level $\lvk$.

Now consider an abelian Chern--Simons theory with action
\be\label{eqn:CShij}
\frac{1}{4\pi}\int\sum_{i,j} h_{ij} \lowB_i\wedge d\lowB_j
\,,
\ee
where $h_{ij}$ are integer elements of a
nonsingular symmetric matrix.
Compactified on $T^2$, the number of states that we get
is the determinant
$$
N_{\text{states}} = \det\{h_{ij}\}
\,.
$$
However, the $4\times 4$ matrix corresponding to
\eqref{eqn:lowCSgen} is singular:
$$
\det\begin{pmatrix}
p_1 & 0 & 0 & 1 \\
0 & p_2 & 0 & 1 \\
0 & 0 & p_3 & 1 \\
1 & 1 & 1 &   1 \\
\end{pmatrix} =
p_1 p_2 p_3 \left(1-\frac{1}{p_1}-\frac{1}{p_2}-\frac{1}{p_3}\right) =
0.
$$
Nevertheless, the zero mode can easily be extracted by
changing variables:
$$
\lowB_1\equiv\lowB_1'
\,,\qquad
\lowB_2\equiv\lowB_2'+\frac{p_1}{p_2}\lowB_1'
\,,\qquad
\lowB_3\equiv\lowB_3'+\frac{p_1}{p_3}\lowB_1'
\,,\qquad
\lowC\equiv\lowC'-p_1\lowB_1'
\,.
$$
Note that this transformation is always in $\SL(4,\Z)$,
since we have arranged the $p_1,p_2,p_3$ in
\eqref{eqn:lowCSsp} so that $\frac{p_1}{p_2}$ and $\frac{p_1}{p_3}$
are integers.

The action \eqref{eqn:lowCSgen} can now be written as
\be\label{eqn:lowCSgenpr}
I = \frac{1}{4\pi}\int\left(
\sum_{i=2}^3 p_i \lowB_i'\wedge d\lowB_i'
+2d\lowC'\wedge\sum_{i=2}^3\lowB_i'
\right)
\,,
\ee
and $\lowB_1'$ does not appear in the action.
This means that when we look for ground states on $T^2$,
we should include a canonical kinetic term proportional to
$\int d\lowB_1'\wedge *d\lowB_1'$
(originating from the gauge kinetic term
for $\lowC$ and other terms).
Such a term will ensure that states coming from excitations
with nonzero $d\lowB_1'$ are not ground states.
If we are only interested in the ground states of the system
it is therefore sufficient to concentrate on the abelian Chern--Simons
theory \eqref{eqn:lowCSgenpr}.
The reduced matrices $h_{ij}$ corresponding to
\eqref{eqn:lowCSgenpr} for the cases $\lvk=1,2,3$ are:
$$
\begin{pmatrix}
3 & 0 & 1 \\
0 & 2 & 1 \\
1 & 1 & 1 \\
\end{pmatrix}
\,,\qquad
\begin{pmatrix}
4 & 0 & 1 \\
0 & 2 & 1 \\
1 & 1 & 1 \\
\end{pmatrix}
\,,\qquad
\begin{pmatrix}
3 & 0 & 1 \\
0 & 3 & 1 \\
1 & 1 & 1 \\
\end{pmatrix}
\,,
$$
and their determinants are $1,2,3$, respectively!
Thus, we have recovered the correct number of ground states.
We conclude this subsection with a few comments.
\begin{enumerate}
\item
We can trace the zero mode $\lowB_1'$
back to the scalar field coming from the component
of the $(2,0)$ anti-self-dual $2$-form $B^{(-)}$ along
the $z,\bz$ directions, i.e., $B_{z\bz}^{(-)}$.
This term corresponds to the $6^{th}$ scalar field of \SUSY{4}
SYM, and in the construction of \secref{sec:Problem}
this zero mode gets lifted by an R-symmetry twist.
In our $(2,0)$-realization this scalar is special,
and while we could not add an R-symmetry twist to lift the zero mode
at the outset,
we can add it to the low-energy theory at the end.
In any case, we do not get any additional multiplicity
of ground states.

\item
Na\"ively, we can attempt to integrate out
$\lowB_1,\lowB_2,\lowB_3$ in \eqref{eqn:lowCSgen}.
The result is obtained by setting $\lowB_i = -\frac{1}{p_i}\lowC$,
but plugging this back into \eqref{eqn:lowCSgen}
we get a vanishing action.
The problem with this prescription is that it ignores the
integral periodicity of the gauge fields.
In fact, this is precisely what Gaiotto and Witten
warned us not to do when dealing with a D$3$-brane ending on a $(p_i,1)$
$5$-brane (see \S8.3 of \cite{Gaiotto:2008ak}).
Here, we see an explicit manifestation of
what can go wrong if we disregard their advice!

\item
There is a connection between the relations among the generators
of the homology group $H_2(T^2\times\MXS)$ and operator relations
in the Hilbert space of ground states.
Consider an abelian Chern--Simons theory of the form
\eqref{eqn:CShij}, with $i,j=1,\dots,d$, compactified on $T^2$.
Let $0\le x_1,x_2\le 2\pi$ be coordinates on this $T^2$,
$\gpifa',\gpifb'$ the $1$-cycles along directions $1$ and $2$
respectively, and
define Wilson lines along the $\gpifa',\gpifb'$
cycles:
$$
\WilV_{1i} = e^{i\oint_{\gpifa'}\lowB_i}
\,,\qquad
\WilV_{2i} = e^{i\oint_{\gpifb'}\lowB_i}.
$$
If $\{h_{ij}\}$
is invertable with inverse $h^{ij}$,
the commutation relations are
$$
\WilV_{1i}\WilV_{2j} = e^{2\pi i h^{ij}}
\WilV_{2j}\WilV_{1i}\,.
$$

The Hilbert space is a representation of this algebra.
We then find that for every $i$,
$$
X_i\equiv \prod_{j=1}^d\WilV_{1j}^{h_{ij}}
\quad\text{and}\quad
Y_i\equiv \prod_{j=1}^d\WilV_{2j}^{h_{ij}}
$$
commute with all $\WilV_{1i},\WilV_{2j}$
and so are central elements of the algebra.
Without loss of generality, we can set their value to $1.$
The Hilbert space can now be constructed by diagonalizing
all $\WilV_{1i}$ ($i=1,\dots,d$) simultaneously.
Let $\ket{\psi}$ be any common eigenstate of all $\WilV_{1i}$.
Then, the full Hilbert space can be constructed by
acting with the $\WilV_{2i}$ on $\ket{\psi}$ and obtaining states
of the form $\prod_{i=1}^d\WilV_{2i}^{N_i}\ket{\psi}$,
where $(N_1,\dots,N_d)\in \Z^d$ is a vector of integers.
The states of the Hilbert space thus correspond to lattice
points in $\Z^d$, but not all lattice points give distinct states.
Since we have identified $Y_i=1$, we find that the lattice
points $(h_{i1},h_{i2},\dots,h_{id})$ correspond to the
same state as $(0,0,\dots,0)$. Let $\Gamma\subset\Z^d$
be the sublattice generated by the $d$ vectors
$(h_{i1},h_{i2},\dots,h_{id})$ ($i=1,\dots,d$).
Then, the basis states of the Hilbert space thus constructed
can be identified with the finite-dimensional set $\Z^d/\Gamma.$
It is not hard to see that $X_i,Y_i$ are central elements
even when $\{h_{ij}\}$ is not invertible.

The point of this note is that in our case,
the relations $X_i=Y_i=1$ have a natural interpretation
in terms of the $(2,0)$ theory.
The Wilson loop
operators $\WilV_{1i},\WilV_{2i}$ descend from
surface operators of the $(2,0)$ theory.
The surface operators $\WilV(\Surf)$ are associated with
closed surfaces $\Surf\subset T^2\times\MXS.$
In our case, since we are working with the abelian theory
and since we are only interested in the ground states,
the surface operators only depend on the homology class
of $\Surf$. This can be argued by noting that in the ground
state the anti-self-dual 3-form flux of the $(2,0)$-theory
vanishes, and that when $\Surf$ and $\Surf'$ are in the same homology
class, we can write $\Surf-\Surf'=\partial\Sigma_3$ so that the difference between
the integral of the $2$-form of the $(2,0)$-theory
on $\Surf$ and on $\Surf'$ is the integral of the
anti-self-dual $3$-form field-strength on $\Sigma_3$.
We can therefore denote the surface operators as $\WilV([\Surf])$
where $[\Surf]$ is the homology class of $\Surf.$

Now, we can match the Wilson lines of
$\lowB_1,\lowB_2,\lowB_3,\lowC$ with surface operators
as follows.
Let $\gamma_i$ be the homology class of the exceptional
fiber at the $i^{th}$ special point ($i=1,2,3$),
and let $\gamma_0$ be the homology class of the generic fiber.
Then, we match
$$
\WilV(\gpifa'\times\gamma_i)\rightarrow
e^{i\oint_{\gpifa'}\lowB_i}
\,,\qquad
\WilV(\gpifa'\times\gamma_0)\rightarrow
e^{i\oint_{\gpifa'}\lowC}
\,,
$$
and similarly,
$$
\WilV(\gpifb'\times\gamma_i)\rightarrow
e^{i\oint_{\gpifb'}\lowB_i}
\,,\qquad
\WilV(\gpifb'\times\gamma_0)\rightarrow
e^{i\oint_{\gpifb'}\lowC}
\,.
$$
The relations $X_i=Y_i=1$ are then seen to be a consequence
of similar relations in homology.
(See \cite{Henningson:2000nu} for a discussion of the commutation
relations for the nonabelian $(2,0)$-theory.)

\end{enumerate}


\subsection{The $U(2)$ theory}
\label{subsec:twozero-U(2)}

We now turn to the nonabelian
gauge group $U(2).$
Schematically, the action is of the form
\bear
I &=& \frac{1}{4\pi}\int\Bigl\{
\sum_{i=1}^3 p_i
\tr (\lowB_i\wedge d\lowB_i
+\tfrac{2}{3}\lowB_i\wedge\lowB_i\wedge\lowB_i)
+\tr (\lowC\wedge d\lowC
+\tfrac{2}{3}\lowC\wedge\lowC\wedge\lowC)
\Bigr\}
\nn\\&&\qquad
+\sum_{i=1}^3 I_i^{[T(U(2))]}(\lowB_i,\lowC)
\,,
\label{eqn:lowCSgenU(2)}
\eear
where $\lowB_1,\lowB_2,\lowB_3$ are the 2+1D $U(2)$
gauge fields coming from the singular points
$\spP_1,\spP_2,\spP_3,$
$\lowC$  is also a 2+1D $U(2)$ gauge field,
and $I_i^{[T(U(2))]}$ is the coupling between $\lowB_i$
and $\lowC$ through the additional $T(U(2))$ degrees of freedom.
Roughly speaking, this coupling
realizes the nonabelian S-duality \cite{Gaiotto:2008ak},
whereby $\lowB_i$ and $\lowC$ are regarded as S-dual variables.
(As noted above, we are ignoring the superpartners
in this discussion.)

Although Gaiotto and Witten have provided an explicit
realization of $T(U(2))$ as the low-energy limit of
a certain \SUSY{4} 2+1D gauge theory,
the full $U(2)\times U(2)$ symmetry, and hence
the coupling to $\lowB_i$ and $\lowC$, relies on an enhanced
symmetry of $T(U(2))$ that is not explicit.
We therefore do not know how to proceed at this moment.
However, we can make some comments about the
$[\perm]$-twisted sector.

What is the interpretation of the $[\perm]$-twisted sectors
in terms of the $(2,0)$-theory construction?
For $U(2)$ gauge group, there is only one $[\perm]$-twisted sector.
In the type-IIA description, the nontrivial $\perm\in S_2$
exchanges the two strings as we go in a loop from $x_3=0$
to $x_3=2\pi\xR$, and so it is reasonable to expect
that in our present M-theory description, we need to exchange
the two M$5$-branes as we go from $x_3=0$ to $x_3=2\pi\xR.$
This exchange of the two branes accompanies the identification
$(z,x_3)\sim (e^{i\pht}z,x_3+2\pi\xR)$
of \eqref{eqn:zx}.

At this point we need to distinguish between two cases --
even $\ord$ and odd $\ord$.
If $\ord$ is odd, then the identification
$(z,x_3)\sim (z,x_3+2\pi\xR\ord)$ is accompanied by
an exchange of the two branes.
If $\ord$ is even this identification is not accompanied
by an exchange of branes.
The two even cases are $\ord=4,6.$
Since $(z,x_3)\sim (z,x_3+2\pi\xR\ord)$ is not accompanied
by exchange of the branes, the reduction to the 4+1D theory
proceeds as in \secref{subsec:Reduction}.

One of the
effects of the nontrivial $\perm$ on this low-energy
4+1D $U(2)$ gauge theory is that as we go around a special point $\spP_i$
with odd $p_i$, we have to also exchange the branches of the
D$4$-branes (that we formally get from the M$5$-branes).
This can be interpreted as a holonomy for $\bulkC$,
which after a suitable conjugation can be written as
\be\label{eqn:holQ}
P\exp\oint_{\spP_i}\bulkC =
\begin{pmatrix}
1 & \\ & (-1)^{\ord/p_i} \\
\end{pmatrix}
\,.
\ee
For odd $\ord/p_i$, this breaks the gauge group
$U(2)\rightarrow U(1)\times U(1).$
In addition, the boundary interaction at $\spP_i$ also
needs to be modified.
Altogether, the action appears quite complicated and
we will not attempt to develop it further in this paper.
It will be interesting to explore this in a future work.


\section{Discussion}
\label{sec:disc}

We have analyzed the Hilbert space of ground states
of the S-duality twisted compactification of \SUSY{4} $U(n)$
SYM on $T^2$, and have seen that in almost all cases,
at least as a representation
of $\SL(2,\Z)$ and the $\Z_\lvk$ symmetry operators $\Psym,\Qsym$,
it breaks up into a direct sum of Hilbert spaces of
Chern--Simons theories with gauge groups of the form
$U(n_1)\times U(n_2)\times\cdots\times U(n_s)$
(with $n=\sum_{j=1}^s n_j$).
Chern--Simons theory with $U(n_j)$ gauge group is described
by specifying the level of $SU(n_j)$ and the
level of the $U(1)$ center, so we use the notation
$$
U(n_j)_{\lvk_j',\lvk_j''}\simeq
[U(1)_{\lvk_j'}\times SU(n_j)_{\lvk_j''}]/\Z_{n_j}.
$$
(There were also two exceptional cases, which involved the Hilbert space $\Hilb_{(2,2)}(\tfrac{\pi}{3})$.) The various decompositions that
we get are listed in \tabref{tab:lvkpp}.
In particular, we saw in \secref{subsec:winding-1}
that in all cases there is a distinguished sector---the $\brperm$-untwisted sector---which is
described by the Hilbert space of $U(n)_{\lvk n,\lvk}.$

In this paper we only studied compactification
on $T^2$. What do our results suggest for
the theory formulated on $\R^{2,1}$?
Is Chern-Simons theory the low-energy theory,
and if so
what is the role of the various sectors
with their different Chern-Simons levels and gauge groups
(as listed in \tabref{tab:lvkpp})?
To make this question more precise,
we need to connect the operators of Chern-Simons
theory to physical operators in our theory.
But the low-energy limit of Wilson loops in the \SUSY{4} 
theory cannot in general be simply a Wilson loop in Chern-Simons
theory, because for $\lvk=2$ for example, the latter is not generally self-adjoint
while the former is, as we have argued in \secref{subsec:Wilson}.
If there is a connection between Chern-Simons
theory and the low-energy
limit of the S-duality twisted compactification on $\R^{2,1}$
it would certainly have to be more complicated than
the ``crude'' conjectures
presented in \cite{Ganor:2008hd}, which at best only captured
the $\brperm$-untwisted sector.
The answer to most of these questions may lie
in the proper description of the low-energy limit of 
Wilson loops. The tools we have developed in this paper
in principle allow the analysis of this problem as well,
by probing the type-IIB D$3$-branes with open strings.
We hope to report on this matter soon \cite{InPreparation}.

\TABLE[t]{
\begin{tabular}{l|l|l}
\hline\hline
$\pht=\frac{\pi}{3}$
& $n=1$ &
$U(1)_1$  \\
\cline{2-3}
$(\lvk=1)$
& $n=2$ &
$U(2)_{2,1}\oplus U(2)_{2,-3}$ \\
\cline{2-3}
& $n=3$ &
$U(3)_{3,1}\oplus [U(1)_1\times U(2)_{2,-3}]\oplus U(3)_{3,-2}$  \\
\cline{2-3}
& $n=4$ &
$U(4)_{4,1}\oplus 2[U(2)_{2,1}\times U(2)_{2,-3}]
\oplus [U(1)_1 \times U(3)_{3,-2}]\oplus\Hilb_{(2,2)}$ \\
\cline{2-3}
& $n=5$ &
$U(5)_{5,1}\oplus U(5)_{5,1}\oplus 2[U(3)_{3,1}\times U(2)_{2,-3}]
\oplus [U(1)_1\times\Hilb_{(2,2)}]\oplus $  \\
& & $[U(2)_{2,1}\times U(3)_{3,-2}]
\oplus [U(2)_{2,-3}\times U(3)_{3,-2}]  $ \\
\hline\hline
$\pht=\frac{\pi}{2}$
& $n=1$ &
$U(1)_2$ \\
\cline{2-3}
$(\lvk=2)$
& $n=2$ &
$U(2)_{4,2}\oplus U(2)_{4,-2}$ \\
\cline{2-3}
& $n=3$ &
$U(3)_{6,2}\oplus [U(1)_2\times U(2)_{4,-2}]\oplus U(3)_{6,-1}$ \\
\hline\hline
$\pht=\frac{2\pi}{3}$
& $n=1$ &
$U(1)_3$ \\
\cline{2-3}
$(\lvk=3)$
& $n=2$ &
$U(2)_{6,3}\oplus U(2)_{6,-1}$ \\
\hline\hline
\end{tabular}
\caption{
The decomposition of the
Hilbert spaces $\Hilb(n,\pht)$  into direct sums
of Hilbert spaces of Chern--Simons theories.
The data in the table is collected from
results in \appref{app:HilbDecomp}.
Trivial $U(n')_{n',1}$ factors were added
to conform to the form \eqref{eqn:HilbSumU}.
Note that for $n=4$ we have two copies of
$U(2)_{2,1}\times U(2)_{2,-3}$. They come
from the sectors $\Hilb_{(2,1,1)}$ and $\Hilb_{(4)}.$
Also, note that the sector $\Hilb_{(2,2)}$ is unresolved.
}
\label{tab:lvkpp}
}

In this paper we have concentrated on gauge groups of
low rank, as we were restricted by the condition $n<\ord.$
In these cases,
as explained in \secref{subsec:LE}, we expect a mass gap.
It would be interesting to extend the analysis to $n\ge\ord.$
Here there are several questions that we can ask.
First, we can still look for a low-energy description
on $\R^{2,1}.$ Since we are dealing with
a 2+1D theory with \SUSY{6} supersymmetry that
we also expect to be conformally invariant
in the low-energy limit, the ABJM theories
\cite{Aharony:2008ug} spring as a natural candidate.
Indeed,
we expect the low-energy limit for $n\ge\ord$
to be an ABJM theory with an appropriate gauge group
that can be determined from the moduli space.
(Note that for $n<\ord$ the low-energy
theories that we have proposed
are supersymmetric in a trivial way ---
as topological theories, all their SUSY generators are zero.)
Second, we can explore the subspace of normalizable ground
states on $T^2.$
Thus, for example, the $\Hilb_{(n_1,n_2,\dots,n_p)}(\pht)$
sector makes sense as long as $n_j<\ord$ (for $j=1,\dots,p$),
even if $n=\sum_1^p n_j\ge\ord.$ The states in this sector
define the normalizable ground states of the $T^2$ compactification
of the theory, even though the full theory has a continuum
of states that start at zero energy.
As a simple case-study, we recall the analysis of
the C-twist with $U(2)$ gauge group
in \secref{subsec:U(2)-C}.
There, we found that even though there is no mass gap,
there is a finite-dimensional Hilbert space of
{\it normalizable} ground states.
Moreover, we matched the ground states with the states
of the type-IIA strings, and the bosonic nature of the strings
had an interesting interpretation in terms of
a restriction on the electric and magnetic fluxes
[see \eqref{eqn:efmfz}].
In the more complicated case of an S-duality twist,
we can also ask whether the subset of normalizable
ground states has a description in terms of a topological
field theory. For example, in section
\secref{subsec:winding-1}
we saw that the subsectors $\Hilb_{(1,1,\dots,1)}(\pht)$
correspond to ground states of $U(n)_{\lvk n,\lvk}$
Chern--Simons theory, even for $n\ge\ord.$
Other sectors might also have extensions for $n\ge\ord.$
This avenue of investigation might be connected to ideas
developed in \cite{Nekrasov:2009ui} about isolating
the ground states of supersymmetric theories and
finding a simpler description for them separately.
We conclude with a summary of a few of the open problems:
\begin{enumerate}

\item
What is the underlying principle that determines the levels
and gauge groups of the $\brperm$-twisted sectors?
Does the decomposition into
sectors survive when the theory is
formulated on $\R^{2,1}$, and if so are they to be 
regarded as ``superselection sectors'', or are there
any physical operators that connect different sectors?
And can the permutation $\perm$ be interpreted as a discrete
$S_n\subset U(n)$ Wilson line? If so, why is it restricted to
$S_n$, i.e., how do fluctuations away from $S_n$ receive a
potential?
Can the unresolved sector $\Hilb_{(2,2)}(\frac{\pi}{3})$
be interpreted as a Chern--Simons Hilbert space?
We note that under certain circumstances flavor symmetry
twists can induce Chern--Simons interactions
\cite{Poppitz:2008hr}, but in our case the R-symmetry twist
alone cannot induce a low-energy Chern--Simons term
because the R-symmetry is nonabelian.
The Chern--Simons couplings are intimately related to the
S-duality twist.

\item
How can these results be recovered directly from
the $(2,0)$-theory? In particular, what is the low-energy
description of the action \eqref{eqn:lowCSgenU(2)}?
What do the Chern--Simons theories that
we found teach us about the S-duality generating $T(SU(n))$
theories that Gaiotto and Witten have found in
\cite{Gaiotto:2008ak}?

\item
The description of the low-energy limit
of Wilson loops as operators on the Hilbert space
of ground states is 
currently under investigation \cite{InPreparation}.

\item
How can the results be extended to $n\ge\ord$?
In this case we can also explore the large $n$ limit
in the context of the AdS/CFT correspondence
\cite{Maldacena:1997re}-\cite{Witten:1998qj}.
If the $\brperm$-untwisted sector
$\Hilb_{(1,1,\dots,1)}$ survives the large $n$ limit,
it would be interesting to find
its holographic dual.
Perhaps the holographic dual of Chern--Simons theory
\cite{Gopakumar:1998ki} will somehow make an appearance.

\item
The analysis of the S-duality and R-symmetry twist
can also be performed
on the topologically twisted \SUSY{4} SYM theories
\cite{Vafa:1994tf}.
(Some preliminary results were discussed in \cite{Ganor:2008hd}.)
For $n\ge\ord$, it might be interesting to look for
connections with the topologically twisted
supersymmetric Chern--Simons theories
\cite{Lee:2008cr}\cite{Kapustin:2009cd}.

\item
Recently,
new ideas about surprising mathematical aspects
of Chern--Simons theory have emerged
(see for instance
\cite{Gukov:2005qp}\cite{Freed:2008jq}).
In this paper, we have suggested that Chern--Simons
theory is  related to S-duality. The latter is also
intimately connected to the Langlands correspondence
\cite{Kapustin:2006pk}. So,
perhaps it is worthwhile to search for a connection
between Chern--Simons theory and the Langlands program.

\end{enumerate}


\acknowledgments
We would like to thank Wu-Yen Chuang,
Jacques Distler, Dan Freed, Eric Gimon, David Gross,
Aki Hashimoto, Simeon Hellerman,
Petr Ho\v{r}ava, Dan Jafferis, Anton Kapustin,
Sungjay Lee, Sonia Paban, Jaemo Park, Soo Jong Rey, Dan Robins,
Mithat \"Unsal, Cumrun Vafa, and Edward Witten
for discussions, correspondence, and comments
during various stages of preparation of this work.
OJG wishes to thank Rob Myers, Mark van Raamsdonk and Wati Taylor,
the organizers of the workshop on
``Emerging Directions in String Theory,''
where preliminary results of this work have been presented in June 2008, and the Banff International Research Station for Mathematical
Innovation and Discovery (BIRS) for their hospitality.
OJG would also like to thank
the Kavli Institute for Theoretical Physics and the organizers of the KITP Mini-conference ``Dualities in Physics and Mathematics'' (March 16-20, 2009) for their hospitality.
YPH would like to thank the Berkeley Center for Theoretical Physics
for hospitality during the final stage of the work.
This work was supported in part by the
Berkeley Center of Theoretical Physics,
in part by the U.S. National Science Foundation
under grant PHY-04-57315,
and in part by the Director,
Office of Science, Office of High Energy Physics,
of the U.S. Department of Energy under Contract
No. DE-AC03-76SF00098. The work of YPH was also supported in part by Basic Science Research Program through the National Research Foundation of Korea (NRF), funded by the Ministry of Education, Science and Technology, under grant  2010-0013526.


\begin{appendix}

\section{Explicit action of $\SL(2,\Z)$ on ground states}
\label{app:SL(2,Z)}

Below we present the operators
$\TdS$ and $\TdT$ that generate
$\SL(2,\Z)$ in the various single-particle Hilbert spaces.
The action of $\SL(2,\Z)$ on
the mutli-particle Hilbert spaces can, of course, be calculated
from the tensor products of the expressions below.
We recall the ambiguity \eqref{eqn:TdGredef}
in the definition of $\TdS$ and $\TdT$,
to which we can also add a phase ambiguity
\be\label{eqn:TdGredefPhase}
\TdT\rightarrow e^{i\phi}\Psym^\txp\Qsym^\txq\TdT
\,,\qquad
\TdS\rightarrow e^{i\phi'}\Psym^\txp\Qsym^\txq\TdS
\,,
\ee
as long as we preserve the group
relations $\TdS^2=(\TdS\TdT)^3=-1.$
Below, we pick arbitrary $\txp,\txq,\phi,\phi'.$
Thus, the expressions below
satisfy the group relations $\TdS^2=(\TdS\TdT)^3=-1$
only up to a phase. This can easily be fixed
by choosing appropriate phases $\phi,\phi'$
in \eqref{eqn:TdGredefPhase},
but we find the formulas easier to read without these
phases, so we have not included them.
We note that only matrix elements of
operators (constructed from $\TdS$ and $\TdT$)
between initial and final states that preserve $\Psym$ and $\Qsym$
are physically meaningful, and the ambiguity
\eqref{eqn:TdGredefPhase} does not affect those matrix elements.


\subsection{Action of $\TdS$ on
single-particle states for $\pht=\frac{\pi}{2}$
($\tau=i$ and $\ord=4$)}


In this case $\lvk=2.$
For $n=1$, $\TdS$ acts as:
\be
\TdS\ket{\fpIIz}=\frac{1}{\sqrt{2}}\left(
\ket{\fpIIz}+\ket{\fpIIc}\right)\,,\qquad
\TdS\ket{\fpIIc}=\frac{1}{\sqrt{2}}\left(
\ket{\fpIIz}-\ket{\fpIIc}\right)\,,
\ee
or, equivalently,
$$
\TdS\ket{[0]}=\frac{1}{\sqrt{2}}\left(
\ket{[0]}+\ket{[\tfrac{1}{2}+\tfrac{1}{2}i]}\right)\,,\qquad
\TdS\ket{[\tfrac{1}{2}+\tfrac{1}{2}i]}=\frac{1}{\sqrt{2}}\left(
\ket{[0]}-\ket{[\tfrac{1}{2}+\tfrac{1}{2}i]}\right)\,.
$$
And for $\TdT$ we have:
\be\label{eqn:TdT-II1}
\TdT\ket{\fpIIz}=\ket{\fpIIz}
\,,\qquad
\TdT\ket{\fpIIc}=e^{\frac{i\pi}{2}}\ket{\fpIIc}
\,,
\ee
or, equivalently,
$$
\TdT\ket{[0]}=\ket{[0]}
\,,\qquad
\TdT\ket{[\tfrac{1}{2}+\tfrac{1}{2}i]}=
e^{\frac{i\pi}{2}}\ket{[\tfrac{1}{2}+\tfrac{1}{2}i]}
\,.
$$


For $n=2$, $\TdS$ acts as:
\be\label{eqn:TdS-II2}
\left[\begin{array}{ll}
\TdS\ket{\fpIIzz}&=
\frac{1}{2}\ket{\fpIIzz}
+\frac{1}{\sqrt{2}}\ket{\fpIIxy}
+\frac{1}{2}\ket{\fpIIcc}
\,,
\\ & \\
\TdS\ket{\fpIIxy}&=
\frac{1}{\sqrt{2}}\ket{\fpIIzz}
-\frac{1}{\sqrt{2}}\ket{\fpIIcc}
\,,
\\ & \\
\TdS\ket{\fpIIcc}&=
\frac{1}{2}\ket{\fpIIzz}
-\frac{1}{\sqrt{2}}\ket{\fpIIxy}
+\frac{1}{2}\ket{\fpIIcc}
\,,
\end{array}
\right.
\ee
or, equivalently,
\be\label{eqn:TdS-IIeq2}
\left[\begin{array}{ll}
\TdS\ket{[0,0]}&=
\frac{1}{2}\ket{[0,0]}
+\frac{1}{\sqrt{2}}\ket{[\tfrac{1}{2},\tfrac{1}{2}i]}
+\frac{1}{2}\ket{[\tfrac{1}{2}+\tfrac{1}{2}i,
\tfrac{1}{2}+\tfrac{1}{2}i]}
\,,
\\ & \\
\TdS\ket{[\tfrac{1}{2},\tfrac{1}{2}i]}&=
\frac{1}{\sqrt{2}}\ket{[0,0]}
-\frac{1}{\sqrt{2}}\ket{[\tfrac{1}{2}+\tfrac{1}{2}i,
\tfrac{1}{2}+\tfrac{1}{2}i]}
\,,
\\ & \\
\TdS\ket{[\tfrac{1}{2}+\tfrac{1}{2}i,
\tfrac{1}{2}+\tfrac{1}{2}i]}&=
\frac{1}{2}\ket{[0,0]}
-\frac{1}{\sqrt{2}}\ket{[\tfrac{1}{2},\tfrac{1}{2}i]}
+\frac{1}{2}\ket{[\tfrac{1}{2}+\tfrac{1}{2}i,
\tfrac{1}{2}+\tfrac{1}{2}i]}
\,.
\\
\end{array}
\right.\nn
\ee
And for $\TdT$ we have:
\be\label{eqn:TdT-II2}
\TdT\ket{\fpIIzz}=\ket{\fpIIzz}
\,,\qquad
\TdT\ket{\fpIIxy}=\ket{\fpIIxy}
\,,\qquad
\TdT\ket{\fpIIcc}=-\ket{\fpIIcc}
\,,
\ee
or, equivalently,
\be\label{eqn:TdT-IIeq2}
\TdT\ket{[0,0]}=\ket{[0,0]}
\,,\quad
\TdT\ket{[\tfrac{1}{2},\tfrac{1}{2}i]}=
\ket{[\tfrac{1}{2},\tfrac{1}{2}i]}
\,,\quad
\TdT\ket{[\tfrac{1}{2}+\tfrac{1}{2}i,
\tfrac{1}{2}+\tfrac{1}{2}i]}=
-\ket{[\tfrac{1}{2}+\tfrac{1}{2}i,
\tfrac{1}{2}+\tfrac{1}{2}i]}
\,.\nn
\ee


For $n=3$, $\TdS$ acts as:
\be
\TdS\ket{\fpIIzzz}=
\frac{1}{\sqrt{2}}\left(\ket{\fpIIzzz}+\ket{\fpIIccc}\right)
\,,
\qquad
\TdS\ket{\fpIIccc}=
\frac{1}{\sqrt{2}}\left(\ket{\fpIIzzz}-\ket{\fpIIccc}\right)\,,
\ee
or, equivalently,
\be
\left[\begin{array}{ll}
\TdS\ket{[0,0,0]} &=
\frac{1}{\sqrt{2}}\left(\ket{[0,0,0]}
+\ket{[\tfrac{1}{2}+\tfrac{1}{2}i,
\tfrac{1}{2}+\tfrac{1}{2}i,\tfrac{1}{2}+\tfrac{1}{2}i]}\right)\,,
\\ & \\
\TdS\ket{[\tfrac{1}{2}+\tfrac{1}{2}i,
\tfrac{1}{2}+\tfrac{1}{2}i,\tfrac{1}{2}+\tfrac{1}{2}i]} &=
\frac{1}{\sqrt{2}}\left(\ket{[0,0,0]}
-\ket{[\tfrac{1}{2}+\tfrac{1}{2}i,
\tfrac{1}{2}+\tfrac{1}{2}i,\tfrac{1}{2}+\tfrac{1}{2}i]}\right)\,.
\\
\end{array}\right.\nn
\ee
And for $\TdT$ we have:
\be\label{eqn:TdTfpII}
\TdT\ket{\fpIIzzz}=\ket{\fpIIzzz}
\,,\qquad
\TdT\ket{\fpIIccc}=e^{-\frac{i\pi}{2}}\ket{\fpIIccc}
\,,
\ee
or, equivalently,
$$
\left[\begin{array}{l}
\TdT\ket{[0,0,0]}=\ket{[0,0,0]}
\,,\\

\\
\TdT\ket{[\tfrac{1}{2}+\tfrac{1}{2}i,
\tfrac{1}{2}+\tfrac{1}{2}i,\tfrac{1}{2}+\tfrac{1}{2}i]}=
e^{-\frac{i\pi}{2}}\ket{\tfrac{1}{2}+\tfrac{1}{2}i,
\tfrac{1}{2}+\tfrac{1}{2}i,\tfrac{1}{2}+\tfrac{1}{2}i]}
\,.
\end{array}\right.
$$


\subsection{Action of $\TdS$ on single-particle states
for $\pht=\frac{\pi}{3}$ ($\tau=e^{\pi i/3}$ and $\ord=6$)}


In this case $\lvk=1.$
For $n=1$, $\TdS$ and $\TdT$ act as:
\be
\TdS\ket{\fpIz}=\ket{\fpIz}\,,
\qquad
\TdT\ket{\fpIz}=\ket{\fpIz}\,,
\ee
or, equivalently,
$$
\TdS\ket{[0]}=\ket{[0]}\,,
\qquad
\TdT\ket{[0]}=\ket{[0]}\,.
$$


For $n=2$, $\TdS$ acts as:
\be\label{eqn:TdS-I2}
\TdS\ket{\fpIzz}=
\tfrac{1}{\sqrt{3}}\ket{\fpIzz}
+\sqrt{\tfrac{2}{3}}\ket{\fpIdd}\,,
\qquad
\TdS\ket{\fpIdd}=
\sqrt{\tfrac{2}{3}}\ket{\fpIzz}
-\tfrac{1}{\sqrt{3}}\ket{\fpIdd}
\,,
\ee
or, equivalently,
\be
\left[\begin{array}{ll}
\TdS\ket{[0,0]}&=
\frac{1}{\sqrt{3}}\ket{[0,0]}
+\sqrt{\frac{2}{3}}\ket{
[\tfrac{1}{\sqrt{3}}e^{\frac{i\pi}{6}},
\tfrac{2}{\sqrt{3}}e^{\frac{i\pi}{6}}]}\,,
\\ & \\
\TdS\ket{[\tfrac{1}{\sqrt{3}}e^{\frac{i\pi}{6}},
\tfrac{2}{\sqrt{3}}e^{\frac{i\pi}{6}}]} &=
\sqrt{\frac{2}{3}}\ket{[0,0]}
-\frac{1}{\sqrt{3}}\ket{
[\tfrac{1}{\sqrt{3}}e^{\frac{i\pi}{6}},
\tfrac{2}{\sqrt{3}}e^{\frac{i\pi}{6}}]}\,.
\\
\end{array}\right.\nn
\ee
$\TdT$ acts as:
\be\label{eqn:TdT-I2}
\TdT\ket{\fpIzz}= \ket{\fpIzz}\,,
\qquad
\TdT\ket{\fpIdd}=e^{-\frac{2\pi i}{3}}\ket{\fpIdd}
\,,
\ee
or, equivalently,
\be
\TdT\ket{[0,0]}=\ket{[0,0]}\,,
\qquad
\TdT\ket{[\tfrac{1}{\sqrt{3}}e^{\frac{i\pi}{6}},
\tfrac{2}{\sqrt{3}}e^{\frac{i\pi}{6}}]} =
e^{-\frac{2\pi i}{3}}\ket{
[\tfrac{1}{\sqrt{3}}e^{\frac{i\pi}{6}},
\tfrac{2}{\sqrt{3}}e^{\frac{i\pi}{6}}]}
\,.\nn
\ee


For $n=3$, we have two $\Z_2$ worldsheet momentum operators
$\wPsym_a,\wPsym_b$ and two $\Z_2$ worldsheet winding number
operators $\wQsym_a,\wQsym_b.$
For the commutation relations we choose
\be\label{eqn:SPQab(k=1)}
\TdS^{-1}\wQsym_a\TdS=\wPsym_b\,,\quad
\TdS^{-1}\wQsym_b\TdS=\wPsym_a^{-1}\,,\quad
\TdS^{-1}\wPsym_a\TdS=\wQsym_b\,,\quad
\TdS^{-1}\wPsym_b\TdS=\wQsym_a^{-1}\,,
\ee
\be\label{eqn:TPQab(k=1)}
\TdT^{-1}\wQsym_a\TdT=\wQsym_a\,,\quad
\TdT^{-1}\wQsym_b\TdT=-\wQsym_b\,,\quad
\TdT^{-1}\wPsym_a\TdT=-\wPsym_a\Qsym_b^{-1}\,,\quad
\TdT^{-1}\wPsym_b\TdT=\wPsym_b\Qsym_a\,.
\ee
Note the $(-)$ sign on the second and third equations of \eqref{eqn:TPQab(k=1)}.
We found that this phase assignment is necessary
so that $\TdT$ will commute with the orbifold
action and keep invariant the subspace spanned by
$$
\ket{\fpIzzz}=\ket{\fpZ_{0,0}}\,,\quad
\ket{\fpIxyd}=\frac{1}{\sqrt{3}}(
\ket{\fpZ_{0,1}}+\ket{\fpZ_{1,0}}+\ket{\fpZ_{1,1}})\,.
$$
We then find that
$\TdS$ acts as:
\be\label{eqn:TdS-I3}
\TdS\ket{\fpIzzz}=
\tfrac{1}{2}\ket{\fpIzzz}
+\tfrac{\sqrt{3}}{2}\ket{\fpIxyd}\,,
\qquad
\TdS\ket{\fpIxyd}=
\tfrac{\sqrt{3}}{2}\ket{\fpIzzz}
-\tfrac{1}{2}\ket{\fpIxyd}\,,
\ee
or, equivalently,
\be
\left[\begin{array}{ll}
\TdS\ket{[0,0,0]}&=
\tfrac{1}{2}\ket{[0,0,0]}
+\tfrac{\sqrt{3}}{2}\ket{
[\tfrac{1}{2},\tfrac{1}{2}\tau,
\tfrac{1}{2}+\tfrac{1}{2}\tau]}\,,
\\ & \\
\TdS\ket{[\tfrac{1}{2},\tfrac{1}{2}\tau,
\tfrac{1}{2}+\tfrac{1}{2}\tau]} &=
\tfrac{\sqrt{3}}{2}\ket{[0,0,0]}
-\tfrac{1}{2}\ket{
[\tfrac{1}{2},\tfrac{1}{2}\tau,
\tfrac{1}{2}+\tfrac{1}{2}\tau]}\,,
\\
\end{array}\right.\nn
\ee
and $\TdT$ acts as
\be\label{eqn:TdT-I3}
\TdT\ket{\fpIzzz}=\ket{\fpIzzz}\,,
\qquad
\TdT\ket{\fpIxyd}=-\ket{\fpIxyd}\,.
\ee
or, equivalently,
\be
\TdT\ket{[0,0,0]}=\ket{[0,0,0]}\,,
\qquad
\TdT\ket{[\tfrac{1}{2},\tfrac{1}{2}\tau,
\tfrac{1}{2}+\tfrac{1}{2}\tau]} =
-\ket{[\tfrac{1}{2},\tfrac{1}{2}\tau,
\tfrac{1}{2}+\tfrac{1}{2}\tau]}\,.\nn
\ee


For $n=4$, $\TdS$ acts (similarly to the $n=2$ case) as:
\be\label{eqn:TdS-I4}
\TdS\ket{\fpIzzzz}=
\tfrac{1}{\sqrt{3}}\ket{\fpIzzzz}
+\sqrt{\tfrac{2}{3}}\ket{\fpIdddd}\,,
\qquad
\TdS\ket{\fpIdddd}=
\sqrt{\tfrac{2}{3}}\ket{\fpIzzzz}
-\tfrac{1}{\sqrt{3}}\ket{\fpIdddd}
\,,
\ee
or, equivalently,
\be
\left[\begin{array}{ll}
\TdS\ket{[0,0,0,0]}&=
\frac{1}{\sqrt{3}}\ket{[0,0,0,0]}
+\sqrt{\frac{2}{3}}\ket{[
\tfrac{1}{\sqrt{3}}e^{\frac{i\pi}{6}},
\tfrac{1}{\sqrt{3}}e^{\frac{i\pi}{6}},
\tfrac{2}{\sqrt{3}}e^{\frac{i\pi}{6}},
\tfrac{2}{\sqrt{3}}e^{\frac{i\pi}{6}}
]}\,,
\\ & \\
\TdS\ket{[
\tfrac{1}{\sqrt{3}}e^{\frac{i\pi}{6}},
\tfrac{1}{\sqrt{3}}e^{\frac{i\pi}{6}},
\tfrac{2}{\sqrt{3}}e^{\frac{i\pi}{6}},
\tfrac{2}{\sqrt{3}}e^{\frac{i\pi}{6}}
]} &=
\sqrt{\frac{2}{3}}\ket{[0,0,0,0]}
-\frac{1}{\sqrt{3}}\ket{[
\tfrac{1}{\sqrt{3}}e^{\frac{i\pi}{6}},
\tfrac{1}{\sqrt{3}}e^{\frac{i\pi}{6}},
\tfrac{2}{\sqrt{3}}e^{\frac{i\pi}{6}},
\tfrac{2}{\sqrt{3}}e^{\frac{i\pi}{6}}
]}\,,
\\
\end{array}\right.\nn
\ee
and $\TdT$ acts as
\be\label{eqn:TdT-I4}
\TdT\ket{\fpIzzzz}=
\ket{\fpIzzzz}\,,
\qquad
\TdT\ket{\fpIdddd}=
e^{\frac{2\pi i}{3}}\ket{\fpIdddd}\,,
\ee
or, equivalently,
\be
\left[\begin{array}{l}
\TdT\ket{[0,0,0,0]}=
\ket{[0,0,0,0]}\,,
\\  \\
\TdT\ket{[
\tfrac{1}{\sqrt{3}}e^{\frac{i\pi}{6}},
\tfrac{1}{\sqrt{3}}e^{\frac{i\pi}{6}},
\tfrac{2}{\sqrt{3}}e^{\frac{i\pi}{6}},
\tfrac{2}{\sqrt{3}}e^{\frac{i\pi}{6}}
]} =
e^{\frac{2\pi i}{3}}\ket{[
\tfrac{1}{\sqrt{3}}e^{\frac{i\pi}{6}},
\tfrac{1}{\sqrt{3}}e^{\frac{i\pi}{6}},
\tfrac{2}{\sqrt{3}}e^{\frac{i\pi}{6}},
\tfrac{2}{\sqrt{3}}e^{\frac{i\pi}{6}}
]}\,.
\\
\end{array}\right.\nn
\ee


For $n=5$, $\TdS,\TdT$ act as the identity:
$$
\TdS\ket{\fpIzzzzz}=\ket{\fpIzzzzz},
\qquad
\TdT\ket{\fpIzzzzz}=\ket{\fpIzzzzz},
$$
or, equivalently,
$$
\TdS\ket{[0,0,0,0,0]}=\ket{[0,0,0,0,0]},
\qquad
\TdT\ket{[0,0,0,0,0]}=\ket{[0,0,0,0,0]}.
$$


\subsection{Action of $\TdS$ on
single-particle states for $\pht=\frac{2\pi}{3}$
($\tau=e^{\pi i/3}$ and $\ord=3$)}


In this case, $\lvk=3$. For $n=1$, $\TdS$ acts as:

\be\label{eqn:TdS-III1}
\left[\begin{array}{ll}
\TdS\ket{\fpIIIz}&=
 \frac{1}{\sqrt{3}}\ket{\fpIIIz}
+\frac{1}{\sqrt{3}}\ket{\fpIIIe}
+\frac{1}{\sqrt{3}}\ket{\fpIIIf}
\,,
\\ & \\
\TdS\ket{\fpIIIe}&=
 \frac{1}{\sqrt{3}}\ket{\fpIIIz}
+\frac{1}{\sqrt{3}}e^{\frac{2\pi i}{3}}\ket{\fpIIIe}
+\frac{1}{\sqrt{3}}e^{-\frac{2\pi i}{3}}\ket{\fpIIIf}
\,,
\\ & \\
\TdS\ket{\fpIIIf}&=
 \frac{1}{\sqrt{3}}\ket{\fpIIIz}
+\frac{1}{\sqrt{3}}e^{-\frac{2\pi i}{3}}\ket{\fpIIIe}
+\frac{1}{\sqrt{3}}e^{\frac{2\pi i}{3}}\ket{\fpIIIf}
\,,
\end{array}
\right.
\ee
or, equivalently
\be
\left[\begin{array}{ll}
\TdS\ket{[0]}&=
 \frac{1}{\sqrt{3}}\ket{[0]}
+\frac{1}{\sqrt{3}}\ket{[\tfrac{1}{\sqrt{3}}e^{\frac{i\pi}{6}}]}
+\frac{1}{\sqrt{3}}\ket{[\tfrac{2}{\sqrt{3}}e^{\frac{i\pi}{6}}]}
\,,
\\ & \\
\TdS\ket{[\tfrac{1}{\sqrt{3}}e^{\frac{i\pi}{6}}]}&=
 \frac{1}{\sqrt{3}}\ket{[0]}
+\frac{1}{\sqrt{3}}e^{\frac{2\pi i}{3}}
  \ket{[\tfrac{1}{\sqrt{3}}e^{\frac{i\pi}{6}}]}
+\frac{1}{\sqrt{3}}e^{-\frac{2\pi i}{3}}
  \ket{[\tfrac{2}{\sqrt{3}}e^{\frac{i\pi}{6}}]}
\,,
\\ & \\
\TdS\ket{[\tfrac{2}{\sqrt{3}}e^{\frac{i\pi}{6}}]}&=
 \frac{1}{\sqrt{3}}\ket{[0]}
+\frac{1}{\sqrt{3}}e^{-\frac{2\pi i}{3}}
  \ket{[\tfrac{1}{\sqrt{3}}e^{\frac{i\pi}{6}}]}
+\frac{1}{\sqrt{3}}e^{\frac{2\pi i}{3}}
  \ket{[\tfrac{2}{\sqrt{3}}e^{\frac{i\pi}{6}}]}
\,.
\end{array}
\right.\nn
\ee
$\TdT$ acts as
\be\label{eqn:TdT-III1}
\TdT\ket{\fpIIIz}=\ket{\fpIIIz}
\,,\qquad
\TdT\ket{\fpIIIe}=e^{-\frac{2\pi i}{3}}\ket{\fpIIIe}
\,,\qquad
\TdT\ket{\fpIIIf}=e^{-\frac{2\pi i}{3}}\ket{\fpIIIf}
\,,
\ee
or, equivalently,
\be
\TdT\ket{[0]}=\ket{[0]}
\,,\qquad
\TdT\ket{[
\tfrac{1}{\sqrt{3}}e^{\frac{i\pi}{6}}]}=
e^{-\frac{2\pi i}{3}}\ket{[
\tfrac{1}{\sqrt{3}}e^{\frac{i\pi}{6}}]}
\,,\qquad
\TdT\ket{[
\tfrac{2}{\sqrt{3}}e^{\frac{i\pi}{6}}]}=
e^{-\frac{2\pi i}{3}}\ket{[
\tfrac{2}{\sqrt{3}}e^{\frac{i\pi}{6}}]}
\,.\nn
\ee


For $n=2$, $\TdS$ acts as:
\be\label{eqn:TdS-III2}
\left[\begin{array}{ll}
\TdS\ket{\fpIIIzz}&=
 \frac{1}{\sqrt{3}}\ket{\fpIIIzz}
+\frac{1}{\sqrt{3}}\ket{\fpIIIee}
+\frac{1}{\sqrt{3}}\ket{\fpIIIff}
\,,
\\ & \\
\TdS\ket{\fpIIIee}&=
 \frac{1}{\sqrt{3}}\ket{\fpIIIzz}
+\frac{1}{\sqrt{3}}e^{-\frac{2\pi i}{3}}\ket{\fpIIIee}
+\frac{1}{\sqrt{3}}e^{\frac{2\pi i}{3}}\ket{\fpIIIff}
\,,
\\ & \\
\TdS\ket{\fpIIIff}&=
 \frac{1}{\sqrt{3}}\ket{\fpIIIzz}
+\frac{1}{\sqrt{3}}e^{\frac{2\pi i}{3}}\ket{\fpIIIee}
+\frac{1}{\sqrt{3}}e^{-\frac{2\pi i}{3}}\ket{\fpIIIff}
\,,
\end{array}
\right.
\ee
or, equivalently,
\be
\left[\begin{array}{ll}
\TdS\ket{[0,0]}&=
 \frac{1}{\sqrt{3}}\ket{[0,0]}
+\frac{1}{\sqrt{3}}\ket{[
\tfrac{1}{\sqrt{3}}e^{\frac{i\pi}{6}},
\tfrac{1}{\sqrt{3}}e^{\frac{i\pi}{6}}]}
+\frac{1}{\sqrt{3}}\ket{[
\tfrac{2}{\sqrt{3}}e^{\frac{i\pi}{6}},
\tfrac{2}{\sqrt{3}}e^{\frac{i\pi}{6}}]}
\,,
\\ & \\
\TdS\ket{[
\tfrac{1}{\sqrt{3}}e^{\frac{i\pi}{6}},
\tfrac{1}{\sqrt{3}}e^{\frac{i\pi}{6}}]}&=
 \frac{1}{\sqrt{3}}\ket{[0,0]}
+\frac{1}{\sqrt{3}}e^{-\frac{2\pi i}{3}}
  \ket{[
\tfrac{1}{\sqrt{3}}e^{\frac{i\pi}{6}},
\tfrac{1}{\sqrt{3}}e^{\frac{i\pi}{6}}]}
+\frac{1}{\sqrt{3}}e^{\frac{2\pi i}{3}}
  \ket{[
\tfrac{2}{\sqrt{3}}e^{\frac{i\pi}{6}},
\tfrac{2}{\sqrt{3}}e^{\frac{i\pi}{6}}]}
\,,
\\ & \\
\TdS\ket{[
\tfrac{2}{\sqrt{3}}e^{\frac{i\pi}{6}},
\tfrac{2}{\sqrt{3}}e^{\frac{i\pi}{6}}]}&=
 \frac{1}{\sqrt{3}}\ket{[0,0]}
+\frac{1}{\sqrt{3}}e^{\frac{2\pi i}{3}}
  \ket{[
\tfrac{1}{\sqrt{3}}e^{\frac{i\pi}{6}},
\tfrac{1}{\sqrt{3}}e^{\frac{i\pi}{6}}]}
+\frac{1}{\sqrt{3}}e^{-\frac{2\pi i}{3}}
   \ket{[
\tfrac{2}{\sqrt{3}}e^{\frac{i\pi}{6}},
\tfrac{2}{\sqrt{3}}e^{\frac{i\pi}{6}}]}
\,.
\end{array}
\right.\nn
\ee
$\TdT$ acts as
\be\label{eqn:TdT-III2}
\TdT\ket{\fpIIIzz}=\ket{\fpIIIzz}
\,,\qquad
\TdT\ket{\fpIIIee}=e^{\frac{2\pi i}{3}}\ket{\fpIIIee}
\,,\qquad
\TdT\ket{\fpIIIff}=e^{\frac{2\pi i}{3}}\ket{\fpIIIff}
\,,
\ee
or, equivalently,
\be
\left[\begin{array}{ll}
\TdT\ket{[0,0]} &= \ket{[0,0]}\,,\\ \\
\TdT\ket{[
\tfrac{1}{\sqrt{3}}e^{\frac{i\pi}{6}},
\tfrac{1}{\sqrt{3}}e^{\frac{i\pi}{6}}]} &=
e^{\frac{2\pi i}{3}}\ket{[
\tfrac{1}{\sqrt{3}}e^{\frac{i\pi}{6}},
\tfrac{1}{\sqrt{3}}e^{\frac{i\pi}{6}}]}\,,
\\ \\
\TdT\ket{[
\tfrac{2}{\sqrt{3}}e^{\frac{i\pi}{6}},
\tfrac{2}{\sqrt{3}}e^{\frac{i\pi}{6}}]} &=
e^{\frac{2\pi i}{3}}\ket{[
\tfrac{2}{\sqrt{3}}e^{\frac{i\pi}{6}},
\tfrac{2}{\sqrt{3}}e^{\frac{i\pi}{6}}]}\,.
\end{array}\right.\nn
\ee

\section{Action of $\SL(2,\Z)$ on Chern--Simons
Hilbert spaces}
\label{app:SL(2,Z)-CS}

The Hilbert space of $U(n)=[U(1)\times SU(n)]/\Z_n$
Chern--Simons theory at level $\lvk$ on $T^2$,
where $U(1)$ is at level $\lvk n$ and $SU(n)$ is at level
$\lvk$, is equivalent to the symmetric product of
$n$ copies of the Hilbert space of $U(1)_\lvk.$
We use this to extract the $\SL(2,\Z)$ representation
of the $SU(n)_\lvk$ Hilbert space.

We can write the states of $U(1)_\lvk$ as $\ket{p}$
with $p=0,\dots,\lvk-1$, and the states of the product
of $n$ copies as $\ket{p_1,\dots,p_n}$
with $0\le p_i\le \lvk-1.$
We then decompose
$$
\sum_{\perm\in S_n}
\ket{p_{\perm(1)},\dots,p_{\perm(n)}}=\sum_{p=0}^{\lvk n-1}
\ket{p_1,\dots,p_n;p}_{SU(n)}\ket{p}_{U(1)}
\,.
$$
As we will soon see, only $n$ out of the $\lvk n$
terms on the right-hand side are nonzero, and the normalization is
$$
\langle p_1,\dots,p_n;p | p_1,\dots,p_n;p\rangle
=\frac{1}{n}N_{p_1\dots p_n}
\,,
$$
where $N_{p_1\dots p_n}$ is calculated as follows.
For $0\le j<\lvk$,
let $m_j$ be the number of indices $i$ for which $p_i=j.$
Then $\sum_{j=0}^{\lvk-1} m_j = n$ and
$$
N_{p_1\dots p_n} =
\frac{n!}{\prod_j m_j!}
\,.
$$

We also need to match the action of large $U(1)$ gauge
transformations that reside entirely inside the $U(1)$ factor
and do not affect the $SU(n)$ degrees of freedom.
They form a $\Z_k\times\Z_k$ group, and act as
\begin{equation*}
\begin{split}
\sum_{\perm\in S_n}
\ket{p_{\perm(1)}+1,\dots,p_{\perm(n)}+1}
&=\PQsymB_1
\sum_{\perm\in S_n}
\ket{p_{\perm(1)},\dots,p_{\perm(n)}}\\
&=\sum_{p=0}^{\lvk n-1}
\ket{p_1,\dots,p_n;p}_{SU(n)}\ket{p+n}_{U(1)}
\,,
\end{split}
\end{equation*}
and
\begin{equation*}
\begin{split}
e^{-\frac{2\pi i}{\lvk}\sum_i p_i}
\sum_{\perm\in S_n}
\ket{p_{\perm(1)},\dots,p_{\perm(n)}}
&=\PQsymB_2
\sum_{\perm\in S_n}
\ket{p_{\perm(1)},\dots,p_{\perm(n)}}\\
&=\sum_{p=0}^{\lvk n-1}
e^{-\frac{2\pi i}{\lvk}p}
\ket{p_1,\dots,p_n;p}_{SU(n)}\ket{p}_{U(1)}
\,.
\end{split}
\end{equation*}
So  $\ket{p_1,\dots,p_n;p}$ is nonzero only if
$p=\sum_i p_i\pmod\lvk$, and
we also get
\be\label{eqn:pppn}
\ket{p_1,\dots,p_n;p-n}=\ket{p_1+1,\dots,p_n+1;p}
\,.
\ee

\subsection{Action of $\TdT$}

For even $\lvk$ we have
\begin{equation*}
\begin{split}
e^{\frac{i\pi}{\lvk}\sum_i p_i^2}
\sum_{\perm\in S_n}
\ket{p_{\perm(1)},\dots,p_{\perm(n)}}
&=
\sum_{\perm\in S_n}
\TdT\ket{p_{\perm(1)},\dots,p_{\perm(n)}}\\
&=\sum_{p=0}^{\lvk n-1}e^{\frac{i\pi}{\lvk n}p^2}
\TdT\ket{p_1,\dots,p_n;p}_{SU(n)}\ket{p}_{U(1)}\,.
\end{split}
\end{equation*}
So,
$$
\TdT\ket{p_1,\dots,p_n;p}_{SU(n)}
=e^{
\frac{i\pi}{\lvk}\bigl(\sum_{i=1}^n p_i^2-\frac{1}{n}p^2\bigr)}
\ket{p_1,\dots,p_n;p}_{SU(n)}
\,.
$$
For odd $\lvk$ and any $n$ we have
\be\label{eqn:TdTp1pn}
\TdT\ket{p_1,\dots,p_n;p}_{SU(n)}
=
(-1)^{p-\sum_{i=1}^n p_i}
e^{
\frac{i\pi}{\lvk}\bigl(
\sum_{i=1}^n p_i^2-\frac{1}{n}p^2\bigr)}
\ket{p_1,\dots,p_n;p}_{SU(n)}
\,,
\ee
where we have used a freedom similar to
\eqref{eqn:TdGredef} to add an extra
factor of $(-1)^p$ so that
for even $\lvk$ we have
$p-\sum_{i=1}^n p_i\equiv 0\pmod{\lvk}$
and therefore $(-1)^{p-\sum_{i=1}^n p_i}=1.$
Note also that for odd $n$ the
extra $(-1)^p$ factor is necessary
for consistency with \eqref{eqn:pppn}.

\subsection{Action of $\TdS$}

For $\TdS$ we have,
\bear
\frac{1}{\lvk^{n/2}}
\sum_{\perm\in S_n}
\sum_{q_1=0}^{\lvk-1}\cdots\sum_{q_n=0}^{\lvk-1}
e^{\frac{2\pi i}{\lvk}\sum_{i=1}^n q_i p_{\perm(i)}}
\ket{q_1,\dots,q_n}
&=&
\sum_{\perm\in S_n}
\TdS\ket{p_{\perm(1)},\dots,p_{\perm(n)}}
\nn\\ &=&
\frac{1}{\sqrt{\lvk n}}
\sum_{q=0}^{\lvk n-1}e^{\frac{2\pi i}{\lvk n}p q}
\TdS\ket{p_1,\dots,p_n;p}_{SU(n)}\ket{q}_{U(1)}
\,.
\nn
\eear
We now take the (partial) inner product of that state with
$\frac{1}{\sqrt{\lvk n}}
\sum_{q=0}^{\lvk n-1}e^{\frac{2\pi i}{\lvk n}p q}
\ket{q}_{U(1)}$, and after some algebra we get
\begin{multline}
\TdS\ket{p_1,\dots,p_n;p}_{SU(n)}
=\\
\frac{1}{\sqrt{\lvk^{n+1} n}}
\sum_{q_1=0}^{\lvk-1}\cdots\sum_{q_n=0}^{\lvk-1}
e^{\frac{2\pi i}{\lvk}\sum_i q_i(p_i-\frac{1}{n}p)}
\sum_{m=0}^{n-1}
e^{-\frac{2\pi i}{n}p m}
\ket{q_1,\dots,q_n;m\lvk + \sum_i q_i}_{SU(n)}\,.
\nn
\end{multline}

\subsection{Action of $Z_n$}
The $U(n)$ states are invariant, so
\begin{align*}
\Omega_1''\ket{p_1,\dots,p_n;p}_{SU(n)}
&=\ket{p_1,\dots,p_n;p+\lvk}_{SU(n)}
\,,
\\
\Omega_2''\ket{p_1,\dots,p_n;p}_{SU(n)}
&=e^{-\frac{2\pi i}{n}p}\ket{p_1,\dots,p_n;p}_{SU(n)}\,.
\end{align*}

\subsection{Example:
$\lbrack U(1)_2\times SU(2)_{-3}\rbrack/\Z_2$}

We have $SU(2)_{-3}$ states of the form
$$
\ket{p_1,p_2;3m+\sum p_i}_{SU(2)}
\,,
\qquad
0\le p_1,p_2<3\,,\quad 0\le m<2
\,.
$$
We add the $U(1)_2$ states to get states of the form
$$
\ket{p_1,p_2;3m+\sum p_i}_{SU(2)}\ket{q}_{U(1)}
$$
with $0\le q<2.$
We then mod out by $\Z_2\times\Z_2$ as follows.
First,
$$
q\equiv 3m+\sum_{i=1}^2 p_i\equiv m+\sum_{i=1}^2 p_i\pmod{2}\,,
$$
so, $q$ is completely determined by $m,p_1,p_2.$
We therefore do not specify $q$ any more.
Next, we need to keep only the $\Omega_1''$-invariant
combinations:
$$
\ket{p_1,p_2}_s\equiv
\sum_{m=0}^1
\ket{p_1,p_2;3m+\sum p_i}_{SU(2)}
\ket{3m+\sum p_i}_{U(1)}\,.
$$
In order for the space spanned by
$\ket{p_1,p_2}_s$ to be closed under
$\TdT$ (and not just $\TdT^2$), we
need to augment \eqref{eqn:TdTp1pn}
by an extra factor of $(-1)^p.$
After some algebra, we then get
\be\label{eqn:TdTpp}
\TdT\ket{p_1,p_2}_s =
(-1)^{\sum p_i}
e^{\frac{\pi i}{3}\bigl\lbrack 2(\sum p_i)^2
-\sum p_i^2\bigr\rbrack}
\ket{p_1,p_2}_s
\,.
\ee
These phases were used for
identifying $\Hilb_{(2)}(\frac{\pi}{3})$
in \appref{app:HilbDecomp}.
In the notation of \eqref{eqn:TdTabcdH2} we have
\bear
\ket{\fpIzz} &=&
\sqrt{2}\ket{0,0}_s
=\sqrt{2}\ket{1,1}_s
=\sqrt{2}\ket{2,2}_s \,,\nn\\
\ket{\fpIdd} &=& \ket{0,1}_s =\ket{1,2}_s=\ket{2,0}_s
           =\ket{1,0}_s =\ket{2,1}_s=\ket{0,2}_s
 \,,\nn
\eear
and up to an overall phase
(see the explanation at the beginning
of \appref{app:SL(2,Z)}),
we find that
\eqref{eqn:TdTpp} agrees with \eqref{eqn:TdT-I2}.

\section{Decomposition of $\Hilb_{(n_1,\dots,n_p)}$
into Chern--Simons Hilbert spaces}
\label{app:HilbDecomp}


\subsection{$\pht=\frac{\pi}{3}$ ($\lvk=1$)}


For $\lvk=1$, $n=2$, we have
$$
\Hilb(2,\tfrac{\pi}{3}) =
\Hilb_{(1,1)}(\tfrac{\pi}{3})
\oplus\Hilb_{(2)}(\tfrac{\pi}{3})\,.
$$
The factor $\Hilb_{(1,1)}$ was discussed in
\secref{subsec:winding-1}, so it only remains to discuss
$\Hilb_{(2)}.$
There are two states which we decompose according to
\eqref{eqn:decpsi}:
\bear
\ket{\fpIzz} &=&
 \ket{a}_{SU(2)}\otimes\ket{0}_{U(1)}
+\ket{b}_{SU(2)}\otimes\ket{1}_{U(1)}
\,,\nn\\
\ket{\fpIdd} &=&
 \ket{d}_{SU(2)}\otimes\ket{0}_{U(1)}
+\ket{c}_{SU(2)}\otimes\ket{1}_{U(1)}
\,,\nn
\eear
where the $U(1)$ is at level $\lvk n=2$.

Up to an unimportant overall phase, we find
\be\label{eqn:TdTabcdH2}
\TdT\ket{a}=\ket{a}
\,,\quad
\TdT\ket{b}=e^{-\frac{\pi i}{2}}\ket{b}
\,,\quad
\TdT\ket{c}=e^{\frac{5\pi i}{6}}\ket{c}
\,,\quad
\TdT\ket{d}=e^{-\frac{2\pi i}{3}}\ket{d}
\,.
\ee
The overall phase can be fixed (up to a cubic root of unity)
by calculating, with the above assignments,
$$
(\TdT\TdS)^3 = e^{-\frac{\pi i}{4}}.
$$
So multiplying $\TdT$ by $e^{\frac{\pi i}{12}}$, for example,
would fix the phase.
We also have
\bear
\TdS\ket{a} &=&
 \frac{1}{\sqrt{6}}\ket{a}
+\frac{1}{\sqrt{6}}\ket{b}
+\sqrt{\frac{1}{3}}\ket{c}
+\sqrt{\frac{1}{3}}\ket{d}
\,,\nn\\
\TdS\ket{b} &=&
 \frac{1}{\sqrt{6}}\ket{a}
-\frac{1}{\sqrt{6}}\ket{b}
-\sqrt{\frac{1}{3}}\ket{c}
+\sqrt{\frac{1}{3}}\ket{d}
\,,\nn\\
\TdS\ket{c} &=&
 \sqrt{\frac{1}{3}}\ket{a}
-\sqrt{\frac{1}{3}}\ket{b}
+\frac{1}{\sqrt{6}}\ket{c}
-\frac{1}{\sqrt{6}}\ket{d}
\,,\nn\\
\TdS\ket{d} &=&
 \sqrt{\frac{1}{3}}\ket{a}
+\sqrt{\frac{1}{3}}\ket{b}
-\frac{1}{\sqrt{6}}\ket{c}
-\frac{1}{\sqrt{6}}\ket{d}
\,,\nn
\eear
and
$$
\Omega_1''\ket{a}=\ket{b}
\,,\quad
\Omega_1''\ket{b}=\ket{a}
\,,\quad
\Omega_1''\ket{c}=\ket{d}
\,,\quad
\Omega_1''\ket{d}=\ket{c}
\,,
$$
$$
\Omega_2''\ket{a}=\ket{a}
\,,\quad
\Omega_2''\ket{b}=-\ket{b}
\,,\quad
\Omega_2''\ket{c}=-\ket{c}
\,,\quad
\Omega_2''\ket{d}=\ket{d}
\,.
$$
These results agree with those of the $SU(2)$ Chern--Simons theory at level $\lvk=-3$ [see \eqref{eqn:tsk=3}]. So we get
$$
\Hilb_{(2)}(\tfrac{\pi}{3})=\Hilb([U(1)_2\times SU(2)_{-3}]/\Z_2)\,.
$$

For $n=3$, we have sectors corresponding to $[\sigma]=(1,1,1),(2,1)$, and $(3)$. The first was discussed in \secref{subsec:winding-1}, and the second is a reducible sector. We now discuss the third case.

We decompose the basis states into the $U(1)$ and $SU(3)$ degrees of freedom as follows:
\bear
\ket{\fpIzzz} &=&
 \ket{a}_{SU(3)}\otimes\ket{0}_{U(1)}
+\ket{b}_{SU(3)}\otimes\ket{1}_{U(1)}
+\ket{c}_{SU(3)}\otimes\ket{2}_{U(1)}
\,,\nn\\
\ket{\fpIxyd} &=&
 \ket{d}_{SU(3)}\otimes\ket{0}_{U(1)}
+\ket{e}_{SU(3)}\otimes\ket{1}_{U(1)}
+\ket{f}_{SU(3)}\otimes\ket{2}_{U(1)}
\,.\nn
\eear
As usual, we extract the action of $\TdT,\TdS,\Omega_1'',\Omega_2''$ on $SU(3)$ degrees of freedom by using the known results for $U(1)$ theory at level $\lvk n=3$. We get
\begin{align*}
\TdT\ket{a}_{SU(3)}&=-\ket{a}_{SU(3)}\,,\quad &
\TdT\ket{b}_{SU(3)}&=-e^{\frac{2\pi i}{3}}\ket{b}_{SU(3)}\,,\quad &
\TdT\ket{c}_{SU(3)}&=-e^{\frac{2\pi i}{3}}\ket{c}_{SU(3)}\,,\\
\TdT\ket{d}_{SU(3)}&=\ket{d}_{SU(3)}\,,\quad &
\TdT\ket{e}_{SU(3)}&=e^{\frac{2\pi i}{3}}\ket{e}_{SU(3)}\,,\quad &
\TdT\ket{f}_{SU(3)}&=e^{\frac{2\pi i}{3}}\ket{f}_{SU(3)}\,,
\end{align*}
up to an overall phase, and
\begin{equation*}
\TdS=\tfrac12
\begin{pmatrix}
\tfrac{1}{\sqrt{3}} & \tfrac{1}{\sqrt{3}} & \tfrac{1}{\sqrt{3}} & 1 & 1 & 1 \\
\tfrac{1}{\sqrt{3}} & \tfrac{\omega}{\sqrt{3}} &  \tfrac{\omega^2}{\sqrt{3}}& 1 & \omega & \omega^2 \\
\tfrac{1}{\sqrt{3}} & \tfrac{\omega^2}{\sqrt{3}} & \tfrac{\omega}{\sqrt{3}} & 1 & \omega^2 & \omega \\
1 & 1 & 1 & -\tfrac{1}{\sqrt{3}} &-\tfrac{1}{\sqrt{3}}  & -\tfrac{1}{\sqrt{3}} \\
1 & \omega & \omega^2 & -\tfrac{1}{\sqrt{3}} & -\tfrac{\omega}{\sqrt{3}} & -\tfrac{\omega^2}{\sqrt{3}} \\
1& \omega^2 & \omega & -\tfrac{1}{\sqrt{3}} &  -\tfrac{\omega^2}{\sqrt{3}}& -\tfrac{\omega}{\sqrt{3}}
\end{pmatrix}\,,
\end{equation*}
where $\omega=e^{\frac{2\pi i}{3}}$, in the basis $\ket{a}_{SU(3)},\ldots,\ket{f}_{SU(3)}$. We also have
\begin{align*}
\Omega_1''\ket{a}_{SU(3)}&=\ket{b}_{SU(3)}\,,\quad &
\Omega_1''\ket{b}_{SU(3)}&=\ket{c}_{SU(3)}\,,\quad &
\Omega_1''\ket{c}_{SU(3)}&=\ket{a}_{SU(3)}\,,\\
\Omega_1''\ket{d}_{SU(3)}&=\ket{e}_{SU(3)}\,,\quad &
\Omega_1''\ket{e}_{SU(3)}&=\ket{f}_{SU(3)}\,,\quad &
\Omega_1''\ket{f}_{SU(3)}&=\ket{d}_{SU(3)}\,,
\end{align*}
and
\begin{align*}
\Omega_2''\ket{a}_{SU(3)}&=\ket{a}_{SU(3)}\,,\quad &
\Omega_2''\ket{b}_{SU(3)}&=e^{\frac{2\pi i}{3}}\ket{b}_{SU(3)}\,,\quad &
\Omega_2''\ket{c}_{SU(3)}&=e^{-\frac{2\pi i}{3}}\ket{c}_{SU(3)}\,,\\
\Omega_2''\ket{d}_{SU(3)}&=\ket{d}_{SU(3)}\,,\quad &
\Omega_2''\ket{e}_{SU(3)}&=e^{\frac{2\pi i}{3}}\ket{e}_{SU(3)}\,,\quad &
\Omega_2''\ket{f}_{SU(3)}&=e^{-\frac{2\pi i}{3}}\ket{f}_{SU(3)}\,.
\end{align*}
The results agree with those of $SU(3)$
Chern--Simons theory at $\lvk=-2$. The latter can be checked,
for example,
by studying the $U(3)_{-2}=[U(1)_{-6}\times SU(3)_{-2}]/\Z_3$
Chern--Simons theory,
using the known results for the $U(1)$ degrees of
freedom and \eqref{eqn:U(n)asU(1)n}.

For $n=4$, we have $\brperm=(1,1,1,1),(2,1,1),(2,2),(3,1),$
and $(4)$ sectors.
The last case is equivalent as representation of $\SL(2,\Z)$
to the $n=2$, $\brperm=(2)$ case
[see \eqref{eqn:TdS-I4}-\eqref{eqn:TdT-I4}].
To see this, one has to change basis
($\ket{\fpIzzzz}\rightarrow -\ket{\fpIdd},
\ket{\fpIdddd}\rightarrow -\ket{\fpIzz},$
and recall that $\TdT,\TdS$ as appear in
\appref{app:SL(2,Z)} are only determined up to an overall phase.)

The only  remaining nontrivial case
(i.e., neither untwisted nor reducible) is the
$\brperm=(2,2)$ sector.
We can write $\Hilb_{(2,2)}(\frac{\pi}{3})$
as a symmetric product
$
\Hilb_{(2,2)}(\frac{\pi}{3}) \simeq
\Hilb_{(2)}(\frac{\pi}{3})^{\otimes 2}/S_2
\,,
$
and using the result $\Hilb_{(2)}(\frac{\pi}{3})\simeq U(2)_{2,-3}$
from above,
we can write $\Hilb_{(2,2)}(\frac{\pi}{3})$
as the symmetric product
of Chern--Simons Hilbert spaces:
$$
\Hilb_{(2,2)}(\frac{\pi}{3})
\simeq
U(2)_{2,-3}^{\otimes 2}/S_2
\,.
$$
This, however, is not good enough for our purposes, because
we would like to present each sector as the Hilbert
space of a gauge theory, and $U(2)^{\otimes 2}/S_2$
is not a group.

The dimension of $\Hilb_{(2,2)}(\frac{\pi}{3})$ is $3$,
so if we attempt to write it as
$[U(1)_4\times\HilbSun_{(2,2)}(\frac{\pi}{3})]/\Z_4$
we find that we need $\dim\HilbSun_{(2,2)}(\frac{\pi}{3})=12.$
The dimension of the Hilbert space of $SU(n')_{\lvk'}$
is $(n'+\lvk'-1)!/\lvk'!(n'-1)!$, so if we assume $n'\le n$
we find only $(n'=2,\lvk'=11)$ and $(n'=12,\lvk'=1)$,
but these are easily ruled out.
We have also explored product gauge groups
such as $U(2)_{2\lvk',\lvk'}\times U(2)_{2\lvk'',\lvk''}$
with $\lvk'+\lvk''=2$, to no avail.

Perhaps
we can obtain a clue to the solution by noting that the symmetric product of $n'$ copies
of $SU(2)_{\lvk'}$ is equivalent to the Hilbert space
of a symplectic group,
\be\label{eqn:SpSym}
\Hilb[SU(2)_{\lvk'}]^{\otimes n'}/S_{n'} \simeq
\Hilb[Sp(n')_{\lvk'}]
\,,
\ee
as can easily be verified by writing
out the explicit wavefunctions,
using the character formulas for affine Lie algebras
\cite{Kac:1990gs} (see also \cite{Bardakci:2009wp}
for a historical review and references).
Another observation is that $\Hilb_{(2,2)}(\frac{\pi}{3})$
is equivalent
as a representation of only $\TdT^2$ and $\TdS$
to the Hilbert space of $Sp(2)/\Z_2\simeq SO(5)$
Chern--Simons theory at level $3.$
To see this, take $n'=2$ and $\lvk'=3$ in \eqref{eqn:SpSym}.
The Hilbert space for $SU(2)$ Chern--Simons theory at level
$3$ was discussed at the end of \secref{subsec:ExU(1)inU(2)}
--- in particular, its dimension is 4,
and the $\Z_2$ large gauge transformation acts as
\eqref{eqn:Z2inSU(2)}. We can therefore obtain the Hilbert space of $Sp(2)/\Z_2$ Chern--Simons theory by first taking the symmetric product of two copies of $\Hilb[SU(2)_3]$, and then requiring invariance under \eqref{eqn:Z2inSU(2)}.

The result is that it is a three-dimensional space spanned by
$$
\ket{a,a}+\ket{b,b}\,,\quad
\ket{a,d}+\ket{b,c}\,,\quad
\ket{c,c}+\ket{d,d}\,,
$$
where, for example,
\begin{equation*}
\begin{split}
\ket{a,a}&\equiv\ket{a}_{SU(2)}\otimes\ket{a}_{SU(2)}
\,,\\
\ket{a,d}&\equiv\tfrac{1}{\sqrt{2}}
(\ket{a}_{SU(2)}\otimes\ket{d}_{SU(2)}
 +\ket{d}_{SU(2)}\otimes\ket{a}_{SU(2)})
\,,\quad
\end{split}
\end{equation*}
with $\ket{a},\ket{b},\ket{c},\ket{d}$ as defined in
\secref{subsec:ExU(1)inU(2)}.
Other states like $\ket{b,b}$ and $\ket{b,c}$
are defined similarly.
We can also read off the action of $\TdT^2$ and $\TdS$
in this basis:
$$
\TdT^2=
\begin{pmatrix}
1&0&0 \\ 0& e^{\tfrac{4\pi i}{3}} & 0 \\ 0&0& e^{\tfrac{2\pi i}{3}}
\end{pmatrix}\,,
\quad
\TdS=\tfrac13
\begin{pmatrix}
1&2&2 \\ 2&1&-2 \\ 2&-2&1
\end{pmatrix}\,.
$$
This matches exactly the action of $\TdT^2$ and $\TdS$ on
$\Hilb_{(2,2)}(\tfrac{\pi}{3})$,
which can be found from \eqref{eqn:TdS-I2} and \eqref{eqn:TdT-I2}.

There is a caveat in this discussion, however, in that we only checked the action of $\TdT^2$, not $\TdT$. The latter is actually not well-defined in the $SO(5)$ Chern--Simons theory Hilbert space, because it does not commute with $\Omega_2''$ of
\eqref{eqn:Z2inSU(2)}. In other words,
$\Hilb_{(2,2)}(\tfrac{\pi}{3})$ is equivalent to the Hilbert space of $SO(5)$ Chern--Simons theory, not as a representation of the full $\SL(2,\Z)$, but as a representation of its subgroup
$\Gamma(2)$. The situation is reminiscent of $U(1)$ Chern--Simons theory at an odd level $\lvk$,
discussed in \secref{subsec:U(1)-wf}, where the theory depends on the choice of spin structure of $T^2$.
Another problem with identifying the $\brperm=(2,2)$ sector
with $Sp(2)/\Z_2$ Chern--Simons theory is that
it is not a subgroup of our gauge group $U(4)$, only its
double-cover $Sp(2)$ is. At this point, therefore,
we are not making any claims about the sector
$\Hilb_{(2,2)}(\frac{\pi}{3}).$

For $n=5$, we have $\brperm=(1,1,1,1,1),(2,1,1,1),(2,2,1),(3,1,1),(3,2),(4,1),$ and $(5)$.
All sectors are either untwisted or reducible,
except for the last one. But the $(5)$ sector is a trivial one-dimensional Hilbert space, so we may set
$$
\Hilb_{(5)}(\tfrac{\pi}{3})=\Hilb[U(5)_{5,1}]\,.
$$


\subsection{$\pht=\tfrac{2\pi}{3}$ ($\lvk=3$)}
For $n=2$, we have
$$
\Hilb(2,\tfrac{2\pi}{3}) =
\Hilb_{(1,1)}(\tfrac{2\pi}{3})
\oplus\Hilb_{(2)}(\tfrac{2\pi}{3}).
$$
The factor $\Hilb_{(1,1)}$ was discussed in
\secref{subsec:winding-1}, so it only remains to discuss
$\Hilb_{(2)}.$
There are three states which we decompose according to
\eqref{eqn:decpsi}:
\bear
\ket{\fpIIIzz} &=&
 \ket{a}_{SU(2)}\otimes\ket{0}_{U(1)}
+\ket{b}_{SU(2)}\otimes\ket{3}_{U(1)}
\,,\nn\\
\ket{\fpIIIee} &=&
 \ket{a}_{SU(2)}\otimes\ket{2}_{U(1)}
+\ket{b}_{SU(2)}\otimes\ket{5}_{U(1)}
\,,\nn\\
\ket{\fpIIIff} &=&
 \ket{a}_{SU(2)}\otimes\ket{4}_{U(1)}
+\ket{b}_{SU(2)}\otimes\ket{1}_{U(1)}
\,,\nn
\eear
where the $U(1)$ is at level $\lvk n=6$. Following the usual procedure, we get
$$
\TdS\ket{a}=\frac{1}{\sqrt{2}}(\ket{a}+\ket{b})
\,,\qquad
\TdS\ket{b}=\frac{1}{\sqrt{2}}(\ket{a}-\ket{b})
\,,
$$
and, up to a phase,
$$
\TdT\ket{a}=\ket{a}
\,,\qquad
\TdT\ket{b}=e^{\frac{\pi i}{2}}\ket{b}
\,.
$$
We also have
$$
\Omega_1''\ket{a}=\ket{a}
\,,\quad
\Omega_1''\ket{b}=-\ket{b}
\,,\qquad
\Omega_2''\ket{a}=\ket{b}
\,,\quad
\Omega_2''\ket{b}=\ket{a}
\,.
$$
These relations agree with the states of $SU(2)_{-1}$, so
$$
\Hilb_{(2)}(\tfrac{2\pi}{3})=\Hilb([U(1)_6\times SU(2)_{-1}]/\Z_2)\,.
$$

\end{appendix}


\bibliographystyle{my-h-elsevier}

\end{document}